%% file: diss.tex
\pdfoutput=1

\documentclass [
DIV10, 
a4paper, pagesize, 
bibliography=totoc, 
final,
BCOR15mm, 
captions=tableheading, 
numbers=noenddot, 
]{scrbook}
\input{definitions}
\begin{document}  

\frontmatter
\input{titlepage}

\onehalfspacing
\chapter{Abstract}
\input{abstract}

\singlespacing
\tableofcontents

\onehalfspacing

\mainmatter

\chapter{Introduction}
        \label{chapt:intro}
        \input{intro.tex}

\chapter{Thermal emission of asteroids}
        \label{chapt:thermal}
        \textit{\input{ThermalIntro}}

        \section{Overview}
                \label{sect:thermal:overview}
                \input{ThermalOverview}

        \section{Relevant physical properties}
                \label{sect:thermal:props}
                \input{ThermalProps}

        \section{Observability}
                \label{sect:thermal:observability}
                \input{Observability}

        \section{Simple models: STM and FRM}
                \label{sect:thermal:STM-FRM}
                \input{simplemodels.tex}

        \section[NEATM]{Near-Earth Asteroid Thermal Model (NEATM)}
                \label{sect:NEATM}
                \input{neatm}

\chapter{Detailed thermophysical modeling}
        \label{chapt:TPM}
        \textit{\input{TPMintro}}

        \section{Overview}
                \label{sect:TPM:overview}
                \input{TPMoverview}
        \section{Thermal physics}
                \label{sect:TPM:modeling}

	        \subsection{Global shape and spin state}
                        \label{sect:TPM:shape-spin}
                        \input{ShapeSpin}
                \subsection{Thermal conduction}
		        \label{sect:TPM:TI}
		        \input{conduction}
                \subsection{Thermal-infrared beaming}
                        \label{sect:TPM:beaming}
                        \input{beaming}

\input{convex}

        \section{TPM fits to asteroid data}
                \label{sect:TPM:fitting}
                \input{fitting}

\chapter[IRTF observations]{Thermal-infrared asteroid observations at the IRTF}
        \label{chapt:IRTF}
        \input{IRTF}

\chapter[Spitzer observations]{Asteroid observations using the Spitzer Space Telescope}
        \label{chapt:SST}
        \textit{\input{SST_intro}}

	\section{Spitzer overview}
		\label{sect:SST:overview}
		\input{SST_overview}
	\section{InfraRed Array Camera (IRAC)}
		\label{sect:IRAC:general}
		\input{IRAC_general}
	\section{InfraRed Spectrograph (IRS)}
		\label{sect:IRS:general}
                \input{IRS_general}

\chapter{Results}
        \label{chapt:results}
\textit{        \input{resultsintro}}

                \section[(433) Eros]{(433) Eros \footnotemark}
\footnotetext{
A preliminary version of the chief result presented herein was prepublished \citep{Ito1}. The analysis of Eros data reported therein was entirely done by me.}

                        \label{sect:Eros}
                       \input{Eros}
		\section[(25143) Itokawa]{(25143) Itokawa \footnotemark}
\footnotetext{
A preliminary version of the results herein were prepublished \citep{Ito1}. The  new ESO observations reported therein were performed and reduced by Marco \Delbo\ and Mario di Martino \citep{Delbo2004}, the new IRTF observations were performed and reduced by myself. The thermophysical modeling of the data set was performed by myself.}
			\label{sect:Itokawa}
                       \input{itokawa}
        
		\section[(1580) Betulia]{(1580) Betulia \footnotemark}
\footnotetext{
Most of the content of this section was prepublished  \citep{Betulia}. I have reduced the presented  data but have not contributed to planning and performing the observations.
 Two independent data analyses are presented in the paper, a NEATM analysis by the first author  and a TPM analysis  by myself. Only the latter analysis is presented here.}
			\label{sect:Betulia}
                       \input{Betulia}

		\section[(54509) YORP]{(54509) YORP \footnotemark}
\footnotetext{
A preliminary version of the results presented herein was presented  in a poster at the IAU General Assembly 2006 in Prague.}
			\label{sect:PH5}
                        \input{PH5}

		\section[(33342) 1998 WT24]{(33342) 1998 WT24 \footnotemark}
\footnotetext{
The content of this section was prepublished \citep{WT24}. I have not contributed to obtaining the  data, but have performed 
a TPM analysis thereof (our TPM  is referred to as ``General Thermophysical Model'' in the paper).}
			\label{sect:WT24}
                       \input{WT24}

	\section[(21) Lutetia,  Rosetta flyby target]{(21) Lutetia, Rosetta flyby target \footnotemark}
\footnotetext{ Most of this section was prepublished  \citep{Lutetia}. I have planned the observations, which were performed by co-author SJB. Data were analyzed by MM and AWH. The remaining three co-authors (JLH, MK, and JDA) represent the MIRSI team which built the mid-infrared imager used in the observations.}
                \label{sect:Lutetia}
               \input{Lutetia}
	\section[(10302) 1989 ML, potential Don-Quijote target]{(10302) 1989 ML, potential Don-Quijote target \footnotemark}
\footnotetext{
The content of this section was prepublished \citep{ML}. Only my contributions to that paper are presented herein.}
			\label{sect:ML}
			\input{ML}

        \section{Eclipses in the binary system (617) Patroclus}
	        \label{sect:Patroclus}
       \input{Patroclus}

 \chapter{Discussion}
         \label{chapt:discussion}
\textit{ \input{discuintro}}
         \input{discussion}
         \input{conclusions}

\appendix
\chapter{TPM for non-convex shapes}
        \label{chapt:TPMconcave}
        \textit{ \input{concaveintro} }

        \input{concave}


\backmatter
\singlespacing

\small{ \input{lib}}

\singlespacing

\input{acknow}

\chapter{Zusammenfassung}
\foreignlanguage{ngerman}{  \input{zusammenfassung}}

\singlespacing


\foreignlanguage{ngerman}{\input{cv_online}}

\end{document}

%% file: definitions.tex
\usepackage{setspace}

\usepackage{amssymb, amsmath, amsthm}

\usepackage[ansinew]{inputenc} 

\usepackage[final]{graphicx} 

\usepackage{varwidth} 
\usepackage{booktabs} 
\usepackage{bbm}      

\usepackage[OT2, OT1]{fontenc}
\usepackage[russian, ngerman, USenglish]{babel}

\usepackage[squaren, textstyle]{SIunits}
\newcommand{\arcsec}{\arcsecond} 
\newcommand{\arcmin}{\arcminute}
\newcommand{\micron}{\micro\metre}

\usepackage{natbib}

\usepackage[font=small,labelfont=bf,singlelinecheck=false]{caption}
\usepackage[breaklinks,bookmarksnumbered]{hyperref} 
\hypersetup{pdftitle={Surface Properties of Asteroids from Mid-Infrared Observations and Thermophysical Modeling},
pdfauthor={Michael Mueller},
pdfsubject={Dissertation, submitted to FU Berlin on 2007/05/21, defended on 2007/07/06},
pdfkeywords={Asteroids, Thermal inertia, Size, Albedo, Yarkovsky effect, Impact hazard, Spitzer observations, IRTF observations} }

\newcommand{\TSS}{\ensuremath{T_{SS}{}}}
\newcommand{\pV}{\ensuremath{p_\text{V}{}}}
\newcommand{\pv}{\pV}
\newcommand{\AU}{\ensuremath{\text{AU}}}
\newcommand{\yr}{\ensuremath{\text{yr}}}
\newcommand{\Jy}{\ensuremath{\text{Jy}}}
\newcommand{\jansky}{\Jy}
\newcommand{\Jansky}{\jansky}
\newcommand{\jy}{\Jy}
\newcommand{\TIunit}{\joule\usk\power{\second}{-1/2}\power{\kelvin}{-1}\power{\metre}{-2}}
\newcommand{\kappaunit}{\watt\usk\power{\kelvin}{-1}\power{\metre}{-1}}
\newcommand{\TIunits}{\TIunit}
\newcommand{\sr}{\steradian}

\newcommand{\sterad}{\steradian}
\newcommand{\gramm}{\gram}
\newcommand{\cm}{\centi\metre}
\newcommand{\km}{\kilo\metre}
\newcommand{\kg}{\kilo\gram}
\newcommand{\hr}{\hour}
\newcommand{\Area}{\ensuremath{\mathcal{A}}}

\newcommand{\cotwo}{\ensuremath{\text{CO}_2}}

\newcommand{\water}{\ensuremath{\text{H}_2\text{O}}}

\newcommand{\textd}{\ensuremath{\text{d}}}
\newcommand{\chitwo}{\ensuremath{\chi^2{}}}

\newcommand{\eS}{\ensuremath{\vec{e_S}{}}}
\newcommand{\eO}{\ensuremath{\vec{e_O}{}}}

\newcommand{\code}[1]{\texttt{#1}}

\newcommand{\AandA} {Astronomy \& Astrophysics}
\newcommand{\aap}{\AandA}
\newcommand{\AdSR} {Adv.\  Space Res.}
\newcommand{\AdSpR}{\AdSR}
\newcommand{\AJ} {Astron.\ Journal}
\newcommand{\AnRevEPS}{Ann.\ Rev.\ Earth Planet.\ Sci.}
\newcommand{\ApJ}  {Astroph.\ Journal}
\newcommand{\ApJS} {The \ApJ\ Supplement Series}

\newcommand{\baas}{Bull.\ AAS}
\newcommand{\BASI}{Bull.\ Astr.\ Soc.\ India}
\newcommand{\EMP}{Earth, Moon, and Planets}
\newcommand{\Icarus} {Icarus}
\newcommand{\JGR}    {J.\ Geophys.\ Res.}
\newcommand{\LPS}{Lunar Planet.\ Sci.}
\newcommand{\MandPS} {M\&PS}
\newcommand{\MNofRAS}{Mon.\ Not.\ of the RAS}
\newcommand{\na}{New A.}
\newcommand{\Nature} {Nature}
\newcommand{\PASJ}{PASJ}
\newcommand{\PASP} {PASP} 
\newcommand{\pss}{Plan.\ and Space Science}
\newcommand{\PSS}{pss}
\newcommand{\SPIE} {Proc.\ SPIE} 
\newcommand{\Science} {Science}

\newcommand{\doi}[1]{\url{http://dx.doi.org/#1}}

\newcommand{\journalref}[6]{#1\ #2. \textit{#3.} #4 \textbf{#5}, #6.}
\newcommand{\journalsubmittedref}[4]{#1\ #2. \textit{#3.} Submitted to #4.}
\newcommand{\journalinpressref}[4]{#1\ #2. \textit{#3.} #4, \textit{in press.}}
\newcommand{\bookref}[5]{#1\ #2. \textit{#3.} #4, #5.}
\newcommand{\bookchapterref}[5]{#1\ #2. \textit{#3.} #5, pp.\ #4.}

\newcommand{\bookforchapter}[4]{In: #1 \textit{#2.} #3, #4}

\newcommand{\editor}[1]{#1\ (Ed.)}
\newcommand{\editors}[1]{#1\ (Eds.)}

\newcommand{\AsteroidsII}{\bookforchapter{\editors{Binzel, R.P., Gehrels, T., Matthews, M.S.}}{Asteroids II}{Univ.\ of Arizona Press}{Tucson}}
\newcommand{\AsteroidsIII}{\bookforchapter{\editors{Bottke, W.F., Paolicchi, P., Binzel, R.P., Cellino, A.}}{Asteroids III}{Univ.\ of Arizona Press}{Tucson}}
\newcommand{\IMPS}{\bookforchapter{\editor{Tedesco, E.D.}}{The IRAS Minor Planet Survey, Tech.\ Rpt.\ PL-TR-92-2049}{Phillips Laboratory}{Hanscom Air Force Base Massachusetts}}
\newcommand{\ACMII}{\bookforchapter{\editor{Warmbein, B.}}{Proc.\ of the Conference: Asteroids, Comets, Meteors ACM 2002. ESA SP-500}{ESA}{Noordwijk, The Netherlands}}
\newcommand{\ACMV}{\bookforchapter{\editors{Lazzaro, D., Ferraz-Mello, S., Fern\'andez, J.A.}}{Asteroids, Comets, and Meteors 2005}{Cambridge University Press}{Cambridge, UK}}

\newcommand{\Broz}{Bro\v{z}}
\newcommand{\Capek}{\v{C}apek}
\newcommand{\Delbo}{Delbo'}
\newcommand{\Durech}{\v{D}urech}
\newcommand{\Fernandez}{Fern\'{a}ndez} 
\newcommand{\Galad}{Gal\'{a}d}
\newcommand{\Hawaii}{Hawai'i}
\newcommand{\Kryszczynska}{Kryszczy\'{n}ska}
\newcommand{\Kuehrt}{Kührt}
\newcommand{\Kusnirak}{Ku\v{s}nir\'{a}k}

\newcommand{\Michalowski}{Micha\l owski}
\newcommand{\Mueller}{Müller} 
\newcommand{\Naranen}{Näränen}
\newcommand{\Nesvorny}{Nesvorn\'{y}}
\newcommand{\Oepik}{Öpik}
\newcommand{\Pravec}{Pravec} 
\newcommand{\Sarounova}{\v{S}arounov\'{a}}
\newcommand{\Vokrouhlicky}{Vokrouhlick\'{y}}
\newcommand{\Zappala}{Zappal\`{a}}

\renewcommand{\eqref}[1]{eqn.\ \ref{#1}}
\newcommand{\Eqref}[1]{Eqn.\ \ref{#1}}
\newcommand{\tableref}[1]{table \ref{#1}}
\newcommand{\Tableref}[1]{Table \ref{#1}}
\newcommand{\figref}[1]{\Figref{#1}}
\newcommand{\Figref}[1]{Fig.\ \ref{#1}}
\newcommand{\sectref}[1]{sect.\ \ref{#1}}
\newcommand{\Sectref}[1]{Sect.\ \ref{#1}}
\newcommand{\chaptref}[1]{chapter \ref{#1}}

\newcommand{\ppageref}[1]{on p.\ \pageref{#1}}
\newcommand{\figrefpage}[1]{\figref{#1} \ppageref{#1}}

\newcommand{\eqrefpage}[1]{\eqref{#1} \ppageref{#1}}

\newcommand{\tablerefpage}[1]{\tableref{#1} \ppageref{#1}}

\newcommand{\sectrefpage}[1]{\sectref{#1} \ppageref{#1}}

\newcommand{\seesect}[1]{(see \sectref{#1})}
\newcommand{\seesectpage}[1]{(see \sectrefpage{#1})}
\newcommand{\seefig}[1]{(see \figref{#1})}
\newcommand{\seefigpage}[1]{(see \figrefpage{#1})}
\newcommand{\seetable}[1]{(see \tableref{#1})}
\newcommand{\seetablepage}[1]{(see \tablerefpage{#1})}
\newcommand{\seechapt}[1]{(see \chaptref{#1})}
\newcommand{\seeeq}[1]{(see \eqref{#1})}
\newcommand{\seeeqpage}[1]{(see \eqrefpage{#1})}


%% file: titlepage.tex


\subject{Dissertation zur Erlangung des akademischen Grades \\
 Doktor der Naturwissenschaften \\
(doctor rerum naturalium)
} 
\title{Surface Properties of Asteroids \\ from Mid-Infrared Observations and \\Thermophysical Modeling}
\author{ 
Dipl.-Phys.\ Michael \Mueller}
\date{2007}

\publishers{
Eingereicht und verteidigt am Fachbereich Geowissenschaften der Freien Universität Berlin. 
Angefertigt am Institut für Planetenforschung des Deutschen Zentrums für Luft- und Raumfahrt e.V.\ (DLR) in Berlin-Adlershof. 
} 

\uppertitleback{%
\minisec{Gutachter} 

Prof.\ Dr.\ Ralf Jaumann (Freie Universität Berlin, DLR Berlin)

Prof.\ Dr.\ Tilman Spohn (Westfälische Wilhems-Universität Münster, DLR Berlin)

\paragraph{Tag der Disputation}
6.\ Juli 2007
}

\dedication{In loving memory of \\ Felix \Mueller\ (1948--2005). 
\\ Wish you were here.}
\maketitle



%% file: abstract.tex
%
The subject of this work is
the physical characterization of asteroids, with an emphasis on the thermal inertia of near-Earth asteroids (NEAs).
Thermal inertia governs the Yarkovsky effect, a non-gravitational force which significantly alters the orbits of asteroids up to $\sim\unit{20}{\km}$ in diameter. Yarkovsky-induced drift is important in the assessment of the impact hazard which NEAs pose to Earth.
Yet, very little has previously been known about the thermal inertia of small asteroids including NEAs.

Observational and theoretical work is reported.
The thermal emission of asteroids has been observed in the mid-infrared (5--\unit{35}{\micron}) wavelength range using the Spitzer Space Telescope and the \unit{3.0}{\metre} NASA Infrared Telescope Facility, IRTF; techniques have been established to perform IRTF observations remotely from Berlin.
A detailed thermophysical model (TPM) has been developed and extensively tested; this is the first detailed TPM shown to be applicable to NEA data.

Our main result is the determination of the thermal inertia of 5 NEAs, increasing the total number of NEAs with measured thermal inertia to 6. For two of our targets, previously available estimates are refined, no reliable estimates have been available for the remaining three.
The diameter range spanned by our targets is 0.1--\unit{17}{\km}.

Our results allow the first determination of the 
typical thermal inertia of NEAs, 
which is around \unit{300}{\TIunit} (corresponding to a thermal conductivity of $\sim\unit{0.08}{\kappaunit}$), 
larger than
the typical thermal inertia of large  main-belt asteroids (MBAs)  by more than an order of magnitude.
In particular, thermal inertia appears to increase with decreasing asteroid diameter.

Our results have been used by colleagues to estimate the size dependence of the Yarkovsky effect, thus explaining the apparent difference in the size-frequency distribution of NEAs and similarly sized MBAs.

Thermal inertia is a very sensitive indicator for the presence or absence of particulate material on the surface, a fact that is widely used in, e.g., Martian geology.
Our estimate of the typical thermal inertia of NEAs is intermediate between values for lunar regolith and bare rock, indicating that even sub-kilometer asteroids are covered with coarse regolith.
This is consistent with spacecraft observations of the \unit{0.32}{\km} wide NEA (25143) Itokawa obtained in 2005.

The correlation of thermal inertia with size indicates a trend of
smaller objects having coarser and/or thinner regolith than larger objects.
This may allow an improved understanding of regolith formation through impact processes.

The first thermal-infrared observations of an eclipsing binary asteroid system are reported.
To this end, the Trojan system (617) Patroclus has been observed using Spitzer.
It is demonstrated that such observations enable uniquely direct thermal-inertia measurements.
In particular, we report the first reliable estimate of the thermal inertia of a Trojan.

Additionally, two targets of future spacecraft encounters, (21) Lutetia and (10302) 1989~ML, have been observed. Their size and albedo has been determined and their surface mineralogy constrained.
Our results for 1989~ML, in particular, are relevant in the current planning of the ESA mission Don Quijote.


%% file: intro.tex
\emph{Asteroids}, also known under the now deprecated name  \emph{minor planets}, are a large population of small Solar System bodies which do not display cometary activity. 
The name asteroid, which was coined by W.\ Herschel in 1802, is derived from the Greek word for star-like---like stars, and unlike planets or comets, asteroids 
appear point-like
in typical telescopic observations.

Small Solar System bodies are the most pristine  material left over from the early days of the Solar System
and have undergone much less processing
than the planets or the Sun throughout the past \unit{4.6}{\giga\yr}.
They therefore preserve crucial information
on the formation and evolution of the Solar System. Asteroids, in particular, are believed to be remnant building material of the inner planets. 
Impacts of asteroids and comets have significantly resurfaced the terrestrial planets and their satellites 
and may have been a significant source of water on Earth \citep[see, e.g.,][for a recent review]{Martin2006}.
Meteorites, the remnants of Earth impactors, are
the major source of extra-terrestrial material available for laboratory studies; 
studies of meteorites and asteroids, the parent bodies of most meteorites,  benefit considerably from one another.
A large impact on Earth
could release sufficient energy to 
cause severe or even fatal damage to our civilization;
the Cretaceous-Tertiary extinction event, during which the dinosaurs died out, is widely believed to have been caused by a catastrophic  impact.

The increasing public awareness of the impact hazard and general scientific interest has stimulated a dramatic increase in asteroid research over the past decade. This includes the dedication of an increasing number of  telescope systems to asteroid discovery.

Nevertheless, the steep increase in asteroid discoveries far outpaces efforts to increase our knowledge about their physical properties.

	\section{Space missions to asteroids}
\label{sect:intro:spacecraft}

\begin{table}
\caption[Overview of asteroids studied by spacecraft]{Overview of asteroids studied by spacecraft, including future targets of  Rosetta (launched in 2004) but not including the two rendezvous targets of Dawn (to be launched in June 2007).}
\label{table:intro:spacecraft}
\centering

\begin{tabular}{llll}
\toprule
Spacecraft & Year & Asteroid target  & \\
\midrule
Galileo & 1991 & (951) Gaspra & Flyby \\
        & 1993 & (243) Ida + Dactyl & Flyby \\
NEAR--Shoemaker    & 1997 & (253) Mathilde & Flyby \\
        & 1998 & (433) Eros     & Flyby \\
        & 2000 & "              & Rendezvous; landed \\
Deep Space 1 & 1999 & (9969) Braille & Flyby \\
Cassini & 2000 & (2685) Masursky & Distant flyby \\
Stardust & 2002 & (5535) Annefrank & Flyby \\
Hayabusa & 2005 & (25143) Itokawa & Rendezvous;  samples taken (?)\\
New Horizons & 2006 & (132524) APL & Distant flyby \\
Rosetta & 2008 & (2867) \v{S}teins & Flyby \\
        & 2010 & (21) Lutetia      & Flyby \\
\bottomrule
\end{tabular}
\end{table}
\begin{figure}[bt]
\centering
  \includegraphics[width=\textwidth]{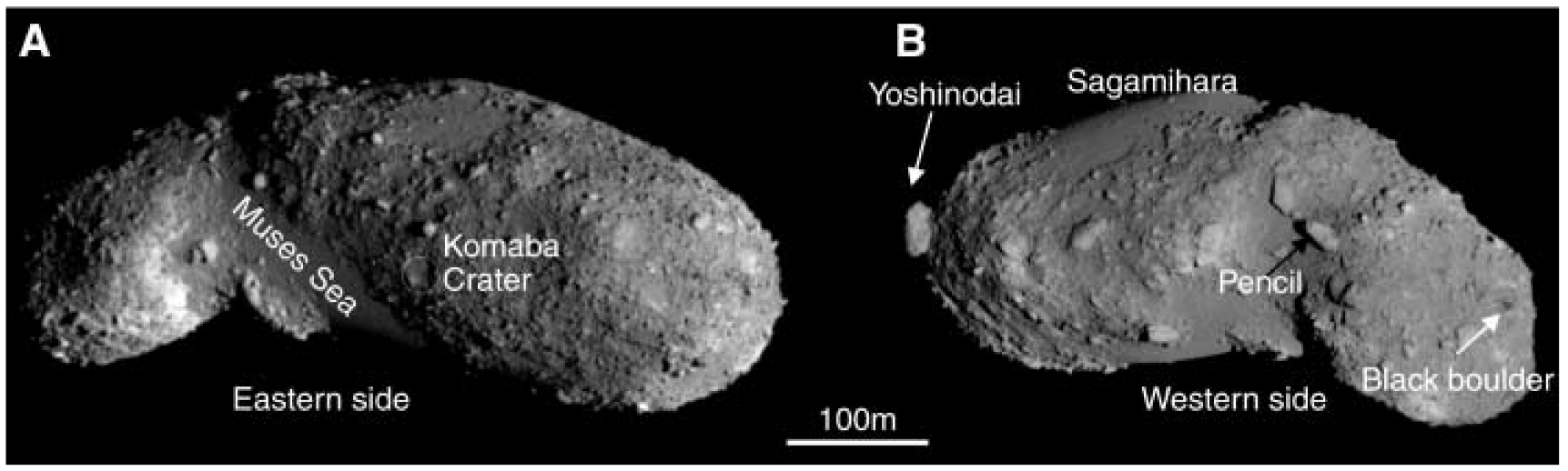}
\caption[Spacecraft imaging of NEA Itokawa]{Global images of near-Earth asteroid Itokawa recorded from the Hayabusa spacecraft. Note the scale!
\citep[Figure from][]{Saito2006}}
\label{fig:intro:ito}
\end{figure}

Since 1991, when the Galileo spacecraft flew by the asteroid (951) Gaspra, asteroids have been targeted by spacecraft several times, see \tableref{table:intro:spacecraft} \citep[see also][for a slightly outdated review]{Farquhar2002}.

Spectacular insights were gained from results of the asteroid rendezvous missions NEAR-Shoemaker and Hayabusa. NEAR-Shoemaker orbited the near-Earth asteroid (433) Eros for about a year until it successfully soft-landed in February 2001, taking further data from ground.
Hayabusa hovered within kilometers from the small (effective diameter around \unit{320}{\meter}) near-Earth asteroid (25143) Itokawa for several months in 2005---note that 
stable spacecraft orbits around such a low-gravity target are hard to find.
After intensively studying the asteroid (see, e.g., \figref{fig:intro:ito}), Hayabusa touched down on the surface twice  to take samples of surface material.
Unfortunately, the spacecraft is experiencing technical difficulties and it is unclear whether samples have been taken. Hayabusa is scheduled to return the sample container to Earth in 2010.
First Hayabusa results were published in a special issue of \textit{Science} on 2 June 2006 (Vol.\ 312, issue 5778).

Asteroid missions  are currently being planned at all major space agencies:
\begin{description}
\item[Dawn] is a NASA mission to rendezvous with the two large main-belt objects (1) Ceres and (4) Vesta, scheduled for launch in June 2007, arrival at Vesta in 2011 and at Ceres in 2015.  Italian and German institutes (including DLR Berlin) contribute two science instruments \citep{Russel2006}.
\item[Don Quijote] is an ESA mission to produce a measurable deflection of a near-Earth asteroid.
The  mission will consist of two spacecraft, a kinetic impactor and an orbiter which will intensively study the asteroid  before and after the deflecting impact. Don Quijote is currently under phase-A study  \citep{HarrisDQ}.
\item[Hayabusa 2] is basically a clone of Hayabusa by the Japan Aerospace Exploration Agency  JAXA. Hayabusa 2 is planned to be launched in 2010 or 2011 and to return samples from another 
near-Earth asteroid. An improved version,
named Hayabusa Mark 2, is also being planned \citep{Yoshikawa2006}.
\item[OSIRIS] 
is a sample-return mission to a near-Earth asteroid currently under consideration at NASA.%
\footnote{ See \url{http://www.nasa.gov/centers/goddard/news/topstory/2007/osiris.html}. OSIRIS is not to be confused with the telescope instruments of the same name, which are located on board the Rosetta spacecraft, at the Keck II telescope, and at the Gran Telescopio Canarias, respectively.}
If selected for further development, the mission may be launched in 2011.
\end{description}

Spacecraft studies of asteroids benefit significantly from  ground-based studies of their targets, and vice-versa.
Mission planning, in particular,
is severely hampered by the general lack of information on the physical properties of potential targets.
\emph{Physical studies of potential spacecraft target asteroids are  of crucial importance in this respect.}

\section{Asteroid populations and their origins}
\label{sect:intro:populations}

Since 1 Jan 1801, when Piazzi discovered (1) Ceres,%
\footnote{ Ceres has been reclassified as a dwarf planet at the IAU General Assembly in August 2006.}
the number of known asteroids has increased dramatically. 
As of 2 May 2007, 374,256 asteroids are known, 
157,788 of them have well-established orbits (see \url{http://cfa-www.harvard.edu/iau/lists/ArchiveStatistics.html}). 
Both numbers are increasing by the thousands per month
due mostly to dedicated  asteroid discovery programs. New telescope systems, which are currently being built \citep[such as Pan-STARRS, see][]{PanSTARRS}, are expected to result in a further increase of the asteroid discovery rate.

\begin{figure}[bt]
\centering
\includegraphics[width=0.6\textwidth]{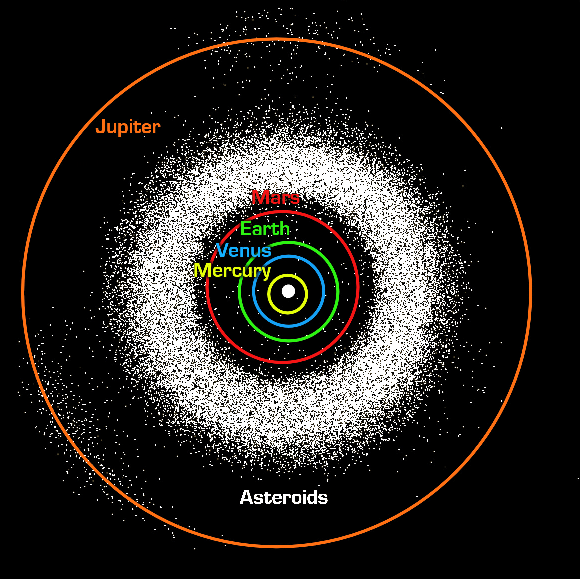}
\caption[Asteroids in the inner Solar System out to Jupiter]{
Asteroids in the inner Solar System out to Jupiter. Note the presence of some asteroids close to the terrestrial planets. Figure credit: NASA/JPL-Caltech/R.\ Hurt (SSC-Caltech).}
\label{fig:intro:populations}
\end{figure}

As can be seen from \figref{fig:intro:populations}, there are three main asteroid populations in the inner Solar System:%
\footnote{
While small bodies beyond Jupiter's orbit without cometary activity, such as Centaurs or trans-Neptunian objects, are given asteroid designations, we shall not consider them as asteroids in the following. They are probably very rich in volatiles and 
 resemble comets more closely  than asteroids.
}
\begin{description}
\item[Main-belt asteroids (MBAs)]
Most known asteroids orbit the Sun in the region between the orbits of Mars and Jupiter, called the asteroid belt or main belt.
The accretion process in the main belt stopped  before a planet could be formed, probably due to dynamical excitation through the gravity of forming Jupiter; present MBA encounter velocities are so high that collisions are more likely to produce fragmentation than accretion \citep{Petit2002}.
MBAs are thus remnant planet building material, left-overs from the formation of the Solar System which have undergone only limited  processing in the past \unit{4.6}{\giga\yr}.

The largest main-belt object is (1) Ceres with a diameter around \unit{950}{\kilo\metre}. 
The observed asteroid size-frequency distribution increases steeply with decreasing size, but drops towards small sizes due to observational incompleteness (in other words: the smallest asteroids have not been discovered, yet).
The smallest newly-discovered MBAs are typically a few \kilo\metre\ in diameter.
\item[Near-Earth asteroids (NEAs)]
Since the discovery of (433) Eros in 1898 it is known that there is an intriguing population of asteroids which approach Earth. 
The Earth-like orbits of some NEAs make them accessible for spacecraft with only a moderate amount of propellant and thus at a relatively low cost.

On average, NEAs are significantly smaller than the known MBAs; the largest NEA is (1036) Ganymede with an estimated diameter around \unit{32}{\kilo\metre}, objects as small as a few tens of meters have been detected. 
As of 1 May 2007, 4619 NEAs have been discovered,
including 712 objects with an estimated diameter of \unit{1}{\km} or larger.%
\footnote{ Source: \url{http://neo.jpl.nasa.gov/stats/}. Note that the number of objects above \unit{1}{\km} in diameter depends on the assumed albedo---see also \sectref{sect:intro:diameter}.}
The total number of the latter is estimated to lie between 700 and 1,100 \citep{Werner2002,Stuart2004}.

A particularly noteworthy group of  NEAs are the Potentially Hazardous Asteroids (PHAs), which approach Earth's orbit
to within \unit{0.05}{\AU} and have  diameters  above \unit{150}{\metre}.%
\footnote{ Note that the diameter of most NEAs is unknown; technically, PHAs are therefore defined as having an absolute optical magnitude $H$ \seesect{sect:intro:diameter} below 22, which corresponds to a diameter above \unit{150}{\metre} for an assumed geometric albedo of $\pv=0.13$.}
As of 7 May 2007, 860 PHAs are known.

\item[Jupiter Trojans] 
There are two large asteroid groups  beyond the main belt, collectively referred to as Jupiter Trojans. They are in  stable 1:1 resonance with Jupiter, 
librating around the $L_4$ and $L_5$ Lagrange points  which lead and trail the planet 
by \unit{60}{\degree} in heliocentric ecliptic longitude, respectively.

The origin of the Trojans is currently under debate.
While they were long believed to have formed near their present position \citep[see, e.g.,][]{Marzari2002}, it has been argued by \citet{Morbidelli2005} that their orbital distribution indicates  they were  captured by Jupiter during the time of the Late Heavy Bombardment, and that they share a volatile-rich parent population with comets and small bodies in the outer Solar System. The latter theory is supported by 
the rather uniform spectral properties and albedos of Trojans similar to cometary nuclei \citep{Barucci2002} and with recent physical studies of large Trojans \citep{Marchis2006, Emery2006}.
\end{description}

\begin{figure}[tb]
\centering
\includegraphics[width=0.7\linewidth]{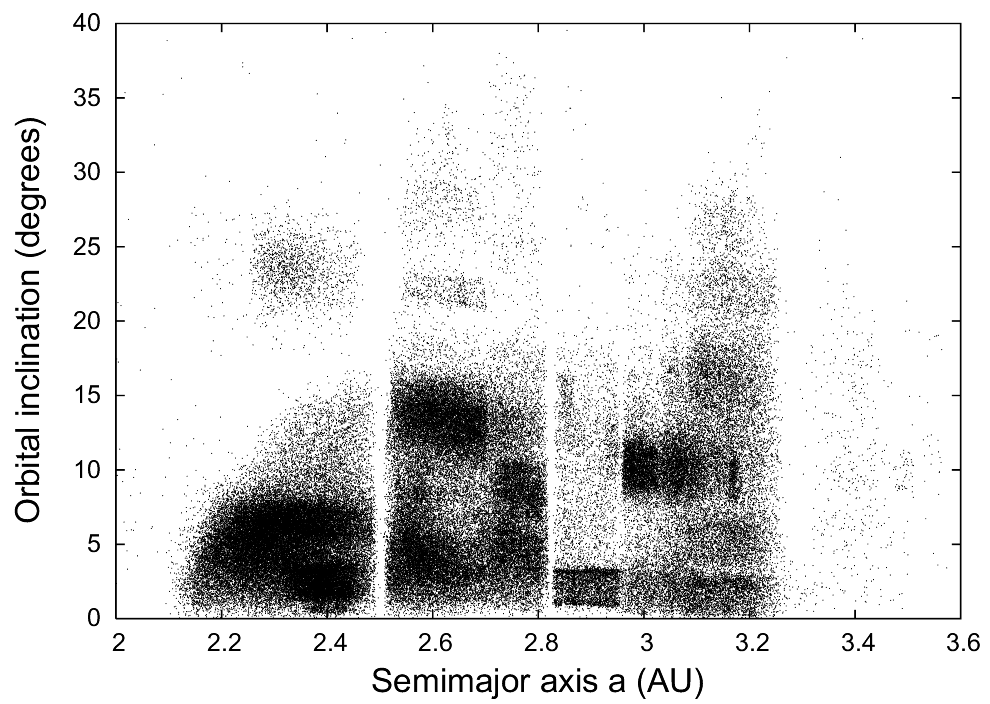}
\caption[Scatter plot of the orbital parameters of numbered MBAs]{Scatter plot of the orbital parameters of numbered MBAs.
The Kirkwood gaps are clearly seen, e.g.\
at semimajor axes of 
$a\sim \unit{2.5}{\AU}$ (3:1 mean-motion resonance with Jupiter), at some \unit{2.8}{\AU} (5:2 resonance), roughly \unit{2.96}{\AU} (7:3 resonance), and the sharp boundary shortward of \unit{3.3}{\AU} (2:1 resonance).
Also some significant clusters corresponding to asteroid families are clearly seen, e.g.\ the Koronis family situated at low inclinations between the 5:2 and 7:3 resonances, seen as a relatively sharp rectangle.
(Orbital parameters were retrieved from the University of Pisa AstDys service, \url{http://hamilton.dm.unipi.it}, on 8 Jan 2007)}
\label{fig:intro:Kirkwood}
\end{figure}

Orbits inside the main belt are highly chaotic, mostly due to the gravitational influence of massive and near-by Jupiter.
In particular, many main-belt orbits are unstable due to  resonance with Jupiter; there is a significant depletion in objects on such orbits, the Kirkwood gaps \citep[see also \figref{fig:intro:Kirkwood}]{Kirkwood1869}.

\paragraph{Asteroid families}
While the largest MBAs are believed to be primordial, most MBAs below a certain threshold size appear to be fragments of larger parent bodies which underwent a catastrophic collisional disruption \citep{Nesvorny2006}.
One might expect fragments of such a breakup event to be on rather similar orbits. Indeed, as can be seen in \figref{fig:intro:Kirkwood}, there are statistically significant clusters in the orbital elements of MBAs referred to as \emph{asteroid families,} which were first noticed and explained by \citet{Hirayama} \citep[see][for a review]{Zappala2002}.
The reflection spectra of asteroids belonging to a family are generally very similar, confirming their common origin \citep{Cellino2002}.

Most known asteroid families appear to be very old, on the order of several \unit{100}{\mega\yr}  \citep{Carruba2003}.
Due to chaotic dynamics,  asteroid families disperse over timescales of roughly \unit{1}{\giga\yr}, making old families hard to detect dynamically \citep{Nesvorny2002b}.
The ages of very young families, on the other hand, can be determined directly, by numerically integrating the orbits of family members backward in time until convergence is reached;
a spectacular case is that of the Karin cluster, the age of which has been determined by \citet{Nesvorny2002} to be only \unit{$5.8\pm 0.2$}{\mega\yr}. The convergence of this backward integration has been shown to improve significantly if the Yarkovsky effect \seesect{sect:intro:Yarko} is taken into consideration \citep{Nesvorny2004}.
Recently, asteroid families
even younger than \unit{1}{\mega\yr} have been reported by \citet{NesvornyVokrouhlicky2006}.

\paragraph{The origin of NEAs}
It is now widely accepted that the dominant NEA source population is the main belt, 
followed by extinct cometary nuclei providing 
\unit{$15\pm 5$}{\%} of the  population \citep[see][and references therein]{BinzelLupishko2006}.
 This is consistent with the diversity in spectral properties and albedo observed among NEAs, which is similar to that of MBAs.

The only known means of delivering sufficient numbers of MBAs into near-Earth space  is through  resonances with Jupiter and later perturbations by the inner planets, which may temporarily trap them in near-Earth orbits, although collisions with the Sun or ejection out of the Solar System are more likely  \citep[see, e.g.,][and references therein]{Morbidelli2002}.

The timescale for resonant ejection out of the main belt is a few \mega\yr, the dynamical lifetime of NEAs is on the order of \unit{10}{\mega\yr}.
However, as apparent from the crater record on terrestrial planets and their satellites, the NEA population has been rather stable over the past \unit{4}{\giga\yr} \citep{Ivanov2002,Werner2002}.
This suggests a steady effect which continuously replenishes the NEA source regions. 

It is now widely believed that this is  accomplished by the  Yarkovsky effect \seesect{sect:intro:Yarko}.
The Yarkovsky-induced drift  inside the main belt takes much longer than the actual resonance-driven transport into near-Earth space, it is therefore the strength of the Yarkovsky effect that determines the timescale and size-dependent efficiency of NEA delivery \citep{Morbidelli2003}.
This is supported by the observed cosmic-ray exposure ages of meteorites, which
average around \unit{10--100}{\mega\yr} for stony meteorites and  an order of magnitude larger for iron meteorites,%
\footnote{ Note that the Yarkovsky effect is generally less effective for objects with very high thermal inertia, such as metallic bodies \seesect{sect:intro:Yarko}.}
significantly longer than the NEA dynamical lifetime and indicative of a substantial drift time spent inside the main belt \citep[see, e.g.,][and references therein]{Bottke2006}.

\section{The Yarkovsky and YORP effects}
\label{sect:intro:Yarko}

\begin{figure}[bt]
\centering
   \includegraphics[width=\textwidth]{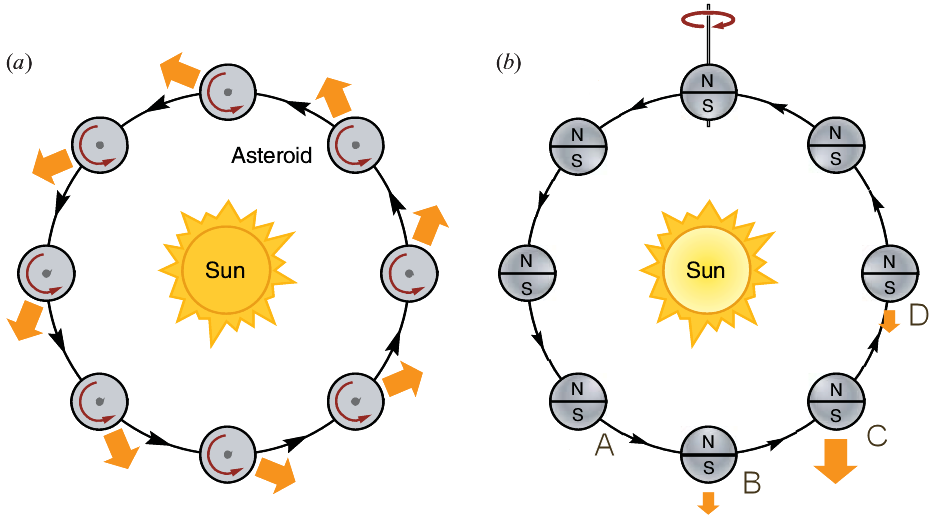}
\caption[Schematic depiction of the diurnal and seasonal Yarkovsky effects]{
\textbf{(a):} Schematic depiction of the \emph{diurnal Yarkovsky effect} for a spin axis perpendicular to the orbital plane. Due to thermal inertia, the trailing afternoon side of a prograde rotator is hotter than the leading morning hemisphere, leading to an emission  surplus  from the former. The resulting net force (arrows) has an accelerating component tangential to the orbit, causing most prominently a secular increase in orbital semimajor axis $a$ (the radial force component is typically negligible against solar gravity).
Analogously, retrograde rotators are decelerated by the Yarkovsky effect,  their $a$  decreases. 
\textbf{(b):} \emph{Seasonal Yarkovsky effect,} with the spin axis inside the orbital plane.
There is an emission surplus from the summer hemisphere. Thermal inertia causes a phase shift between seasons and orbital revolution (orange arrows), leading to a net tangential force component after averaging over one orbit (see positions A--D). The seasonal effect always decreases $a$.
\citep[Figure adapted from][]{Bottke2006}.}
\label{fig:intro:Yarko}
\end{figure}

It has been realized over the past decade that asteroid dynamics  is governed not only by gravity and mutual collisions, but also by the non-gravitational Yarkovsky and YORP effects, both caused by the recoil force from thermally emitted photons.
As with ion spacecraft propulsion, the resulting momentum transfer is slight but steady, and therefore capable of slowly but substantially altering the orbits (Yarkovsky effect) and spin states (YORP effect) of small asteroids or meteoroids.%
\footnote{ Orbital drift due to thermal emission was first considered by \citet{Yarkovsky} in a private publication \citep[which was long lost, but has recently been rediscovered; see][for a reprint]{BrozThesis}. 
\Oepik, having read Yarkovsky's paper,  reproposed and named the Yarkovsky effect much later  \citep{OepikYarko}, but until the 1990s it was widely considered irrelevant. 
The YORP effect was proposed by \citet{Rubincam2000}, and named after Yarkovsky and also O'Keefe, Radzievskii, and Paddack, who had considered similar effects between 1954 and 1976.}
Both effects have been observed (see below). See \citet{Bottke2006} for a recent review.

\paragraph{Yarkovsky effect}
As depicted in \figref{fig:intro:Yarko}, 
surface temperature asymmetries due to thermal inertia \seesect{sect:intro:TI} lead to a gradual increase or decrease in orbital semimajor axis $a$, depending on the spin axis orientation and obliquity.
There are diurnal and  seasonal components of the Yarkovsky effect, which are respectively most efficient in the situations depicted in \figref{fig:intro:Yarko}.

\textit{
Since the Yarkovsky effect is driven by  surface temperature asymmetries, it depends crucially on the  thermal inertia.} 
Specifically, it vanishes in the limiting cases of zero and infinite thermal inertia where the temperature distribution is symmetric about the subsolar point. Obviously, spin rate and heliocentric distance are also relevant.

Very importantly, the Yarkovsky effect is size dependent:
For objects much larger than the penetration depth of the heat wave (typically at the \centi\metre-scale) and all other parameters kept constant,
the photon recoil force scales with $D^2$ (with diameter $D$), while the mass scales with $D^3$, so the acceleration scales with $D^{-1}$. 
Smaller objects become increasingly isothermal, weakening the Yarkovsky effect; 
their interaction with the solar radiation field is dominated by the Poynting-Robertson effect or the radiation pressure.

There is now ample evidence that the Yarkovsky effect  strongly influences the orbital dynamics of asteroids below  \unit{$\sim20$}{\kilo\metre} in diameter:
\begin{itemize}
\item The small NEA (6489) Golevka was shown by \citet{YarkoGolevka} to have undergone an orbital drift in the years 1991--2003 which cannot be explained by gravitational perturbations alone, but is fully consistent with an additional Yarkovsky-induced drift.
The Yarkovsky effect had previously been found to alter the orbit of the LAGEOS satellite  \citep{Rubincam1987, Rubincam1988, Rubincam1990}.
\item 
As seen above, there must be a steady mechanism bringing small MBAs into powerful resonances which deliver them into near-Earth space---only the Yarkovsky effect is known to do so in a way consistent with the observed NEA distributions in size, spectral type, and spin axis obliquity.
In particular, its size dependence explains the apparently different  size-frequency distributions of NEAs and MBAs
\citep[see][for a detailed discussion]{Delbo2007}.
\item 
The Yarkovsky effect is required to explain the observed orbital distribution inside asteroid families.
Most spectacularly, the orbits of asteroids belonging to the very young Karin family (see above) were seen to have evolved under the Yarkovsky effect \citep{Nesvorny2004}.
Furthermore, Yarkovsky-induced drift is required to make the observed orbital dispersion in evolved asteroid families compatible with that of young families and also with 
model calculations of 
the initial fragment ejection velocity distribution \citep{Carruba2003,Bottke2006}.
\item The Yarkovsky effect is crucial to assess the impact hazard from individual asteroids. Specifically, it determines whether the NEA 1950~DA, the object with the highest currently  known impact probability \seesect{sect:intro:impact}, will hit Earth in 2880 or not \citep{Giorgini2002}.
\end{itemize}

\paragraph{YORP effect}

The photon recoil force combined with the radiation pressure of absorbed sunlight may also cause a net torque,
altering  the spin axis obliquity and the rotation rate of small objects.
The YORP torque depends critically on the object's shape, in particular it vanishes for spherical or ellipsoidal objects \citep[see][for a recent definition of a shape-dependent parameter describing the strength of the YORP torque]{Scheeres2007}.
While small asteroids are known to have highly irregular shapes in general, the shape of individual objects is usually unknown, although that situation is likely to improve significantly in the next decade \seesect{sect:intro:shape}.
The YORP effect is therefore less well studied than the Yarkovsky effect.

Nevertheless, the first direct observations of YORP-induced shifts in rotational period have very recently been  reported for the NEAs (1862) Apollo \citep{Kaasalainen2007} and  (54509) YORP \citep[][54509 was known as 2000~PH5 before 2 April 2007; see also \sectref{sect:PH5}]{Lowry2007, Taylor2007}.
The YORP effect was seen by \citet{Vokrouhlicky2003} to determine the distribution of spin axis obliquities in asteroid families. The YORP effect may also explain the observed size-dependence of asteroid spin-rate distributions \seesect{sect:intro:spin} and may be important in the forming of binary asteroid systems \seesect{sect:intro:binaries}.

	\section{Asteroids impacting Earth: Hazard and link to meteorites}
\label{sect:intro:impact}

The terrestrial planets and their satellites have been resurfaced by impact cratering. This is quite evident on bodies such as the Moon or Mars, where erosion processes are relatively slow, but also on Earth a significant number of impact craters has been preserved---the most widely known in Germany is  the Nördlinger Ries.

Most Earth  impactors  are very small in size and are completely destroyed upon atmospheric entry causing only a ``falling star''. 
Among  objects that reach the ground, the majority is barely large and robust enough to do so---these produce meteorites which represent the major source of extraterrestrial material available for study in Earth laboratories.
Large impactors some thousand tons in mass or above,
 however, are not significantly decelerated by the atmosphere. 
They hit the ground at velocities above the Earth escape velocity of \unit{11.2}{\kilo\metre\per\second}
and release their correspondingly large kinetic energy in a crater forming process.
While our understanding of the latter is still highly incomplete \citep[see, e.g.,][and references therein]{Holsapple2002,deNiem2005} it is clear that impactors with diameters around \unit{1}{\kilo\metre} release a significantly higher amount of energy than a nuclear warhead; such impacts would cause global catastrophes \citep[see, e.g.,][for reviews]{Morrison2002,Chapman2004b}.
The Cretaceous-Tertiary extinction event, during which the dinosaurs died out, is widely believed to have been caused by the impact of an object around \unit{10}{\kilo\metre} in diameter \citep{Alvarez1980}.

The US Congress held hearings to investigate the impact hazard and charged NASA with the task of discovering
\unit{90}{\%} of all near-Earth objects (NEOs) larger than \unit{1}{\kilo\metre} in diameter within ten years; this \emph{Spaceguard Survey} was initiated in 1998, the due date for the spaceguard goal is  end of 2008. 
Several successful asteroid discovery programs have been initiated leading to a steep and ongoing increase in asteroid discoveries.
A follow-up discovery program, possibly requiring NASA to discover \unit{90}{\%} of all NEOs above \unit{140}{\metre} in diameter until 2020, is currently under discussion.%
\footnote{ A law requiring NASA to report to Congress about the feasibility of such a program was signed into law in December 2005, NASA's report to Congress was published in March 2007; see \url{http://neo.jpl.nasa.gov/neo/report2007.html}.}
It might include the deployment of mid-infrared space telescopes for asteroid discovery, such as the proposed NASA mission NEOCam \citep{Mainzer2006}.

As of 7 May 2007, the highest known Earth impact probability for an individual object is \unit{0.33}{\%} for a potential impact of the \unit{1.1}{\kilo\metre} wide NEA (29075) 1950 DA in 2880 \citep{Giorgini2002}---the only known impact probability larger than the accumulated background risk due to unknown objects of comparable size, thus leading to a positive hazard rating on the Palermo scale by \citet{PalermoScale}.
It is worth pointing out that the uncertainty in the risk assessment by \citeauthor{Giorgini2002}\ is dominated by the lack of knowledge of  physical properties of the asteroid which govern the magnitude of the Yarkovsky effect.

The NEA (99942) Apophis \citep[then known as 2004 MN4,  \unit{$270\pm 50$}{\metre} in diameter, see][]{Delbo2007b} 
held, for a brief period after its rediscovery in December 2004, an unprecedentedly large probability for an impact in 2029, peaking at \unit{2.7}{\%} and severely disconcerting the NEA community over the Christmas holidays.
On 27 Dec 2004 the 2029 impact could be ruled out on the basis of newly obtained astrometric data; the miss distance from the geocenter in 2029 is currently estimated to be 
$5.89\pm 0.35$ Earth radii \citep[$3\sigma$ uncertainty;][]{Chesley2006}.
The subsequent orbit, however, will be severely perturbed by 
Earth's gravity, possibly onto impact course. The corresponding risk is dominated by a potential impact in 2036, with a probability of
$2.2\cdot 10^{-5}$.
Again, for accurate risk assessment the Yarkovsky effect must be taken into consideration \citep{Chesley2006}.

The design of
 asteroid deflection missions, which would become necessary if an impactor were to be discovered, is an active area of engineering research \citep[see, e.g.,][]{Kahle2006}.
ESA is planning a precursor mission to an asteroid deflection mission, \emph{Don Quijote}, which is currently under phase-A study \seesect{sect:intro:spacecraft}.

\section{Physical properties of asteroids}

There is a growing body of information on the physical properties of asteroids, although the rapid discovery rate leaves most known objects uncharacterized.
Some asteroids, however, have been scrutinized with spacecraft or have been studied by ground-based observers in great detail.

The emerging picture is still rather incomplete and highly diverse.

\subsection{Diameter and albedo}
\label{sect:intro:diameter}

For most asteroids, the size, arguably the most basic physical property, is only poorly known. Note that asteroids are typically far too small to be spatially resolved with current telescopes.
In only a few cases could asteroid sizes  be determined by means of direct imaging from near-by spacecraft, the Hubble Space Telescope, or ground-based telescopes equipped with adaptive optics.
Another rather direct way of determining asteroid sizes is from observations of stellar occultations.

For most asteroids, only optical photometric data are available, typically from astrometric measurements with limited photometric accuracy. The amount of reflected sunlight is proportional to the projected area and the albedo, allowing coarse conclusions on the size to be drawn.
An important quantity is the absolute optical magnitude $H$, 
which is defined as the visual magnitude corrected to heliocentric and observer-centric distances of \unit{1}{\AU} and a solar phase angle of \unit{0}{\degree} \citep{HG}.
$H$ is related to diameter $D$ and geometric albedo \pv\ by
\citep{FowlerChillemi}:
\begin{equation}
D = 10^{-H/5}\frac{\unit{1329}{\kilo\metre}}{\sqrt{\pv}}.
\label{eq:FowlerChillemi}
\end{equation}
Asteroid albedos range from some $\pv=0.02$ up to around 0.6, thus  diameters estimated in this way are very uncertain.

A widely used method to determine asteroid sizes is from observations of their thermal emission, which is proportional to the projected area but only a weak function of albedo (see \chaptref{chapt:thermal} for a detailed discussion). This method, pioneered by \citet{Allen1970}, is the source of most known asteroid diameters \citep{SIMPS}. Other methods of determining asteroid sizes include  observations at radar wavelengths \citep{Ostro2002}.

Alternatively, the diameter can be determined if \pv\ is known.
Methods for determining asteroid albedos include studies of the optical brightness and also of the polarization of reflected sunlight  as a function of solar phase angle \citep[see][for a review; note that the latter method is so far based on a purely empirical correlation between albedo and certain polarization properties]{Muinonen2002}.

\subsection{Taxonomy}
\label{sect:intro:mineralogy}

Conclusions on the mineralogical composition of asteroid surfaces  can be drawn from reflection properties at visible and near-IR wavelengths, chiefly from spectral features and  albedo measurements.
Asteroid reflection properties  are routinely compared to those of meteorites. This way, much could be learned about the composition of asteroids and about the  origin of most meteorites.

Different kinds of taxonomic systems are used in order to describe observed asteroid reflection properties, and also to link them with analogue meteorites.
The most widely used taxonomic systems are those by \citet{Tholen1984} and \citet{BusBinzel}.
While taxonomic classification relies chiefly on spectroscopic or spectrophotometric observations, it can be greatly constrained with albedo measurements alone.

A large number of taxonomic classes have been proposed, but  most asteroids belong to one of the following classes \citep[or ``complexes'' in the notation of][]{BusBinzel}
with generally mnemonic names:
\begin{description}
\item[C] is for carbonaceous:
C-type asteroids display spectra and albedos consistent with a composition similar to that of carbonaceous chondritic meteorites. They are very dark, generally $\pv < 0.1$. Most objects in the outer main belt appear to be C types \citep[see][Fig.\ 19]{BusBinzel}.
\item[S] is for silicaceous: 
S-type asteroids show spectral features indicative of a silicate composition similar to stony meteorites, with \pv\ normally in the range from 0.10 to 0.25.

S types dominate the inner main belt and the near-Earth population.
S-type asteroids are therefore commonly associated with the most frequent meteorite type, the ordinary chondrites.
However, the spectral features and albedos of the similar but less frequent \emph{Q-type} asteroids fit those of ordinary chondrites much better.
It is widely believed that S and Q-type asteroids are of identical bulk mineralogy, and that their surfaces age due to impacts by micro-meteorites and/or the solar wind (note that meteorites have lost their original surface during atmospheric entry), this process is called \emph{space weathering}. In this picture, S and Q-type asteroids are respectively the old and young endmembers of a continuum; spacecraft imaging of the S-type asteroids Ida and Eros appears to support this idea \citep[see][for reviews]{Clark2002,Chapman2004}. Recently, \citet{Lazzarin2006} found evidence that asteroids of other spectral types are also space weathered.
%
\item[X or EMP] 
Most remaining asteroids have rather featureless spectra at visible wavelengths, they are called X-type asteroids. There are three distinct groupings of X types differing in  albedo:
   \begin{description}
   \item[E] with high albedo ($\pv>0.3$), probably related to  enstatite achondrite meteorites
   \item[M] with moderate albedo ($0.1<\pv<0.2$), some of which appear to be related to iron meteorites, but others appear to be non-metallic 
   \item[P] with very low albedo ($\pv<0.1$). P-type asteroids are believed to be composed of silicates very high in organic material, there are no known meteorite analogues.
Together with the equally dark D-type asteroids (not listed here), P types are very abundant among the Jupiter Trojans. 
Some D and P-type asteroids in  the near-Earth population are believed to be extinct cometary nuclei (depending on their orbital properties).
   \end{description}
For X types, an albedo determination is particularly diagnostic of mineralogical composition. The study of composition and possible subtle spectral features of X-type asteroids is a very active area of research. 

\end{description}

\subsection{Spin rate}
\label{sect:intro:spin}

Asteroid spin rates differ significantly from object to object, from only a few minutes up to several weeks. 
The rotation states of  MBAs are mostly determined from mutual collisions: The observed spin rates of MBAs larger than \unit{50}{\kilo\metre} in diameter follow a Maxwellian distribution as predicted by this model, with a mean rotation period around \unit{10}{\hour}  \citep{HarrisPravecACM2005}.
Smaller asteroids deviate, increasingly so with decreasing size, from a Maxwellian distribution.
In comparison, both very low and very high rotation rates are over-represented, indicating the presence of an effect capable of spinning small asteroids up or down. 
The YORP effect is widely believed to be responsible for this \citep{HarrisPravecACM2005}.

There is an intriguing dichotomy in asteroid spin rates:
While the  periods of all known asteroid larger than \unit{1}{\kilo\metre} in diameter are larger than \unit{2.2}{\hour}, most smaller objects spin significantly faster, at spin rates of only a few minutes in extreme cases.
This is widely seen as indicative of their internal structure \seesect{sect:intro:internal}.

\subsection{Shape and spin axis}
\label{sect:intro:shape}

The shape and spin axis of most asteroids are unknown. It must be kept in mind that asteroids are typically too small to be spatially resolved. 
There are two well-established techniques to determine physical models of asteroid shape and spin state from ground-based observations, 
namely from time-resolved photometric observations at optical wavelengths \citep[see][]{Kaasalainen2002} or 
from radar observations of their rotationally induced Doppler frequency shift \citep{Ostro2002}.
Both methods typically require a large amount of input data taken at various  aspect angles.
Note that shape models obtained from the inversion of optical photometry are typically convex, concavities are virtually impossible to resolve using that technique.
 
Asteroid shapes found so far vary significantly: Larger MBAs are typically  nearly spherical, although there are notable exceptions such as the ``dog-bone shaped'' \unit{$217\times 94\times 81$}{\kilo\metre} asteroid (216) Kleopatra \citep{Kleopatra}.
The shape diversity of smaller asteroids, most of which are expected to be collisional shards, is significantly larger.
Upcoming asteroid discovery programs such as Pan-STARRS promise to provide an extensive database of well-calibrated optical photometric data, which will allow the shapes and spin states of at least several thousands of asteroids to be determined in the next decade \citep{Durech2005}.

\subsection{Internal structure---are asteroids piles of rubble?}
\label{sect:intro:internal}

The internal structure of asteroids is an important and very active area of research. 
In particular it is not clear whether asteroids have significant tensile strength or whether some (or most) are loose gravitational aggregates, called \emph{rubble piles} \citep[see][for a review]{Richardson2002}.
This is of particularly practical importance in the case of potential Earth impactors:
Rubble piles may be significantly harder to deflect than monolithic objects.

\paragraph{Surface morphology}
The two asteroid rendezvous missions carried out so far revealed that Eros does not appear to be a rubble pile \citep{Cheng2004} while Itokawa does \citep{Fujiwara2006}.
Flyby imaging revealed very large craters, comparable to the objects' radii, on some asteroids (Gaspra, Ida, and Mathilde) and on Mars' satellite Phobos, presumably a captured asteroid. Those large craters are largely seen as indicative of a very weak, rubble-pile-like internal structure \citep[see, e.g.,][and references therein]{Chapman2002,Cheng2004} because monolithic bodies would be expected to be disrupted by shock waves such as those generated during the crater-forming impacts.
Weak, porous bodies dissipate shock waves efficiently and can therefore sustain much larger impacts \citep{Holsapple2002}.
This is supported by the presence of adjacent and undisturbed large craters on Mathilde; on a solid target, the formation of the later crater would have significantly affected the former.

\paragraph{Mass density}
Inferences on internal structure can be made from the mass density, which is known for a small set of asteroids consisting of spacecraft targets, some binary systems \seesect{sect:intro:binaries}, and a few asteroids which made recent close encounters with other asteroids \citep{Hilton2002}.
While the bulk mass densities of the largest MBAs above \unit{500}{\kilo\metre} in diameter match those of analogue meteorites very well, smaller objects are significantly under-dense.
There is a cluster of objects, including Eros, with macroporosities around \unit{20}{\%} (i.e.\ \unit{20}{\%} of the volume is apparently void) which are believed to contain major cracks formed by shattering through sub-catastrophic impacts.
Other objects
display much higher macroporosities, beyond \unit{50}{\%} in extreme cases; these objects must contain major voids
and are widely believed to be rubble piles (Mathilde falls into this category, which is consistent with its craters mentioned above) \citep[see][for a review]{Britt2002}.

It is worth noting that the uncertainty in mass density is typically dominated by the diameter uncertainty \citep[see][]{Merline2002,Richardson2006}.

\paragraph{Spin rate}
Further information can be potentially obtained from the observed dichotomy in asteroid spin rates \seesect{sect:intro:spin}: As was first noted by \citet{Harris1996}, the apparent ``spin barrier'' around \unit{2.2}{\hour} coincides with the spin rate at which the centrifugal force at the equator of a spherical body equals gravity. Rubble piles at faster spin rates would therefore disrupt, the conspicuous lack of such fast rotators larger than \unit{1}{\kilo\metre} appears to indicate that most, if not all, of them are rubble piles.
Small fast rotators, on the other hand, are held together by tensile strength and are expected to be monolithic.

Both conclusions have recently been challenged by  \citet{Holsapple2007} who argues that for bodies larger than some \unit{10}{\kilo\metre} in diameter, tensile strength is negligible relative to gravity.As a result, such asteroids are subject to the quoted spin barrier even if they possess significant tensile strength.
\citeauthor{Holsapple2007}  estimated the  tensile strength required to stabilize  known small fast rotators to be on the order of only \unit{10--100}{\kilo\pascal},  ``the strength of moist sand'' in the words of the author.

\subsection{Binarity}
\label{sect:intro:binaries}

An intriguing asteroid sub-population is that of the binary asteroids, i.e.\ asteroids with a gravitationally bound satellite. The first asteroid satellite, Dactyl, was found orbiting the MBA (243) Ida through imaging from the \textit{Galileo} spacecraft \citep{Belton1996}.
Since then, some 70 binary asteroid systems have been discovered, using different observational techniques. Their orbits range from near-Earth space out  into the trans-Neptunian region, with primary body diameters between just a few kilometer and several hundred \citep{Richardson2006, Noll2006}.

The mass, which for asteroids is only rarely known \citep[see][]{Hilton2002}, can be calculated for binary systems from Newton's form of Kepler's Third Law if the mutual orbit is well constrained.

The total angular momentum of most small NEA binaries is just above the critical limit at which a single body of equal mass would  disrupt. This indicates that they were formed from a parent body which was  spun up (e.g.\ through the YORP effect), ultimately leading  to its fission into a binary system \citep{HarrisPravecACM2005}.
This is seen as an indication for their  being rubble piles \seesect{sect:intro:internal}, which is consistent with the low lightcurve amplitude of most binaries.

\subsection{Regolith}
\label{sect:intro:regolith}

Large atmosphereless bodies such as Mars, Mercury, or our Moon are  known to be covered with a thick layer of regolith, which formed from retained impact ejecta.
Large asteroids are known to display at least some regolith on their surfaces,  smaller objects are known to be increasingly depleted in fine dust grains \citep[see, e.g.,][]{Dollfus1989}.%
\footnote{ Throughout this thesis, regolith is loosely defined  as a layer of particulate material covering the surface or parts of it.}
This latter finding was explained with the low asteroid gravity which allows fine impact ejecta (with  higher average thermal velocities) to escape.
Asteroids below a certain threshold size were thus expected to be basically regolith free, the threshold diameter was estimated to be between 10 and \unit{70}{\kilo\metre} \citep[see][for a review]{Scheeres2002}.
This was consistent with the findings of \citet{Lebofsky1979} and \citet{Veeder1989}, who indirectly estimated the  thermal inertia  \seesect{sect:intro:TI} of a number of small NEAs to be very high, indicating a lack of thermally insulating regolith on their surfaces.

To the big surprise of the asteroid community, however, spacecraft imaging in 1991 \seesect{sect:intro:spacecraft} revealed indications for a substantial regolith layer on the MBA (951) Gaspra with an effective diameter of only \unit{12}{\kilo\metre}.
The NEA (433) Eros, around \unit{17}{\kilo\metre} in diameter, was unambiguously seen to be thickly covered with regolith, although an intriguingly large number of boulders are present on its surface.
Even the small NEA (25143) Itokawa (effective diameter around \unit{320}{\metre}) is not entirely regolith free: recent spacecraft imaging showed a thought-provoking surface dichotomy between boulder-strewn surface patches, apparently  devoid of powdered material, and very flat, regolith-dominated regions \seefigpage{fig:intro:ito}.

While the origin of regolith on asteroids is far from being completely understood, it is plausible that the regolith found on relatively small asteroids indicates  a weak surface structure, i.e.\ low material strength and/or high porosity \citep[see, e.g.,][]{Asphaug2002,Chapman2002,Holsapple2002,Scheeres2002}.
In this case, crater formation is dominated by gravity rather than material strength, which reduces ejecta velocities and enables the gravitational retention of a non-negligible fraction thereof.
This conforms with first results from the \emph{Deep Impact} mission \citep{AHearn2005}, where a \unit{370}{\kilo\gram} projectile hit the nucleus of comet 9P/Tempel 1 at a relative velocity of \unit{10.3}{\kilo\metre\per\second}---apparently a significant fraction of the ejected dust  was very slow and later reaccumulated, indicative of a very low material strength of the nucleus.
We caution, however, that the formation and later dynamics of ejecta is likely to be very different among comets and asteroids.

No detailed spacecraft imaging is available for asteroids between \unit{0.32}{\kilo\metre} (Itokawa) and \unit{12}{\kilo\metre} (Gaspra) in diameter; it is therefore unclear if they display regolith or not. 
In particular it is unclear whether there is a clear transition size above which asteroids are fully covered with regolith, while smaller objects are not.
In the light of the previous paragraph, studies of asteroid regolith coverage may allow conclusions to be drawn on the elusive but important material strength and may further our understanding of regolith formation through impact processes.


\subsection{Thermal inertia}
\label{sect:intro:TI}

Thermal inertia is a measure of  the resistance to changes in surface temperature: The surface of a body with low thermal inertia heats up or cools down readily, while bodies with high thermal inertia tend to keep their surface temperature for longer (see \sectref{sect:thermal:TI} for a more detailed discussion).

\textit{Thermal inertia governs the important Yarkovsky and YORP effects}
 \seesect{sect:intro:Yarko};  thermal inertia estimates are crucial for model calculations of both. 
Thermal inertia also determines the temperature environment in which lander missions \citep[see, e.g.,][for some considerations]{Binzel2003} have to operate: Low thermal inertia causes harsh temperature contrasts between the day and the night side, while in the case of high thermal inertia the diurnal temperature profile is much smoother.
Furthermore,  thermal inertia is a very sensitive indicator for the presence or absence of regolith on the surface \seesect{sect:intro:regolith}: The thermal inertia of lunar regolith is some 50 times lower than that of bare rock, which in turn is nearly an order of magnitude below that of metal \seetablepage{table:thermalproperties}.
This is widely used in planetary science;  several Mars orbiters, e.g., carried temperature sensitive instruments in order to derive global thermal-inertia maps by means of which exposed bedrock can be distinguished from  regolith \citep[see, e.g.,][and references therein]{Christensen2003, Putzig2005}.

Little is known so far about the thermal inertia of asteroids, virtually nothing is known about the thermal inertia of small asteroids including NEAs.
Ground-based determinations of asteroid thermal inertia are challenging, both in terms of observing and modeling: 
Extensive spectrophotometric or spectroscopic observations in the difficult mid-infrared wavelength range ($\sim5$--\unit{35}{\micron}) are needed \seesect{sect:thermal:observability}. Such observations are hampered by the large atmospheric opacity throughout most of this wavelength range, combined with a large level of background radiation stemming from the atmosphere, clouds, and the telescope itself which emits thermal radiation peaking at a wavelength around \unit{10}{\micron}.
On the modeling side, difficulties arise because  crucial parameters such as the object's shape and spin state are typically not known. Furthermore, sufficiently detailed thermophysical models had so far only been tested for application to large MBAs, which differ from small NEAs in many important ways \citep[see][for a review and \chaptref{chapt:TPM} for a detailed discussion]{HarrisLagerros2002}.

It was realized already in the 1970s that the typical thermal inertia of large asteroids must be small, comparable to that of the Moon \citep[see, e.g.,][for a review]{Morrison1977}. 
However, no quantitative results were available, with the notable exception of (433) Eros \citep[][based on a very approximate shape model]{LebofskyRieke1979} and (1) Ceres and (2) Pallas \citep{Spencer1989}.
The first large-scale thermal-inertia study was by \citet{MuellerLagerros1998} who quantitatively determined the thermal inertia of 5 large MBAs.

There has been some controversy about the typical thermal inertia of NEAs, which had not been measured directly so far. \Citet{LebofskyBetulia, Lebofsky1979} and \citet{Veeder1989} found indirect evidence that a significant fraction of NEAs should have a very high thermal inertia indicative of a surface consisting of bare rock. \citet{Delbo2003}, on the other hand, performed a thermal spectrophotometric survey of NEAs and stated that the  majority of their targets must possess a thermally insulating layer of regolith.
Note that the thermal inertia of small asteroids is particularly relevant since they are substantially influenced by the Yarkovsky effect \seesect{sect:intro:Yarko} which is governed by thermal inertia.

\section{Scope of this work}

\textit{The primary aim of this work is to augment the number of asteroids with known thermal inertia.} 
Emphasis is put on  NEAs for which practically no reliable information is available so far. We have also determined the size and albedo of two asteroid targets of upcoming spacecraft visits.

The following questions are addressed:
\begin{itemize}
\item What is the typical thermal inertia of NEAs?
\item What can be learned about their regolith coverage?
\item Does thermal inertia depend on size, as might be expected from models of regolith retention?
\item What is the size and albedo of our targets, and how can we constrain their surface mineralogy?
\end{itemize}
This requires
extensive observations of the thermal emission of our targets in the mid-infrared wavelength range ($\sim5$--\unit{35}{\micron}), combined with observations of the reflected sunlight and a suitable model of the thermal emission.

An adequate thermophysical model for NEAs has been developed and tested \seechapt{chapt:TPM}. 
Previously available models of NEA thermal emission are not sufficiently detailed for the quantitative determination of thermal inertia, while available thermophysical models for atmosphereless bodies (on which the model described herein is based) were neither designed nor tested for application to NEAs.

Observations  were made with the NASA Infrared Telescope Facility on Mauna Kea~/ \Hawaii\ (\chaptref{chapt:IRTF}) and the Spitzer Space Telescope (\chaptref{chapt:SST}).

We present detailed studies of individual objects rather than a general survey; our results for individual asteroids are presented in \chaptref{chapt:results}. Nevertheless, our results allow the first firm conclusions to be drawn on the NEA distribution in thermal inertia. These and other results are discussed in \chaptref{chapt:discussion}.


In the final \chaptref{chapt:conclusions}  our main conclusions are summarized, possible future work is discussed in \chaptref{chapt:future}.


%% file: ThermalIntro.tex
In this chapter we  briefly summarize some goals and methods of the study of asteroid thermal emission.
After a brief overview section (\sectref{sect:thermal:overview}),
those physical properties which are most relevant for the thermal emission of asteroids will be introduced in \sectref{sect:thermal:props} (a more detailed discussion of some  will be given  in \chaptref{chapt:TPM}). The observing conditions in the mid-infrared wavelength-range, in which the thermal emission of asteroids typically peaks, will be discussed in \sectref{sect:thermal:observability}. 

There are two different ways of interpreting thermal data, depending on the available database and other previous knowledge about the particular asteroid. If only little information is available, the diameter and albedo can be estimated, but assumptions must be made about thermal properties such as thermal inertia. If, however, much information is available, thermal properties can be \emph{derived} from the thermal data, in addition to  potentially more accurate estimates of diameter and albedo. 

Two ``simple'' thermal models which are widely used to determine asteroid diameters and albedos are presented in \sectref{sect:thermal:STM-FRM}.
Throughout this work, we make frequent use of the Near-Earth Asteroid Thermal Model (NEATM) described in \sectref{sect:NEATM}, which allows qualitative information on thermal properties to be obtained in addition to diameter and albedo. 
A detailed thermophysical model is required for quantitative determination of thermal inertia; see \chaptref{chapt:TPM}.


%% file: ThermalOverview.tex
The thermal emission of asteroids contains many important clues about their physical properties; indeed, the study of asteroid thermal emission (often referred to as thermal radiometry) is the dominant source of known diameters and albedos \seesect{sect:intro:diameter} and the only established ground-based means of determining the crucial thermal inertia \seesect{sect:intro:TI}.

The principle of thermal radiometry is simple:
Asteroids are heated up by absorption of sunlight, the absorbed energy is radiated off as thermal emission.
The total emitted thermal radiation at different wavelengths can be calculated by convolving the temperature distribution over the asteroid surface with the temperature-dependent thermal emission of single facets (using, e.g., the Planck black-body law).

While the optical brightness of an asteroid is proportional to its albedo (which can vary between roughly 2 and \unit{60}{\%}; see \sectref{sect:intro:diameter}), its thermal emission is only a weak function of albedo and therefore a much better proxy for size; this approach has been pioneered by \citet{Allen1970}, with important early contribution by, e.g., \citet{Matson1971} and \citet{Morrison1973}.
However, complications arise because  other important physical properties (such as thermal inertia, surface roughness, shape, and spin state, all of which are typically unknown) significantly influence the thermal emission of asteroids.
On one hand this imposes difficulties for the determination of diameters, but on the other hand the thermal flux contains more information than about diameter alone.

Thermal observations of asteroids are hampered by
the fact that typical asteroid temperatures are not too different from those of most objects on Earth, leading to a huge background radiation in the mid-infrared wavelength range in which asteroid thermal emission peaks.
Furthermore, the Earth's atmosphere is mostly opaque in this wavelength range, with the exception of a few ``atmospheric windows;'' this will be further discussed in \sectref{sect:thermal:observability}.

\paragraph{Early days of asteroid thermal studies}
Throughout the 1970s and 1980s, thermal-infrared observations of asteroids, then typically performed at a single thermal wavelength,
proved to be very fruitful for determining asteroid diameters and gave quite good agreement with other techniques,
culminating in the advent of the IRAS Minor Planet Survey \citep{IRAS, SIMPS} which provided thermal measurements of about two thousand  asteroids and resulted in the largest currently available catalog of asteroid diameters and albedos.

The most widely used thermal model was the Standard Thermal Model (STM) discussed in \sectref{sect:thermal:STM-FRM}.
It was developed to determine the diameters of large, bright MBAs from single-wavelength observations at low phase angle (on which observations had to focus for reasons of instrument sensitivity).
The STM is based on a spherical shape;  observations are assumed to take place at opposition, thermal inertia is neglected. This fixes the temperature distribution on the asteroid surface and hence the \emph{color temperature}%
\footnote{ The color temperature is determined from the spectral distribution of thermally emitted flux. By virtue of Wien's displacement law, the thermal emission of colder bodies is skewed towards larger wavelengths compared to hotter bodies. \label{foot:colortemperature}}
The generally good agreement of STM-derived diameter estimates for large, bright MBAs with estimates determined using other techniques (e.g.\ through stellar occultations or polarimetry) provided indirect evidence for a low thermal inertia of these objects, consistent with their apparent regolith cover \seesect{sect:intro:regolith}.

The first asteroid for which the STM diameter differed significantly from diameters obtained using other techniques was the NEA (1580) Betulia \citep[][see also \ref{sect:Betulia}]{LebofskyBetulia} with an estimated  diameter (as of 1978) around \unit{7}{\kilo\metre}.
The apparent discrepancy could be resolved by using a different thermal model, the Fast-Rotating Model (FRM; see \sectref{sect:FRM}), in which effectively an infinite thermal inertia is assumed, leading to a different color temperature.
\citet{LebofskyBetulia} concluded that Betulia was regolith free, perfectly consistent with the ideas about regolith retention prevailing at that time.

\citet{Veeder1989} report \unit{10}{\micron} observations of 22 NEAs. They derive STM and FRM diameters, with a typical discrepancy around \unit{30}{\%}. Results obtained using other techniques favor STM results in some cases and FRM results in others, leading to significant systematic uncertainties for the remaining targets.
It must be emphasized that observations at a  single thermal wavelength provide no information on the color temperature. Hence they do not allow one to discriminate between concurrent thermal models on the basis of thermal data alone, whereas multi-wavelength observations do.

\paragraph{Modern NEA observations}
Thanks to advances in detector technology, multi-wavelength thermal-infrared spectrophotometry of asteroids is now quite possible, even for small NEAs, given favorable circumstances.

In general, neither the STM nor the FRM provide a good fit to the thermal spectrum of an NEA.
Typically,  better fits can be reached by using the 
Near-Earth Asteroid Thermal Model (NEATM; see \sectref{sect:NEATM} for a detailed discussion).
In contrast to simpler models, the NEATM does not \emph{assume} an apparent color temperature but is used to \emph{derive} the color temperature; this requires
 observations obtained at two or more thermal wavelengths.
The NEATM can be used to determine NEA diameters, which were previously prone to significant systematic uncertainties,  to an accuracy typically within \unit{15}{\%} \seesect{sect:NEATM:accuracy}. Furthermore, conclusions on the thermal inertia can be drawn from the NEATM fit parameter $\eta$  \seesect{sect:NEATM:eta}.

\citet{Delbo2003} performed thermal-IR observations of a large sample of small NEAs around \unit{1}{\kilo\metre} in diameter using the \unit{10}{\metre} Keck-2 telescope and determined the sizes and albedos of their targets. While \citet{Delbo2003} could not make quantitative statements about the typical thermal inertia of their targets from a NEATM analysis alone, they could exclude a large thermal inertia indicative of bare rock. This appears to be incompatible with the afore-mentioned findings by \citet{Lebofsky1979}, who claimed their NEA targets to be predominantly rocky.
%

\paragraph{Thermophysical modeling}
For reliable determinations of the thermal inertia, a detailed thermophysical model (TPM) is required in which the effect of thermal inertia is explicitly taken into account.
Additionally, TPM-derived diameter and albedo estimates promise to be more accurate than those derived using highly idealized ``simple'' thermal models such as those alluded to above.

Meaningful application of a TPM, however, requires information on the object's shape and spin axis orientation \seesect{sect:thermal:TI}, which is often not available \seesect{sect:intro:shape}.
Moreover, a large set of high-quality thermal-infrared data is typically required in order to constrain the inevitably larger number of fit parameters in a meaningful way.
For these reasons, TPM-based research has so far focused on bright MBAs, for which the required observational data are more readily available.

The ongoing technological progress now enables high-quality thermal-infrared observations of faint asteroids including NEAs. Furthermore, the number of NEAs with well-determined shape and spin state is growing rapidly.
This thesis contains a description of the first TPM shown to be applicable to NEAs \seechapt{chapt:TPM}.

%% file: ThermalProps.tex
The thermal emission of an asteroid is determined from the temperature distribution on its surface convolved with the temperature-dependent emission of the surface elements.
In practice, the most relevant parameters are the easiest to model: the object diameter $D$ (see below for a diameter definition for non-spherical objects), the heliocentric distance $r$, and the observer-centric distance $\Delta$: Fluxes are proportional to $\left(D/\Delta\right)^2$, temperatures are proportional to $r^{-2}$.
The apparent color temperature is determined from the physical temperature distribution, which also affects the absolute flux level (Stefan-Boltzmann law).

Among the parameters that determine the temperature are:%
\footnote{ This chapter contains a qualitative discussion of these parameters, see \chaptref{chapt:TPM} for a quantitative discussion.}
the albedo \seesect{sect:thermal:size-albedo}; thermal inertia \seesect{sect:thermal:TI}; surface roughness \seesect{sect:thermal:beaming}; and shape and spin state \seesect{sect:thermal:shape}. 

Observable fluxes also depend on the observation geometry \seesect{sect:thermal:obsgeometry}, chiefly on the solar phase angle, $\alpha$; they also depend on the temperature-dependent spectral characteristics of the thermal emission \seesect{sect:thermal:emission}.

\begin{figure}
\centering
   \includegraphics[angle=-90, width=0.6\linewidth]{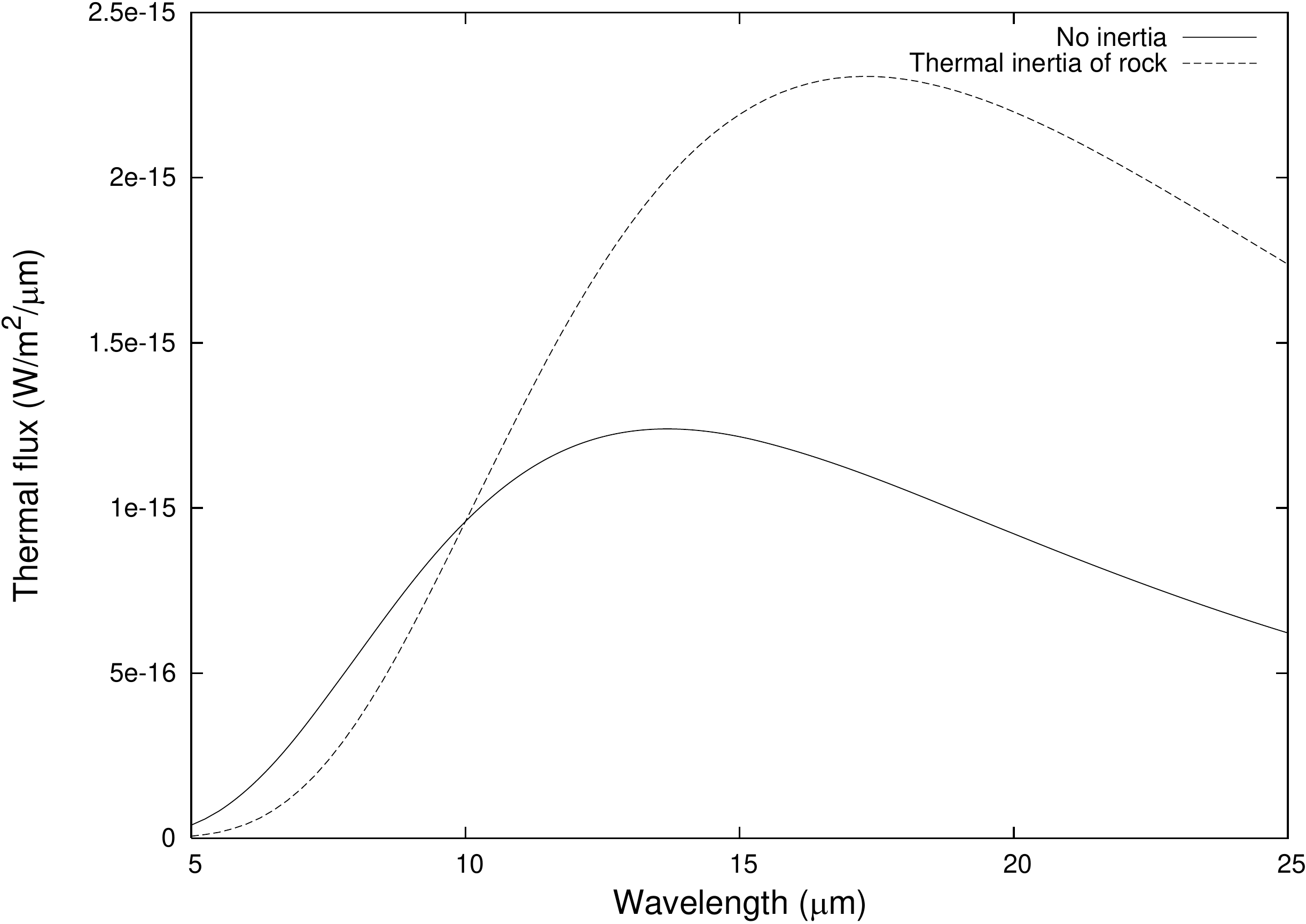}
\caption[Thermal emission of two model asteroids, one without thermal inertia, one with that of bare rock]{Thermal emission of two  spherical model asteroids, one with a diameter of \unit{100}{\kilo\metre} and vanishing thermal inertia (solid line), one with a diameter around \unit{237}{\kilo\metre} and the thermal inertia of bare rock (dotted line).
They cannot be distinguished through single-wavelength observations made at \unit{10}{\micron}, whereas observations at a second wavelength (e.g.\ \unit{20}{\micron}) enable the ambiguity to be resolved. Model fluxes were calculated using the model code described in \chaptref{chapt:TPM} for observations at opposition ($\alpha = 0$), with heliocentric distance of \unit{3}{\AU} and observer-centric distance of \unit{2}{\AU}. The spin axis is perpendicular to the line of sight, the spin period equals \unit{6}{\hour}. No surface roughness was assumed.}
\label{fig:thermal:singlewavelength}
\end{figure}

Observations at a single thermal wavelength contain no information on the color temperature  \seefig{fig:thermal:singlewavelength}, leading to significant diameter uncertainties in the interpretation of such measurements. Measurements at two or more thermal wavelengths combined with a suitable thermal model allow a cold and large asteroid to be distinguished from a hot and small object, reducing systematic diameter uncertainties.
Furthermore, the color temperature bears information on the physical parameters which determine the temperature, chiefly the thermal inertia.

\subsection{Size and albedo}
        \label{sect:thermal:size-albedo}

\paragraph{Diameter}
All other parameters being constant, the thermal emission is proportional to the projected area $\Area$
and hence to $D^2$, where $D$ denotes the diameter. 

The ``diameter'' of a non-spherical object is not uniquely defined.
For the reason above, diameters obtained from simple models based on spherical geometry are area-equivalent diameters: $\pi/4\cdot D^2 = \Area$.
This definition is  inconvenient to use when an asteroid shape model is available, since it depends on the observing geometry.
Whenever our thermophysical model (see \chaptref{chapt:TPM}) is used, diameters are defined as  volume-equivalent diameters, i.e.\ that of a sphere with identical volume $V$
\begin{equation}
  \label{eq:Deff}
  \frac{\pi}{6}D^3 = V.
\end{equation}
In practice, the difference among the two definitions is negligible except for extremely elongated shapes.

\paragraph{Albedo}

The amount of solar flux absorbed by an asteroid is proportional to $(1-A)$ with the bolometric Bond albedo $A$.
$A$ is defined as the ratio of reflected or scattered flux over incoming flux, scattering into all directions is considered.
$A$ is therefore restricted to lie between 0 and 1. 
For Solar-System objects, $A_V$ (i.e.\ the Bond albedo in the $V$ band) is a good approximation to $A$.

The geometric albedo \pv\ is 
 defined as the ratio of the visual brightness of an object observed at zero phase angle to that of a perfectly diffusing Lambertian disk of the same projected area and at the same distance as the object. For planetary bodies, \pv\ is more readily measurable than $A$ and a widely quoted parameter. The ratio $q := A_V/\pv$ is called the phase integral.
In the standard HG system \citep{HG}, 
\begin{equation}
  \label{eq:q}
q=0.290 + 0.684\times G 
\end{equation}
with the slope parameter $G$. $G$ can be determined from  optical photometric measurements made at different  phase angles but is often not available; a default value of $G=0.15$ is then typically assumed.
Note that objects with $\pv > 1$, while unusual, are by no means unphysical; highly backscattering objects such as mirrors may have $\pv > 1$, the measured geometric albedos of some Kuiper belt objects exceed unity \citep[][and references therein]{Stansberry2007}.

The amount of sunlight scattered by an asteroid, and hence its optical brightness, is proportional to its albedo and its projected area \seeeqpage{eq:FowlerChillemi}. The absorbed flux, which is later thermally reemitted, is proportional to $1-A$. For typical asteroids, $A$ is much closer to 0 than to 1, hence thermal fluxes do not critically depend on albedo. 
From thermal-emission data, it is therefore possible to determine the diameter nearly independently from the albedo. Combining the diameter result with optical photometric data, it is then possible to determine the albedo.
Note that while this statement holds for nearly all asteroids due to their relatively low $A$, it would be wrong for very-high-albedo objects such  as the Kuiper belt objects alluded to above.

\subsection[Thermal inertia]{Thermal inertia}
        \label{sect:thermal:TI}

\begin{figure}
  \centering
\includegraphics[angle=-90, width=0.75\linewidth]{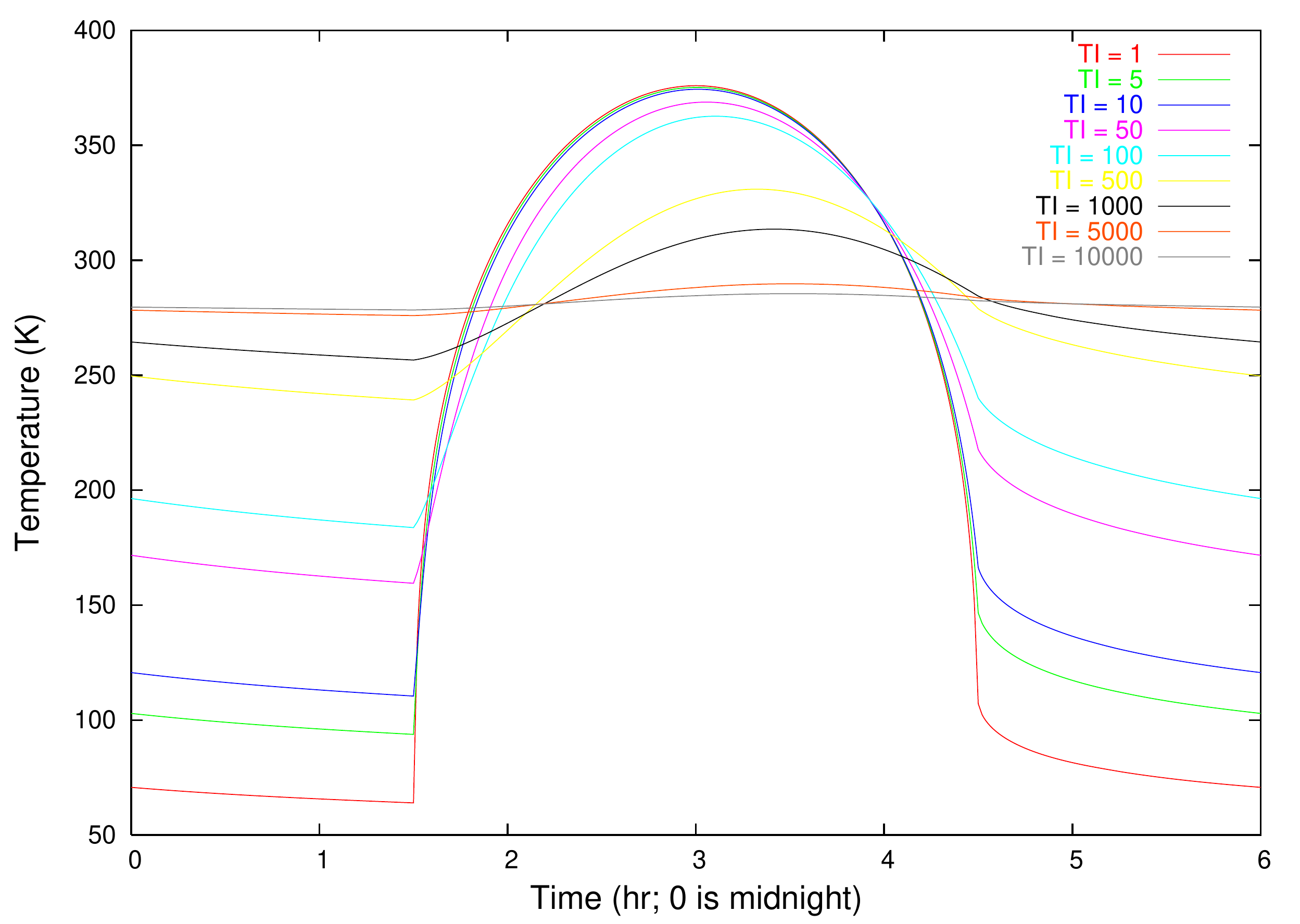}
  \caption[Synthetic diurnal temperature curves for different values of thermal inertia]{Synthetic diurnal temperature curves on the equator of a model asteroid for different values of thermal inertia (in units of \TIunit). Increasing thermal inertia smooths  temperature contrasts and additionally causes the temperature peak to occur after the insolation peak at \unit{3}{\hour}.  The asteroid is situated at a heliocentric distance of $r=\unit{1.1}{\AU}$, has a spin period of \unit{6}{\hour}, a Bond albedo of $A=0.1$, and its spin axis is  perpendicular to the orbital plane.}
  \label{fig:thermal:TI}
\end{figure}

Thermal inertia is a measure of the resistance to changes in surface temperature and is closely related to thermal conductivity.
The surface of an object with zero thermal inertia would be in instantaneous thermal equilibrium with external heat sources; the surface temperature on an asteroid with zero thermal inertia, in particular, would drop to zero immediately after sunset, hence no thermal emission would originate from the non-illuminated hemisphere.
All physical objects have some thermal inertia, such that surface elements require a certain amount of time to heat up or cool down. On planetary objects, this induces a phase lag between insolation and surface temperature.
Night-time temperatures no longer vanish; by virtue of energy conservation day-side temperatures are reduced \seefig{fig:thermal:TI}.

\begin{table}[tb]
\fbox{
\begin{minipage}{0.93\columnwidth} 
\caption{Thermal inertia: Some familiar examples.}
\label{table:thermal:TI}
\paragraph*{Noon}
The hottest time of day is generally \emph{after} noon, 
and the hottest time of the year is generally after the summer solstice
due to thermal inertia.


\paragraph*{Oceanic and continental climate}
The thermal inertia of water greatly exceeds that of soil. 
Consequently, the climate close to  oceans or large lakes is generally mild, with  moderate temperature differences between both day and night or summer and winter.
This is contrasted by continental climate as in, e.g., central Siberia, with hot summers but notoriously extreme winters.

\paragraph*{Earth and Moon}
Although the Moon is at the same heliocentric distance as the Earth, lunar temperatures oscillate between around \unit{100}{\celsius} at day time and some \unit{-150}{\celsius} at night. This is caused by the extremely low thermal inertia of its regolith-dominated surface together with the low spin rate.

\end{minipage}
} 
\end{table}

See \sectref{sect:TPM:TI} for a formal definition of thermal inertia and a mathematical discussion, \tableref{table:thermal:TI} for some familiar examples and \sectref{sect:intro:TI} for a discussion of what is known about asteroid thermal inertia and physical implications thereof.

In asteroid observations at low phase angles (as  typical for MBAs and objects in the outer Solar System), the chief effect of thermal inertia is a reduction of the day-time temperature relative to a low-thermal-inertia object, hence a reduction in absolute flux level and also in apparent color temperature (i.e.\ the observed flux is skewed towards longer wavelengths). The enhanced emission from the night side does not contribute significantly to the observable flux since most parts of the non-illuminated hemisphere are not visible at low phase angles.
The effect of thermal inertia on large-phase-angle observations, such as typical NEA observations, is less straightforward to predict and generally requires careful modeling.

On asteroids, thermal inertia is caused by thermal conduction into and from the subsoil. Large asteroids are well known to be covered with dusty regolith, which is a poor thermal conductor, hence their thermal inertia is very low (see \sectref{sect:TPM:TI} for a discussion).
Generally, neglecting their thermal inertia does  not introduce large systematic diameter uncertainties.

Little is known, however, about the thermal inertia of small asteroids including NEAs \seesect{sect:intro:TI}, therefore one is ill advised to neglect their thermal inertia. Also, NEAs are regularly observed at much larger solar phase angles \seesect{sect:thermal:obsgeometry}, where the effects of thermal inertia become more pronounced.
In the interpretation of thermal NEA data, it is therefore crucial to take thermal inertia into account to avoid significant systematic diameter uncertainties.

The effect of thermal inertia is tightly coupled to the rotational properties:
A slow rotator with high thermal inertia may mimic the diurnal temperature curve of an otherwise identical fast rotator of low thermal inertia (see \eqrefpage{eq:def_thermalparameter}).
Also, the spin axis orientation is important: Obviously, the diurnal temperature distribution on an object whose spin axis points towards the Sun shows no effect of thermal inertia whatsoever; thermal inertia has the most profound influence on the diurnal temperature distribution if the subsolar point is at the equator.

\subsection[Beaming due to surface roughness]{Beaming due to surface roughness}
        \label{sect:thermal:beaming}

By comparing thermal diameters with occultation diameters, \citet{STM} and \citet{LebofskySpencer1989} found what appeared to be a systematic thermal-flux surplus at low phase angles:
Thermal emission is ``beamed'' into the sunward direction, such that at low phase angles a larger-than-expected flux level is observable at an elevated apparent color temperature.
This effect is referred to as \emph{thermal-infrared beaming}.

Like the well-known optical opposition-effect \citep[see, e.g.,][]{BelskayaShevchenko2000}, thermal-infrared beaming is thought to be caused by surface roughness. 
Imagine
a hemispherical crater at the subsolar point, 
where the solar incidence vector coincides with the crater symmetry axis.
Inside the crater, surface elements can exchange energy radiatively leading to \emph{mutual heating} due both to sunlight scattered inside the crater and due to reabsorption of thermal emission.
Relative to an equally-sized, flat surface patch, the crater therefore absorbs a larger amount of energy and thermally emits at an elevated effective temperature.
Furthermore, craters situated off the subsolar point contain surface elements which point towards the Sun (and, at low phase angle, to the observer). This reduces the amount of thermal \emph{limb darkening} relative to a Lambertian emitter.

Due to conservation of energy, one would  expect a reduced color temperature and reduced flux level at larger phase angles. In particular,  beaming would be expected to lead to a phase-angle dependence of the apparent color temperature (see \sectref{sect:NEATM:eta} for a further discussion).

\subsection[Shape and spin state]{Shape and spin state}
        \label{sect:thermal:shape}

While large MBAs generally tend to be nearly spherical, smaller asteroids display a large diversity of shape \seesect{sect:intro:shape}
which, combined with their spin, typically causes their projected visible area  to vary with time.
This induces a time variability in optical brightness (optical lightcurve) and also in  thermal emission (thermal lightcurve), typically with two peaks per asteroid revolution, such that the lightcurve period  equals half the spin period. The lightcurve amplitude  depends on the asteroid shape, on the phase angle of the observations, and also on the aspect angle \citep{Zappala1990}. It may be larger than one magnitude for extremely elongated objects, corresponding to a minimum-to-maximum flux variability of a factor around 2.5.

It is important to take account of the rotational thermal-flux variability when deriving diameters. Failure to do so  not only introduces an unpredictable diameter offset but might also cause the estimated color temperature to be flawed (typically, spectrophotometric observations in different filters are not taken simultaneously).
If the shape of the object is known, a detailed thermophysical model \seechapt{chapt:TPM} can be used to exploit this information. Typically, however, the shape is unknown and only a few thermal data-points are available, typically insufficient to trace the thermal lightcurve.
In that case, thermal data are often ``lightcurve-corrected'' on the basis of optical lightcurve data, which are typically more readily available. 


However, optical and thermal lightcurve may differ in phase and/or structure due to the effects of shape, surface structure, thermal inertia, or albedo variegation.
While this may cause uncertainties for lightcurve correction of thermal flux values to derive diameters using thermal models based on spherical shape, these lightcurve effects can often be exploited to determine the thermal inertia using a thermophysical model. \citet{LebofskyRieke1979}, e.g., observed a phase shift between thermal and optical lightcurve data of (433) Eros which they explained in terms of a temperature lag due to thermal inertia (see \sectref{sect:Eros}); \citet{Lellouch2000} report a similar phase lag in the thermal lightcurve of (134340) Pluto.

\subsection{Observation geometry}
        \label{sect:thermal:obsgeometry}

          

It is clear that observed thermal fluxes depend critically on the observing geometry:
Fluxes scale with $\Delta^{-2}$ (observer-centric distance $\Delta$), the absorbed solar energy scales with $r^{-2}$ (heliocentric distance $r$).
Objects in the outer Solar System are therefore much colder than bodies in near-Earth space and consequently their thermal emission peaks at much longer wavelengths.

\paragraph{Solar elongation}
The practical observability of asteroids is typically determined by their solar elongation, i.e.\ their angular distance from the Sun as seen by the observer, together with their declination.
Objects on the celestial equator  
at solar elongations below \unit{90}{\degree} culminate on the day sky and are consequently difficult to observe from ground
(see \sectref{sect:SST:solarsystem} for solar-elongation constraints on observations with the Spitzer Space Telescope).

\paragraph{Solar phase angle}
Closely related to the solar elongation is the solar phase angle, $\alpha$, which is of great importance for thermal modeling. 
Distant objects such as MBAs typically reach their peak brightness at opposition, when $\alpha\sim 0$ and the solar elongation is maximized (typically close to \unit{180}{\degree})---this is different for near-Earth objects with reach their peak brightness around the date of closest approach.

Objects at large observer-centric distances can only be observed at relatively small phase angles 
whereas NEAs often sweep large ranges of phase angle within a few weeks during close approaches to Earth.

At low phase angles, the observed thermal flux is vastly dominated by the hot subsolar region; thermal inertia reduces both the observed flux level and the color temperature  by transporting heat to the night side, from where it is not observable. For low thermal inertia, the temperature distribution is nearly symmetric about the subsolar point, which nearly coincides with the sub-observer point; due to this approximate symmetry, radiometric diameters are largely insensitive to the  temperature distribution.

At large phase angles, however, large portions of the non-illuminated side become observable, rendering the observable thermal emission more sensitive to the details of the temperature distribution. The latter is determined by shape, thermal inertia, spin state, and surface roughness.

At low phase angles, the cooling effect of thermal inertia may be countered by the beaming effect due to surface roughness, which increases the apparent color temperature. 
At larger phase angles, however, also beaming is expected to lead to a cooling effect, making it difficult to disentangle the two effects.

Accurate derivation of thermal properties such as thermal inertia hence typically requires observations at several phase angles.

\paragraph{Aspect angles}
Also important are the aspect angles, chiefly the subsolar and sub-observer latitude on the asteroid. These depend on the asteroid spin axis, which is often unknown. 
Lightcurve effects and the effect of thermal inertia are maximized when the spin axis is perpendicular to the viewing plane (i.e.\ when both the subsolar and sub-observer point are on the equator), both effects are minimized if the asteroid is viewed pole-on.

\subsection{Thermal emission model}
        \label{sect:thermal:emission}

The temperature-dependent spectral characteristics of asteroid thermal emission is typically modeled using a gray-body law, i.e.\  a Planck black body with a spectrally constant  emissivity $\epsilon$.
The latter assumption is only approximately valid. 
There are well-known spectral features in the thermal infrared; those with the largest spectral contrast are due to silicates and located at wavelengths of 8--10 and \unit{15--25}{\micron}.
Not many thermal-infrared spectroscopic observations of asteroids have been published \citep[see][and references therein---see also \sectref{sect:Patroclus}]{Lim2005,Emery2006}, but the typical spectral contrast of detected features is only a few percent if spectra were detectable at all. 

While the emissivity is roughly spectrally constant over the relevant wavelength range, the exact value of the bolometric emissivity is less well constrained.
As is common practice, we assume $\epsilon = 0.9$, which is a typical value for silicate powders known from laboratory measurements \citep[see, e.g.,][]{HovisCallahan1966}.
Bolometric emissivities cannot exceed 1, and for most common materials, $\epsilon$ is within \unit{10}{\%} of 0.9; a notable  exception are polished metal surfaces which can have emissivities down to a few percent \citep[see, e.g.,][Tab.\ 29]{Berber1999}.
To first order, asteroid thermal fluxes are proportional to $\epsilon D^2$, hence the emissivity uncertainty induces a fractional diameter uncertainty of \unit{5}{\%} at most---except for polished metallic objects, a very implausible asteroid surface model.


%% file: Observability.tex

\begin{figure}
  \centering
\includegraphics[angle=-90, width=0.6\linewidth]{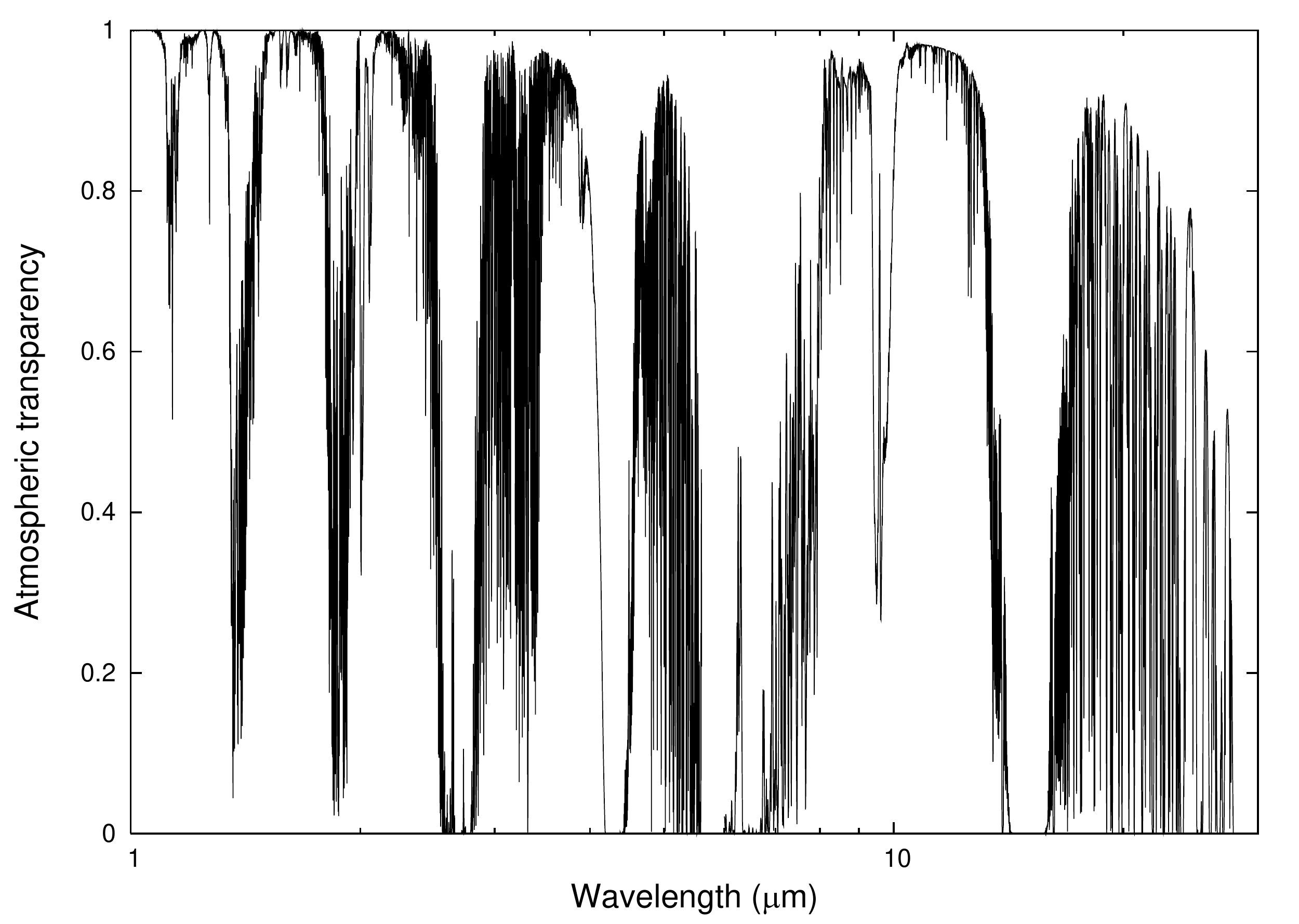}
\caption[Atmospheric transmissivity as a function of wavelength]{Atmospheric transmissivity as a function of wavelength over the summit of Mauna Kea/\Hawaii, one of the best sites for infrared observations. The clearly visible transmissive regions are called ``atmospheric windows'' and are named (with increasing wavelength) J, H, K, L, M, N, and Q; thermal-infrared windows are the M window ($\sim\unit{5}{\micron}$), the N window ($\sim\unit{10}{\micron}$), and the Q window ($\sim\unit{20}{\micron}$). The atmospheric transparency model was developed by \citet{Lord1992}, data were  made available through the GEMINI website \url{http://www.gemini.edu/sciops/ObsProcess/obsConstraints/ocTransSpectra.html} (assuming \unit{1}{\milli\metre} of precipitable water and an airmass of 1.5).}
\label{fig:thermal:MK}
\end{figure}


The thermal emission of asteroids peaks in the mid-infrared wavelength range (also referred to as \emph{thermal infrared}), typically between 10 and \unit{20}{\micron}.
The thermal emission of outer-Solar-System objects peaks at larger wavelengths, beyond \unit{50}{\micron} in the case of  Kuiper belt objects.

The atmosphere is mostly opaque in the thermal-infrared wavelength range and practically totally opaque at larger infrared wavelengths, chiefly due to absorption from \cotwo\ and \water.
Ground-based observations of the thermal emission of outer-Solar-System bodies are therefore virtually impossible, while asteroid observations  are restricted to ``atmospheric windows''  \seefig{fig:thermal:MK}.

Further problems stem from the fact that thermal emission of, e.g., the atmosphere, clouds, or even the telescope mirrors cause high levels of rapidly varying background radiation;
thermal-infrared detectors are typically cooled with liquid helium in order to minimize their own thermal emission.
The large  background level makes special observation techniques necessary, such as those discussed in \sectref{sect:IRTF:strategy}.

In the thermal infrared, it is therefore particularly advantageous to observe from a vantage point above most of the atmosphere (e.g.\ at the summit of a high mountain or in an airborne telescope) or above all of the atmosphere using a space telescope.
The currently most sensitive imaging instruments in the thermal infrared  are on board the Spitzer Space Telescope, although its  aperture of  \unit{85}{\centi\metre} is much smaller than that of, e.g., the \unit{10}{\metre} Keck-2 telescope.

A widely used unit of mid-infrared flux (monochromatic flux density) is \watt\usk\metre\rpsquared\reciprocal\micron, also widely used is the  Jansky (\Jy); \unit{1}{\Jy} equals \unit{$10^{-26}$}{\watt\usk\power{\metre}{-2}\power{\hertz}{-1}}. Fluxes are converted from one unit into the other as follows:
\begin{equation}
  \label{eq:Jy}
  F_{\Jy}\left(\lambda\right) = F_{\watt\rpsquare\metre\reciprocal\micron}\left(\lambda\right) \times \lambda_{\micron}{}^2 \times 0.33356 \times 10^{12}
\end{equation}
where the latter factor contains the reciprocal of the speed of light required to convert from flux per wavelength to flux per frequency.

%% file: simplemodels.tex
%
%

In this section, two widely used, yet highly idealized, thermal models are described: 
The Standard Thermal Model (STM, see \sectref{sect:STM}), which neglects the combined effect of rotation and thermal inertia, and the Fast Rotating Model (FRM, see \sectref{sect:FRM}), which effectively assumes an infinitely large thermal inertia.

Both models were developed in the 1970s, 
when thermal-infrared observations of asteroids were effectively limited to a single wavelength.
The color temperature is  fixed by the respective model assumptions, hence data at a single thermal wavelength are sufficient to estimate the diameter.


\subsection{Standard Thermal Model (STM)}
\label{sect:STM}

In the Standard Thermal Model \citep[STM, see][and references therein]{STM}, the asteroid is assumed to be spherical, to have a vanishing thermal inertia (hence its spin state is irrelevant), and to be observed at opposition, i.e.\ at a phase angle of \unit{0}{\degree}. 
Under these assumptions, conservation of energy determines the temperature at the subsolar point \TSS\ of a smooth asteroid:
\begin{equation}
  \label{eq:TSS:pre_STM}
\epsilon\sigma  \TSS^4 = \left(1-A\right) \frac{S}{r^2}
\end{equation}
where $\epsilon$ denotes the bolometric emissivity, $\sigma$ the Stefan-Boltzmann constant, $A$ the bolometric Bond albedo \seesect{sect:thermal:size-albedo}, $S$ the solar constant, and $r$ is the heliocentric distance in \AU.
In the absence of thermal inertia, temperatures are in instantaneous equilibrium with insolation, and hence the temperature distribution on the surface solely depends on the angular distance from the subsolar point (or, equivalently, the angle formed by the solar incidence vector and local zenith) $\Phi$:
\begin{equation}
  \label{eq:STM:Tphi}
  T\left(\Phi\right) =
  \begin{cases}
    \TSS \cos^{\frac{1}{4}} \Phi & \text{if}\ \Phi \leq \unit{90}{\degree} \\
    0         & \text{otherwise (i.e.\ Sun is below local horizon)}
  \end{cases}
\end{equation}
Using the Planck function
\begin{equation}
  \label{eq:Planck}
  B\left(\lambda, T\right) = \frac{2\pi h c^2}{\lambda^5} \frac{1}{\exp\left[h c /\left(\lambda k T \right)\right] - 1 }
\end{equation}
with the Planck constant $h$, velocity of light $c$, and Boltzmann constant $k$, the total flux $f(\lambda)$ at wavelength $\lambda$ then equals
\begin{equation}
  \label{eq:STM:flux}
  f(\lambda) = \frac{\epsilon D^2}{2\Delta^2} \int\limits_0^{\pi/2} B\left(\lambda, T\left(\Phi\right)\right) \sin\Phi\cos\Phi \textd\Phi.
\end{equation}
The symmetry about the subsolar point renders the azimuthal integral trivial, leaving only a one-dimensional integral to be performed numerically.

The STM assumes observations to take place at opposition, whereas real observations typically occur at $\alpha > 0$, requiring a phase-angle correction.
\citet{LebofskySpencer1989} employ an empirical phase coefficient of \unit{0.01}{\text{mag}\per \text{degree}}, which was found by \citet{Matson1971} to be a good approximation to the phase curve of asteroids observed at N-band wavelengths and at phase angles up to \unit{30}{\degree}.
It must be emphasized that the STM is not applicable at larger phase angles. 

\citet{STM} and \citet{LebofskySpencer1989} found that diameters estimated using this ``naive'' STM were systematically larger than estimates derived using other techniques, which they attributed to thermal-infrared beaming \seesect{sect:thermal:beaming}. As a first-order correction, the so-called \emph{beaming parameter $\eta$} was introduced into the energy balance (\eqref{eq:TSS:pre_STM})%
\footnote{ They effectively follow  \citet{Jones1974}.}
\begin{equation}
  \label{eq:TSS:STM}
\epsilon\sigma\boldsymbol{\eta}  \TSS^4 = \left(1-A\right) \frac{S}{r^2}.
\end{equation}
$\eta < 1$ enhances the model temperature and thus the expected flux level (thereby reducing model diameters required to match measured fluxes) while $\eta > 1$ reduces both the temperature and flux level.
By comparing occultation diameters of the few largest MBAs to radiometric diameters, they determined a best-fit ``canonical'' value of $\eta = 0.756$.
The thus modified STM is widely used, and frequently referred to as the ``refined'' STM to distinguish it from the case $\eta=1$.

The STM was designed to interpret single-wavelength measurements; there is only one free parameter, namely the diameter $D$ (the albedo $A$ which appears in \eqref{eq:TSS:STM} is linked to $D$ through the optical magnitude $H$, see \sectref{sect:intro:diameter}). When, however, observations at more than one wavelength are available, it is common practice to derive one diameter value per data point and to compare the diameters. They are then often referred to by the used wavelength, e.g.\ as ``N-band diameter'' or ``Q-band diameter.''

For large MBAs, it was found that STM-derived diameters typically agree well with estimates derived using other techniques. Much less is known about smaller asteroids including NEAs.
The STM was used to derive the largest currently available catalog of asteroid diameters and albedos \citep{IRAS, SIMPS} from data obtained with the InfraRed Astronomy Satellite (IRAS).

\subsection{Fast Rotating Model (FRM)}
\label{sect:FRM}

An alternative, equally simple model was devised by \citet{LebofskyBetulia}, called the Fast Rotating Model (FRM) or Isothermal Latitude Model (ILM).
In this model, the diurnal temperature distribution is constant for regions of constant geographic latitude. This corresponds to the assumption of infinitely fast rotation about a spin axis perpendicular to the observing plane spanned by the Sun, the observer and the asteroid; the often-made assertion that the FRM assumes an infinite thermal inertia is not strictly valid since it neglects lateral heat conduction, which  would cause the asteroid to become isothermal.
Nevertheless, the FRM should be more appropriate for high-thermal-inertia asteroids than the STM.

The FRM was developed in the late 1970s to explain the discrepancy in different diameter estimates for the NEA (1580) Betulia, for which the STM diameter was found to be much lower than estimates resulting from radar and polarimetric observations \citep[][see also \sectref{sect:Betulia}]{LebofskyBetulia}. Using the FRM rather than the STM resolved the apparent discrepancy, hence it was concluded that Betulia had a very high thermal inertia consistent with a surface of bare rock.


Under the FRM assumptions, a strip at geographic latitude $\theta$ (width $D/2\ \textd\theta$) is in thermal equilibrium with the absorbed sunlight averaged over one rotation:
\begin{equation}
\label{eq:energybalance_FRM}
\left(\pi \frac{D^2}{2} \cos\theta\textd\theta\right)\epsilon\sigma T^4
=
\frac{(1-A)S}{r^2} \cos\theta \left(\frac{D^2}{2} \cos\theta\textd\theta\right)
\end{equation}
(energy is emitted from a total area of $\pi D^2/2\ \cos\theta\textd\theta$ and absorbed on a total projected area of $D^2/2\ \cos\theta\textd\theta$, the second factor of $\cos\theta$ on the right-hand side of \eqref{eq:energybalance_FRM} is for the solar incidence angle).
The temperature distribution equals
\begin{align}
\label{eq:FRM:T}
  T\left(\theta\right) &= \TSS\cos^{\frac{1}{4}}\theta \\
\label{eq:FRM:TSS}
  \TSS                 &= \left(\frac{S}{r^2} \frac{1-A}{\pi\epsilon\sigma} \right)^\frac{1}{4}.
\end{align}
Note that  \eqref{eq:FRM:TSS} formally corresponds to \eqref{eq:TSS:STM} with $\eta=\pi$;  thermal inertia carries energy from the day side towards the night side, hence the former is much cooler than in the STM case.
The total model flux equals
\begin{equation}
  \label{eq:FRM:flux}
  f(\lambda) = \frac{\epsilon D^2}{\pi\Delta^2} \int\limits_0^{\pi/2} B\left[\lambda, T\left(\theta\right)\right] \cos^2\theta \textd{\theta}.
\end{equation}

Due to the rotational symmetry of the temperature distribution, FRM fluxes do not depend on solar phase angle, hence no phase-angle correction is required. One might therefore expect the FRM to be a more appropriate model than the STM for observations of high-thermal-inertia objects at large phase angles.


%% file: neatm.tex
In the STM and the FRM described in \ref{sect:thermal:STM-FRM}, the temperature distribution on the asteroid surface as well as the maximum temperature are completely determined by the model, where very different assumptions are made on thermal properties.
In general, neither model provides a good fit to the measured spectral emission properties of small asteroids such as NEAs. 
\citet{NEATM} proposed a modified variant of the STM, the Near-Earth Asteroid Thermal Model (NEATM),
in which the model temperature scale is adjusted to fit the observed data, enabling a first-order correction to effects of thermal inertia, surface roughness, or shape.
This requires spectrophotometric data at two or more thermal wavelengths and enables some information on thermal properties to be obtained \seesect{sect:NEATM:eta} in addition to estimates of diameter and albedo which are generally more accurate than those based on simpler thermal models such as the STM or FRM
(this will be discussed in \sectref{sect:NEATM:accuracy}).
In \sectref{sect:NEATM:fitting} we describe our fitting routine, which is used \emph{mutatis mutandis} also for thermophysical model fitting.

Despite its name, the NEATM is suitable for application to thermal data from any atmosphereless body, not only NEAs.
In this work, thermal-infrared data of various asteroids are fitting using the NEATM (most data are also fitted using the thermophysical model described in \chaptref{chapt:TPM}).



                \subsection{Model description}
                        \label{sect:NEATM:description}

Like the STM, on which it is based, the NEATM assumes a spherical asteroid shape with an STM-like temperature distribution (\eqref{eq:STM:Tphi} and \eqref{eq:TSS:STM}). In contrast to the STM, however, the parameter $\eta$ no longer has a fixed value. Rather, $\eta$ is varied in order to match the spectral distribution of the observed data, i.e.\ the apparent color temperature; the physical significance thereof will be discussed in \sectref{sect:NEATM:eta}.
Furthermore, in contrast to the STM, where observations are assumed to take place at opposition and observed data are phase-angle corrected using an empirical phase coefficient, NEATM fluxes are calculated at the phase angle $\alpha$ at which the observations took place assuming Lambertian emission.
To this end, a two-dimensional integral is performed over that part of the sphere which is illuminated and visible to the observer. In a spherical coordinate system with the subsolar and sub-observer points at the equator ($\theta=0$) and azimuth angle $\phi=0$ at the subsolar point (the sub-observer point is at $\phi=\alpha$):
\begin{equation}
  \label{eq:NEATM:flux}
  f\left(\lambda\right) = 
\frac{\epsilon D^2}{\Delta^2} 
\int\limits_0^{\pi/2} \textd\theta 
\int\limits_{-\frac{\pi}{2}+\alpha}^{\frac{\pi}{2}+\alpha} \textd\phi\
 B \left[\lambda, T\left(\theta,\phi\right)\right]\
\cos^2\theta\ \cos\left(\phi-\alpha\right)
\end{equation}
(the factor $\cos\theta\cos(\phi-\alpha)$ is the cosine of the observer angle, a second factor $\cos\theta$ stems from the surface element $\textd A = \cos\theta \textd\theta \textd\phi$)
with
\begin{equation}
  \label{eq:NEATM:T}
  T(\theta,\phi) = 
  \begin{cases}
     \TSS \cos^\frac{1}{4}\theta \cos^\frac{1}{4}\phi & \text{for} \cos\theta\cos\phi\geq0 \\
     0 & \text{otherwise (Sun below local horizon)}
  \end{cases}
\end{equation}
and (see \eqref{eq:TSS:STM})
\begin{equation}
  \label{eq:NEATM:eta}
  \epsilon\sigma\eta\ \TSS^4 = \frac{(1-A)S}{r^2}.
\end{equation}

The NEATM contains two free parameters to fit the data: the diameter $D$ (from which the bolometric Bond albedo $A$ is calculated) and the model parameter $\eta$. Thermal measurements $f_i(\lambda_i)$ at two or more well spaced thermal wavelengths are required for meaningful NEATM fits. 

                \subsection[Physical significance of $\eta$]{Physical significance of $\boldsymbol{\eta}$}
                        \label{sect:NEATM:eta}

\begin{figure}
  \centering
\includegraphics[angle=-90,width=0.6\linewidth]{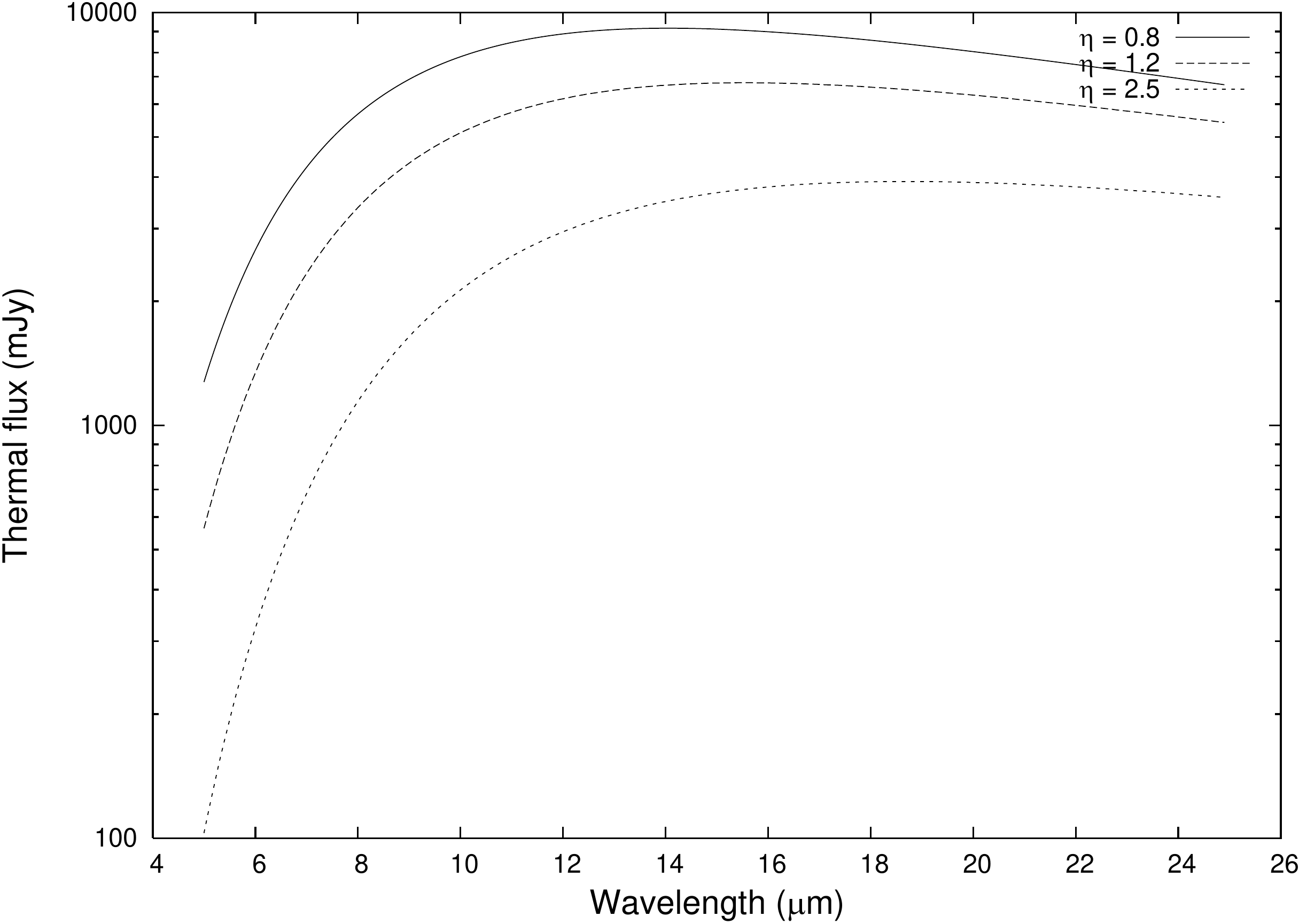}
  \caption[NEATM spectra for three different $\eta$ values]{NEATM spectra of a model asteroid with $\eta=0.8$, $\eta=1.2$, and $\eta=2.5$. Remaining model parameters are: $H=16$, $G=0.15$, $\pv=0.2$, $r$ = \unit{1.1}{\AU}, $\Delta=\unit{0.1}{\AU}$,  $\alpha=\unit{20}{\degree}$, and $\epsilon=0.9$.}
  \label{fig:NEATM:spectra}
\end{figure}

The value of $\eta$ scales the subsolar temperature \TSS\ (\eqref{eq:NEATM:eta}) and hence the temperature throughout the surface (\eqref{eq:NEATM:T}), where $\eta=1$ corresponds to a smooth zero-thermal-inertia (Lambertian) surface, 
$\eta<1$ corresponds to a globally elevated temperature, while $\eta>1$ corresponds to a reduced temperature.
Varying $\eta$ therefore allows the model spectrum to be fitted to the spectral distribution (apparent color temperature) of the observed data
(see \ref{fig:NEATM:spectra} for a schematic example).
No such mechanism to match the observed color temperature is available in simpler models such as the STM or FRM, hence NEATM-derived diameters may be expected to be more accurate \seesect{sect:NEATM:accuracy}.
In particular, NEATM spectra with low $\eta$ values around the ``canonical'' STM value of $\eta=0.756$ mimic STM spectra (the agreement is identical at $\alpha=0$), while NEATM values for large $\eta$ values above $\sim 2.5$ tend towards FRM spectra \seefig{fig:NEATM:spectra}; in a way, therefore, the NEATM interpolates between those two extreme models.

$\eta$ is a measure of the apparent color temperature, from which conclusions can be drawn on the physical temperature and hence on thermal properties.
As described in \sectref{sect:thermal:obsgeometry}, thermal inertia and surface roughness  alter the apparent color temperature, depending on the solar phase angle $\alpha$ at which the observations took place. 
One may therefore expect  $\eta$ determinations at different phase angles to contain information about thermal inertia and thermal-infrared beaming.

\begin{figure}
  \centering
\includegraphics[angle=-90,width=0.75\linewidth]{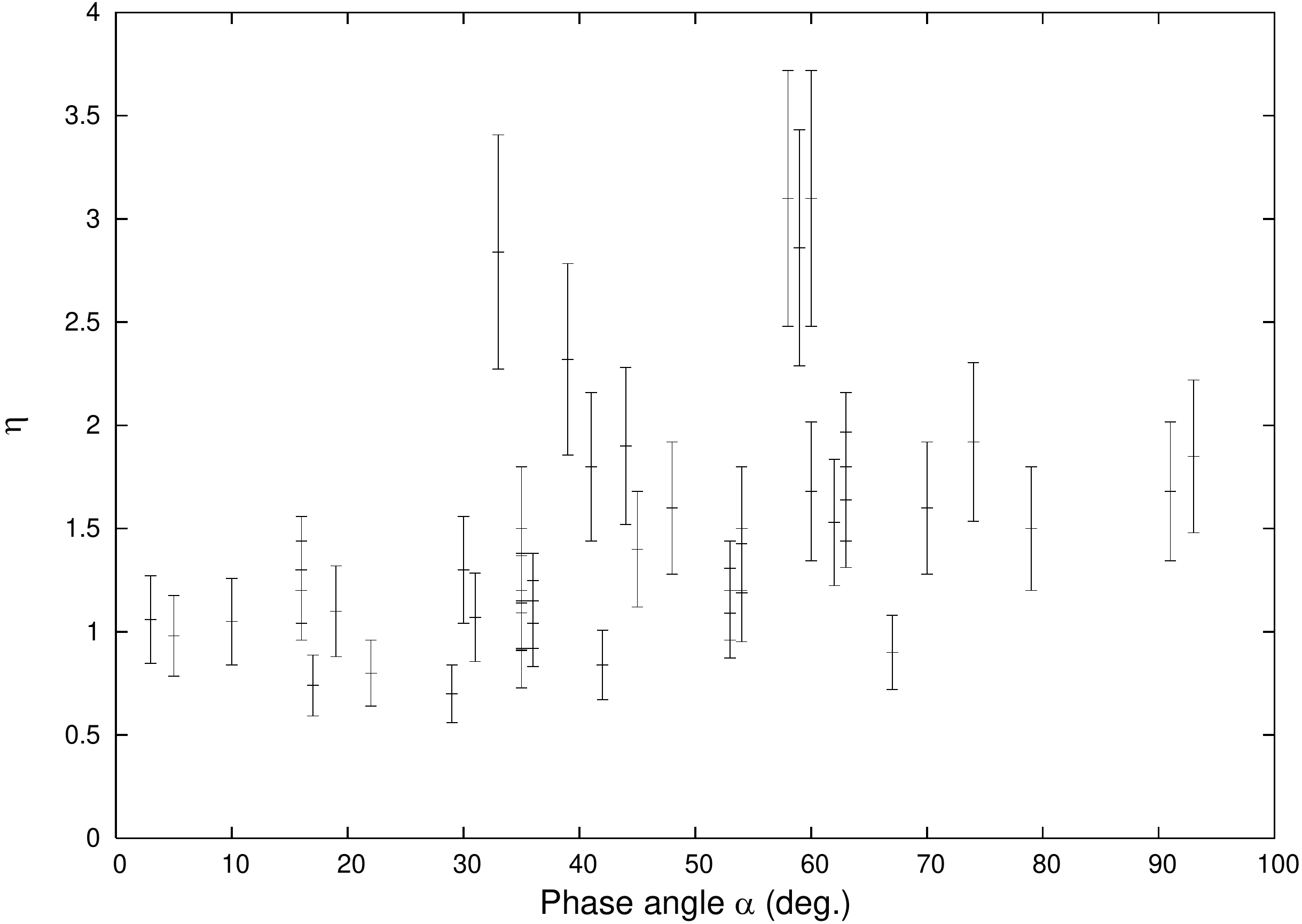}
  \caption[Observed NEATM model parameter $\eta$ as a function of phase angle $\alpha$ for an ensemble of NEAs.]{Observed NEATM model parameter $\eta$ as a function of phase angle $\alpha$ for an ensemble of NEAs. Most asteroids have been observed only once, but there are, e.g., 5 data points for (5381) Sekhmet. 
\citep[Data compiled in][see references therein]{Delbo2007}}
  \label{fig:NEATM:etas}
\end{figure}

\citet{Delbo2007} compiled $\eta$ values from all currently published multi-wave\-length spectrophotometric NEA observations of sufficient data quality \seefig{fig:NEATM:etas}.
There is a clear correlation between $\eta$ and $\alpha$, which is probably due to infrared beaming. The conspicuous lack of large $\eta$ values at low phase indicates that most NEAs in this ensemble do not have a large thermal inertia indicative of a bare-rock surface \citep{Delbo2003}, with the four (or five) objects displaying the largest $\eta$ values possibly being exceptions. 
An ensemble-average thermal inertia of these NEAs was determined by fitting the distribution of $\eta$ values using a detailed thermophysical model similar to that described in \chaptref{chapt:TPM} \citep{Delbo2007}.


                \subsection{Accuracy of NEATM-derived diameters and albedos}
                        \label{sect:NEATM:accuracy}

The systematic uncertainty about the validity of the model assumptions inherent to the NEATM translate into systematic uncertainties of determined diameters and albedos, where the systematic fractional albedo uncertainty is twice that of the fractional diameter uncertainty by virtue of \eqref{eq:FowlerChillemi}.
In many practical cases, other sources of uncertainty such as statistical and systematic flux uncertainties or uncertainties caused by the lightcurve-correction procedure (or lack thereof) can be neglected relative to the systematic diameter uncertainty. Additionally, the uncertainty in absolute optical magnitude $H$ contributes towards the albedo uncertainty.

In the case of large MBAs, STM-derived diameters are known to be in good agreement with occultation diameters, with a relative deviation of \unit{11}{\%} for a sample of 12 MBAs of low lightcurve amplitude \citep[where the two diameter estimates can be compared readily; see][and references therein]{HarrisLagerros2002}.
At the small phase angles at which MBAs are observed, the NEATM and the STM produce generally very similar model predictions provided that the best-fit $\eta$ value is not large \cite[which is generally the case for large objects, see][]{Walker2003}.

        \subsubsection{Comparison of NEATM diameters and albedos of NEAs with other results}
                \label{sect:NEATM:comparison-data}

In the case of small asteroids including NEAs, 
assessment of the systematic diameter and albedo accuracy is hampered by the scarcity of 
 ``ground truth'' data for comparison.
Since the work of \citet{Veeder1989} it is known that there is significant systematic uncertainty in NEA diameters if only the STM and FRM are considered.
Multi-wavelength spectrophotometry analyzed using the NEATM effectively allows one to interpolate between those two extreme models but is still sufficiently simple to be applicable in the absence of detailed \emph{a priori} information on the target (given which more detailed modeling is preferable).

\citet{HarrisLagerros2002} quote STM, FRM, and NEATM diameters and albedos  for a sample of 20 NEAs and conclude that generally NEATM results are in better agreement with albedos expected on the basis of taxonomic classification  \seesect{sect:intro:mineralogy} or spacecraft imaging (in the case of 433 Eros).

While no occultation diameters of NEAs have been published,
radar measurements can provide very accurate estimates of NEA diameters.%
\footnote{ However, it must be kept in mind that radar does not necessarily provide a more direct diameter estimate than thermal data. 
See also \sectref{sect:NEA:D}.}
\citet[pp.\ 100--104 in chapt.\ 5]{Delbo2004} compare radiometric diameters with radar-derived diameters for all NEA data available at that time.
He finds  the rms.\ deviation between NEATM and radar diameters of the considered NEAs to be around \unit{20}{\%}.

It is often quoted \citep[see, e.g.,][]{WT24} that comparison of NEATM results with results from other sources, such as radar, indicate that the overall systematic uncertainty is less than \unit{15}{\%} in diameter and \unit{30}{\%} in albedo, but uncertainties may be larger in the case of very elongated or highly irregularly shaped objects and/or for observations at very large phase angles when night-side emission (which is neglected in the NEATM) becomes more relevant.

        \subsubsection{Mutual comparison of thermal and thermophysical models}
                \label{sect:NEATM:comparison-mutual}

It is instructive to check the NEATM against other thermal models, particularly against detailed thermophysical models. Although it must be cautioned that also the latter carry systematic uncertainties which are currently not well explored, such studies may reveal systematic offsets of NEATM diameters as a function of, e.g., thermal inertia or phase angle of observations.

\citet{Harris2006} used NEATM to fit synthetic thermal flux values, which were generated using a thermophysical model including the effects of thermal inertia but without surface roughness. He showed that for $\alpha<\unit{50}{\degree}$ and thermal inertia below some \unit{500}{\TIunit}, NEATM-derived diameters were within \unit{15}{\%} of the input diameter.
In a similar analysis,  \citet{Wright2007} found that the NEATM  reproduces the input diameter to within \unit{10}{\%} (r.m.s.) for $\alpha < \unit{60}{\degree}$ for a particular model of surface roughness and different thermal inertia values.
None of these analyses considered the effect of irregular shape beyond cratering, where the latter is neglected in Harris' analysis.

                \subsection{NEATM fitting routine}
                \label{sect:NEATM:fitting}

In the course of this thesis work, a NEATM model code was implemented in C++, independent of that by \citet{NEATM} or \citet{Delbo2004}. Model fluxes from all three implementations were found to agree with one another except for negligible numerical noise. Also a fitting routine was developed, different from those by Harris or \Delbo\ but leading to identical results; since this fitting routine is used \emph{mutatis mutandis} also for our thermophysical model, it shall here be described in some detail.

Given $n$ data points $d_i(\lambda_i)$ observed at wavelength $\lambda_i$ and their uncertainties $\sigma_i$, we aim at minimizing%
\footnote{
We make frequent use of the \emph{reduced} $\chi^2$ which is defined as $\chi^2$ as defined in \eqref{eq:NEATM:chi2} divided by $n-m$, with the number of data points, $n$, and the number of free fit parameters, $m$. For NEATM fits, $m=2$ (diameter and $\eta$). For fitting purposes, it is irrelevant which definition is used; for gaging the goodness of fit, reduced $\chi^2$ is the more meaningful quantity.}
\begin{equation}
  \label{eq:NEATM:chi2}
  \chi^2 = \sum_{i=1}^n \frac{\left[m_i(\lambda_i)-d_i(\lambda_i)\right]^2}{\sigma_i^2}
\end{equation}
where the model fluxes $m_i(\lambda_i)$ are 
determined from \eqref{eq:NEATM:flux} with parameters
\begin{itemize}
\item Observing geometry: $r$, $\Delta$, $\alpha$
\item Asteroid constants: $H$, $G$, $\epsilon$
\item NEATM fit parameters: $\eta$ and either $D$ or \pv.
\end{itemize}

In our fitting routine,  $\eta$ and the geometric albedo \pv\ are varied; 
the factor $D^2$ in \eqref{eq:NEATM:flux} is proportional to $1/\pv$ by virtue of 
\eqref{eq:FowlerChillemi}. 
The $\eta$ axis of the parameter space is searched for the minimum \chitwo\ longward of $\eta=0.6$ (implicitly assuming that $\eta$ values below 0.6 are unphysical) at a step width of 0.4 in $\eta$. For each $\eta$ value, the best-fit \pv\ value and corresponding \chitwo\ are determined until the minimum in $\chi^2$ is boxed, i.e.\ until a series of three consecutive $\eta$ values is found for which the middle value leads to the lowest of the three $\chi^2$ values.
Then the best-fit $\eta$ value is determined using a bisection algorithm.

The best-fit \pv\ for constant $\eta$ is determined iteratively.
To this end, NEATM fluxes are calculated using a seed value of $\pv_0$ (usually 0.2, but this value can be changed for debugging).
Using linear regression, we determine a scale factor $\kappa$ \seeeq{eq:NEATM:fitting} which leads to the best fit of the model fluxes with the data without changing the color temperature. 
Following iterations assume the \pv\ value of their predecessor divided by the thus obtained $\kappa$ value.
This is repeated until \pv\ stabilizes to a user-defined fractional accuracy goal.

If temperatures were independent of \pv,
the integrand in \eqref{eq:NEATM:flux} would be independent of \pv\ and thermal fluxes would be proportional to $1/\pv\propto\kappa$ for constant $H$, hence the iteration above would converge to the exact result in the first step.
In reality, temperatures are a weak function of \pv, hence the iteration typically converges after a small number of iterations.
The convergence velocity improves with increasing number of data points and with decreasing albedo.
We observed that in most cases we  over-correct \pv, i.e.\ that $\kappa>1$ is typically followed by $\kappa<1$ in the next iteration, and vice-versa.
Sometimes this behavior leads to stable oscillations so  the algorithm does not converge.
To prevent this, we  update \pv\ by dividing by $\sqrt{\kappa}$ rather than $\kappa$, and by $\sqrt[4]{\kappa}$ if $\pv\geq 0.5$.
This has led to stable convergence in all cases so far.

\paragraph[Determining the best-fit \pv\ for constant $\eta$]{Determining the best-fit $\mathbf{\pv}$ for constant $\boldsymbol{\eta}$}
It is assumed that fluxes are nearly proportional to $1/\pv$ and thus to $\kappa$.
\Eqref{eq:NEATM:chi2} then reads:
\begin{equation}
  \label{eq:NEATM:fitting}
  \chi^2 = \sum_i^n \frac{\left[\kappa\  m_i(\lambda_i) - d_i(\lambda_i)\right]^2}{\sigma_i^2} =: \kappa^2 MM - 2\kappa MD + DD
\end{equation}
with 
\begin{subequations}
\label{eq:NEATM:MM}  
\begin{align}
  MM &:= \sum_i^n \left(\frac{m_i(\lambda_i)}{\sigma_i}\right)^2 \\
  MD &:= \sum_i^n \frac{m_i\left(\lambda_i\right) d_i\left(\lambda_i\right)}{\sigma_i^2} \\
  DD &:= \sum_i^n \left(\frac{d_i(\lambda_i)}{\sigma_i}\right)^2.
\end{align}
\end{subequations}
The sums $MM$, $MD$, and $DD$ can be calculated in a numerically efficient way after model fluxes have been calculated; in particular, model fluxes need not be stored in memory.
The best-fit value of $\kappa$ satisfies  $\textd \chi^2/\textd \kappa = 0$, i.e.\ 
 $\kappa = MD/MM$.


%% file: TPMintro.tex
A detailed thermophysical model (TPM) is presented which is applicable to all asteroids including NEAs.
The effects of thermal inertia, spin state, irregular shape, and thermal-infrared beaming are explicitly taken into account. Arbitrary convex shapes are allowed for, a generalization to non-convex shapes is under development (see \chaptref{chapt:TPMconcave} in the appendix).

Realistic thermophysical modeling is required in order to reach the primary goal of this thesis, namely to determine the thermal inertia of NEAs through analysis of thermal-infrared observations; we wish to emphasize that there is no other well-established method of measuring the thermal inertia of asteroids.
Furthermore, 
since the thermal physics of asteroid surfaces is here modeled in a more realistic way than in the highly idealized thermal models described in the previous chapter, 
TPM-derived estimates on diameter and albedo are potentially more accurate.

The thermal emission of NEAs is more challenging to model than that of MBAs due to their larger thermal inertia and because they are typically observed at much larger solar phase angles.
While TPMs applicable to MBA data have previously been available, we here report the first  TPM shown to be applicable to NEA data.

After an overview section, the thermal physics of asteroid surfaces is discussed in \sectref{sect:TPM:modeling}. The model implementation is presented in \sectref{sect:TPMconvex:implementation}, validation test are presented in \sectref{sect:TPMconvex:validation}. 
\Sectref{sect:TPM:fitting} is devoted to fitting techniques.



%% file: TPMoverview.tex
Various models have been proposed  to overcome the limitations of simple thermal models. E.g., \citet{Hansen1977} proposed a physical model for asteroid surface roughness along with an approximative treatment of conduction on a spherical asteroid.
Non-spherical asteroid shapes have been modeled by \citet{Brown1985}, who  proposed a variant of the STM with ellipsoidal asteroid shape. \citet{Spencer1990} proposed an improved variant of \citeauthor{Hansen1977}'s model in which thermal conduction is modeled in more detail. A variant of  \citeauthor{Spencer1990}'s model has been used by \citet{Delbo2004}.
The most realistic asteroid TPM currently available is that by  \citet{LagerrosI,LagerrosIII,LagerrosIV}  in which thermal conduction and surface roughness are explicitly modeled on an asteroid of arbitrary shape.
The Lagerros TPM has been widely used and is well tested for applications to large MBAs. It enabled their thermal emission to be studied to such a high accuracy  that they are used as calibration standards for space telescopes \citep[see, e.g.,][]{MuellerLagerros1998, MuellerLagerros2002}.

TPMs for asteroids were largely inspired by 
observations  and models of other atmosphereless bodies.
The first extraterrestrial body with well understood thermal properties was the \emph{Moon}.
It was found to display a thermal-infrared beaming effect by \citet{PettitNicholson1930} who also concluded that the thermal conductivity of lunar regolith was extremely low \citep[see also][]{Wesselink1948}.
Around the Apollo era the thermal properties of the Moon were studied in great detail from  ground-based observations, in-situ measurements, and laboratory analysis of returned lunar samples
\citep[see, e.g.,][and references therein]{Buhl,WinterKrupp1971,Saari1972,Jones1975,Langseth1976}.
While Martian results are not directly applicable to asteroids due to Mars' atmosphere, lessons can be learned from spacecraft observations of \emph{Martian satellites} \citep[see, e.g.,][]{Lunine1982,Kuehrt1992}, which are thought to be captured asteroids.
Also \emph{Mercury} is an atmosphereless body with a well observable and well modeled beaming effect \citep{Emery1998}. 
Due primarily to its slow spin, only the vicinity of Mercury's terminator is expected to be influenced  appreciably by thermal inertia.
No effects of thermal conduction on Mercury could be found from ground-based observations,  but thermal-infrared spacecraft observations using the MERTIS instrument \citep{Benkhoff2006} on BepiColombo are to be expected in the  future.

Our TPM is based on that by \citet{LagerrosI, LagerrosIII, LagerrosIV}.
Minor improvements to the physical modeling are proposed.
The implementation is completely independent of Lagerros'.
The model code was verified to produce physical results for very large solar phase angles and for a thermal inertia up to that of bare rock; extensive tests are reported in \sectref{sect:TPMconvex:validation}.
No such tests of the Lagerros TPM have been reported, but it is clear that his model was primarily aimed at application to MBAs.
For geometric reasons, ground-based observations of MBAs cannot be performed at phase angles largely exceeding \unit{30}{\degree}. Furthermore, the typical thermal inertia of MBAs appears to be comparable to that of lunar regolith. For both reasons, modeling the thermal emission of MBAs is numerically less challenging than that of NEAs.

Our  model is implemented in an object-oriented way (in C++) which makes it easy to add new features.

\subsection{Model description}
\label{sect:TPM:description}

The asteroid is modeled as a convex mesh of typically a few thousand triangular facets. 
It is assumed to rotate about a fixed axis, i.e.\ non-principal-axis rotation (``tumbling'')  is not supported.
All physical surface properties, such as albedo, emissivity, thermal inertia, and surface roughness are assumed to be constant over the surface.
Local surface temperatures are calculated from the local insolation geometry. One-dimensional thermal conduction into and from the subsoil is taken into account, all relevant parameters are assumed to be constant. To model surface roughness, model craters in the form of subdued hemispheres are added. The crater density and  opening angle can be varied. Inside craters, shadowing, multiple scattering of optical and thermal flux, and reabsorption of both are fully taken into account, leading to thermal-infrared beaming. 

\paragraph{Global-scale convexity}
The asteroid shape model is assumed to be convex, only small-scale concavity is modeled in terms of craters.
Global-scale convexity implies that facets cannot shadow one another, neither can they exchange energy radiatively which would lead to mutual heating.
Since facets are typically very large compared to the penetration depth of the diurnal heat wave \seesect{sect:TPM:TI}, lateral heat conduction between facets can be neglected.
Hence, the thermal flux emanating from an individual facet can be calculated independent of all other facets, which significantly simplifies the numerical treatment.
A more general model variant, in which non-convex shape is allowed for, is described in \chaptref{chapt:TPMconcave} in the appendix.

Most available  models of asteroid shapes are convex by design  \citep[see][for a review]{Kaasalainen2002}. Lacking such a shape model, one typically assumes a spherical or ellipsoidal shape, which are also convex.

\subsection{Model parameters}
\label{sect:TPM:parms}

Required asteroid parameters are the shape and spin state (see \sectref{sect:TPM:shape-spin} for details), the absolute optical magnitude $H$, slope parameter $G$, and emissivity $\epsilon$. Model variables are the geometric albedo \pv\ (which determines the diameter $D$ through $H$, and furthermore the Bond albedo $A$), the thermal inertia in SI units (\TIunit), and the crater density and opening angle (in degrees).
Model fluxes are calculated for a given Julian date, wavelength (in \micron), and observing geometry, where the latter is defined by the heliocentric and observer-centric asteroid position in ecliptic coordinates (ecliptic longitude and latitude in degrees, distance in \AU).
Data taken at different epochs should be stored in separate fit files, but they can be read in and fitted simultaneously.
Our techniques to fit model parameters to a given set of data are described in \sectref{sect:TPM:fitting}.

\subsection{Implementation overview}
\label{sect:TPM:impl_overview}

The TPM program takes \code{.convex} files as input. Those are generated using auxiliary programs based on computer-readable shape models \seesect{sect:TPM:shape-spin}.
On each facet, the diurnal temperature distribution is calculated as described in \sectref{sect:TPM:TI}. Thermal-infrared beaming is modeled as described in \sectref{sect:TPM:beaming}. Disk-integrated model fluxes are calculated by summing up contributions from all visible facets.


%% file: ShapeSpin.tex
In our TPM, the asteroid shape is  modeled as a mesh of typically a few thousand planar triangular facets, the typical  format of available asteroid shape models.
The shape is  defined by the coordinates of $n$ vertices which form $2n-4$ facets, each defined by the three indices of its vertices.
As is common practice, a body-centric coordinate system is used in which the $z$ axis corresponds to the spin axis; non-principal-axis rotation (``tumbling'') is not supported so far.
To model spherical and ellipsoidal shapes, an auxiliary program was developed which produces models of triangulated spheres of user-specified resolution and stretches them into ellipsoids if desired.

Geometric and thermal model tasks are separated in the code as far as possible, such that time-consuming geometric calculations need to be performed only once per shape.
To this end, auxiliary programs have been developed to convert shape model files provided by colleagues
(currently, the quasi-standard OBJ wavefront format and Mikko Kaasalainen's variant thereof are supported)
 into a specifically defined type called \code{.convex}.
For the thermal emission of convex asteroids the vertex positions are irrelevant, hence \code{.convex} files contain  solely
 a list of $2n-4$ outbound surface-normal vectors \seesect{sect:TPM:dA} and the model's intrinsic diameter \seesect{sect:TPM:intrinsicD}. Shape models are not checked for convexity, they are assumed to be convex.

\subsubsection{Outbound surface-normal vector}
\label{sect:TPM:dA}

Each facet is defined by three vertex vectors $\vec{v_{1,2,3}}$ and has an outbound surface-normal vector (normal to the facet with a modulus equal to the facet size)
\begin{equation}
  \label{eq:TPM:dA}
  \vec{\textd\Area} = \pm \frac{1}{2} \left(\vec{v_2}-\vec{v_1}\right) \times \left(\vec{v_3}-\vec{v_1}\right)
\end{equation}
where $\times$ denotes the vector product.
We calculate $\vec{\textd\Area}$ with the positive sign and then check the orientation; the sign is flipped  if the resulting vector is inbound.
Assuming that the origin of the coordinate system is inside the object%
\footnote{ Strictly speaking, the more stringent requirement holds that all straight lines connecting the origin to the vertices must entirely lie within the object. By the definition of convexity, this is implied if the origin is inside the object. \label{foot:convex_concave}}
 (which is true for all common distributions of shape models), $\vec{\textd\Area}$ is outbound if 
\begin{equation*}
  \vec{\textd\Area} \cdot \frac{\vec{v_1}+\vec{v_2}+\vec{v_3}}{3} > 0
\end{equation*}
(note that the division by 3 is not required to perform this test and is therefore not done in the code).

\subsubsection{Intrinsic diameter}
\label{sect:TPM:intrinsicD}

So far, all linear dimensions are in unspecified units. To convert them into physical units, they must be multiplied with a scale factor
\begin{equation}
  \label{eq:TPM:intrinsicD}
  s := \frac{D_\text{phys}}{D_\text{intr}}
\end{equation}
where $D_\text{phys}$ is the physical diameter determined from the constant $H$ and the variable \pv\ (\eqref{eq:FowlerChillemi}), and $D_\text{intr}$ is the intrinsic diameter of the shape model in unspecified units. %
As is common practice,  the effective diameter is defined as that of the sphere of identical volume (see \eqrefpage{eq:Deff}).
The total volume $V$ of the polyhedral shape model equals the sum of  all tetrahedral volumes $V_i$ defined by the origin and the three vertices $\vec{v_{i,123}}$ belonging to facet $i$
\begin{subequations}
  \label{eq:TPM:volume}
  \begin{align}
    V &= \sum_i \frac{1}{6} \left| \left(\vec{v_{i,1}}\times\vec{v_{i,2}}\right) \cdot \vec{v_{i,3}} \right| \\
   D_\text{intr} &= \sqrt[3]{\frac{6V}{\pi}}.
  \end{align}
\end{subequations}
The scale factor $s$ is  updated inside the TPM code whenever the variable \pv\ is updated.

\subsubsection{Transformation into a co-rotating system}
\label{sect:TPM:bodyfix}

For the calculation of model fluxes, the coordinates of the Sun, \eS, and of the observer, \eO, in a co-rotating asteroid-centric coordinate system are required.
They are calculated from the epoch of the observation, $t$, and from the  ecliptic coordinates of the asteroid in heliocentric $\left(\lambda_S, \beta_S\right)$ and observer-centric $\left(\lambda_O, \beta_O\right)$ frames, respectively, which are typically taken  from ephemeris generators.
Further required input parameters are the epoch of zero rotational phase $\text{JD}_0$, the spin period $\mathcal{P}$, and the ecliptic coordinates of the spin axis $\left(\lambda_A, \beta_A\right)$.
The transformation into the bodycentric co-rotating system reads: ($R_a\left(\phi\right)$ denotes the counter-clockwise rotation about the $a$ axis by the angle $\phi$ in radians)
\begin{equation}
  \label{eq:TPM:geom:transfo}
- \vec{e_x} = 
R_z\left( -\left(t-\text{JD}_0\right)  
 \frac{2\pi}{P} 
  \right)
R_y\left(\beta_A - \frac{\pi}{2} \right)
\left(
  \begin{array}{c}
\cos\left(\lambda_x-\lambda_A\right) \cos\beta_x \\
\sin\left(\lambda_x-\lambda_A\right) \cos\beta_x \\
\sin\beta_x
  \end{array}
\right)
\end{equation}
using the usual Euler rotations (substitute $S$ or $O$ for $x$; the first Euler rotation is performed explicitly,  leading to the longitude $\lambda_x-\lambda_A$); the minus sign in front of $\vec{e_x}$ reflects the fact that ephemeris coordinates denote vectors pointing towards the asteroid, whereas in the following we require  vectors pointing away from it.


%% file: conduction.tex

Thermal conduction into and from the subsoil causes asteroids to display thermal inertia \seesect{sect:thermal:TI}, such that their surface temperatures not only depend on the instantaneous insolation but also on the thermal history.
As will be discussed in \sectref{sect:conduction:physics}, the thermal inertia of asteroid surfaces may vary by some two orders of magnitude, depending on various surface properties, leading to significant differences in surface temperatures and hence in thermal fluxes.

As is common practice, lateral heat conduction is neglected;  the length scale of thermal conduction phenomena, the skin depth $l_S$, is typically in the \centi\metre-range, much below the resolution of available asteroid shape models.

Like most authors of recent asteroid thermophysical models \citep[see, e.g.,][]{LagerrosI,LagerrosIV,Delbo2004,Wright2007}
we  follow the example of \citet{Spencer1989} and assume all relevant thermal parameters to be constant with depth and hence with temperature (see discussion below).

\subsubsection{Mathematical description}
\label{sect:conduction:math}

Thermal conduction can be described in terms of the vector-valued heat flux, $\vec{\Phi}$, which is defined as the amount of heat transfer per cross-sectional area. 
$\vec{\Phi}$ points into the direction of the heat transfer, which is proportional to the gradient of the temperature $T$ 
\begin{equation}
\label{eq:thermalconductivity}
\vec{\Phi} = - \kappa \vec{\nabla} T.
\end{equation}
$\kappa$ is a material specific constant, the \emph{thermal conductivity.}
Inside a thermal conductor, the thermal energy per unit volume reads $\rho c T$, with the surface bulk mass density, $\rho$,%
\footnote{ $\rho$ should not be confused with the total bulk mass density of the body which can be very different from that of the surface, e.g.\ in the case of loose regolith covering solid rock.}
 and the specific heat, $c$. By conservation of energy, local changes in thermal energy act as sources of heat flux, i.e.:
\begin{equation}
\label{eq:general_heat_transfer}
\frac{\partial}{\partial t} \rho c T = \vec{\nabla}\cdot \kappa \vec{\nabla} T.
\end{equation}

In the case of opaque atmosphereless bodies such as asteroids, a boundary condition of this partial differential equation of second order stems from the absorption of solar flux at the surface \seeeqpage{eq:STM:Tphi}%
\footnote{ Note that non-opaque materials, such as water ice, allow sunlight to be absorbed at non-negligible depth leading to, e.g., the solid-state greenhouse effect \citep[see][and references therein]{Kaufmann2006,Kaufmann2007}.}
\begin{equation}
\label{eq:energybalance_conduction}
\epsilon\sigma T^4 = (1-A)\frac{S}{r^2} \mu_S + \Phi_N
\end{equation}
where $\Phi_N$ is the heat flux projection onto the outbound surface normal, and $\mu_S$ is
the cosine of the local zenith distance of the Sun ($\mu_S$ is defined to vanish when the Sun is below local horizon).%
\footnote{
For the sake of compactness, we consider direct insolation as the only source of absorbed incoming flux throughout this section; other sources (originating, e.g., from other facets inside concavities such as craters) are straightforward to add to the solar-radiation term. Similarly, shadowing can easily be incorporated by defining $\mu_S$ to vanish whenever the facet is shadowed.}
Thermal conduction thus couples surface temperatures to the sub-surface temperature profile.

Throughout this work, three approximations are made:
\begin{description}
\item[Constant thermal conductivity] 
The thermal conductivity $\kappa$ is assumed to be spatially constant, which tacitly includes that $\kappa$ be temperature independent (otherwise $\vec{\nabla}\kappa = \partial \kappa/\partial T \, \vec{\nabla} T$, 
see the discussion in \sectref{sect:conduction:physics}).
Under this assumption, \eqref{eq:general_heat_transfer} reduces to the well-known diffusion equation
\begin{equation}
\label{eq:heatdiffusion}
\frac{\partial}{\partial t} T = \frac{\kappa}{ \rho c}\Delta T.
\end{equation}
with the Laplace operator $\Delta = \vec{\nabla}\cdot\vec{\nabla}$.
\item[One-dimensional heat flow]
As we will see below, thermal conduction on asteroids is effective over typical length scales in the \centi\metre-range, significantly below the resolution of known shape models. 
We can therefore neglect lateral heat conduction and only  consider  one-dimensional heat flow into and from the subsoil.
Throughout this work, a coordinate system will be used in which the $Z$ axis coincides with the local surface normal,  $Z=0$ at the surface and $Z>0$ below the surface. 
\item[No seasonal effects] 
The insolation (and hence the temperature) on asteroids has typically two fundamental periods, the spin period $\mathcal{P}$ and the orbital period $\mathcal{T}$, leading to diurnal and seasonal effects, respectively. Typically, $\mathcal{T}\gg\mathcal{P}$. As will be seen below, seasonal effects are typically negligible on asteroids.
Insolation and  temperature are then periodic in time with period $\mathcal{P} =: 2\pi/\omega$. 
\end{description}

Under these assumptions, it is possible to express all relevant quantities in a dimensionless way, following \citet{Spencer1989}:
\begin{subequations}
\label{eq:uztau}
\begin{align}
\tau &= \omega t \\
z    &= Z/l_S \\
u    &= T/\TSS
\end{align}
\end{subequations}
with the following definitions of skin depth $l_S$, subsolar temperature \TSS, thermal inertia $\Gamma$, and thermal parameter $\Theta$:
\begin{subequations}
\begin{align}
\label{eq:skindepth}
l_S  &= \sqrt{\frac{\kappa}{\omega\rho c}} \\
\label{eq:TSS}
\TSS &= \sqrt[4]{\frac{(1-A)S/r^2}{\epsilon\sigma}}.
\\
\Theta &= \frac{\kappa/l_S}{\epsilon\sigma\TSS{}^3} = \sqrt{\omega}\frac{\Gamma}{\epsilon\sigma\TSS{}^3}
\label{eq:def_thermalparameter} \\
\Gamma &= \sqrt{\kappa\rho c}
\label{eq:def_TI}
\end{align}
\end{subequations}
\Eqref{eq:heatdiffusion} and \eqref{eq:energybalance_conduction} acquire the form
\begin{subequations}
\label{eq:heatconduction_dimensionless}
\begin{align}
\frac{\partial}{\partial \tau} u (z,\tau) &= \frac{\partial^2}{\partial z{}^2} u (z,\tau) 
\label{eq:heat_dimensionless} \\
u(0,\tau)^4 &= \mu_S(\tau) + \Theta
\frac{\partial}{\partial z}u(0,\tau).
\label{eq:boundarycondition_dimensionless} 
\end{align}
\end{subequations}
The  heat transfer problem depends solely on the thermal parameter, $\Theta$,  which is proportional to the thermal inertia, $\Gamma$, and otherwise independent of thermal properties.
It is easily seen that $\Theta=0$ corresponds to vanishing thermal inertia (c.f.\ \eqref{eq:STM:Tphi}) while $\Theta\to\infty$ corresponds to  an FRM-like asteroid  (c.f.\ \eqref{eq:energybalance_FRM}). $\Theta$ is proportional to $1/\sqrt{\mathcal{P}}$, so the FRM limit is approached by fast rotators.

Equation \eqref{eq:heat_dimensionless} is a partial differential equation of second order which requires two suitable boundary conditions to be solvable: \Eqref{eq:boundarycondition_dimensionless} and the requirement that the temperature be spatially constant at infinite depth
\begin{equation}
\label{eq:boundarycondition2_dimensionless}
\lim_{z\to\infty}\frac{\partial}{\partial z}u(z,\tau) = 0.
\end{equation}
Since the surface boundary condition is $\tau$ periodic with  period $2\pi$, so must the solution $u(z,\tau)$, and there 
 is a unique  solution to the diffusion equation \eqref{eq:heatconduction_dimensionless} in combination with \eqref{eq:boundarycondition2_dimensionless}, namely the wave equation
\footnote{
Derivation: A solution of the partial differential equation \eqref{eq:heatconduction_dimensionless} $u(z,\tau)\in\mathbbm{R}$ with time period $2\pi$ can be Fourier decomposed  \citep[see, e.g.,][]{Wesselink1948}
\begin{equation*}
u(z,\tau) = \sum_{n\in \mathbbm{Z}} a_n \exp(i n \tau) g_n(z)\qquad \forall_n\forall_z: a_n g_n(z) = \overline{a_{-n} g_{-n}(z)}
\end{equation*}
where $\overline{z}$ denotes the complex conjugate of $z$. 
Denoting derivatives w.r.t.\ $z$ with primes
we get:
\begin{align*}
n=0:\ \ &g_0^{\prime\prime}(z) = 0        &\longrightarrow g_0(z) &= \lambda_0\ z+u_0 \\
n>0:\ \ &g_n^{\prime\prime}(z) = ing_n(z) &\longrightarrow g_n(z) &= \lambda_n \exp \left[\sqrt{n/2}\left(1+i\right) z \right] + \mu_n \exp \left[-\sqrt{n/2}\left(1+i\right) z\right] \\
n<0:\ \ &g_n^{\prime\prime}(z) = ing_n(z) &\longrightarrow g_n(z) &= \lambda_n \exp \left[\sqrt{-n/2}\left(1-i\right)z\right] + \mu_n \exp \left[-\sqrt{-n/2}\left(1-i\right)z\right]
\end{align*}
with constants $\lambda_n$ and $\mu_n$. 
The boundary condition \eqref{eq:boundarycondition2_dimensionless}
requires all $\lambda_n$ to vanish. 
The proof is  finished by substituting $a_n \mu_n(z)$ with $u_n \exp(i\phi_n)/2$ ($\forall_n: u_n\in\mathbbm{R}\  \text{and}\ \phi_n\in\mathbbm{R}$).
} 
\begin{equation}
\label{eq:heatdiffusion_abstractsolution}
u\left(z, \tau\right) = u_0 + \sum_{n=1}^\infty u_n \exp\left(-z\sqrt{n/2}\right) \cos \left(n \tau - z\sqrt{n/2} - \phi_n \right).
\end{equation}
The parameters $u_n$ and $\phi_n$ remain to be determined from the surface boundary condition \eqref{eq:boundarycondition_dimensionless}.
In practice, this is less convenient than a straightforward numerical integration of the differential equation, so the heat-wave solution is of little practical use.
It is, however, important to realize that the amplitude of the heat wave decays exponentially with depth.

\subsubsection{Physical discussion}
\label{sect:conduction:physics}

\begin{figure}
  \centering
  \begin{minipage}[t]{0.48\linewidth}
    \includegraphics[angle=-90, width=\linewidth]{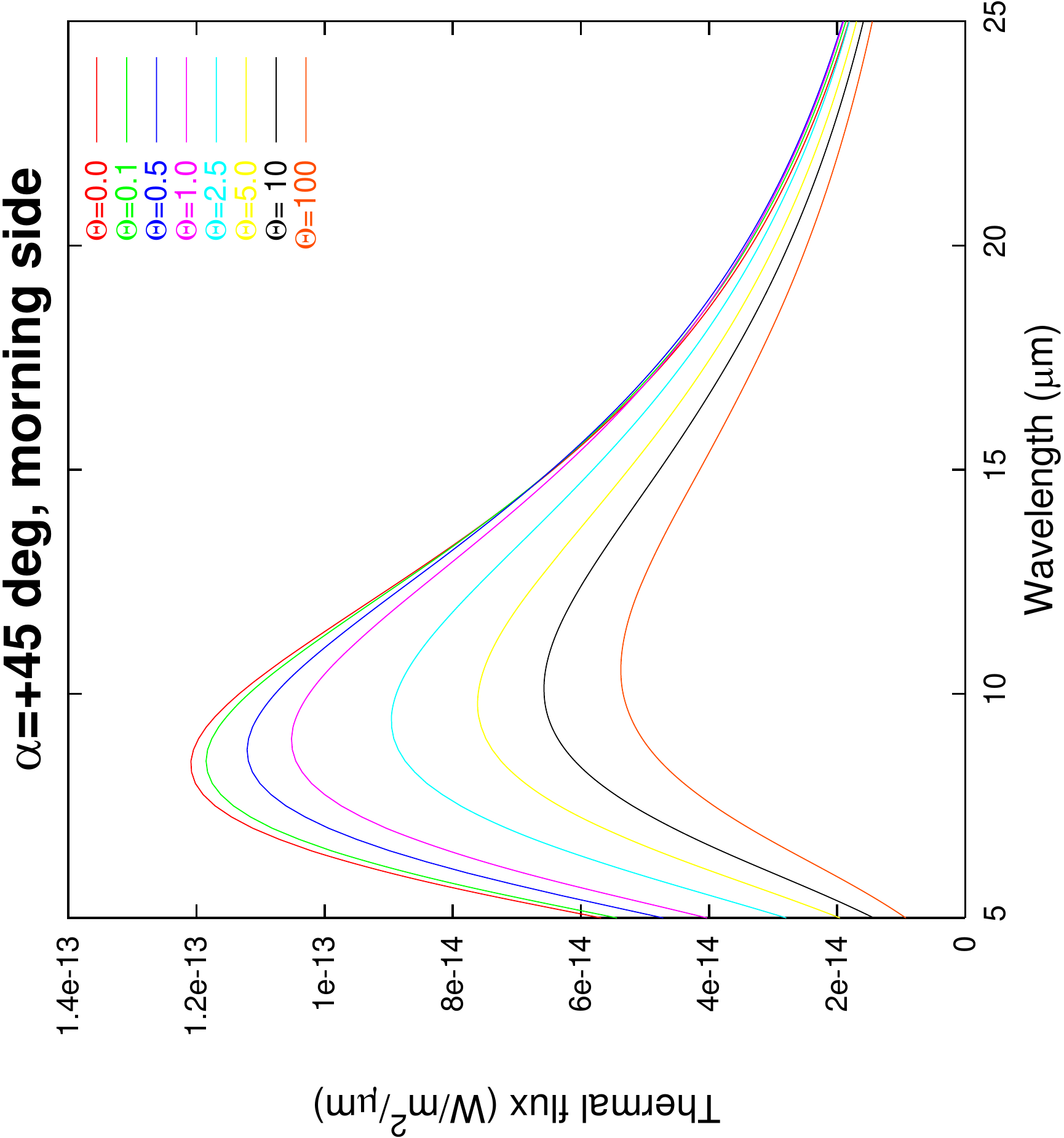}
  \end{minipage}
  \begin{minipage}[t]{0.48\linewidth}
    \includegraphics[angle=-90, width=\linewidth]{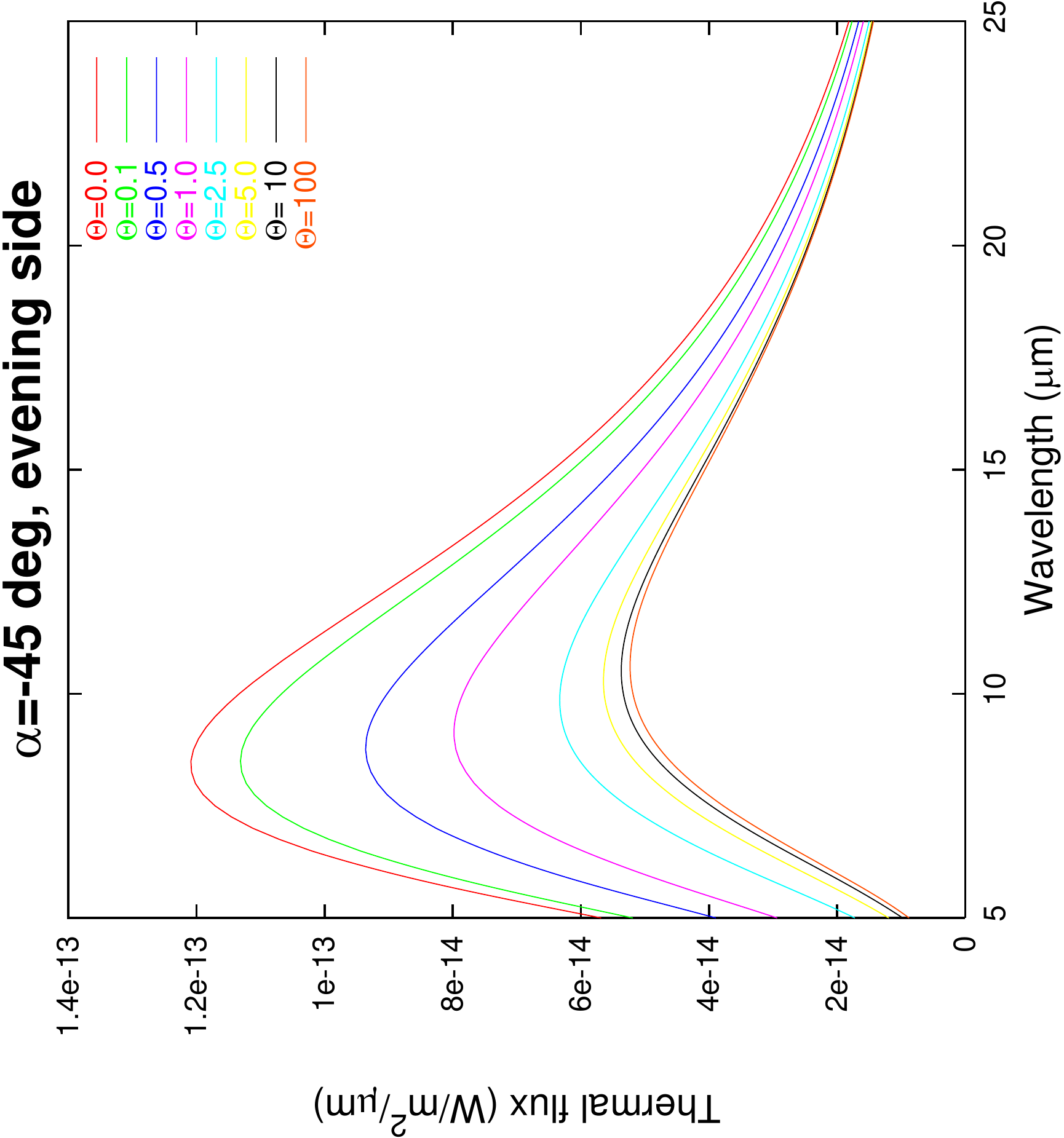}    
  \end{minipage}
  \caption[Thermal model spectra of a spherical smooth asteroid for different values of $\Theta$ and $\alpha=\pm\unit{45}{\degree}$.]{Thermal model spectra of a spherical smooth NEA for different values of $\Theta$, which is proportional to thermal inertia $\Gamma$, and for phase angles of $\alpha=\pm\unit{45}{\degree}$, placing the sub-observer point at local times of 9~AM and 3~PM, respectively (Sun and observer are above the equator).
Note the large \emph{morning-evening}  asymmetry for intermediate $\Theta$ values, which asymptotically vanishes for low and  large $\Theta$ values. This asymmetry drives the important Yarkovsky effect \seesect{sect:intro:Yarko} and facilitates determinations of thermal inertia from thermal data despite the competing effect of  thermal-infrared beaming which does not display a morning-evening asymmetry.
Model parameters are $r=\unit{1.1}{\AU}$, 
$\Delta=\unit{0.1}{\AU}$, $\mathcal{P}=\unit{6}{\hour}$, $A=0.1$, and $\epsilon=0.9$.
Hence, $\Theta=1$ corresponds to $\gamma\sim\unit{160}{\TIunit}$. A typical value for NEAs is \unit{300}{\TIunit} \seesect{sect:NEA:TI}.
}
  \label{fig:conduction:Theta}
\end{figure}

When observing the effect of thermal conduction on asteroid thermal fluxes, the primary observable quantity is $\Theta$ (see \eqref{eq:def_thermalparameter}).
As can be seen from \figref{fig:conduction:Theta}, $\Theta$ is difficult to measure in the limiting cases when it approaches 0 or $\infty$, and is most readily measurable for intermediate values of the order of 1.

Depending on their angular spin velocity $\omega$ and the subsolar temperature \TSS\ (which depends chiefly on the heliocentric distance $r$), objects with identical thermal inertia may nevertheless have very different thermal parameters.
While it is intuitively clear that the surface temperatures of fast rotators are more significantly influenced by thermal conduction than those of cold rotators, the dependence of $\Theta$ on \TSS\ is less intuitively obvious; to explain the latter, it is instructive to think of thermal emission into space (proportional to $\TSS^4$) as a heat transfer mechanism which competes with conduction into the subsoil (proportional to \TSS).
This explains, 
among other things, why the STM (which assumes $\Theta=0$) appears to be a poor thermal model for Kuiper belt objects \citep[see][and references therein]{Stansberry2007} despite the  low thermal inertia typically expected for such objects: Due to their large heliocentric distances, Kuiper belt objects are cold, and hence $\Theta$ can no longer be neglected even for low $\Gamma$ values.

The three physical parameters $\rho$, $c$, and $\kappa$ cannot be determined separately, but only the thermal inertia $\Gamma=\sqrt{\rho c\kappa}$  \seeeq{eq:def_TI}.
As can be seen from \tablerefpage{table:thermalproperties}, 
for plausible asteroid-surface materials $\kappa$ is much more variable than $\rho$ and $c$, hence $\Gamma$ is largely determined from $\kappa$.
$\Gamma$ is often transformed into $\kappa$ and vice-versa assuming  typical values of $\rho$ and $c$.
Confusingly, it is common practice among modelers of the Yarkovsky and YORP effects to consider $\kappa$ as the primary variable \citep[see, e.g.,][and references therein; ``thermal inertia'' is often not mentioned in such papers]{Bottke2006}, while thermal observers and modelers tend to highlight $\Gamma$ rather than $\kappa$.

\paragraph{Geological interpretation}
\begin{table}[tb]
\caption[Thermal properties of some typical materials]{Thermal properties of some typical materials: thermal conductivity $\kappa$, mass density $\rho$, specific heat capacity $c$, all for temperatures of \unit{20}{\celsius} unless otherwise stated.
Thermal inertia $\Gamma$ is  calculated from \eqref{eq:def_TI}, skin depths $l_S$ from \eqref{eq:skindepth}, assuming periods of $\mathcal{P} =\unit{24}{\hour}$ or \unit{365}{\dday}.
References: Text books and standard tables (\citealp{Dubbel}; \citealp{VolgerLaasch1989}; \citealp{Berber1999}; \citealp{Huette}), 
and \citet{WinterKrupp1971} for lunar regolith. 
Values in parentheses were estimated based on similar materials.
}
\label{table:thermalproperties}

\footnotesize{ \centering
\begin{tabular}{lcccccc}
\toprule
Material & $\kappa$ & $\rho$ & $c$ & $\Gamma$ & $l_S$ (day) & $l_S$ (year) \\
     & \kappaunit & \kilogram\usk\rpcubic\metre & \joule\usk\reciprocal\kilogram\reciprocal\kelvin & \TIunit &\centi\metre &\centi\metre \\     
\midrule
Nickel        & 91   & 8850 & 448 & $19\cdot10^3$ & 56 & $1000$ \\
Iron          & 81   & 7860 & 452 & $17\cdot10^3$ & 56 & $1000$ \\
Granite       & 2.9  & 2750 & 890 & 2600 & 13  & 250 \\
Marble        & 2.8  & 2600 & 800 & 2400 & 14  & 260 \\
Water ice, \unit{0}{\celsius} 
              & 2.25 & 917 & 2000 & 2040 & 13  & 252 \\
Water, \unit{0}{\celsius}
              &0.56  & 1000& 4200 & 1500 & 4.3 & 82 \\
Snow (compact)
              & 0.46 & 560  & 2100& 740  & 7.3 & 140\\
Sandy soil    & 0.27 & 1650 & 800 & 600  & 5.3 & 100\\
Coal          & 0.26 & 1350 & 1260& 665  & 4.6 & 88 \\
Pumice        & 0.15 & 800  &(900)& 330  & 5.4 & 100 \\
Paper         & 0.12 & 700  & 1200& 320  & 4.4 & 85 \\
Polystyrene foam 
              & 0.03 & 50   &1500 & 47   & 7.4 & 140\\
Air, \unit{20}{\celsius}
              & 0.026& 1.2  & 1000&5.6   & 55  & 1000\\
Lunar regolith&0.0029& 1400 & 640 & 51   & 0.7 & 13 \\
\bottomrule
\end{tabular}
}

\footnotesize{\textit{
\emph{Note:} 
The entry for air only applies to small volumes of air, where convective heat transfer (which dominates in large volumes) is inefficient. For very porous bodies on Earth (such as polystyrene foam), heat conduction through air trapped inside the pores is the dominant heat transfer mechanism, placing a lower limit on thermal conductivity. On atmosphereless bodies such as the Moon or asteroids, lower conductivities are possible.}}
\end{table}

As can be seen in \tablerefpage{table:thermalproperties}, the thermal inertia of lunar regolith is around \unit{50}{\TIunit}, roughly that of very light polystyrene foam.
Other plausible soil materials, such as coal or sand, have much larger thermal inertia, and bare  rock (granite or marble) reaches \unit{$\Gamma\sim2500$}{\TIunit}, some 50 times larger than that of lunar regolith.
Metals are excellent thermal conductors, metallic meteoroids may therefore display 
a very large thermal inertia some 350 times larger than that of lunar regolith.

The thermal inertia of a given particulate material decreases with decreasing grain size unless grains are much larger than the thermal skin depth \citep[see, e.g.,][]{Jakosky1986, Clauser1995,Presley1997}.
Thermal inertia is thus a very sensitive indicator for the presence or absence of loose surface material. This is widely used in Martian geology  \citep[see, e.g.,][and references therein]{Mellon2000,Christensen2003,Putzig2005}.
It must be noted, however, that even the thin Martian atmosphere greatly enhances the thermal conduction among fine grains \citep[see][]{Presley1997} relative to purely radiative heat transfer. 
To the best of our knowledge, the dependence of thermal inertia on grain size in a vacuum has not been studied so far, therefore it is presently not straightforward to interpret thermal-inertia values on asteroids in terms of grain size.

Another trend apparent in \tableref{table:thermalproperties} is that thermal inertia decreases with porosity (see the difference between ice and snow or the low thermal inertia of volcanic pumice).
It is therefore hard to tell \emph{a priori} what thermal inertia an asteroid composed of ``bare rock'' should display, but the commonly quoted value of \unit{2,500}{\TIunit} (see the values for granite and marble in \tableref{table:thermalproperties}) may be expected to be an upper limit.

It is also apparent from \tableref{table:thermalproperties} that neglecting lateral heat conduction on asteroid surfaces is unlikely to introduce significant systematic uncertainties: Even metallic objects with their huge thermal inertia and with a relatively slow rotation rate of \unit{24}{\hr} would have a thermal skin depth of only \unit{56}{\centi\metre}, such that for all objects larger than a few tens of meters in diameter lateral heat conduction can be safely neglected.
However, lateral heat conduction is very important for thermal modeling of the precursor bodies of metallic meteoroids and their  Yarkovsky drifts. 
Since such objects cannot be observed with current mid-IR telescopes, they are beyond our scope.

\paragraph{Heat transfer mechanisms}
There are three major heat transfer mechanisms: conduction, convection, and thermal radiation.
While convection is irrelevant on atmosphereless bodies, both conductive and radiative heat transfer could plausibly occur on asteroids.
Conduction would be expected to occur within surface grains, while radiation should dominate the heat transfer between grains. In the limit of point-like grains, conduction vanishes while for compact bodies radiative heat transfer can be neglected.
The relative importance of the two heat transfer processes thus depends on the typical surface-grain size, which is unknown for asteroids.

For conductive heat transfer, $\kappa$ is largely independent of temperature $T$
(for the materials and the temperature range relevant for our purposes)
whereas radiative heat transfer is well described with $\kappa\propto T^3$.
For lunar regolith, it is known from Apollo \emph{in-situ} measurements and laboratory analysis of returned lunar samples 
that both conductive and radiative heat transfer  are relevant, such that $\kappa = a + b T^3$  with constants $a$ and $b$ \citep[see, e.g.,][and references therein; in their model, $a$ and $b$ are functions of depth]{Jones1975}.

\citet{Kuehrt1989} proposed a complex TPM in which lunar results are rescaled to the conditions prevailing on the Martian satellites.
When attempting to use their model to fit observational data of Deimos and Phobos, however, they reverted to a simplified model, in which only the radiative $T^3$ term is considered \citep{Giese1990, Kuehrt1992}. Note that in this case heat transfer is no longer described by the diffusion equation \eqrefpage{eq:heatdiffusion}, but an additional term occurs, which stems from the derivative of $\kappa$.

Most TPMs for asteroids proposed so far assume  $\Gamma$ to be independent of depth and temperature \citep[see][and references therein]{Spencer1989,LagerrosI,Delbo2004,Wright2007}. This implicitly prefers conductive over radiative heat transfer. Another, widely quoted, interpretation is that the exact dependence of thermal parameters on depth and temperature is too poorly constrained by available data to be modeled explicitly, hence one reverts to constant values which are effective averages over the relevant length scales.

Due to their generally lower heliocentric distance, NEAs are typically hotter than MBAs, hence one might expect the radiative $T^3$ term to be more important for their thermal emission than for that of MBAs.
However, the relative importance of the $T^3$ radiative term and the $T^0$ conductive term are not clear \emph{a priori}.
An ``Apollo-like'' thermal conductivity model with two or more fit parameters ($a$ and $b$ given above) would probably be most realistic, but the ratio $a/b$ would be very hard to constrain with typical asteroid data.
We adopted a model in which all thermal properties including $\kappa$ are assumed to be constant, but we note that it may be worthwhile  to consider a model in which $\kappa$ is proportional to $T^3$. 
This is left to future work.


%% file: beaming.tex
As introduced in \sectref{sect:thermal:beaming}, the emission characteristics of asteroids surfaces are different to those of  smooth Lambertian surfaces, 
with an observed relative temperature and flux enhancement at low phase angles which, due to conservation of energy, must correspond to relative flux losses at large phase angles.
This phenomenon is referred  to as \emph{thermal-infrared beaming} 
and is well known from thermal observations of the Moon
\citep[see, e.g.,][for an overview]{Saari1972}
and Mercury
\citep[e.g.][]{Emery1998}.
The surfaces of these bodies are well known to be densely covered with impact craters;
``cratered'' thermophysical models
were seen to reproduce the observed beaming well \citep[see, e.g.,][]{Buhl,WinterKrupp1971,Emery1998}.
In these models, craters are modeled as sections of hemispheres and 
 the beaming effect stems from \emph{mutual heating} of facets due to reabsorption of scattered and thermally emitted flux inside the crater, and furthermore from \emph{shadowing} effects which become relevant at large phase angles, both leading to sharp temperature contrasts on small length scales.

Similar crater models have been successfully  applied to planetary satellites \citep[e.g.][]{Giese1990,Kuehrt1992} and are frequently employed to model  asteroid surface roughness
\citep[see][and references therein]{Hansen1977,Spencer1990,LagerrosIV,Delbo2004}.

These models differ in the degree to which multiple scattering inside craters is taken into account:
While \citet{Hansen1977} neglect multiple scattering altogether but include shadowing and mutual heating due to reabsorption of thermal flux, \citet{Spencer1990} additionally considers multiple scattering of sunlight but not of thermal flux (equivalently, $\epsilon=1$ is assumed inside craters); \citet{Delbo2004} follows \citeauthor{Spencer1990}'s approach. 
The crater model by \citet{Kuehrt1989} is virtually identical to Spencer's but differs in the treatment of thermal conduction inside craters (see below).
The model by \citet{LagerrosIV} is the most complete crater model currently available: Direct and multiply scattered sunlight, shadowing, and reabsorption and multiple scattering of thermal radiation are taken into account.
We here present an improvement over the model by \citet{LagerrosIV} where multiple thermal scattering is fully considered to all orders (see \sectref{sect:beaming:fluxes} for our improvement to Lagerros' model).

Several methods have been proposed in the literature to model thermal conduction inside hemispherical craters. 
\citet{LagerrosIV} and \citet{Delbo2004} 
explicitly solve the one-dimensional heat conduction problem for each surface tile inside the crater \citep[see above for limitations of the][model]{Delbo2004} which is, however, computationally very expensive.
Although a more complex model had been proposed by \citet{Kuehrt1989}, the same authors used a simplified variant thereof for fitting observations of the Martian satellites \citep{Giese1990,Kuehrt1992}, 
in which effectively only one subsoil depth is considered---while this satisfactorily reproduces the effect of thermal inertia on the diurnal lightcurve amplitude, it fails to reproduce the phase lag introduced by thermal inertia.
We use an approximation proposed and validated by \citet{LagerrosIV} (see \sectref{sect:beaming:approximation}) in which the numerical treatment of thermal conduction and cratering decouple, which is numerically highly advantageous.

\subsubsection{Model assumptions}
\label{sect:beaming:assumptions}

Beaming is modeled  by adding  craters to each surface facet. Variable model parameters are the crater opening angle (equivalent to the relative crater depth) and the crater density,  i.e.\ the  surface fraction covered in craters.
Very small craters with diameters comparable to or below the thermal skin depth $l_S$ (in the \centi\metre-range for typical asteroid surfaces, see  \tablerefpage{table:thermalproperties}) do not contribute significantly to the observable beaming because temperature contrasts are reduced by lateral heat conduction. We shall only consider much larger craters, such that lateral heat conduction can be neglected.
Under this assumption, thermal fluxes are independent of the crater size distribution  for a given opening angle and crater density.
One-dimensional heat conduction inside craters is considered in an approximate way, which decouples the treatment of craters from that of thermal conduction.
Multiple scattering and reemission of both sunlight and thermal emission are fully taken into account,  our approach is an improvement over that by \citet{LagerrosIV}, the most complete available in the literature so far.
So far, the only considered source of input flux is the Sun; for globally non-convex asteroid shapes, where facets may receive additional flux from one another, a generalized model may be required (see \sectrefpage{sect:TPMconcave:beaming}).

Two different radiation fields inside the crater are considered, one at optical wavelengths with total energy density $J_V(\vec{r})$ and corresponding absorptivity 
$1-A$ ($A$ denotes the bolometric Bond albedo); and another radiation field $J_{IR}(\vec{r})$ containing 
thermal radiation (integrated over all thermal wavelengths), with 
spectrally constant emissivity = absorptivity $\epsilon$ and reflectivity $1-\epsilon$. 
These two fields are independent from one another, with the exception that absorption of optical energy is a source of thermal energy.

\subsubsection{Geometry}
\label{sect:beaming:geometry}

Since the crater size distribution is irrelevant, the crater radius is set to unity without loss of generality.
The crater shape then solely depends on the opening angle $\gamma$, where low $\gamma$ corresponds to shallow craters and $\gamma=\unit{180}{\degree}$ corresponds to craters shaped as full hemispheres, the deepest craters considered. The slope of surface facets at the crater rim relative to neighboring smooth facets equals $\gamma/2$.
Throughout the following, we use the following parametrization of the crater surface
\begin{equation}
  \label{eq:beaming:r}
\vec{r} = \left(
  \begin{array}{c}
    \sin\theta\cos\phi \\ \sin\theta\sin\phi \\ -\cos\theta 
  \end{array}
\right)
\end{equation}
where $\phi$ runs from 0--$2\pi$ and $\theta$ from 0--$\gamma/2$. The crater floor is at depth $-1$, the crater rim at depth $-\cos(\gamma/2)$, hence the total crater depth is $1-\cos(\gamma/2)$.
Note that in our notation the value of the opening angle is twice
as large as in the notations of \citet{Kuehrt1989,Spencer1990,Emery1998,Delbo2004}: in their notation, e.g., the full hemisphere has an opening angle of \unit{90}{\degree} rather than \unit{180}{\degree}.
\citet{Hansen1977} and \citet{LagerrosIV} parametrize crater shape in terms of depth over diameter $S^\prime$
\footnote{ They use the symbol $S$ which we replace by $S^\prime$ to avoid confusion with the solar constant.}
 rather than opening angle $\gamma$; for comparison the following expressions are helpful:
\begin{subequations}
  \label{eq:beaming:S}
  \begin{align}
  S^\prime= \frac{1-\cos(\gamma/2)}{2} &= \sin^2\left(\frac{\gamma}{4}\right) \\
1-S^\prime &= \cos^2\left(\frac{\gamma}{4}\right).
  \end{align}
\end{subequations}
The outbound area element $\vec{\textd\Area}(\vec{r})$ inside the crater is given by
\begin{equation}
  \label{eq:beaming:dA}
  \vec{\textd\Area}\left(\vec{r}\right) = -\sin\theta\ \vec{r}\ \textd\theta\ \textd\phi,
\end{equation}
in particular the outbound unit normal vector $\vec{n}$ equals
\begin{equation}
  \label{eq:beaming:n}
  \vec{n}\left(\vec{r}\right) = -\vec{r}.
\end{equation}

We make frequent use of the local directional cosines of the directions towards the Sun, $m_S(\vec{r})$, and the observer, $m_O(\vec{r})$; and of the cosines of the angular distances of Sun, $\mu_S$, and observer, $\mu_O$, from local zenith. All these quantities are clipped to be non-negative.
Denoting the unit direction-vector towards the Sun or, respectively, the observer as $r_x$ (substitute $S$ or $O$ for $x$) and the unit vector in $z$ direction (i.e.\ local zenith) as $\vec{e_z}$, $\mu_x$ equals $\vec{r_x}\cdot\vec{e_z}$ (or 0 if $\vec{r_x}$ is below local horizon) and 
\begin{equation}
  \label{eq:beaming:m}
m_x\left(\vec{r}\right) =
\begin{cases}
  -\vec{r}\cdot\vec{r_x} & \text{if $\vec{r_x}$ is visible from $\vec{r}$} \\
 0  & \text{otherwise (i.e.\ $\vec{r}$ is eclipsed/occulted)}
\end{cases}
\end{equation}
$\vec{r_x}$ is visible from $\vec{r}$ if $\mu_x$ is positive and if the straight line containing $\vec{r}$ with tangent vector $\vec{r_x}$ intersects the sphere circumscribing the crater above the crater rim (in addition to the trivial intersection at $\vec{r}$ itself):
\begin{equation}
  \label{eq:beaming:sees}
  \vec{r}\cdot\vec{e_z} - 2\left(\vec{r}\cdot\vec{r_x}\right) \left(\vec{r_x}\cdot\vec{e_z}\right) > -\cos\left(\frac{\gamma}{2}\right).
\end{equation}

The vectors $\vec{r_S}$ and $\vec{r_O}$ in the crater coordinate system can be constructed from scalar products performed in the asteroid-centric coordinate system (scalar products are invariant under rotations) which is computationally advantageous:
\begin{align}
  \label{eq:beaming:SO_noinertia}
\vec{r_S} &= \left(
  \begin{array}{c}
    \sqrt{1-\mu_S{}^2} \\  0 \\ \mu_S
  \end{array} \right) \\
\vec{r_O} &= \left(
  \begin{array}{c}
    \sqrt{1-\mu_O{}^2}\cos\phi_A \\ \sqrt{1-\mu_O{}^2}\sqrt{1-\cos^2\phi_A} \\ \mu_O
  \end{array}
\right).
\end{align}
The $\mu_x$ can be calculated from the scalar products of $\vec{r_x}$ and $\vec{\textd \Area}$  and
\begin{equation}
  \label{eq:beaming:azi}
  \cos\phi_A = \frac{\vec{r_S}\cdot\vec{r_O} - \mu_S\mu_O}{\sqrt{1-\mu_S{}^2}{\sqrt{1-\mu_O{}^2}}}.
\end{equation}
(note that this approach fails when thermal conduction inside the crater is explicitly modeled, since then knowledge of $\vec{r_x}$ is required as a function of rotational phase).

Another important quantity in the following discussion is the \emph{view factor} $V_{r r^{\prime}}$ from $\vec{r}$ to $\vec{r^\prime}$. It is defined as the fraction of radiative energy per area leaving the facet centered at $\vec{r}$ and directly striking that at $\vec{r^\prime}$. Assuming Lambertian emission, the view factor is a purely geometric quantity symmetric in $\vec{r}$ and $\vec{r^\prime}$; it equals
\begin{equation}
  \label{eq:beaming:viewfactor_general}
  V_{r r^{\prime}} = \frac{
\left(\vec{n}\cdot\left( \vec{r^{\prime}} - \vec{r}  \right) \right)
\left(\vec{n^{\prime}}\cdot\left( \vec{r} - \vec{r^{\prime}}  \right) \right)
}
{\pi \left|\vec{r}-\vec{r^{\prime}}\right|^4}
\end{equation}
with the unit outbound surface-normal vectors $\vec{n}$ and $\vec{n^\prime}$. In particular, the solid angle under which a facet is visible at another is proportional to the product of its size and the view factor.
Inside a sphere, $\vec{n}=-\vec{r}$ (see \eqref{eq:beaming:n}), hence (using $\vec{r}\cdot\vec{r}=\vec{r^\prime}\cdot\vec{r^\prime} = 1$)
 \begin{equation}
 V_{r r^{\prime}} = \frac{
 (\vec{r}\cdot( \vec{r^{\prime}} - \vec{r}  ) )
 (\vec{r^{\prime}}\cdot( \vec{r} - \vec{r^{\prime}}  ) )
 }
 {\pi \left|\vec{r}-\vec{r^{\prime}}\right|^4}
= \frac{1}{4\pi}.  
   \label{eq:beaming:viewfactor}
 \end{equation}
The fact that the view factor is constant is a peculiarity of the sphere owing to its high symmetry; it will prove to be crucial in the following.
In particular, it enables the analytic evaluation of two important surface integrals:
\begin{equation}
  \label{eq:beaming:integral1}
  \int\limits_\Area V_{r r^{\prime}} \textd \Area = \frac{2\pi}{4\pi}\int\limits_0^{\gamma/2} \sin\theta\textd \theta = \frac{1-\cos(\gamma/2)}{2}=\sin^2\left(\frac{\gamma}{4}\right)
\end{equation}
and 
\begin{equation}
  \label{eq:beaming:integral2}
  \int\limits_\Area  V_{r r^{\prime}} m_x\left(\vec{r}\right) \textd \Area
= \mu_x \sin^2\left(\frac{\gamma}{4}\right)
\cos^2\left(\frac{\gamma}{4}\right).
\end{equation}
\begin{proof}[Proof of \eqref{eq:beaming:integral2}]
By construction, the integral over the entire crater $\Area$ equals the integral over that part $\Area^\prime$ from which $\vec{r_x}$ is not obstructed because $m_x(\vec{r})$ vanishes elsewhere. On $\Area^\prime$, the integral can be written as $\frac{1}{4\pi} \int_{\Area^\prime} \vec{r_x}\cdot\vec{\textd \Area^\prime}$ (see \eqref{eq:beaming:n} and \ref{eq:beaming:m}) which, by virtue of Gauss' theorem, equals the sum of three contributions:
\begin{enumerate}
\item
the integral over the crater ``cap'' $= \frac{2\pi}{4\pi}\mu_x \int_0^{\sin(\gamma/2)}\rho \textd\rho = \mu_x/4 \sin^2(\gamma/2) = \mu_x\sin^2(\gamma/4)\cos^2(\gamma/4)$
\item
the vanishing integral over the ``terminator,'' i.e.\ the boundary between volume elements that see or do not see $\vec{r_x}$. 
The surface normal vector of that region is perpendicular to $\vec{r_x}$ by construction, hence this integral vanishes (if the entire crater sees $\vec{r_x}$, this integral vanishes trivially) 
\item
the volume integral of the (vanishing) divergence of $\vec{r_x}$ inside the region circumscribed by the three areas.
\end{enumerate}
\end{proof}

\subsubsection{Optical flux}
\label{sect:beaming:optical}

A necessary prerequisite to the determination of the temperature distribution inside the crater is a complete knowledge of the radiation field at optical wavelengths $J_V(\vec{r})$.
Apart from direct insolation, surface elements  receive scattered light from other facets, where multiple scattering  occurs, i.e.\ light scattered from one facet to another may be scattered again at the latter.

There are two approaches to model multiple scattering, either by summing up individual scattering orders (which we will do at the end of this section for illustrative purposes) or by self-consistency, i.e.\ by solving the integral equation
\begin{equation}
  \label{eq:beaming:optical1}
  J_V(\vec{r}) = A \left[ \frac{S}{r^2}m_S(\vec{r})  + \int\limits_{\Area^{\prime}} J_V(\vec{r^{\prime}}) V_{rr^{\prime}} \textd \Area^{\prime} \right],
\end{equation}
with solar constant $S$ and heliocentric distance $r$ (in \AU). 
\Eqref{eq:beaming:optical1} can be solved analytically for hemispherical craters because the view factor $V_{rr^{\prime}}$ is constant \seeeq{eq:beaming:viewfactor}, hence the integral in 
\eqref{eq:beaming:optical1} is independent of $\vec{r}$, i.e.\ a mere constant $K_1$
\begin{subequations}
  \label{eq:beaming:K1a}
\begin{align}
  J_V(\vec{r}) &= A \left[ \frac{S}{r^2}m_S(\vec{r})  + K_1 \right]. \\
  K_1          &= \int\limits_{\Area^\prime} J_V(\vec{r^\prime}) \frac{\textd \Area^\prime}{4\pi} 
= A \left[ \frac{S}{r^2} \int\limits_{\Area^\prime} m_S(\vec{r}) \frac{\textd \Area^\prime}{4\pi}
 + K_1 \int\limits_{\Area^\prime} \frac{\textd \Area^\prime}{4\pi}
\right]
\end{align}
\end{subequations}
by reinserting \eqref{eq:beaming:optical1}. Using the integrals \eqref{eq:beaming:integral1} and \ref{eq:beaming:integral2}:
\begin{subequations}
  \label{eq:beaming:K1}
  \begin{align}
    K_1 &= A \left[
\frac{S}{r^2}\mu_S\sin^2\left(\frac{\gamma}{4}\right)\cos^2\left(\frac{\gamma}{4}\right) + K_1 \sin^2\left(\frac{\gamma}{4}\right)
\right] \\
    K_1 &= A \mu_S \frac{S}{r^2} \frac{\sin^2\left(\gamma/4\right) \cos^2\left(\gamma/4\right)}{1-A\sin^2\left(\gamma/4\right)}
  \end{align}
\end{subequations}
such that
\begin{equation}
  \label{eq:beaming:optical}
  J_V(\vec{r}) = A  \frac{S}{r^2} 
\left[
m_S\left(\vec{r}\right) + \mu_S \frac{\sin^2\left(\gamma/4\right) \cos^2\left(\gamma/4\right)}{1-A\sin^2\left(\gamma/4\right)} 
\right].
\end{equation}

\paragraph{Correction of the Bond albedo}

The presence of craters lowers the albedo relative to that of a flat surface patch.
Due to the absorption processes  associated with multiple scattering, the crater scatters less optical flux outward than a flat surface of equal area would. 
The corrected Bond albedo $A_\text{corr}$ equals  $V_\text{out}/V_\text{in}$ with total in- and outgoing optical flux $V_\text{in}$ and $V_\text{out}$, respectively.%
\footnote{ In principle, this ratio must be averaged over the hemisphere of all possible incidence directions. Below, however, we will find the albedo to be independent of the incidence angle.}
The total solar flux $V_\text{in}$ entering through the crater rim equals $\mu_S S/r^2$ times the projected area of the crater cap $\pi\sin^2(\gamma/2) = 4\pi\sin^2(\gamma/4)\cos^2(\gamma/4)$.
The total outgoing flux $V_\text{out}$ equals $\int_{\Area^\prime}\int_\Area J_V(\vec{r^\prime}) V_{rr^{\prime}}\textd \Area \textd \Area^\prime$, where the integral is performed over the crater interior $\Area^\prime$ and the crater cap $\Area$.
By virtue of Gauss' theorem, one can deform $\Area$ into the  complement of the sphere circumscribing the crater (i.e.\ the range $\gamma/2<\theta<\pi$) without changing the result, such that
\begin{align}
\nonumber
  V_\text{out} &= \int\limits_0^{2\pi}\textd \phi \int\limits_{\gamma/2}^{\pi}\sin\theta\textd\theta K_1 \qquad \text{(see \eqref{eq:beaming:K1a})} \\
\nonumber
  &= 2\pi \left(2\cos^2\left(\frac{\gamma}{4}\right)\right) A\mu_S \frac{S}{R^2} \frac{\sin^2(\gamma/4)\cos^2(\gamma/4)}{1-A\sin^2(\gamma/4)} \\
\nonumber
 &= V_\text{in} A \frac{1-\sin^2(\gamma/4)}{1-A\sin^2(\gamma/4)}
\end{align}
such that the corrected Bond albedo reads
\begin{equation}
  \label{eq:beaming:correctedBond}
  A_\text{corr} = A \frac{1-\sin^2(\gamma/4)}{1-A\sin^2(\gamma/4)}.
\end{equation}
\begin{figure}
  \centering
\includegraphics[width=0.5\linewidth, angle=-90]{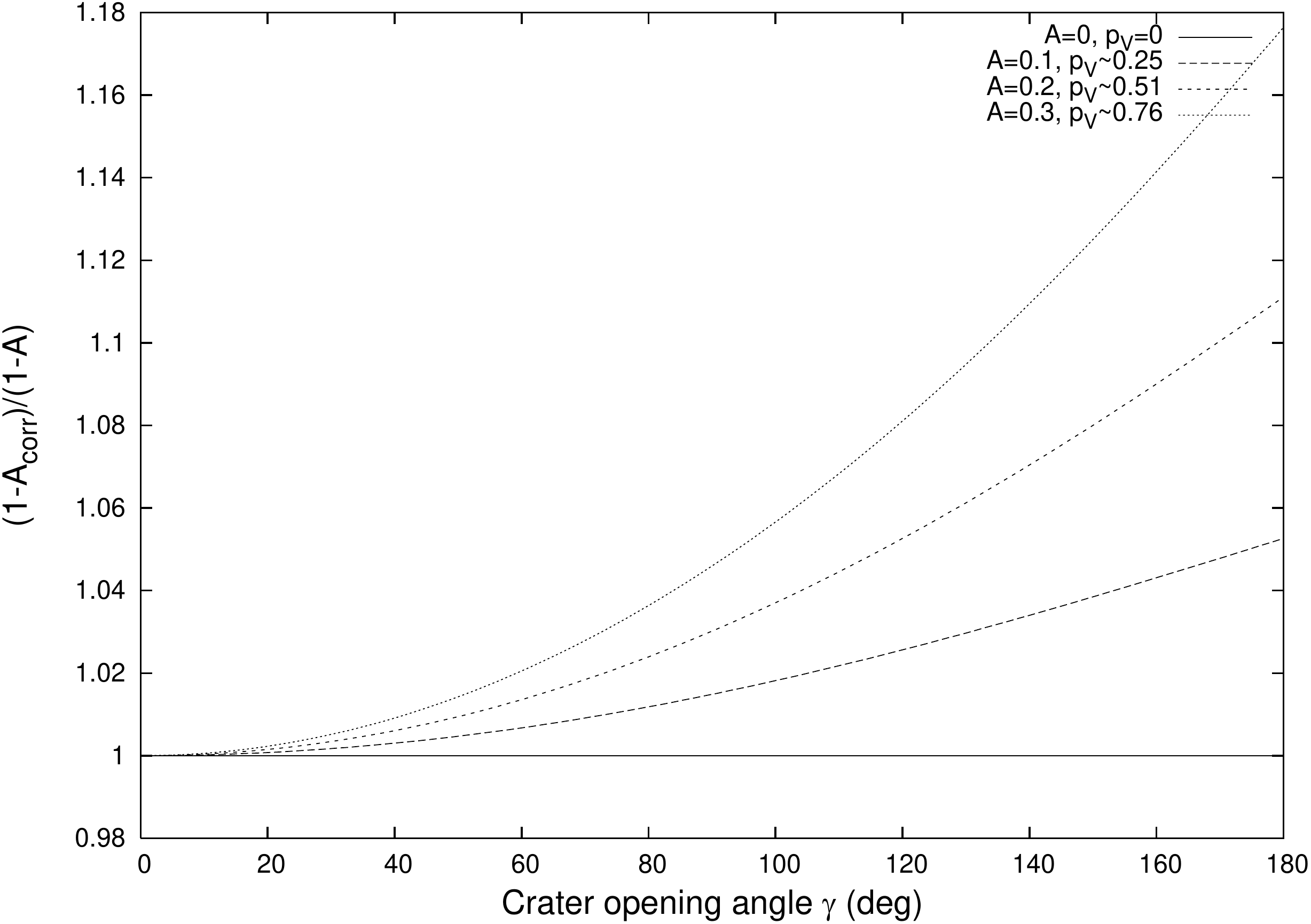}
  \caption{Dependence of absorptivity $1-A$, which determines temperatures, on the albedo correction due to the presence of craters (see \eqref{eq:beaming:correctedBond}) as a function of crater opening angle $\gamma$ for different values of flat-surface Bond albedo $A$. To convert $A$ into geometric albedo \pv, $G=0.15$ is assumed \seesect{sect:thermal:size-albedo}.}
  \label{fig:beaming:Bond}
\end{figure}
See \figref{fig:beaming:Bond} for a depiction of the relative changes in absorptivity $1-A$. For realistic asteroid albedos $\pv \leq 0.6$ (see \sectref{sect:intro:diameter}; $\pv=0.6$ corresponds to $A=0.24$ assuming $G=0.15$, see \sectref{sect:thermal:size-albedo}), the relative change in absorptivity cannot greatly exceed \unit{10}{\%}, leading to only  moderature changes in temperature since $T\propto \sqrt[4]{1-A}$.
It is, however, very important to use the corrected Bond albedo when verifying that the ingoing optical flux equals the total optical and thermal flux, which is an important test of model consistency  used in the following.

\paragraph{Direct summation of scattered light to all orders}
It is instructive to verify \eqref{eq:beaming:optical} by  explicitly summing up  scattered contributions to all orders, which can be performed analytically in this case:
For the sake of this discussion, let us denote the amount of directly scattered sunlight emanating from $\vec{r}$ as $J_0 (\vec{r})$ and the $n$-times scattered component as $J_n(\vec{r})$. Then
\begin{equation}
\label{eq:beaming:optical:direct}
J_V(\vec{r}) =  \sum_{n=0}^{\infty} J_n(\vec{r})
\end{equation}
and the following relations hold:
\begin{align}
\nonumber
J_0(\vec{r}) &= A\frac{S}{r^2} m_S(\vec{r})
\\ \nonumber
J_1(\vec{r}) &= A \int\limits_{\Area^\prime} J_0(\vec{r^{\prime}}) V_{rr^{\prime}} \textd \Area^\prime = A^2 \frac{S}{r^2} \mu_S \sin^2\left(\frac{\gamma}{4}\right) \cos^2\left(\frac{\gamma}{4}\right)
\\ \nonumber
\forall_{n\geq 1}: J_{n+1}(\vec{r}) &= A \int\limits_{\Area^\prime} J_n(\vec{r^{\prime}}) V_{rr^{\prime}} \textd \Area^\prime = A \sin^2\left(\frac{\gamma}{4}\right) J_n(\vec{r})
\end{align}
(the middle equation holds because of \eqref{eq:beaming:integral2} and, inductively, implies the third equation together with \eqref{eq:beaming:integral1}---note that all $J_n(\vec{r})$ are independent of $\vec{r}$ for $n\geq1$).
\Eqref{eq:beaming:optical:direct}  acquires the form of a geometric series  and can be summed up analytically ($\sum_{n=0}^\infty q^n = 1/(1-q)$ for $|q|<1$; here, $q=A\sin^2(\gamma/4)<1$), yielding \eqref{eq:beaming:optical}.

\subsubsection{Temperature distribution}
\label{sect:beaming:temperature}

Similar to the total optical flux $J_V(\vec{r})$, the total thermal flux emanating from $\vec{r}$, $J_{IR}(\vec{r})$,  equals the sum of the locally emitted flux  $\epsilon\sigma T^4(\vec{r})$ plus the flux scattered away from $\vec{r}$
\begin{align}
\label{eq:beaming:T1general}
J_{IR} ( \vec{r}) &=
\epsilon\sigma T^4(\vec{r}) 
+ (1-\epsilon) \int\limits_{\Area^{\prime}}J_{IR} ( \vec{r^{\prime}}) V_{rr^{\prime}}\textd \Area^{\prime}
\\
&= \epsilon\sigma T^4(\vec{r}) + (1-\epsilon)K_2,
  \label{eq:beaming:T1}
\end{align}
where, again, $K_2$ is constant because the view factor $V_{rr^{\prime}}$ is constant. As above
\begin{align}
K_2 &= \int\limits_{\Area^\prime}J_{IR}\left(\vec{r^\prime}\right) V_{rr^{\prime}}\textd \Area^{\prime}
 = \epsilon\sigma \int\limits_{\Area^\prime} T^4(\vec{r^\prime}) \frac{\textd \Area^{\prime}}{4\pi}
  + (1-\epsilon)K_2\sin^2\left(\frac{\gamma}{4}\right)
\nonumber \\
K_2 &= \frac{\epsilon\sigma}{1-(1-\epsilon)\sin^2(\gamma/4)} \int\limits_{\Area^\prime} T^4(\vec{r^\prime}) \frac{\textd \Area^{\prime}}{4\pi}.
  \label{eq:beaming:K2}
\end{align}

Sources of the temperature field $T(\vec{r})$ are absorption of direct sunlight, of  scattered optical flux (see \eqref{eq:beaming:optical}), of  direct and scattered emission $J_{IR}(\vec{r})$ received from other facets,  and thermal conduction from the subsoil:
\begin{align}
\label{eq:beaming:T2general}
\left(\frac{T(\vec{r})}{\TSS}\right)^4
 &= m_S(\vec{r}) 
+ \int\limits_{\Area^\prime}\left[ 
\left(1-A\right) \frac{J_V (\vec{r^\prime})}{\epsilon\sigma\TSS^4}
+ \epsilon \frac{J_{IR} (\vec{r^\prime})}{\epsilon\sigma\TSS^4}
\right] V_{rr^{\prime}}\textd \Area^{\prime}
+ \frac{\Theta}{\TSS} \frac{\partial T}{\partial z}
\\   \label{eq:beaming:T2}
&= m_S \left(\vec{r}\right) 
+ \frac{1-A}{\epsilon\sigma\TSS^4}K_1
+ \frac{\epsilon}{\epsilon\sigma\TSS^4}K_2
+ \frac{\Theta}{\TSS} \frac{\partial T}{\partial z}
\end{align}
where the temperature derivative in the last term is with respect to the dimensionless depth coordinate of the one-dimensional heat-conduction problem considered in \sectref{sect:TPM:TI}, which is not to be confused with the $z$ coordinate in the crater coordinate system.
See \eqref{eq:TSS} for the definition of the subsolar temperature \TSS.

If thermal inertia is explicitly modeled inside the crater, \eqref{eq:beaming:T2}  takes the role of the surface boundary-condition \eqref{eq:boundarycondition_dimensionless} of the heat-conduction problem, which must be solved in conjunction with \eqref{eq:beaming:K2}.

If thermal conduction is neglected inside craters, the last term in \eqref{eq:beaming:T2}  vanishes, such that $K_2$ can  be determined by inserting \eqref{eq:beaming:T2} and \eqref{eq:beaming:K1} into \eqref{eq:beaming:K2}:
\begin{equation}
  \label{eq:beaming:K2_noinertia}
  K_2 = \epsilon\sigma\TSS^4 \mu_S \frac{\sin^2(\gamma/4)}{1-A\sin^2(\gamma/4)} \qquad \text{neglecting thermal conduction}
\end{equation}
and hence 
\begin{equation}
  \label{eq:beaming:T}
\left(\frac{T\left(\vec{r}\right)}{\TSS}\right)^4
=   m_S(\vec{r}) + \frac{\mu_S \sin^2\left(\gamma/4\right)}{1-A\sin^2\left(\gamma/4\right)} \left(\epsilon + A\cos^2\frac{\gamma}{4} \right).
\end{equation}
This represents an analytic solution to the temperature distribution inside the crater, where multiple scattering of both sunlight and thermal asteroid radiation are taken into account to all orders \citep{LagerrosIV}.

\subsubsection{Fluxes}
\label{sect:beaming:fluxes}

The observable thermal flux $F(\lambda)$ at wavelength $\lambda$ 
is the sum of the directly emitted flux component $F_0(\lambda)$ and an infinite number of scattered components $F_i(\lambda)$, where the index $i>0$ denotes the scattering order 
\begin{equation*}
  F(\lambda) = \sum_{i=0}^\infty F_i(\lambda).
\end{equation*}
Each of these components equals an integral of the respective local flux component $F_i(\lambda,\vec{r})$ over the visible portion of the crater.
In particular
\begin{subequations}
  \label{eq:beaming:F0}
  \begin{align}
  F_0\left(\lambda\right) &= \int\limits_\Area 
\frac{m_O\left(\vec{r}\right)}{\pi\Delta^2} F_0\left(\lambda, \vec{r}\right) \textd\Area 
\\   
  F_0\left(\lambda, \vec{r}\right) &= \epsilon B\left(\lambda, T\left(\vec{r}\right)\right)
  \end{align}
\end{subequations}
with the Planck function $B(\lambda, T)$ (see \eqref{eq:Planck}), the observer-centric distance $\Delta$, and the local directional cosine towards the observer $m_O$ (see \eqref{eq:beaming:m}).
Scattered orders are determined recursively
\begin{subequations}
  \label{eq:beaming:Fi}
  \begin{align}
  F_{i+1}\left(\lambda\right) &= \int\limits_\Area \frac{m_O\left(\vec{r}\right)}{\pi\Delta^2} F_{i+1}\left(\lambda, \vec{r}\right) \textd\Area 
\\   
  F_{i+1}\left(\lambda, \vec{r}\right) &= (1-\epsilon) \int\limits_\Area F_{i}\left(\lambda, \vec{r}\right) \frac{\textd\Area}{4\pi},
  \end{align}
\end{subequations}
where we have used $V_{rr^{\prime}}=1/4\pi$ (\eqref{eq:beaming:viewfactor}).
In particular
\begin{subequations}
  \label{eq:beaming:F1}
  \begin{align}
  F_1\left(\lambda, \vec{r}\right)
&= (1-\epsilon)\int\limits_\Area \epsilon B\left(\lambda, T\left(\vec{r}\right)\right) \frac{\textd\Area}{4\pi}
\\
 F_1\left(\lambda\right)
&= (1-\epsilon)  \int\limits_\Area \frac{m_O(\vec{r})}{\pi\Delta^2}  \frac{\textd\Area}{4\pi} \int\limits_{\Area^\prime} \epsilon B\left(\lambda, T\left(\vec{r^\prime}\right)\right) \textd\Area^\prime
\\ &= (1-\epsilon) \mu_O 
\frac{\sin^2(\gamma/4)\cos^2(\gamma/4)}{\pi\Delta^2}
\int\limits_\Area F_0\left(\lambda,\vec{r}\right) \textd\Area.
  \end{align}
\end{subequations}

\paragraph{Observable flux according to \citet{LagerrosIV}}
In \citet{LagerrosIV}, the total observable flux is approximated as the sum of directly emitted flux $F_0$ and singly scattered flux $F_1$ \citep[c.f.][eqn.\ 19]{LagerrosIV}, 
\begin{subequations}
  \label{eq:beaming:FLagerros}
  \begin{align}
  F(\lambda)_\text{Lagerros} 
&= F_0(\lambda)+F_1(\lambda)
\\
&= \frac{\epsilon}{\pi\Delta^2} \int\limits_\Area 
\left[ m_O(\vec{r}) + \mu_O\sin^2\left(\frac{\gamma}{4}\right)\cos^2\left(\frac{\gamma}{4}\right)(1-\epsilon) \right]
B\left(\lambda, T\left(\vec{r}\right)\right) \textd\Area.
  \end{align}
\end{subequations}
It is not discussed therein why higher scattering orders are neglected.

\paragraph{Observable flux to all orders}
While, to the best of our knowledge, this has not been discussed in the literature so far, it is feasible to consider \emph{all} scattering orders in the calculation of observable flux.
To this end it is crucial to realize that, because the view factor $V_{rr^{\prime}}=1/4\pi$ is constant, all scattered flux components $F_{i>0}(\lambda,\vec{r})$ are actually independent of $\vec{r}$
as is easily seen in the recursion \eqref{eq:beaming:Fi}.
Hence
\begin{equation}
\label{eq:beaming:Frecursion1}
  F_{i+1}\left(\lambda, \vec{r}\right) = (1-\epsilon)\int\limits_{\Area} F_i \left(\lambda, \vec{r}\right) \frac{\textd\Area}{4\pi}
= (1-\epsilon)\sin^2\left(\frac{\gamma}{4}\right) F_i \left(\lambda, \vec{r}\right)
\end{equation}
for all $i\geq 1$ (using \eqref{eq:beaming:integral1}). The sum of all scattered flux components acquires the form of a geometric series
\begin{equation}
  \label{eq:beaming:Frecursion2}
  \sum_{i=1}^\infty F_i\left(\lambda\right)
= F_1\left(\lambda\right) \sum_{i=0}^\infty \left[(1-\epsilon)\sin^2 \left(\frac{\gamma}{4}\right) \right]^i
= \frac{F_1\left(\lambda\right)}{1-(1-\epsilon)\sin^2(\gamma/4)}
\end{equation}
and the flux to all orders reads
\begin{align}
\nonumber
  F(\lambda)
&= \sum_{i=0}^\infty F_i(\lambda)
\\ \label{eq:beaming:Fallorders}
&= \frac{\epsilon}{\pi\Delta^2} \int\limits_\Area 
\left[ m_O(\vec{r}) + \frac{\mu_O(1-\epsilon)\sin^2(\gamma/4)\cos^2(\gamma/4)}{1-(1-\epsilon)\sin^2(\gamma/4)}
\right]
B\left(\lambda, T\left(\vec{r}\right)\right) \textd\Area.
\end{align}
Note, in particular, that the only difference between this expression and Lagerros' approximation (\eqref{eq:beaming:FLagerros}) is the redefinition of a constant factor, hence the effort to evaluate them through numerical integration is identical.

\begin{figure}
  \centering
\includegraphics[angle=-90, width=0.6\linewidth]{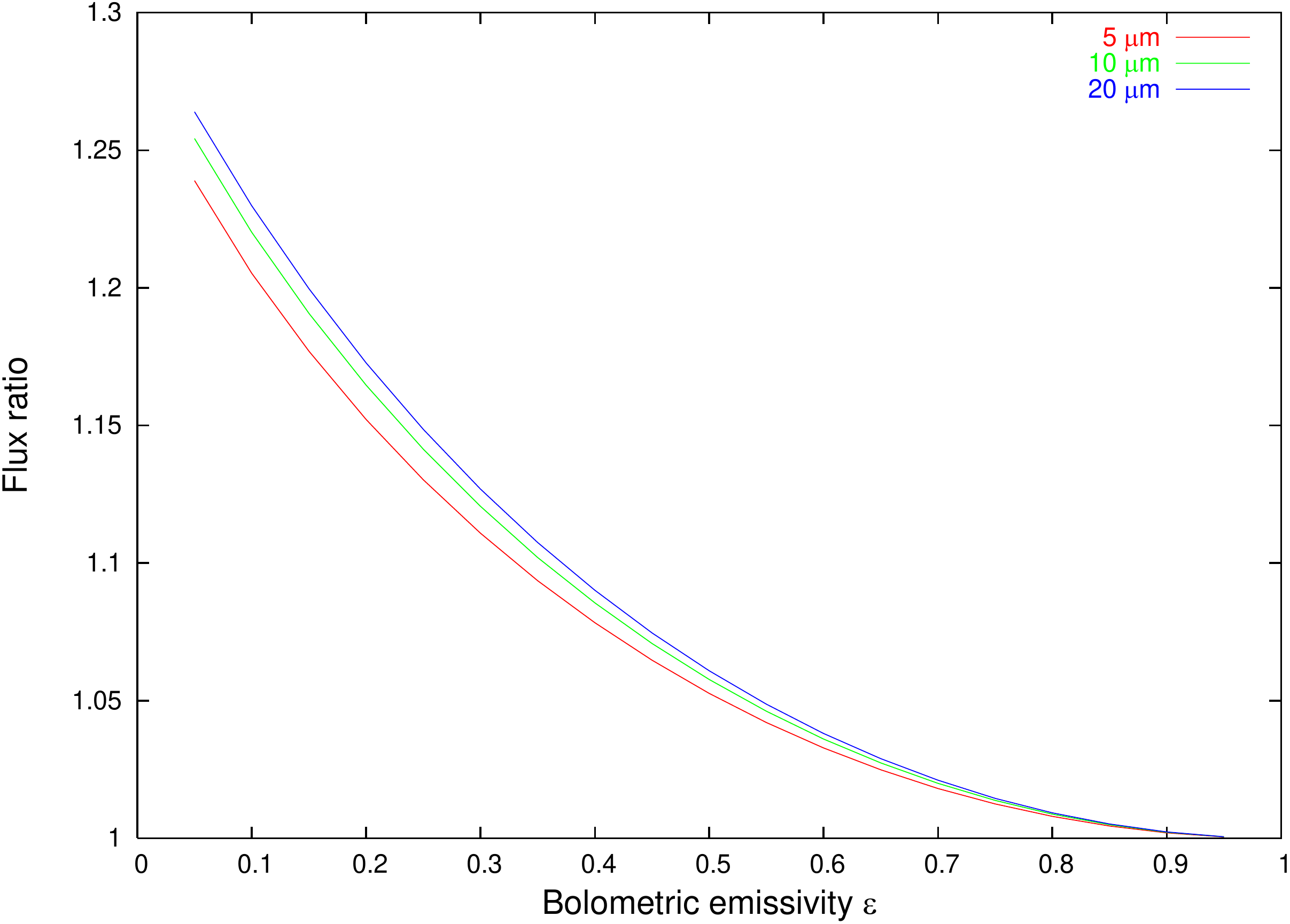}
  \caption[Ratio of crater model fluxes resulting from the solution to all scattering orders relative to Lagerros' approximation as a function of $\epsilon$.]{Ratio of crater model fluxes resulting from the solution to all scattering orders \eqref{eq:beaming:Fallorders} relative to Lagerros' approximate treatment  \eqref{eq:beaming:FLagerros} as a function of $\epsilon$. Both Sun and observer are at local zenith, used parameters are $\gamma=\unit{180}{\degree}$, $r=\unit{1.1}{\AU}$, $A=0.1$. Model fluxes were determined using the code described in \sectref{sect:TPMconvex:beaming}.}
\label{fig:beaming:Lagerros_all}
\end{figure}

The difference between the two expressions is in second and higher scattering orders, i.e.\ of the order $(1-\epsilon)^2$.
For asteroids, $\epsilon\sim0.9$, this would be expected to lead to differences at the  percent level, only. This has been verified numerically \seefigpage{fig:beaming:Lagerros_all}; only for rather unrealistically small $\epsilon$ values do the two expressions differ by more than \unit{10}{\%}.%
\footnote{ We have checked both solutions for conservation of energy as described in \sectref{sect:TPMconvex:validation:beaming}. It was seen that the full solution does conserve energy to within numerical noise, while Lagerros' approximation does not---as expected, the mismatch increases systematically with decreasing $\epsilon$ and is insignificant for $\epsilon=0.9$.}
However, since the effort in numerical evaluation is identical, there is no good reason \emph{not} to use the full solution.

\subsubsection{Approximative treatment of thermal conduction inside  craters}
\label{sect:beaming:approximation}

We do not explicitly model thermal conduction inside the crater, but rather use an approximation proposed by \citet[Eqn.\ 23]{LagerrosIV}.
There, it is proposed that the relative temperature change due to thermal inertia inside the crater equals the relative temperature change outside the crater:
\begin{equation}
  \label{eq:beaming:approx1}
  \frac{T_\text{crater}(\Theta)}{T_\text{crater}(0)} = \frac{T_\text{smooth}(\Theta)}{T_\text{smooth}(0)}.
\end{equation}
Under this approximation, the numerical treatment of cratering decouples from that of thermal conduction.
In particular, advantage can be taken of the analytic expression  for the temperature distribution inside the crater neglecting thermal conduction (\eqref{eq:beaming:T}).

\citet{LagerrosIV} generated model fluxes using both his full-blown model and a simplified model making the  approximation \eqref{eq:beaming:approx1}.
He found that the latter systematically overestimates fluxes relative to the former, increasingly so with increasing crater opening angle and thermal parameter, but 
 mutual agreement stayed within  \unit{1}{\%} for the specific circumstances of his test.

While this is not discussed in \citet{LagerrosIV}, the approximation \eqref{eq:beaming:approx1} can not be used on the night side, where temperatures would vanish without thermal inertia. We therefore neglect craters on the night side altogether. 
For the same reason, temperature ratios diverge close to the terminator. Together with the finite spatial resolution of the shape models used by us, this leads to random occurrences of
unphysically large flux contributions from facets close to the terminator. To prevent such overshoots, we clip the temperature ratio on the right-hand side of \eqref{eq:beaming:approx1} to be $\leq 1.3$;  we have verified that model fluxes are largely independent of the precise value of this threshold value within reasonable limits.



%% file: convex.tex
   	\section{Implementation}
                \label{sect:TPMconvex:implementation}

Our TPM model code has been implemented in C++. 
It compiles on  a Windows XP platform using the compiler which is part of the  Visual Studio .NET 2003 suite and furthermore under Linux using gcc.
All debugging and testing has been performed under Windows.

In the code development, emphasis was put on a 
transparent, generic, and object-oriented
code structure which makes it easy to add new features.
Numerical efficiency was not a primary implementation goal. If in doubt, we erred on the side of simplicity and understandability 
as opposed to sophistication and obscurity.
Not only does this save on development time (including time required for potential further development) but it also helps in the physical validation of the code.
On the other hand, the  model code is numerically quite inefficient, such that fits to large databases can require several nights of CPU time on a \unit{2.66}{\giga\hertz} PC.

                        \subsection{Class structure}
                                \label{sect:TPMconvex:philosophy}

The most important objects inherit from the abstract, generic base classes 
\code{asteroid} or \code{ThermalModelConvex} (the latter inherits from the more abstract \code{ThermalModel}).

\code{asteroid} is a purely abstract base class, only objects belonging to the sub-class \code{TriangulatedConvex} can be instantiated. 
Within its scope, all required information on the asteroid shape and spin state is stored, as well as 
the constants $H$, $G$, and $\epsilon$. \pv\ is a \code{protected} variable within the scope of \code{asteroid}, its value can be changed using a routine which also updates the Bond albedo $A$ and the diameter scale factor $s$ (since $H$ is constant, changing \pv\ is equivalent to changing the diameter).
\code{TriangulatedConvex} contains several routines to calculate disk-integrated thermal fluxes. These routines rely on \code{ThermalModelConvex} objects (see below) to calculate flux contributions from single facets; the actual model to be used is specified by passing a reference to the appropriate \code{ThermalModelConvex} object.
The transformation of heliocentric and observer-centric asteroid coordinates into asteroid-centric directions towards the Sun or the observer \seesect{sect:TPM:bodyfix} is performed within the scope of \code{TriangulatedConvex}.

The chief interface between \code{asteroid} and \code{ThermalModelConvex} objects are the functions \code{fluxModFactors} and \code{ThermalLightCurveModFactors} within the scope of \code{ThermalModelConvex}. They return a dimensionless flux value or a dimensionless thermal lightcurve for a given time, observing geometry, and a single facet.
All returned flux values must be multiplied by a constant factor of $2\pi\epsilon h c^2/ (\Delta^2\lambda^5) \times s^2$ (with the scale factor $s$ defined in \eqrefpage{eq:TPM:intrinsicD}) to convert them into units of \watt\per\metre\squared\per\micron.
This multiplication is done by the calling routine within the scope of \code{asteroid} after adding contributions from all facets.
\code{fluxModFactors} is overloaded to allow calculation of fluxes for either a single wavelengths or simultaneously for a vector of wavelengths, the latter significantly increases the efficiency if spectra are calculated for non-vanishing values of thermal inertia.

Important auxiliary classes include \code{ConvexFile}, \code{SpinState}, and \code{fitFileSI}. The latter serves to read in an ASCII file containing thermal flux values along with the epoch and observing geometry of the observations.
All model flux values are in units of \watt\per\meter\squared\per\micron. The auxiliary class \code{fitFileMJy} is used to read in fit files with fluxes in units of \milli\Jy\ and to convert them into units of \watt\per\metre\squared\per\micron; \code{fitFileMJy} inherits from \code{fitFileSI}.

In order to fit thermal data, the TPM code is used to output ASCII files containing $\chi^2$ for different combinations of thermal parameters which are then manually analyzed. 
Output files are in a suitable format for  plotting using the open-source software \code{gnuplot}.
For each combination of thermal parameters, a best-fit \pv\  is determined analogous to our NEATM-fitting approach \seeeqpage{eq:NEATM:fitting}. If the crater density is considered as a fit parameter, linear regression is used to  analytically determine the best-fit crater density.
$\chi^2$ values are determined by calling an appropriate function at the \code{main}-routine level.
Prior to this, the data file (or files) are read in and 
a \code{TriangulatedConvex} object is instantiated with all asteroid constants including the shape model (passed as a \code{ConvexFile} object) and the \code{SpinState}.
Then, for each combination of relevant model parameters, a \code{ThermalModelConvex} object is instantiated,%
\footnote{ If thermal inertia is considered and if data were taken at more than one epoch, data from each epoch should be stored within a separate fit file, and separate \code{ThermalModelConvex} objects should be instantiated for each.
Differences in heliocentric distance $r$ translate into differences in thermal parameter $\Theta$ for otherwise constant parameters.}
which is used to calculate model fluxes for the times and observing geometry given in the fit files. 

                        \subsection{Thermal conduction}
                                \label{sect:TPMconvex:TI}

The subclass of \code{ThermalModelConvex}
for numerical modeling of thermal conduction on smooth surface elements 
 is \code{ThermalInertiaOnlyConvex}.
Thermal conduction is modeled as described in \sectref{sect:TPM:TI}.
The dimensionless heat diffusion equation (\eqref{eq:heatconduction_dimensionless}) in the subsoil and the boundary condition at the surface (\eqref{eq:boundarycondition_dimensionless}) are discretized in a straightforward fully explicit way with equidistant sampling points in space and time. This is numerically stable provided the time resolution is not too coarse, specifically  \citep[see][sect.\ 19.2]{NR} the parameter \code{dTdZ2} (dimensionless time resolution divided by the square of the dimensionless depth resolution) must not exceed 0.5. The numerical integration is truncated at a certain depth, at which the infinite-depth boundary condition \eqref{eq:boundarycondition2_dimensionless} is taken into account.

In the constructor, the thermal parameter and the desired fractional accuracy goal are specified as well as the discretization parameters \code{nTime} (number of time steps), \code{nZ} (number of depth steps), and \code{zMax} (maximum depth in units of skin depths). The constructor checks whether the stability criterion mentioned above is met and throws an exception otherwise.

To calculate surface temperatures,  the cosine of the solar angular zenith distance $\mu_S$, clipped to be non-negative, is determined for each time step and stored into an array. If all entries are essentially zero, zero flux is returned (the facet is not illuminated for any time of day).
Then an array containing the temperature profile is initialized with a constant temperature distribution,%
\footnote{ It has been verified that the final surface temperature is independent of that initialization.}
and the asteroid is spun until the surface temperature at the desired time has remained constant to within the user-specified fractional accuracy goal. For all time steps, the temperature profile is kept in the computer's RAM, which proved helpful for debugging and validating the code.
An asteroid revolution consists of \code{nTime} time steps. For each, the new subsoil temperature profile \code{newP[i]} (with $1\leq i \leq \code{nZ}-2$) is determined from the old profile \code{oldP[i]} using
\begin{equation}
  \label{eq:TPMconvex:conduction1}
\code{newP[i]} = \code{oldP[i]} + \code{dTdZ2} \left( \code{oldP[i-1]} + \code{oldP[i+1]} - 2\code{oldP[i]}\right),
\end{equation}
with the parameter \code{dTdZ2} defined above.
The surface temperature \code{newP[0]} is determined from a  discretization of the boundary condition (\eqref{eq:boundarycondition_dimensionless})
\begin{equation}
  \label{eq:TPMconvex:boundary@surface}
\code{newP[0]}^4 = \mu_S + \Theta\frac{\code{newP[1]}-\code{newP[0]}}{\code{dZ}}  
\end{equation}
with the dimensionless spatial resolution \code{dZ}. This nonlinear equation is solved using Newton's method, with \code{oldP[0]} as a first guess for \code{newP[0]}.
The new temperature at the maximum depth considered, \code{newP[nZ-1]}, is calculated from \eqref{eq:TPMconvex:conduction1} assuming that the fictitious \code{oldP[nZ]} equals \code{oldP[nZ-1]}, thus approximating the  boundary condition at infinite depth (\eqref{eq:boundarycondition2_dimensionless})
\begin{equation}
  \label{eq:TPMconvex:boundary2}
  \code{newP[nZ-1]} = \code{oldP[nZ-1]} + \code{dTdZ2}*\left(\code{oldP[nZ-2]} - \code{oldP[nZ-1]}\right).	
\end{equation}

Typically used values for \code{nZ}, \code{zMax}, and \code{nTime} are 25, 6, and 300, respectively, a typical value for the fractional accuracy goal is 0.0001.

The fully explicit discretization scheme used to solve the heat diffusion equation and the discretization of the boundary conditions are fully encapsulated within the scope of the class \code{ThermalInertia} and its member class \code{TimeStep}. Other discretization schemes, such as Crank-Nicholson \citep[see, e.g.,][sect.\ 19.2]{NR} can be implemented easily without changes to the remaining code.

                        \subsection{Beaming}
                                \label{sect:TPMconvex:beaming}
                                
\paragraph{Without thermal inertia}

Fluxes are calculated by 
\code{HemisphericNoInertia} objects, which inherit from \code{ThermalModelConvex}.
Returned fluxes are for a crater density of \unit{100}{\%} and
must be combined with ``non-cratered'' fluxes in the calling function to account for lower crater densities.

The temperature distribution inside the crater follows from \eqref{eq:beaming:T}, fluxes are obtained by numerically integrating \eqref{eq:beaming:Fallorders}.%
\footnote{ A class that integrates Lagerros' approximation \eqref{eq:beaming:FLagerros} has also been implemented---it differs from that considered herein by a mere redefinition of a constant parameter.}

The two-dimensional integral over the crater surface is performed separately for $\phi$ and $\theta$.  For each,  a Simpson quadrature algorithm has been implemented 
which adaptively refines the step width until the user-defined fractional accuracy goal is reached \citep[see, e.g.,][chapt.\ 4]{NR}.

Care was taken to minimize the number of calls to the numerically expensive trigonometric functions. E.g., in the polar integral the variable $\theta$ is transformed into $z=\cos\theta$ such that $\int_0^{\gamma/2}\sin\theta\textd\theta$ becomes $-\int_{\cos\gamma/2}^1\textd z$.

\paragraph{With thermal inertia}
As has been discussed in \sectref{sect:beaming:approximation}, thermal conduction inside craters is modeled in an approximative way.
Model fluxes are calculated by objects belonging to the class \code{InertiaTimesCraterLagerros}, which own  respectively one instance of \code{HemisphericNoInertia} and of \code{ThermalInertiaOnlyConvex}.

The \code{ThermalInertiaOnlyConvex} object is used to calculate the temperature at a smooth surface patch with thermal conduction.
If the Sun is below local horizon ($\mu_S\leq0$), the crater routine is not called, and fluxes are returned corresponding to that temperature.
Otherwise, the ratio of the temperature with and without thermal conduction is determined (without thermal inertia, $T/\TSS=\sqrt[4]{\mu_S}$)
and clipped to be $\leq 1.3$ (see discussion in \sectref{sect:beaming:approximation}).
``Cratered'' model fluxes are calculated with a rescaled temperature distribution. 
This is accomplished without 
changes to the implementation of \code{HemisphericNoInertia} by using a trick which is based on the fact that \code{fluxModFactors} is a function of $\lambda T$, hence rescaling the temperature $T$ is equivalent to rescaling the wavelength $\lambda$ by the same factor (note that the coefficient of the Planck function containing $\lambda^{-5}$ is not multiplied in \code{fluxModFactors} but in the calling function).
Different fractional accuracy goals can be chosen for the \code{HemisphericNoInertia} and \code{ThermalInertiaOnlyConvex} objects.

        \section{Validation}
                \label{sect:TPMconvex:validation}

In this section,  tests for internal model consistency are reported, with an emphasis  on validating the model for application to NEA data, i.e.\ for large phase angles ($>\unit{30}{\degree}$) and for thermal inertias up to \unit{2500}{\TIunit}. 
Another important model validation was its application to thermal-infrared observations of a well-studied reference NEA, (433) Eros, which will be reported in \sectref{sect:Eros}.

                        \subsection{Thermal conduction}
                        \label{sect:TPMconvex:validation:TI}

A first qualitative validation of our  thermal-conduction model is from visual inspection of \figrefpage{fig:thermal:TI}, which was generated using the TPM code. As required, increasing thermal inertia reduces the diurnal temperature contrast and shifts the temperature peak towards the afternoon side.
Additionally, the peak temperature at low thermal inertia approaches the theoretical value of \TSS\ around \unit{376}{\kelvin} (for the parameters specified), while for large thermal inertia the essentially constant temperature  approaches $\TSS/\sqrt[4]{\pi}$ (see \sectref{sect:FRM}) or roughly \unit{283}{\kelvin} as required.

Due to conservation of energy, the thermally emitted power integrated over one asteroid revolution must match the total absorbed solar power.
This was numerically checked, using the discretization parameters stated at the end of \sectref{sect:TPMconvex:TI}, a fractional accuracy goal of 0.001, and assuming
the situation depicted in \figref{fig:thermal:TI}---i.e., facet at the equator, $r=\unit{1.1}{\AU}$, $A=0.1$, $\mathcal{P}=\unit{6}{\hour}$.
For a thermal inertia of \unit{10,000}{\TIunit}, the total model emission was found to be too low by roughly \unit{16}{\%}, around \unit{4}{\%} too low for \unit{2500}{\TIunit}, and around \unit{1}{\%} for \unit{50}{\TIunit}.
Tightening the fractional accuracy goal to 0.0001 incurs a penalty of largely increased program run time, but leads to a conservation of energy  to within \unit{1.5}{\%} for thermal-inertia values up to \unit{10,000}{\TIunit}, to \unit{0.4}{\%} for thermal inertia up to \unit{2,500}{\TIunits}, and much better for lower thermal inertia. 
No asteroid studied to date displays a thermal inertia larger than \unit{1000}{\TIunit}, hence a fractional accuracy goal of 0.001 is typically sufficient.

We wish to stress that the  numerical effort for sufficiently accurate modeling of thermal conduction increases with thermal inertia; TPM codes suitable for MBAs may therefore be unsuitable for NEAs with typically much larger thermal inertia.

                        \subsection{Beaming without thermal conduction}
                        \label{sect:TPMconvex:validation:beaming}

As detailed in \sectref{sect:thermal:beaming}, thermal-infrared beaming is an enhancement relative to a Lambertian surface both in absolute flux level and in apparent color temperature for observations taken at low phase angles.

\begin{figure}
\centering
\includegraphics[angle=-90, width=0.7\linewidth]{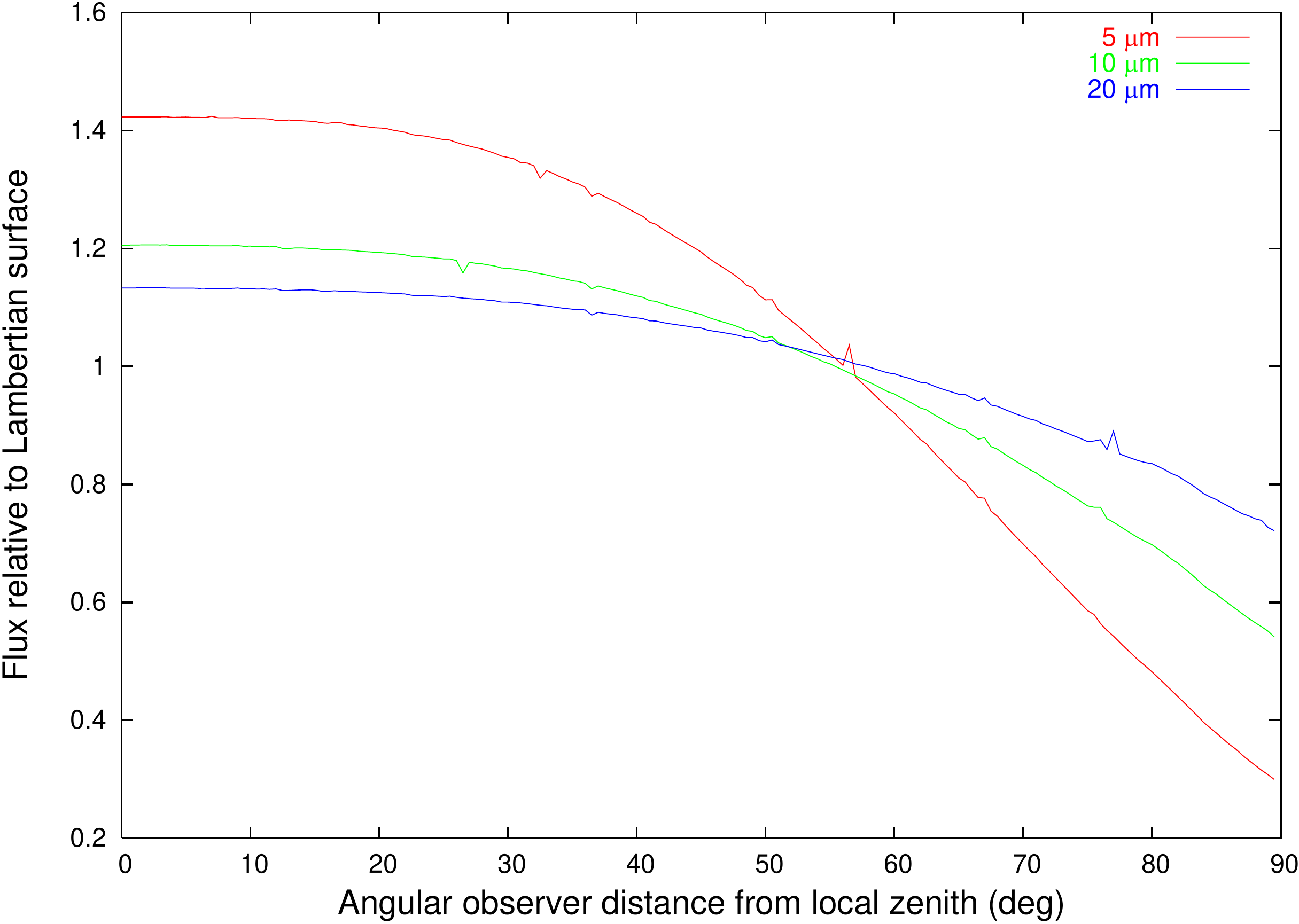}
  \caption[Relative flux enhancement due to a hemispheric crater ($\gamma=\unit{180}{\degree}$) as a function of observer angular distance from local zenith; the Sun is at local zenith.]{Relative flux enhancement due to a hemispheric crater ($\gamma=\unit{180}{\degree}$) as a function of observer angular distance from local zenith; the Sun is at local zenith. Further model parameters: $r=\unit{1.1}{\AU}$, $A=0.1$, $\epsilon=0.9$. The slight ``wiggles'' in the lines are due to numerical noise, they disappear for more stringent accuracy goals (we here use 0.001, a typical value).}
  \label{fig:beaming:Sun@zenith}
\end{figure}

\begin{figure}
  \centering
\includegraphics[angle=-90, width=0.7\linewidth]{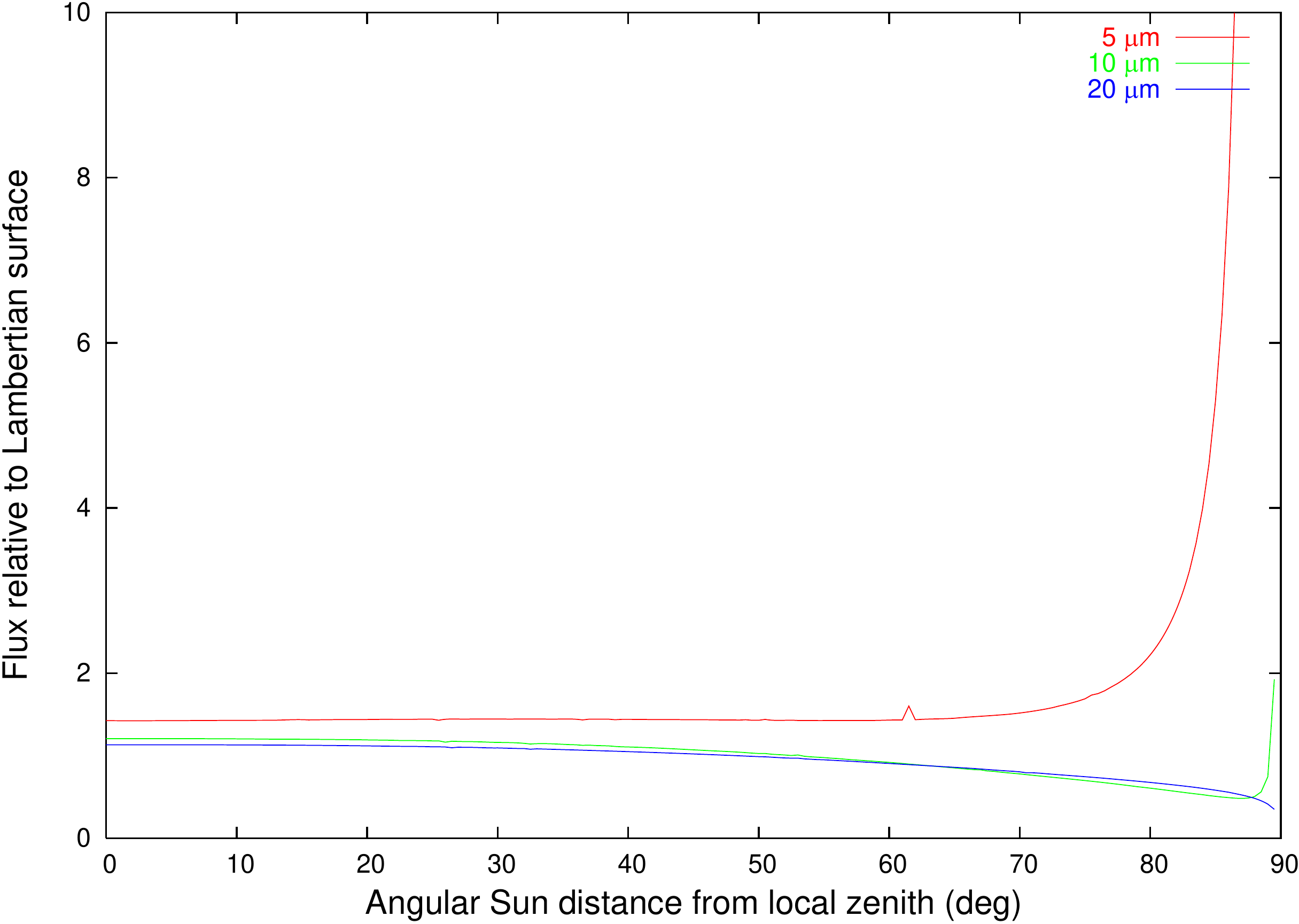}
  \caption{Relative flux enhancement as in \figref{fig:beaming:Sun@zenith}, but as a function of solar angular distance from local zenith, with observer at local zenith.}
  \label{fig:beaming:Obs@zenith}
\end{figure}

\begin{figure}
  \centering
\includegraphics[angle=-90, width=0.7\linewidth]{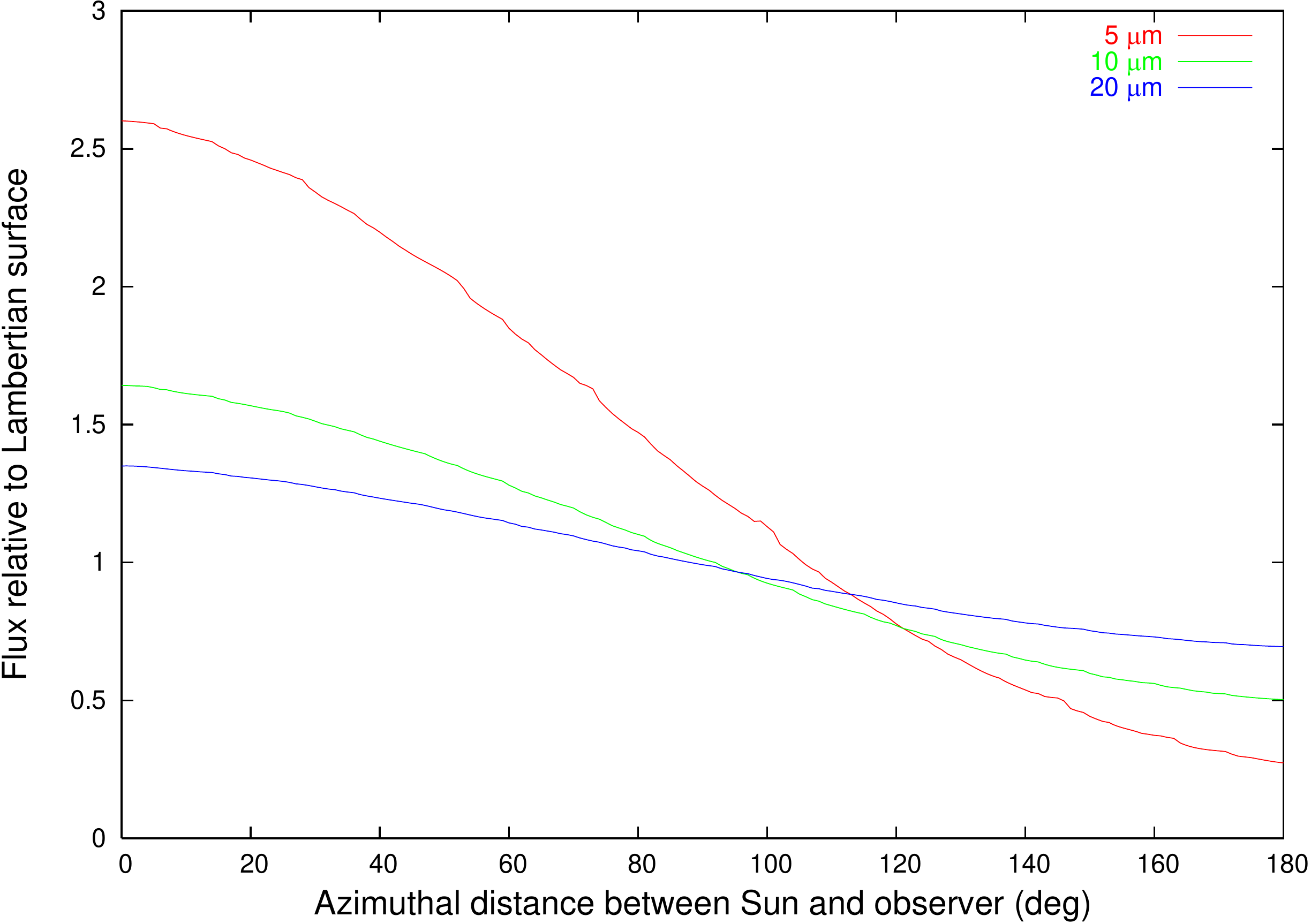}
  \caption{Relative flux enhancement as in \figref{fig:beaming:Sun@zenith} with both Sun and observer  placed at a zenith distance of \unit{45}{\degree}, the azimuthal distance between the two is varied.}
  \label{fig:beaming:azi}
\end{figure}

Figures \ref{fig:beaming:Sun@zenith}, \ref{fig:beaming:Obs@zenith}, and \ref{fig:beaming:azi} show plots of crater fluxes relative to a flat Lambertian surface for hemispherical craters ($\gamma=\unit{180}{\degree}$) and for different viewing geometries (many other opening angles were tested, leading to qualitatively equi\-valent results).
As expected, beaming not only enhances the flux at low phase angles (i.e.\ low zenith distances in Figures \ref{fig:beaming:Sun@zenith} and \ref{fig:beaming:Obs@zenith}, low azimuthal distance in \figref{fig:beaming:azi}) but also the apparent color temperature: the flux ratio increases with decreasing wavelength. This is compensated for at large phase angles, when fluxes are reduced relative to a Lambertian emitter---more so at short wavelengths such that the color temperature is reduced relative to a Lambertian emitter.

An important consistency check is to verify the conservation of energy, i.e.\ to  compare ingoing solar flux with the total outgoing energy. For a hemispherical crater of opening angle $\gamma=\unit{180}{\degree}$ and projected area of $\pi$ this means
\begin{equation}
\label{eq:TPMconvex:energy}
\pi\mu_S\left(1-A_\text{corr}\right)\frac{S}{r^2} = \int\limits_0^\infty \textd\lambda \int\limits_\Area\textd\Area\ F(\lambda)  
\end{equation}
for, e.g., a hemisphere \Area\ above the crater, with the observable flux \eqref{eq:beaming:Fallorders} and the corrected Bond albedo \eqref{eq:beaming:correctedBond}.

Conservation of energy was tested 
 by numerically integrating \eqref{eq:TPMconvex:energy} for different values of solar zenith distance (assuming $r=\unit{1.1}{\AU}$, $A=0.1$, and $\epsilon=0.9$). To this end, the integral over $\textd\lambda$ was discretized and truncated with 80 equidistant $\lambda$ steps between 0.5 and \unit{79.5}{\micron} using the Simpson integration scheme \citep[see, e.g.,][]{NR}; for the integral over $\textd\Area$, a Monte-Carlo scheme was implemented with 2,500 uniformly distributed random directions towards the observer,%
\footnote{ Note that the distribution of vectors resulting from uniformly distributed polar coordinates  would \emph{not} be uniform but rather biased towards the poles. Instead, we have drawn three  Cartesian coordinates per vector uniformly distributed between $-1$ and 1 (0 and 1 for the $z$ coordinate, so vectors are above local horizon) and rescaled the resulting vectors to unit length.}
each Monte-Carlo integration was executed four times with identical parameters in order to gage the statistical noise.
The typical scatter among the four runs is  \unit{1}{\%}.
In order to gage the absolute accuracy,  disk-integrated
thermal fluxes were generated for a spherical Lambertian emitter without thermal inertia and integrated those fluxes over $\textd\lambda$ and \textd\Area. The resulting total thermal power emitted by the asteroid equals the total absorbed power to within the numerical noise ($\sim\unit{1}{\%}$).

To check the crater routine for conservation of energy, it turned out to be convenient to check it against a Lambertian emitter.
To this end, two integrals are calculated using the discretization above, where once the integrand is  the output of the crater routine and once that of the Lambertian routine,%
\footnote{ In both instances, the \code{fluxModFactors} routines of the respective objects are called, hence returned values must be multiplied by $\lambda^{-5}$.}
then the two resulting integrals are compared. 
Note that the crater induces a reduced albedo $A_\text{corr}$ relative to the smooth surface \seesect{sect:beaming:optical}
due to the increased optical absorption inside the crater.
Hence, the ``crater integral'' should be larger than the Lambertian integral by a factor of 
$(1-A_\text{corr})/(1-A)$, which equals roughly 1.05 for the parameters considered here \seefigpage{fig:beaming:Bond}.
For  a fractional accuracy goal of 0.01, energy was seen to be conserved within \unit{1}{\%} (the Sun was placed at zenith distances of 0, 30, and \unit{60}{\degree}).

\subsection{Beaming with thermal conduction}
\label{sect:TPMconvex:validation:beaminginertia}

When thermal inertia is considered,  ingoing and outgoing radiation are no longer in instantaneous equilibrium
but only after integration over a asteroid revolution.
The latter integral reduces to a mere sum of contributions from each of the \code{nTime}  time steps, where for each the integral over $\textd\lambda$ and \textd\Area\ is performed as above.
As in the case without thermal inertia, the integral over fluxes returned by an \code{InertiaTimesCraterLagerros} object are compared to the integral over  Lambertian fluxes multiplied by $(1-A_\text{corr})/(1-A)$.

Integration runs were performed for $\mathcal{P}=\unit{6}{\hour}$, and the remaining parameters as in the rest of this section.
The Sun was placed at an angular distance of \unit{90}{\degree} from the spin pole (i.e.\ over the  equator), facets pointing 30, 60, and \unit{90}{\degree} away from the pole were considered, for thermal-inertia values of 50, 500, and \unit{2500}{\TIunit}.
For a fractional accuracy goal of 0.0001 for the treatment of thermal inertia (as in the case of thermal inertia alone) and of 0.01 for cratering (as for cratering alone), energy was found to be conserved to within a few percent in all cases.


%% file: fitting.tex
In principle, 
our TPM contains four parameters which 
can be varied independently to obtain the best fit to the data: the diameter $D$ (which is related to \pv\ through the optical magnitude $H$), the thermal inertia $\Gamma$,  the crater opening angle $\gamma$, and the crater density $\rho_c$.
It has been shown by \citet{Emery1998} and \citet{LagerrosIV} that 
the modeled effect of surface roughness is practically a function of a single parameter, the mean surface slope $\bar{\Theta}$, which is a function of crater opening angle and crater density, reducing the dimensionality of the parameter space by one.

Originally \citep[see][]{Ito1}, our approach was to fix the crater opening angle at a constant value and to determine the combination of diameter, thermal inertia, and crater density that best fit the data.
In this approach we exploited the fact that, as discussed in \sectref{sect:TPM:parms}, model fluxes depend linearly on the assumed crater density. For a given combination of diameter and thermal inertia it is therefore possible to analytically determine the best-fit crater density through linear regression (with the constraint that the density must be non-negative and \unit{$\leq100$}{\%}).

Later, however, we found that the crater roughness is often not significantly constrained by the data.
We find it more instructive and more transparent to use four preset combinations of roughness parameters which span the range of possible surface roughness,
and to determine the best-fit diameter and thermal inertia for each.
The results  are then compared with one another, potentially showing that some roughness model fits the data significantly better than others, and otherwise illustrating the range of roughness-induced uncertainty in thermal inertia and diameter.
The four roughness models considered throughout this thesis are
\begin{description}
\item[No roughness] $\gamma=0$, $\rho_c=0$
\item[Low roughness] $\gamma=\unit{117.70116}{\degree}$, $\rho_c=0.4$
\item[Default roughness] $\gamma=\unit{144.59046}{\degree}$, $\rho_c=0.6$
\item[High roughness] $\gamma=\unit{151.75834}{\degree}$, $\rho_c=1.0$
\end{description}
The parameters for ``low'', ``default'', and ``high'' roughness 
have been  defined by \citet[see also references therein]{Mueller2004} on the basis of thermal-infrared analyses of MBAs.

Synthetic thermal fluxes are generated for the observing geometry and at the wavelengths of the observational data.
To fit diameter and thermal inertia for a given set of roughness parameters, 
model fluxes are calculated on an equidistant grid of thermal-inertia values from the largest value considered down to 0. 
For each value of thermal inertia, a best-fit diameter is determined
from the assumed input diameter and the scale factor $\kappa$
which minimizes
\begin{equation}
  \label{eq:NEAfitting:chi2}
  \chi^2 = \sum_{i=1}^n\left(\frac{\kappa\cdot m_i-d_i}{\sigma_i}\right)^2
\end{equation}
(for data points $d_i$ with uncertainties $\sigma_i$ and synthetic model fluxes $m_i$; see \eqref{eq:NEATM:fitting}). The  best-fit diameter and the corresponding $\chi^2$ are stored.
We neglect that changing the diameter is, for constant $H$, equivalent to changing the albedo, which in principle changes model temperatures and thermal fluxes in a nontrivial way. A full recomputation of model fluxes would, however, be computationally very expensive.
We would expect that it is safe to neglect the influence of albedo on temperature if \pv\ is reasonably constant, i.e.\ if the resulting $\kappa$ value is close to unity (see also \sectref{sect:NEATM:fitting}).

For the first (largest) thermal-inertia value considered in each series, an arbitrary default-value of \pv\ is used.
For all following steps, model fluxes are calculated assuming the best-fit \pv\ determined in the previous thermal-inertia step.
For not too large steps in thermal inertia, 
$\kappa$ typically stabilizes to values close to unity within a small number of iterations (see, e.g., \sectref{sect:Eros:results} for a discussion)
such that our approximation above becomes uncritical.

For each roughness model, an ASCII file is output which can be used as a data file for GNUplot. Those files contain, among other things, data lines with entries for thermal inertia, the best-fit $\chi^2$ obtained, and the corresponding best-fit \pv.
Best-fit roughness models, thermal inertias, and diameters are then determined from an analysis of the obtained plots (see, e.g., \sectref{sect:Eros:results}).
The run time scales with the number of data points and  the number of facets in the shape model, typical values range between a few hours and a few days on a \unit{2.66}{\giga\hertz} PC.


%% file: IRTF.tex
\textit{
The Infrared Telescope Facility (IRTF) is a \unit{3.0}{\metre} telescope on Mauna Kea~/ \Hawaii, which is optimized for infrared observations of Solar-System objects. It is operated by the Institute for Astronomy (University of \Hawaii) under a cooperative agreement with NASA.
Observing time with the IRTF is made available twice a year in an international open competition.
}

\textit{
We have been awarded IRTF time for thermal-infrared observations of asteroids in the semesters 2004A, 2004B, 2005A, and 2005B. Techniques have been established to perform IRTF observations remotely from Berlin.
We have used the Mid-InfraRed Spectrometer and Imager (MIRSI) and  the optical CCD camera ``Apogee,'' the latter for support observations.
This thesis also contains IRTF thermal-infrared data which have been obtained with the JPL Mid-InfraRed Large-well Imager (MIRLIN). MIRLIN
was replaced at the IRTF by MIRSI in 2003.
}

\textit{
This chapter contains a brief description of the  capabilities of IRTF, MIRSI, and Apogee as needed for our purposes, followed by a section about the remote observing technique established by us (\sectref{sect:IRTF:remote-control}). We then explain our observing strategies (\sectref{sect:IRTF:strategy}) and detail the data reduction techniques used by us (\sectref{sect:IRTF:reduction}).
}

	\section{IRTF setup and instrumentation}
                \label{sect:IRTF:setup}

The IRTF is part of the Mauna Kea Observatory~/ \Hawaii\ (IAU observatory code 568). Mauna Kea is among the most favorable locations for infrared telescopes worldwide owing to its  altitude of some \unit{4200}{\metre} above sea level, low atmospheric humidity, and good seeing.

The IRTF is a Cassegrain telescope optimized for infrared astronomy.
Its primary mirror has an aperture of \unit{3.0}{\metre}.
It can be pointed to declinations between \unit{+67}{\degree} and \unit{-56}{\degree} at an hour angle within \unit{$\pm5$}{\hour}.
Differential tracking can be used for for observations of Solar-System objects including  NEAs with fast apparent motion.

The Cassegrain secondary mirror provides a $f/35$ beam at the Cassegrain focus where the science instruments are typically situated (an additional Coud\'{e} focus is available).
IRTF imaging is diffraction limited at mid-infrared wavelengths.
The Cassegrain secondary mirror can be oscillated at a frequency up to \unit{40}{\hertz} at a maximum projected amplitude of \unit{6}{\arcmin}, which is important for, e.g.,  MIRSI observations (``chopping''; see \sectref{sect:MIRSI}).

Up to four science instruments can be mounted simultaneously at the Cassegrain focus, instrument change-over can normally be accommodated within \unit{30}{\min}; significantly faster in the case of 
changes between MIRSI and Apogee, which require only a filter to be moved and require only a few seconds.

The telescope shutter consists of two parts for airmass ranges of up to 1.5 (maximum zenith distance \unit{$\sim48$}{\degree}, only one shutter part open) or up to 2.9 (both parts open) (see \eqrefpage{eq:IRTF:airmass} for a definition of airmass).

\begin{figure}
  \centering
\includegraphics[width=0.8\linewidth]{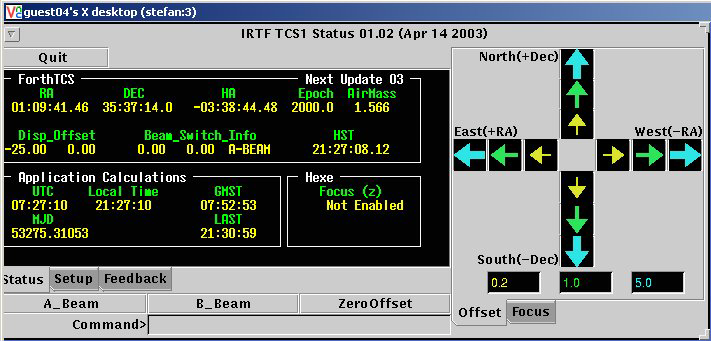}
  \caption{tcs1\_status, the IRTF tool for focusing and  fine tuning of  pointing (screenshot from a DLR machine during an IRTF observing run on 27 Sept.\ 2004).}
  \label{fig:IRTF:tcs1}
\end{figure}

Telescope  pointing and tracking is done  by the Telescope Operator (TO) who also sets up the instruments. Focusing and fine tuning of the pointing is in the responsibility of the observer;  the IRTF software tcs1\_status is used for this purpose \seefig{fig:IRTF:tcs1}.%
\footnote{ See \url{http://irtfweb.ifa.hawaii.edu/observing/remote_obs/tcs1_status.pdf}.}

		\subsection{MIRSI}
                        \label{sect:MIRSI}
The Mid-Infrared Spectrometer and Imager MIRSI is a mid-infrared camera with imaging capabilities between $\sim5$ and \unit{26}{\micron} and additional spectroscopic capabilities over the 8--14 and \unit{17--26}{\micron} atmospheric windows.
MIRSI has been built at the Boston University \citep{MIRSI} and 
 has been commissioned at the IRTF in 2003, replacing MIRLIN.
It is not an IRTF facility instrument but has been made available by the instrument PI (the late L.\ Deutsch; now J.\ Hora) on a collaborative basis. The MIRSI detector is operated at a temperature between \unit{6--12}{\kelvin} inside a helium cooled dewar.
The MIRSI control computer system is connected to the IRTF mirror control unit such that it can control chop and  nod movements (see below).

\begin{table}
     \caption[Overview of MIRSI filters used in our programs: wavelengths and sensitivities]{Overview of MIRSI filters used in our programs, together with their estimated $1\sigma$ point-source sensitivity in \unit{1}{\minute} on-source integration time if mounted on the IRTF \citep[source:][for the filter specifications and \url{http://cfa-www.harvard.edu/mirsi/mirsi_spec.html} for the sensitivities]{MIRSI}.}
     \label{table:MIRSI:filters}
     \centering
     \begin{tabular}{r|ll}
\toprule
Central wavelength & Width & $1\sigma$ sensitivity for \unit{1}{\minute} integration \\
(\micron) & (\%) & (\milli\Jy) \\
\midrule
4.9 & 21  & 16.2 \\
7.8 & 9.0 & 46.3 \\
8.7 & 8.9 & 59.1 \\
9.8 & 9.4 & 19.7 \\
11.7 & 9.9& 24.0 \\
12.3 & 9.6& (no value given)\\
18.4 & 8.0& 57.3 \\
24.8 & 7.9& (no value given)\\
\bottomrule
\end{tabular}
\end{table}

Our programs make use of MIRSI in imaging mode, the spectroscopy mode is disregarded in the following.
The Si:As impurity-band-conduction detector has $320\times240$ pixels at a projected pixel scale of \unit{0.27}{\arcsec} when mounted on the IRTF, corresponding to a field-of-view (FOV) size around $\unit{85}{\arcsec}\times \unit{64}{\arcsec}$. At MIRSI wavelengths, IRTF imaging is diffraction limited.
See \tableref{table:MIRSI:filters} for a list of MIRSI filters used in our observations.

Due to the high level of background radiation, single MIRSI exposures (frames) typically saturate within fractions of a second, depending on the filter and on atmospheric conditions.
MIRSI observations therefore consist of long series of  frames, where the desired frame time and total on-source integration time are set by the observer.
Depending on the observing mode, observations are taken at different positions to enable correction for background radiation.
For our observations, MIRSI was used in chop/nod mode. In chop/node mode, MIRSI commands the telescope's Cassegrain secondary mirror to ``chop'', i.e.\ to oscillate in a square wave pattern of a user-defined frequency and amplitude (\unit{4}{\hertz} and \unit{20}{\arcsec}, in our case). At each chop position, an integer number of frames is taken. 
Half of the total integration time is chopped at the specified location, then the primary mirror is slewed (``nodded'') to an offset position where a second chop set is taken.
Combining chopping and nodding enables more accurate subtraction of background radiation \citep[see, e.g.,][sect.\ 3.3, for a detailed discussion]{Delbo2004}. 
The data obtained
in a chop/nod exposure are stored as a single \code{FITS} file with three extensions where each of the four contained images is the sum of all frames taken at one particular beam position.

The directions and amplitudes of the chop and nod movements are set by the TO upon the observer's request. Timing parameters are specified by the observer through the MIRSI control software.

Due to overheads for, e.g., detector read-out or wait times for the secondary mirror to settle after a chop movement, there is a significant difference between the integration time (i.e.\ the total time spent collecting measured photons) and the image acquisition time (integration time plus all overheads). All other parameters kept constant, shortening the frame time increases the ratio of acquisition time over integration time. 

\begin{figure}
  \centering
\includegraphics[width=0.8\linewidth]{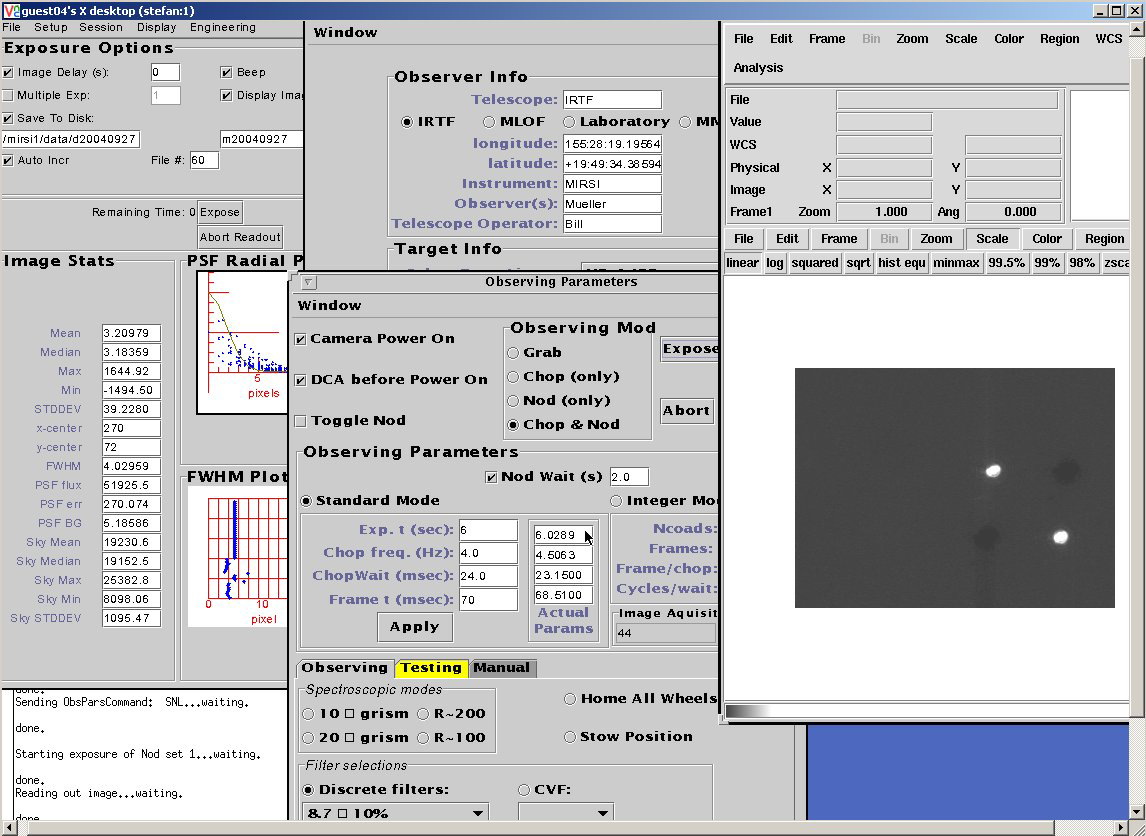}
  \caption{Screenshot of the MIRSI control software during a remote IRTF observing run on 27 Sept.\ 2004.}
  \label{fig:IRTF:MIRSIsoft}
\end{figure}

In the windows opened by the MIRSI control software (see \figref{fig:IRTF:MIRSIsoft}, see also \url{http://www.cfa.harvard.edu/mirsi/MIRSI_Obs_Guide.pdf} for a user guide),
the observer specifies the observing mode and corresponding timing parameters, and controls the filter wheel.
Information about the target can be entered; this is used by the software to store the target name and the airmass in the header of the resulting \code{FITS} file (in addition to the filter used, observation time and duration, etc.). 
Note that the airmass is \emph{not} taken directly from the telescope control system (such as that displayed in tcs1\_status; see \figref{fig:IRTF:tcs1}) but is rather calculated from the system date, the telescope coordinates, and the target RA and dec.\ entered in the target information form---for the latter, it is important to strictly keep the format described in the observer guide because otherwise the program will crash, typically requiring a lengthy reboot of the control computer and  instrument re-initialization.

		\subsection{Apogee}
                        \label{sect:apogee}

Apogee is an optical-wavelength camera using a commercial $1024\times1024$ pixel CCD with a field of view of \unit{66}{\arcsec}. Several filters are available, we used a standard V filter in all our observations.
Apogee is thermoelectrically cooled to an operational temperature of \unit{$-30$}{\celsius} to reduce the noise level.
We typically used Apogee in $4\times4$--binning mode (thus shrinking the FOV to $256\times256$ pixels) to reduce the required detector readout times. 
Changeover between MIRSI and Apogee or vice-versa takes only a few seconds, enabling nearly-simultaneous optical photometry of MIRSI targets to be obtained.

Apogee is significantly more sensitive to asteroid flux than MIRSI.
Using the V filter, signal-to-noise ratios exceeding 100 are reached within a few tens of seconds for all asteroid targets which are sufficiently bright  for MIRSI observations.

The Apogee control software automatically saves all obtained data in \code{FITS} format.

        \section{Remote control of observations}
                \label{sect:IRTF:remote-control}

The IRTF supports remote control of observations using MIRSI and Apogee. We have established techniques to remotely control IRTF observations from DLR Berlin and have successfully observed remotely  a number of times.%
\footnote{ See also \url{http://solarsystem.dlr.de/HofW/nr/245/}.}

During remote observations, the  control programs for both MIRSI and Apogee are executed locally at the IRTF, but the screen output and input from mouse and keyboard can be transferred to/from any computer with access to the internet. For this purpose the VNC protocol is used, which is encrypted and password-protected. A VNC viewer program is required to run on the observer's machine.%
\footnote{ Freeware VNC viewers are available for  UNIX/Linux and Windows platforms at \url{http://www.realvnc.com/}.}
MIRSI and Apogee require one virtual VNC desktop each.
Additionally, observers require access to the program tcs1\_status for focusing and pointing fine-tuning.%
\footnote{ See \url{http://irtfweb.ifa.hawaii.edu/observing/remote_obs/tcs1_status.pdf}.}
It is advisable to run tcs1\_status on another computer within the Institute for Astronomy network. 
To remotely connect to the machine running tcs1\_status from a Windows machine in Berlin, it is most convenient to open a third VNC server on the \Hawaii\ machine running tcs1\_status.

Each VNC server on a machine is identified with its desktop number, which are counted upward 
starting with 1 (desktop 0 is the native desktop which cannot usually be forwarded).
VNC communication between two computers is performed through designated ports, where the default port for a virtual desktop with number $n$ is $5900+n$. Remote control of IRTF observations thus requires 
the ports 5901, 5902, and 5903 by default.
These ports are blocked by the DLR firewall, and the DLR IT administration is reluctant to open them.
Instead, these ports were ``tunneled'' using an encrypted SSH connection.

In addition to the VNC connections, a channel for direct communication with the telescope operator is required. The IRTF default for remote observers is a PolyCom system or, alternatively, a PC running NetMeeting.
Since neither possibility was easily feasible from within the DLR network (and because telephone contact over several hours of observing time would have been rather expensive and inconvenient), we proposed an alternative solution which is now among the default options offered by the IRTF, namely voice-over-IP communication. 

The Mauna Kea Weather Center (\url{http://hokukea.soest.hawaii.edu/index.cgi}) holds many useful resources to remotely check the 
 atmospheric conditions above Mauna Kea.
Astronomy-minded weather forecasts for the summit are available (which generally differ vastly from forecasts for the rest of the island) in addition to real-time webcam images from several telescopes at Mauna Kea, real-time meteograms from several telescopes, and nearly-real-time satellite imagery. 
A particularly useful tool is the CFHT ``skyprobe''
(\url{http://www.cfht.hawaii.edu/Instruments/Skyprobe/})
for estimates of the atmospheric extinction. Significant changes in the extinction typically indicate the presence of clouds.
After each observing run, we typically save a copy of the skyprobe plot for the respective night, in addition to a Mauna-Kea meteogram, in order to enable later assessment of the atmospheric conditions.

        \section{IRTF observing strategy}
                \label{sect:IRTF:strategy}

                \begin{figure}
                  \centering
\includegraphics[width=0.8\linewidth]{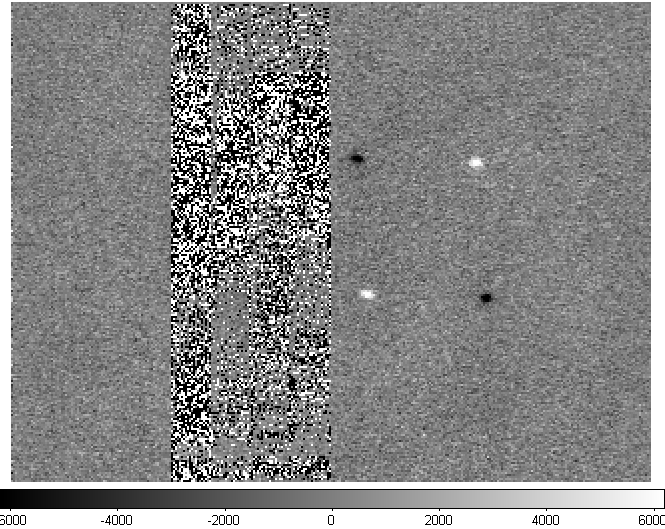}
                  \caption{Coadded and subtracted MIRSI image with the four beams not quite at the nominal position. Note the artefact pattern in the left half of the detector. (Observation of Itokawa, 10 July 2004, 12:53 UT---this frame could not be used to measure the target flux, see \tablerefpage{table:ito:MIRSI})}
                  \label{fig:IRTF:MIRSIgarbled}
                \end{figure}

\paragraph{General considerations}
All MIRSI observations are performed in chop/nod mode to enable subtraction of the celestial background. 
Since our targets are not spatially resolved, the directions of the chop and nod throws are uncritical; we used  \unit{20}{\arcsec} East-West and North-South, respectively.
We used a chop frequency of \unit{4}{\hertz}, a chop wait time of \unit{24}{\milli\second}, and a nod wait time of \unit{2}{\second}.

No flatfield for MIRSI is available and it would be very time consuming to obtain one; to minimize the possible effect of pixel-to-pixel variations in detector sensitivity, we have aimed at keeping all sources at a constant position on the chip using tcs1\_status. We chose a nominal position such that the center of the four beam positions is centered in the detector's $y$ coordinate and at \unit{75}{\%} in the $x$ direction, where the latter is motivated by the fact that the detector's ``left'' half is occasionally malfunctioning (see \figrefpage{fig:IRTF:MIRSIgarbled}).
The FOV is large enough to have all four beam positions on the chip, nevertheless.

Apogee observations are interspersed between the MIRSI observations. 
As for MIRSI, no flatfield is available for Apogee. The telescope is not moved between MIRSI and Apogee observations, such that also for Apogee the on-chip target position is reasonably constant, minimizing the possible effect of pixel-to-pixel variations in sensitivity. Apogee is used with a standard V filter in all our observations.

\paragraph{Calibration observations and observation planning}
To facilitate the derivation of absolutely calibrated infrared fluxes and optical V-magnitudes, suitable calibration standard stars are chosen from the lists of \citet[for MIRSI]{CohenVII,CohenX} and \citet[for Apogee, we picked calibration standards of similar spectral class to the Sun]{Landolt1973,Landolt1992} for each observing run.
Calibration standards are typically observed at the beginning and at the end of each observing run. If the atmospheric conditions appear slightly unstable, further calibration observations are interspersed between the target observations to gage the atmospheric stability.
To facilitate accurate correction for atmospheric extinction (also called airmass correction, see \eqrefpage{eq:IRTF:extinction}), calibration standards are typically observed at airmasses which bracket those of the primary scientific targets. If the target airmasses span a large range, a calibrator at an intermediate airmass is observed to gage the quality of the airmass correction. Whenever possible, the same calibration standard star is used for all required airmasses (i.e.\ it is reobserved at a later time) to minimize any possible cross-calibration problems.

The observations are planned such that the science targets are observed at the minimum possible airmass.
We furthermore aim at minimizing the number of required changes to the telescope shutter
(see \sectref{sect:IRTF:setup}) which occur when the telescope pointing transgresses an airmass boundary value of 1.5.

\paragraph{Starting an observing run}

After  starting and initializing all required control units for the telescope itself and the instruments, and, in the case of remote observations, after establishing all required communication channels, the instruments must be initialized (most of these tasks are performed by the TO but may require observer input).

As soon as the instruments are initialized, the telescope should be slewed to a bright MIRSI target, typically a calibration standard star, to focus the telescope using tcs1\_status (see \figrefpage{fig:IRTF:tcs1}).
Telescope focus is a function of local temperature, therefore it is required to check occasionally during an observation run whether the telescope is still  focused, particularly in the first half of the night when the telescope dome cools down significantly.

At the same time, suitable frame times have to be determined for each MIRSI filter to be used during the night:
Frame times are limited by the requirement that the sky background, which vastly dominates the flux at MIRSI wavelengths, must not bring the detector close to saturation, so that the detector response to the source flux remains linear.
On the other hand, frame times should be as long as possible to optimize the time efficiency of MIRSI observations (see \sectref{sect:MIRSI}).
Background levels can vary greatly on a night-to-night basis, so the optimum frame time must be determined for each MIRSI filter and each observing run.
The MIRSI software  analyses observational data ``on the fly'' and displays,  among other things,
 the average, median, and maximum value of sky background radiation (see \figrefpage{fig:IRTF:MIRSIsoft}).
Frame times should be chosen which bring the average sky level to values not greatly exceeding 20,000 as displayed by the MIRSI control software (Bus, 2004, private communication). 
Once determined, the filter-dependent frame times should be
used  throughout the observing run. 
Significant changes in the sky background during the night typically indicate non-photometric conditions, 
with the exception that sky levels  grow moderately with airmass and furthermore during twilight.

As soon as Apogee has reached its operational temperature of \unit{$-30$}{\celsius}, we take a number of bias frames. Dark frames are not usually taken.
During the first Apogee observations it is verified that  Apogee is on focus, otherwise the TO is asked to adjust the relative focus between MIRSI and Apogee.

\paragraph{During the observations}

When observing a faint science target, it is usually first observed through the \unit{11.6}{\micron} filter which, among MIRSI's filters, is most sensitive to asteroid thermal emission.
From the asteroid brightness in this filter it is estimated how long observations in other filters might take and whether challenging filters, such as the Q-band \unit{18.4}{\micron} filter, are tried at all.

During longer observations of a single target, we typically monitor its thermal lightcurve in the \unit{11.6}{\micron} filter and intersperse observations in other MIRSI filters and Apogee V-band observations.
It is particularly convenient to perform a quick Apogee observation
while MIRSI is busy changing filters, which may take up to a minute.%
\footnote{ This is a subjective estimate obtained during hectic observing nights.}
For asteroid targets sufficiently bright to be observed with MIRSI, required Apogee integration times with the V filter are of the order of tens of seconds at most, while instrument changeover between Apogee and MIRSI is done within seconds.

During observations, all observation parameters are logged using the log sheets contained at the end of the MIRSI observer guide.%
\footnote{ See \url{http://www.cfa.harvard.edu/mirsi/MIRSI_Obs_Guide.pdf}.}
Most importantly, the observing time, duration, target name, filter, and airmass are logged in the designated columns.
The ``comments'' column is used, among other things,  for logging  manual telescope offsets and changes to focus settings.

	\section{Data reduction}
                \label{sect:IRTF:reduction}

The measured asteroid flux is proportional to the detector response per integration time, the proportionality constant is determined from observations of calibration standard stars.
The detector response to exposures is 
determined through synthetic aperture photometry  after instrument-specific image-cleaning procedures.

Measured flux values must be corrected for the effect of atmospheric extinction (``airmass correction''). Corrections for the effect of finite filter breadth (``color correction'') are discussed.

		\subsection{MIRSI}

MIRSI data reduction is aided by a set of 
IDL routines developed 
 by Eric Volquardsen, available on-line at 
\url{http://irtfweb.ifa.hawaii.edu/~elv} and described at \url{http://irtfweb.ifa.hawaii.edu/~elv/mirsi_steps.txt}.
We use the routine \code{mirsi\_coad}
to  remove  instrument-specific artefacts and to coadd the observations taken at the four chop/nod beam positions with the appropriate signs.
%
Note that cosmic ray hits are not a source of concern for MIRSI observations because their effect is dwarfed by the celestial background.

After coadding, MIRSI images display the source four times with two positive  and two negative detections (see, e.g., \figrefpage{fig:IRTF:MIRSIgarbled}).
These four detections are copied pixel-wise into a single, ``registered'' detection after determining four 
non-overlapping regions centered on each of the four detections and inverting the sign of the regions surrounding the negative detections.
The edge length of these regions was chosen  large enough to facilitate  accurate background subtraction during synthetic aperture photometry, while guaranteeing that the  region does not exceed the chip boundary.
We typically use an edge length of 50 MIRSI pixels corresponding to \unit{13.5}{\arcsec}.

                \subsubsection{Synthetic aperture photometry}
                        \label{sect:IRTF:aperture}

Detector counts are determined by performing synthetic aperture photometry on the registered images.
To this end, the pixel content of a synthetic aperture, i.e.\ a circular region surrounding the source centroid, is added up and corrected for the amount of background radiation contained inside the synthetic aperture.
The background level is usually estimated from the pixel content of an annular region centered at the detection centroid. 
The background annulus should be far enough from the source to be uncontaminated with source flux but should be close enough to provide a good estimate of the local background level (which may be spatially variable).
Accurate flux estimation from faint detections (low $S/N$) depends critically on appropriate background estimates.
A practical criterion for the appropriateness of a background estimate
is whether or not it leads to a 
stable photometric growthcurve \citep{Howell1989}---see also the detailed discussion in \citet[chapter 3]{Delbo2004}.

Image registration and synthetic aperture photometry on MIRSI images can be automatized using the  IDL routines by E.\ Volquardsen. 
We felt it was safer to  perform these steps interactively, using the IDL-based registration and aperture-photometry tool by M.~\citet[sect.\ 3.6]{Delbo2004} which is a modified version of the ATV package by Aaron Barth (\url{http://www.astro.caltech.edu/~barth/atv/}).
Image registration is done manually as described above. The synthetic aperture photometry routine automatically determines the centroid of the registered source.
Care was taken to determine appropriate background values for each source by studying the photometric growthcurve.
To avoid systematic uncertainties, the same aperture radius was chosen for all observations obtained during one observing run in a specific filter.
To accommodate for possible temporal variations in the width of the point spread function, which may arise due to changes in seeing but also due to focus drifts, conservatively large aperture radii were chosen.

\subsubsection{Airmass correction}
        \label{sect:IRTF:MIRSIairmass}

Fluxes determined from ground-based astronomical observations must be corrected for the effect of atmospheric extinction, which reduces the amount of flux received at the ground.
Atmospheric extinction scales with the angular zenith distance of the observed target $\zeta$, most conveniently expressed in terms of the airmass AM
\begin{equation}
  \label{eq:IRTF:airmass}
  \text{AM}:=\frac{1}{\cos\zeta}.
\end{equation}
To first order in airmass, the received fluxes $f$ of a constant source observed at 
 different airmasses are related by
\begin{equation}
  \label{eq:IRTF:extinction}
-2.5\log  \frac{f(\text{AM}_1)}{f(\text{AM}_2)}= E \left(\text{AM}_1-\text{AM}_2\right)
\end{equation}
with the extinction coefficient $E$ in units of mag/airmass.
Extinction coefficients for the Mauna Kea site depend greatly on the wavelength considered and may furthermore vary significantly from night to night \citep[see, e.g.,][and references therein]{Krisciunas1987}; at infrared wavelengths, extinction coefficients depend chiefly on the atmospheric water content.
A night during which extinction coefficients vary significantly is non-photometric by definition.

Practically, we determine extinction coefficients for each MIRSI filter and observing run by comparing the background-corrected detector response to observations of calibration standard stars at different wavelengths. 
While it is possible to relate observations of different calibration standard stars in this way, we tried to reobserve the same calibration standard whenever possible, in an attempt to minimize systematic uncertainties in the airmass correction
\seesect{sect:IRTF:strategy}.
Thus determined extinction coefficients were compared to the list compiled by \citet{Krisciunas1987} for the Mauna-Kea site, which serves as a useful cross-check.

Once the extinction coefficients are determined, all background-corrected detector counts are referred to the same airmass, where we typically use $\text{AM}=1$.
Larger values were occasionally chosen for observations which were obtained at large airmass.

\subsubsection{Flux calibration}

After correction for background and airmass, 
detector counts $C$ are proportional to the measured source flux $f$ and the total on-source integration time $t$. 
Denoting quantities referring to the asteroid target with subscript $a$ and quantities referring to the calibration star with subscript $s$:
\begin{equation}
  \label{eq:IRTF:calib}
f_a = \frac{C_a}{t_a} f_s \frac{t_s}{C_s}.
\end{equation}
Flux values for calibration stars were taken from \citet{CohenVII,CohenX}.

                        \subsubsection{Color correction}
                                \label{sect:IRTF:CC}

As can be seen in \tablerefpage{table:MIRSI:filters}, 
MIRSI together with the filters used by us is a broad-band photometer, hence it would be expected that fluxes must be color corrected 
taking account of the target spectrum and the total spectral response of the combined  telescope--filter--detector system (see \sectref{sect:IRAC:CC} for a general discussion).

Determination of color-correction factors is here hampered by the lack of published filter transmission curves.
Assuming a rectangular transmission curve centered at the  central wavelength and with the  spectral breadths given in 
\tableref{table:MIRSI:filters}, \citet[][appendix B]{Delbo2004} determined  color-correction factors for all  MIRSI filters used by us and a number of black-body temperatures; the accuracy of his results is hard to estimate lacking information on the filter profile.
For black-body temperatures between \unit{200 and 400}{\kelvin}, which are appropriate for NEAs and MBAs, the thus determined approximate color-correction factors do not exceed a few percent, much below the remaining systematic and statistical uncertainties. 
Color corrections were therefore not applied to MIRSI fluxes.

		\subsection{Apogee}

Like for MIRSI, the IRTF provides a set of IDL routines for the reduction of Apogee data 
developed by Eric Volquardsen (see \url{http://irtfweb.ifa.hawaii.edu/~elv} and  \url{http://irtfweb.ifa.hawaii.edu/~elv/apogee_steps.txt}).
The automatized data reduction steps include 
correction for bias and dark current,
detection and removal of cosmic ray hits,
synthetic aperture photometry,
and airmass correction.
We have carefully checked all pipeline elements (e.g.\ against our independent synthetic aperture photometry routine) and found them to work reliably for our high-signal-to-noise Apogee data.
Only the detection and removal of cosmic ray hits requires visual inspection on a case-by-case basis to avoid flagging and ``cleaning'' parts of the source.
The final product of the Apogee data reduction pipeline is an ASCII file containing instrumental magnitudes for all observations.
Source magnitudes are determined by adding a constant  magnitude offset determined from  the instrumental magnitudes of the calibration standards \citep[with V magnitudes listed in][]{Landolt1973,Landolt1992}. 
By default, Volquardsen's IDL routines determine magnitudes of the brightest source in the field. Occasionally, Apogee's field of view contained  field stars which were brighter (in the V band) than the asteroid target. In resulting image frames, the field star was masked semi-automatically prior to the proper data reduction, for which purpose we have developed an IDL routine.


%% file: SST_intro.tex
The Spitzer Space Telescope (known as Space InfraRed Telescope Facility, SIRTF, until after launch) is the fourth and final member of NASA's series of space telescopes called \emph{Great Observatories,}
following the Hubble Space Telescope, the Compton Gamma Ray Observatory, and the Chandra X-Ray Observatory.
Spitzer 
has
imaging and spectroscopy capabilities in the mid-infrared wavelength range from 3.6--\unit{160}{\micron}. 
Spitzer is the currently most sensitive mid-infrared telescope, owing mostly to its  vantage point far away from the Earth atmosphere and thermal environment.

Spitzer was launched in 2003 and is expected to stay fully operational until spring 2009. 
Observing time with Spitzer is made available annually
since summer 2004  through an international competition. We were awarded a total of four asteroid observing programs in open competition, two in cycle II (2005--06) and further two in cycle III (2006--07). As of 20 April 2006, i.e.\ up to Spitzer cycle III,
 there is a total of 16 asteroid observing programs with Spitzer, including our four.%
\footnote{ Awards of observing time for cycle IV were announced to successful proponents on 2 May 2007. We were awarded one cycle-IV program. Statistics is not yet publicly available.}

In \sectref{sect:SST:overview}, some general information on Spitzer is provided, including its optical design.
Sections \ref{sect:IRAC:general} and \ref{sect:IRS:general} are devoted to the two focal plane instruments used in our observing programs, the InfraRed Array Camera (IRAC) and the InfraRed Spectrometer (IRS), respectively.
In either section,  the instrument capabilities are presented and  our observing strategies and data reduction techniques are detailed. 
The scientific goals and achievements of our observations will be discussed in the following chapters.

More information can be found in the Spitzer Space Telescope Observer's Manual (referred to as \citet{SOM} in the following) provided on-line by the Spitzer Science Center: \url{http://ssc.spitzer.caltech.edu/documents/som/}. See also \citet{SST}.


%% file: SST_overview.tex

The Spitzer spacecraft carries a \unit{0.85}{\metre} reflecting telescope with  three  focal plane science instruments performing imaging and spectroscopy in the wavelength range from 3.6--\unit{160}{\micron}. 
In flight, telescope and focal plane instruments are cryogenically cooled with liquid helium.
The spacecraft was launched on a Delta rocket from Cape Canaveral, USA on 25 Aug 2003. Its orbit is heliocentric, trailing behind Earth, thus avoiding the adverse affects of the Earth atmosphere and thermal environment. After cooling down the telescope, followed by initial  validation and calibration measurements,  science observations began on 1 Dec 2003; they are expected to last 
as long as cryogen is available, probably until spring 2009.

%
%
\begin{figure}
\centering
\includegraphics[width=0.6\textwidth]{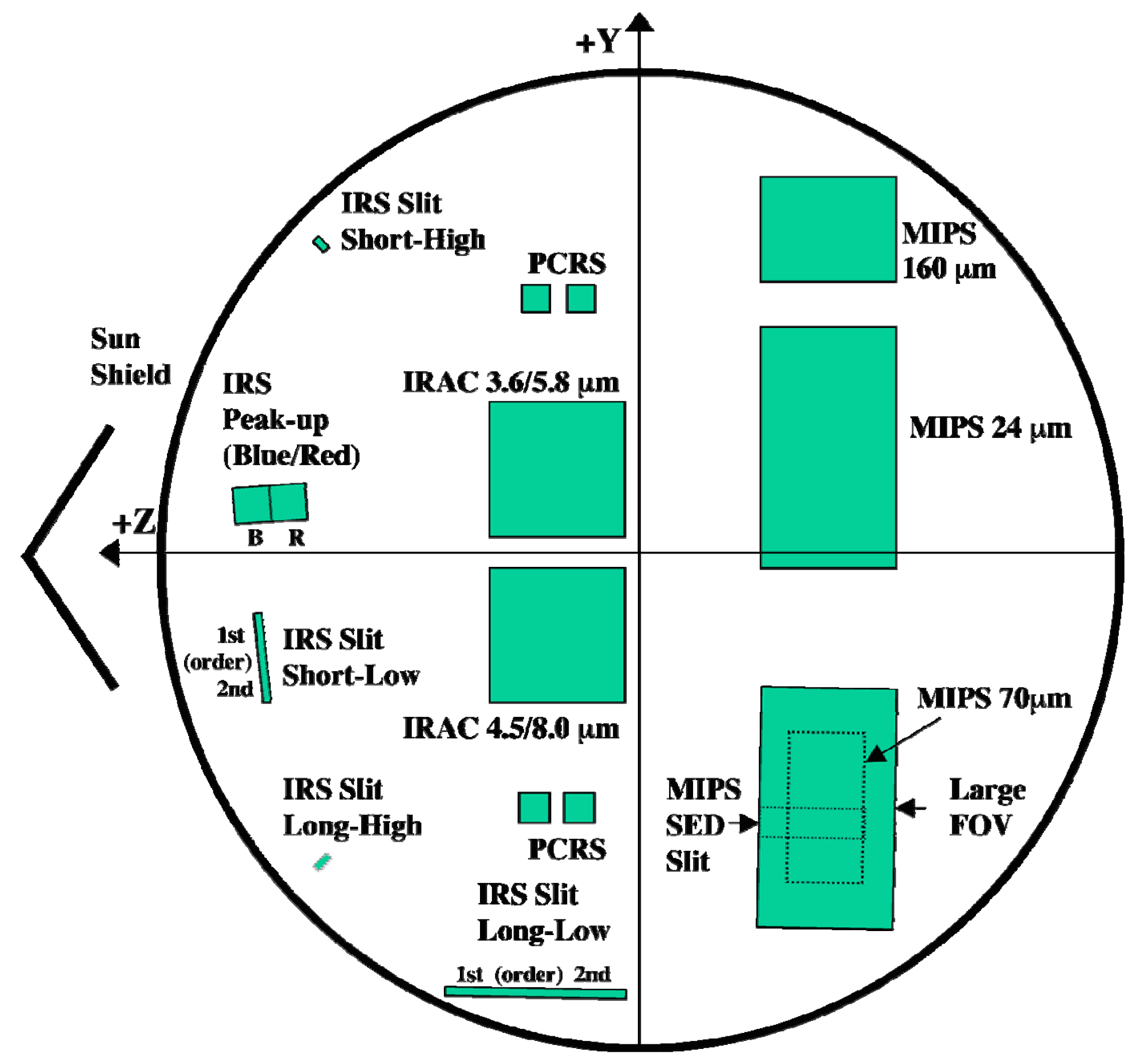}
\caption[Aperture positions of the Spitzer science instruments on the focal plane]{Aperture positions of the Spitzer science instruments on the focal plane.
There are several apertures per instrument, corresponding to different instrument operation modes. Those for IRAC will be discussed in \sectref{sect:IRAC:general}, those for IRS in \sectref{sect:IRS:general} (MIPS is not used in our observations). Also on the focal plane are pointing control reference sensors (PCRS).
The projected focal plane radius is \unit{16}{\arcmin}.
\citep[Fig.\ from][p.\ 60]{SOM}. } 
\label{fig:SST:focalplane}
\end{figure}

The three science instruments are (see \figref{fig:SST:focalplane} for their position on the focal plane):
\begin{itemize}
\item \textbf{InfraRed Array Camera (IRAC)}, four imaging detectors operating at central wavelengths of 3.6, 4.5, 5.8, and \unit{8.0}{\micron}, respectively, each with a square field-of-view of about \unit{5}{\arcmin} at a pixel scale of \unit{$\sim$ 1.2}{\arcsec} (PI G.G.\ Fazio, Smithsonian Astrophysical Observatory, Harvard-Smithsonian Center for Astrophysics);
\item \textbf{InfraRed Spectrograph (IRS)}, a  spectrometer covering the wavelength range from  5.2--\unit{38.0}{\micron} at relative spectral resolutions $\lambda/\Delta\lambda$ between 64 and 128 (low-resolution mode) or $600$ (high-resolution mode). IRS also has  imaging capabilities at wavelengths of about 16 and \unit{22}{\micron} with a field of view of about $\unit{1}{\arcmin}\times \unit{1.2}{\arcmin}$ at a pixel scale of \unit{$\sim$ 1.8}{\arcsec} (PI J.R.\ Houck, Cornell University);
\item \textbf{Multiband Imaging Photometer for Spitzer (MIPS)}, an imager with three broad-band filters centered at wavelengths of 24, 70, and \unit{160}{\micron}. MIPS has an additional low-resolution ($\lambda/\Delta\lambda \sim 20$) spectroscopy mode at wavelengths of 55--\unit{95}{\micron} and a Total Power Mode for measuring the absolute sky brightness in that wavelength range (PI G.\ Rieke, Steward Observatory, University of Arizona). MIPS is not used in our observing programs and is thus disregarded in the following.
\end{itemize}

The spacecraft is operated by the Spitzer Science Center (SSC), located at the campus of the California Institute of Technology in Pasadena, CA, USA. 

\subsection{Spacecraft design and orbit}
\label{sect:SST:design}
\begin{figure}[tb]
  \centering
       \includegraphics[width=0.7\textwidth]{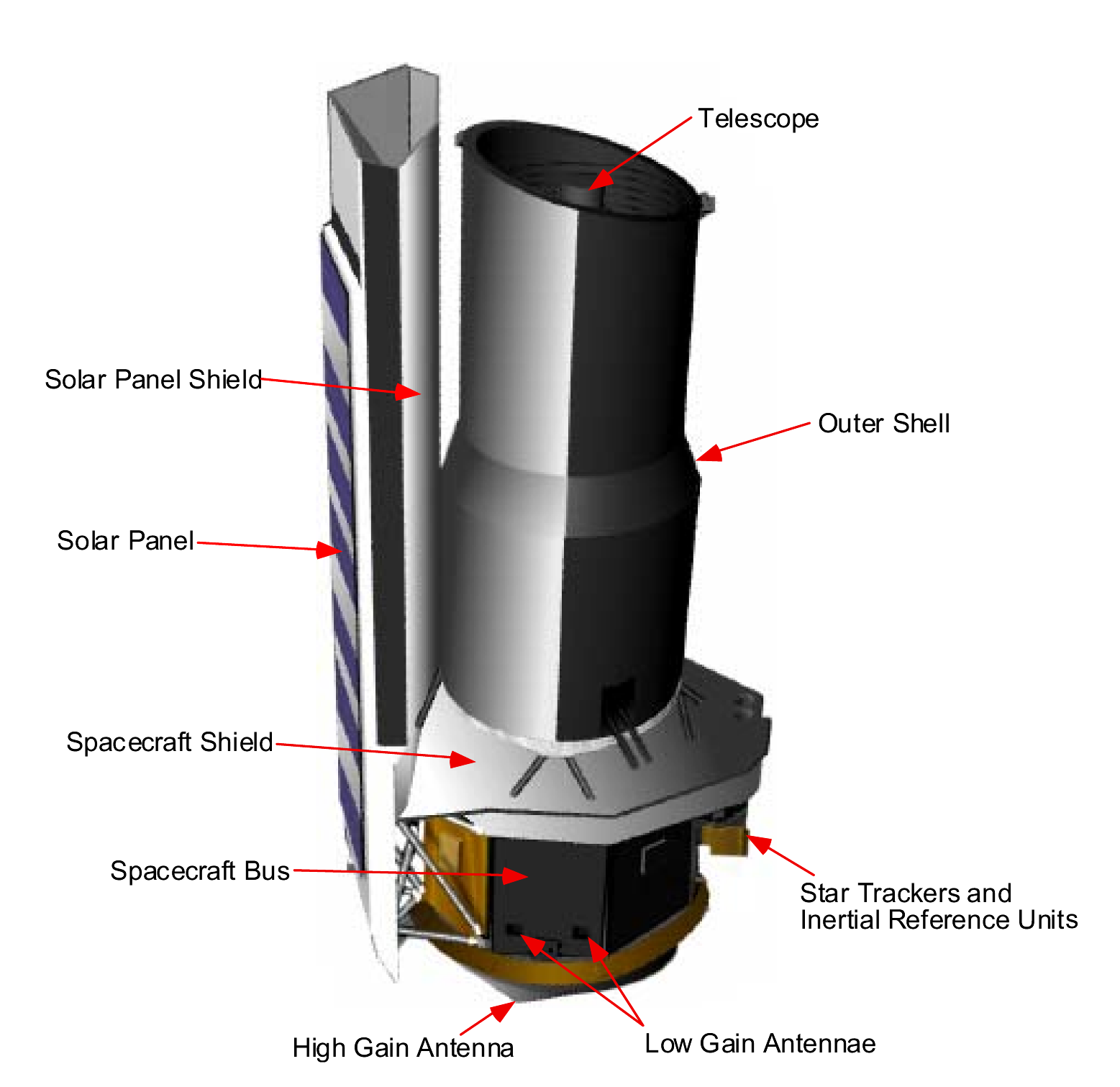}
   \caption[Design of the Spitzer spacecraft]{Design of the Spitzer spacecraft; it is \unit{$\sim 4$}{\metre} tall and \unit{$\sim 2$}{\metre} in diameter. The longest axis coincides with telescope boresight. \citep[Fig.\ from][p.\ 35]{SOM}.}
   \label{fig:SST:design}
\end{figure}

See \figrefpage{fig:SST:design} for an overview of  the spacecraft design. Its main components are:
\begin{itemize}
\item 
Cryogenic Telescope Assembly: A cryostat attached to the helium tank; it contains the cold parts of the focal plane science instruments at a helium bath temperature of \unit{1.4}{\kelvin}. The telescope itself is 
cooled through helium vapor vented off the cryostat to a  temperature  of 6--\unit{12}{\kelvin}.
\item Spacecraft Bus,
hosting all command and data handling units, along with the systems for attitude control and telecommunication. The spacecraft bus is not cooled, and thermally shielded from the cryogenic telescope assembly.
\item Solar Panel and Solar Panel Shield, which provide electric power and at the same time shield the spacecraft (particularly the cryogenic assembly) from direct sunlight.
\end{itemize}

\begin{figure}[tb] 
\centering
\includegraphics[width=0.7\textwidth]{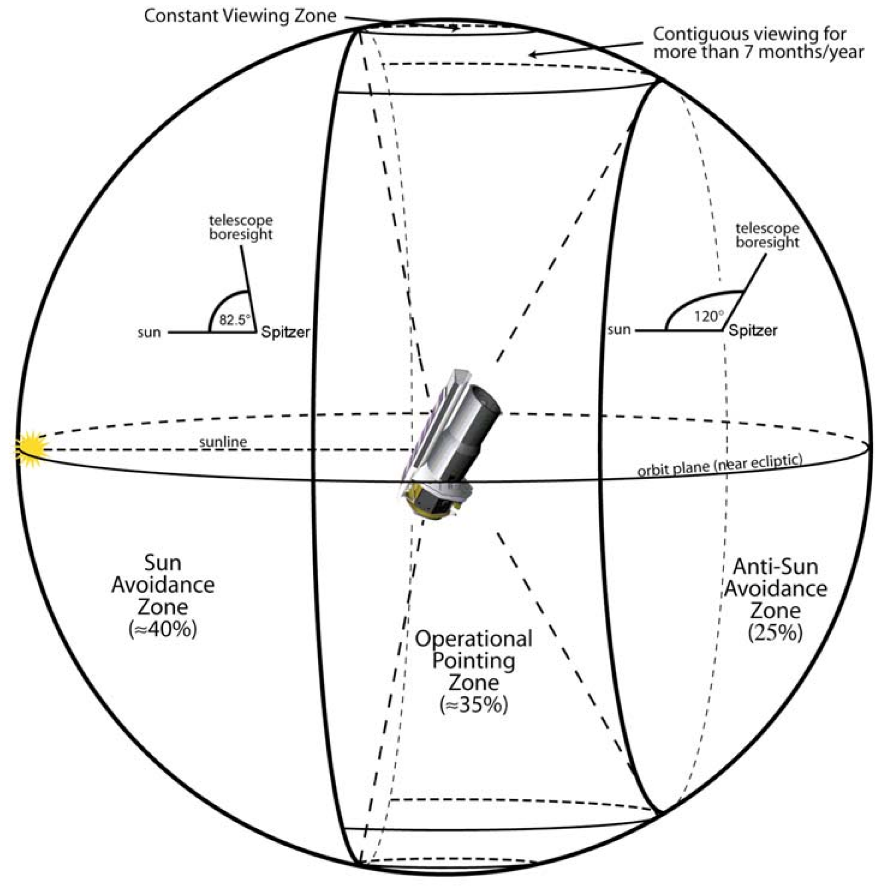}
\caption{Depiction of the region into which Spitzer can point \citep[Fig.\ from][p.\ 26]{SOM}.}
\label{fig:SST:viewing}
\end{figure}

The permissible range of spacecraft attitude is severely constrained:
The requirement to shield the spacecraft from direct sunlight virtually forbids ``roll'' rotations around telescope boresight (the roll angle is fixed to its nominal value \unit{$\pm 2$}{\degree}).
Furthermore, telescope pointing is restricted to solar elongations above \unit{82.5}{\degree} in order to prevent sunlight from entering the telescope, and to solar elongations below \unit{120}{\degree} in order to guarantee sufficient electric power output from the solar panels.
Rotations about the axis pointing towards the Sun, however, are unconstrained, resulting in a viewing zone of annular shape, at solar elongations between 82.5 and \unit{120}{\degree}; see \figref{fig:SST:viewing}.
This has a profound influence on observations of Solar System objects, which will be discussed in \sectref{sect:SST:solarsystem}.

Spitzer is on an Earth-trailing heliocentric orbit, drifting away from Earth at a rate of about \unit{0.12}{\AU\per\yr}. It is thus not only far away from the Earth atmosphere, but also from its thermal environment, which would otherwise substantially heat up the spacecraft and thus limit the cryogenic lifetime. Also, this orbit limits the projected size of the Earth and Moon avoidance zones for telescope observations.


\subsection{Optical design}
\label{sect:SST:telescope}

The telescope has an aperture of \unit{0.85}{\metre} and a focal ratio of $f/12$ (at \unit{5.5}{\kelvin}; focal length \unit{10.2}{\metre})  and follows the Ritchey-Chr\'etien variant of the Cassegrain design.
It is entirely made of beryllium, a light metal of  low heat capacity at low temperatures.
Imaging is diffraction limited at wavelengths above \unit{5.5}{\micron}.

There are no moving parts on the spacecraft, with the exception of the camera shutter of IRAC (which is, however, not being planned to be used in flight) and a mirror inside MIPS.
The absence of (movable) camera shutters hampers calibration measurements, in particular dark current measurements, which will be discussed in the appropriate instrument sections.

Only one science instrument is powered up at any time, in order to minimize cryogen consumption due to dissipation of electric power. Instrument change-over entails 
a substantial amount of engineering and calibration activities. 
Instruments usually stay  on for roughly one week, called an instrument campaign.

Spitzer has different mechanisms for changing and controlling telescope  pointing which, in the absence of moving parts, is tantamount to spacecraft attitude.
The  accuracy of ``blind'' slews  is \unit{0.5}{\arcsec} ($1 \sigma$), which is sufficient to acquire imaging targets but not always sufficient for spectroscopy. 
Relative pointing (for slews of up to \unit{30}{\arcmin}) is more accurate.
Pointing is stable to within  \unit{0.1}{\arcsec} ($1\sigma$) over \unit{1000}{\second}.
Spitzer is capable of linearly tracking moving targets of apparent velocities of up to \unit{1}{\arcsec\per\second}, which is sufficient for all our programs.

\subsection{Planning and proposing observations}
\label{sect:SST:planning}

The majority of Spitzer observing time is open to international observers in an open competition. 
General Observer (GO) proposals are invited by the Spitzer Science Center (SSC) on an annual basis in mid February, 
programs awarded observing time are scheduled within a 12-month period starting the following summer.
Director's Discretionary Time (DDT) is available
for observations of unexpected phenomena of particularly high scientific merit
which cannot be accommodated by GO programs.
DDT can be applied for at any time on relatively short notice. Three out of our four Spitzer programs were GO programs, we were awarded DDT for one program. 

All proposals must be submitted electronically to the SSC,
along with a full specification of all proposed observations, using the SSC-supplied software SPOT (Spitzer Planning Observations Tool). 
SPOT can be used to determine
times at which prospective targets are observable with Spitzer; 
Spitzer pointing is constrained in solar elongation \seesect{sect:SST:design}, neither may the telescope be pointed close to bright moving sources (planets, dwarf planets, the Moon, and the brightest asteroids) which may temporarily damage the detectors.

In SPOT, different types of constraints can be imposed on the relative or absolute timing of proposed observations, we will discuss some of these in \sectref{sect:SST:solarsystem}. Any timing constraint imposed must be given a scientific justification.

If an observing program is accepted, the parameters entered into SPOT are automatically transformed into commands uplinked to the spacecraft.


Due to Spitzer's high sensitivity and accuracy, 
in the planning of observations 
effects must be considered which are  unfamiliar from ground-based mid-infrared observations.
For example, contamination due to faint background sources becomes a problem (see \sectref{sect:SST:solarsystem}), potentially leading to confusion noise. 
Furthermore, possible detector dysfunctionalities  (e.g.\ due to cosmic ray hits or defective pixels) 
together with the non-interactive nature of Spitzer observations require imaging observations to be highly dithered, i.e.\ several consecutive observations of the target should be taken at small spatial offsets from one another. SPOT provides several instrument-specific templates for that.
Dithering has the additional advantage of mitigating against any  residual pixel-to-pixel variations in sensitivity not caught in the flat-fielding process.


\subsection{Asteroid-specific aspects to observation planning}
\label{sect:SST:solarsystem}

\paragraph{Parallax}
Spitzer is \emph{not} on a geocentric orbit, but trails behind the Earth on a heliocentric orbit at a steadily increasing geocentric distance. This causes a parallax between the Spitzer-centric and geocentric  apparent positions of solar-system targets, particularly for near-Earth objects.
 \textit{Horizons,} the JPL on-line ephemeris service (\url{http://ssd.jpl.nasa.gov/?horizons}), can be used to generate Spitzer-centered ephemerides of solar-system objects---the telescope code for Spitzer is $-79$. 

\begin{figure} 
\centering
   \includegraphics[angle=-90,width=0.7\textwidth]{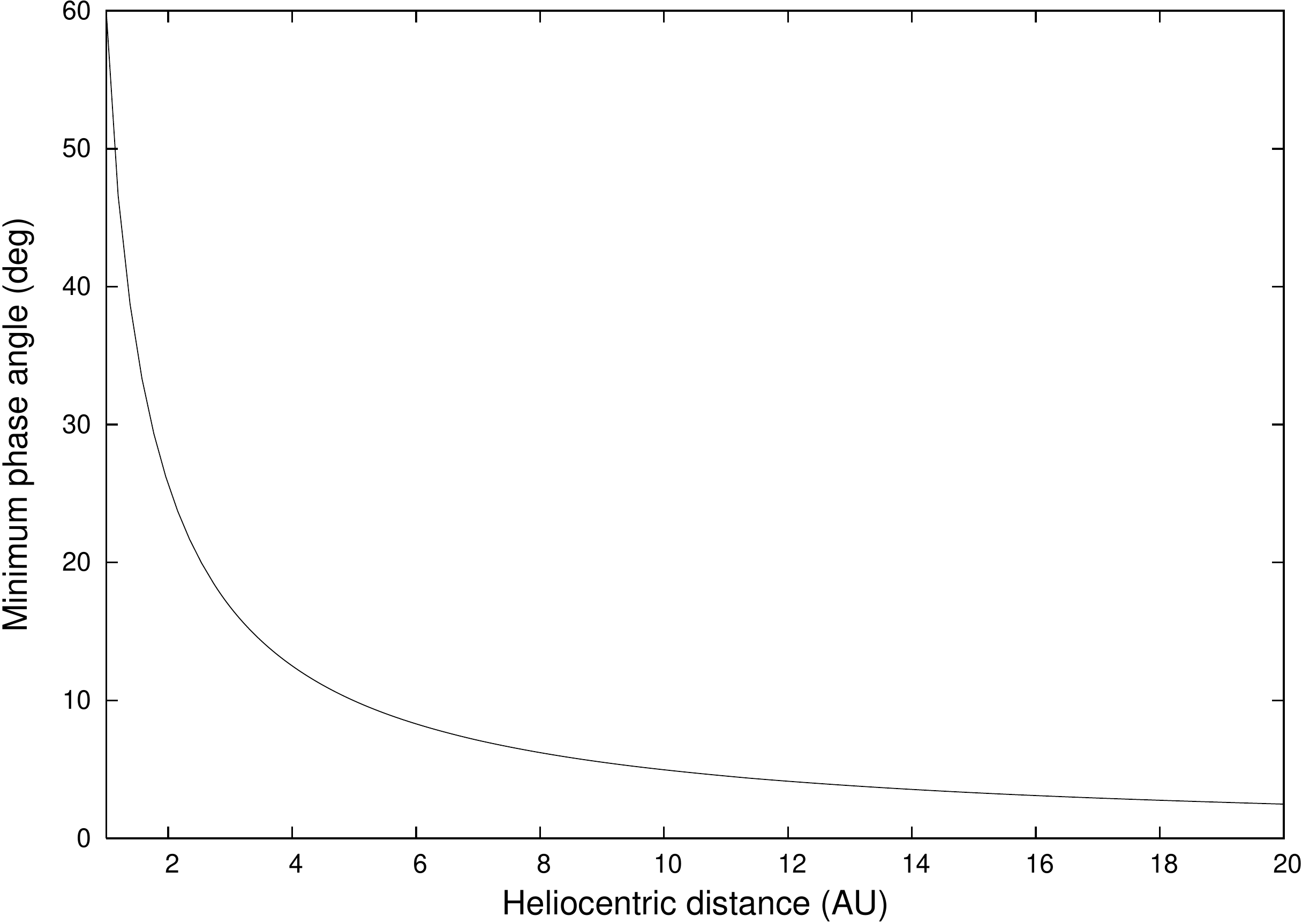}
\caption[Minimum solar phase angle for Spitzer observations as a function of heliocentric distance]{Minimum solar phase angle, $\alpha_{\text{min}}$ for Spitzer observations of Solar System objects  as a function of heliocentric distance, $r$ \seeeq{eq:alphamin_Spitzer}. Spitzer's heliocentric distance $r_\text{SST}$ is assumed to equal \unit{1}{\AU}.}
\label{fig:SST:alpha}
\end{figure}

\paragraph{Solar elongation constraints}
As discussed in \sectref{sect:SST:design}, spacecraft design constrains telescope pointings to solar elongations between 82.5 and \unit{120}{\degree}. While for inertial targets this only constrains the \emph{time} of their observability, for solar-system objects this prevents observations close to opposition (solar elongation of \unit{180}{\degree}), where it would be preferable to observe them.
Not only do solar-system objects generally reach their peak brightness at opposition, but they are also best observable from ground during that time. The impossibility of close-to-opposition Spitzer observations thus hampers simultaneous ground-based support observations with, e.g., optical telescopes, which would have been advantageous for some of our programs.
Most importantly, however, the solar elongation constraint provides a lower limit on the solar phase angle, $\alpha$, at which objects are observable---minimum phase angles occur at opposition. The importance of the phase angle for thermal modeling is discussed in \sectref{sect:thermal:obsgeometry}.
The minimum  solar phase angle $\alpha_{\text{min}}$ for  Spitzer observations of solar-system objects is a function of their heliocentric distance $r$ and of Spitzer's heliocentric distance $r_{\text{SST}}$:
\begin{equation}
\label{eq:alphamin_Spitzer}
\alpha_{\text{min}} (r) = \arcsin{\frac{\sqrt{3}}{2}\frac{r_\text{SST}}{r}},
\end{equation}
where  it is assumed that the minimum phase angle occurs at an elongation of \unit{120}{\degree} (true for $r>r_\text{SST}$).
See \figrefpage{fig:SST:alpha} for a plot.

\paragraph{Instrument change-over}
Only one Spitzer instrument is powered up at any time.
Therefore, multi-wavelength observations using different instruments entail a time gap between observations of a few hours at least, which may result in an additional flux uncertainty due to lightcurve effects.

\paragraph{Background sources}
With ground-based mid-infrared telescopes such as the IRTF (see \chaptref{chapt:IRTF}), only very bright targets can be observed, which are virtually guaranteed to be the only observable source inside the FOV. This is different with Spitzer.
Its high sensitivity
leads to several potential problems, which need to be mitigated against in both observation planning and data analysis:
\begin{itemize}
\item Very bright background sources inside (or close to) the FOV heavily saturate the detector and compromise  observations due to, e.g., blooming. The timing of asteroid observations must therefore be constrained to times where no such bright background source is near-by. This time-consuming task can be accomplished by overlaying star charts with the asteroid path using SPOT (see, e.g., \figrefpage{fig:IRAC:FOVs}). The most useful star charts for this purpose are those derived from the IRAS survey at wavelengths of 12 and \unit{25}{\micron}. Most IRAS sources would heavily saturate Spitzer cameras in deep integrations.
\item
For deep imaging observations, the brightness of (spatially resolved or unresolved) background sources may be comparable to that of the asteroid target, causing difficulties for reliable target flux extraction in their vicinity.
Although this latter effect is mitigated  by the apparent motion of asteroids, it severely compromised several of our asteroid observations. Note that deep imaging observations regularly consist of a time series of individual pointings, during which the asteroid moves.
Data can be partially recovered by rejecting frames with close asteroid-star encounters.
\end{itemize}

Other reasons for imposing timing constraints on asteroid observations may be target brightness (particularly for near-Earth objects, where brightness is a strong function of time) or phased observations of the asteroid lightcurve.

\subsection{Spitzer data reduction}
\label{sect:SST:data}

The SSC discourages observers from performing calibration observations; instead, calibration observations are regularly performed by members of the instrument teams along with maintenance operations.
All science data are partially reduced at the SSC. To this end, the SSC hosts instrument-specific
sets of automated partial data reduction routines, 
referred to as the \emph{BCD pipelines,} where BCD stands for Basic Calibrated Data. The BCD pipelines remove most known instrument artefacts from the data and also perform absolute flux calibration. While raw data are made available to the observer, most parts of the BCD pipelines and most required calibration files are not.

The remaining data reduction steps to be taken by the observer include removal of residual instrument artefacts, stacking and coadding appropriate data subsets, extraction of target fluxes, and correction of target fluxes for, e.g., the effect of filter breadth. See the appropriate subsections of \sectref{sect:IRAC:general} and \ref{sect:IRS:general}.


%% file: IRAC_general.tex

IRAC (acronym for InfraRed Array Camera) is an imaging camera
at the low-wavelength end of Spitzer's spectral coverage:
It consists of four detectors operating 
at central wavelengths around  3.6, 4.5, 5.8, and \unit{8.0}{\micron} (referred to as channels 1--4 in the following). 
Each  detector has a square field-of-view (FOV)  256~pixels wide, corresponding to roughly \unit{5.2}{\arcmin} at a pixel scale around \unit{1.2}{\arcsec}.

At IRAC wavelengths, reflected sunlight and thermal emission contribute to the observable asteroid flux. 
At \unit{3.6}{\micron} (channel 1), thermal flux is generally negligible relative to reflected sunlight, the converse applies to channel 4 (\unit{8.0}{\micron}).
The IRAC wavelength range is at the Rayleigh-Jeans tail of the emission of stellar background sources, therefore their brightness relative to that of asteroids decreases with wavelength.

This section starts with a short overview of the instrument layout and its capabilities as used in our observing programs (\sectref{sect:IRAC:layout}). 
Our asteroid observation strategies are described in \sectref{sect:IRAC:strategy}.
In the following two sections, we describe the employed data reduction techniques: First the partial, automated IRAC BCD pipeline
hosted by the Spitzer Science Center  (\sectref{sect:IRAC:BCD_pipeline}), then  the data reduction techniques employed by us (\sectref{sect:IRAC:asteroids}).
\Sectref{sect:IRAC:validation} is devoted to tests of our data reduction pipeline.

More information on  instrument layout and nominal data quality can be found in \citet{Fazio2004}, \citet{Hora2004}, and in the IRAC data handbook provided on-line \citep{IRACdatahandbook} (we made use of version 3.0 from 20 Jan.\ 2006, \url{http://ssc.spitzer.caltech.edu/irac/dh/}).

\subsection{Instrument layout}
\label{sect:IRAC:layout}

\begin{table}
  \caption[IRAC filter wavelengths and spectral breadths]{IRAC filter wavelengths and spectral breadths. See \eqrefpage{eq:IRAC:effectivelambda} for a definition of the effective wavelength.}
  \label{table:IRAC:channels}
\centering
  \begin{tabular}{l|llll}
    \toprule
     & Channel 1 & Channel 2 & Channel 3 & Channel 4 \\
    \midrule
    Mnemonic wavelength (\micron) & 3.6 & 4.5 & 5.8 & 8.0 \\
    Effective wavelength (\micron)  & 3.550 & 4.493 & 5.731 & 7.872 \\
    FWHM (\micron)                & 0.74  & 1.02  & 1.41  & 2.88 \\
    Relative filter width         & 21\%  & 23\%  & 25\%  & 36\% \\
    \bottomrule
  \end{tabular}
\end{table}

\begin{figure}
\centering
   \includegraphics[width=0.5\textwidth]{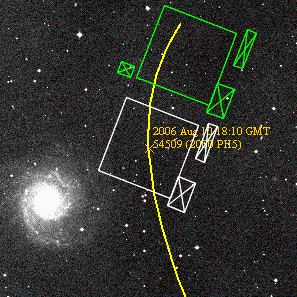}
   \caption[Projection of the two IRAC FOVs onto the sky]{Projection of the two IRAC FOVs onto the sky: The FOV shared by channels 2 and 4 is depicted white, that of channels 1 and 3 is green. The crossed-out rectangles next to the FOVs represent the respective straylight-avoidance zones (see text). Depicted is a $\unit{20}{\arcmin}\times \unit{20}{\arcmin}$ patch of Digital Sky Survey (DSS) data
centered at the position of (54509) YORP (then known under its provisional designation 2000~PH5) on 10 Aug 2006, 18:10 UT (yellow line and cross). The large spiral galaxy  is Messier object 74 = NGC 628.
}
  \label{fig:IRAC:FOVs}
\end{figure}

IRAC consists of four CCD detectors sharing a common electronic and cryogenic framework. Channels 1 and 2 are In:Sb detectors,  channels 3 and 4 are Si:As detectors.
Filter wavelengths and spectral breadths for all channels are summarized in  \tableref{table:IRAC:channels}.
The detectors are situated inside the Cryogenic Telescope Assembly (see \figrefpage{fig:SST:design}) at a helium bath temperature of \unit{1.4}{\kelvin}. All instrument control units are in the Spacecraft Bus.
Each detector is 256~x~256 pixels in size, with a projected pixel size around \unit{1.2}{\arcsec}~x~\unit{1.2}{\arcsec}. Each field of view (FOV)  is about \unit{5.2}{\arcmin}~x~\unit{5.2}{\arcmin} wide.
Channels 1 and 3
share a common aperture and FOV using a dichroic beam-splitter,  the same applies to channels 2 and 4.
The edges of the two  FOVs are separated by about \unit{1.5}{\arcmin}, so there is no overlap on the sky \seefig{fig:IRAC:FOVs}.

IRAC imaging is  limited by diffraction with mean diffraction widths of \unit{1.66}{\arcsec}, \unit{1.72}{\arcsec}, \unit{1.88}{\arcsec}, and \unit{1.98}{\arcsec} (FWHM) for channels 1--4, resp. With a \unit{$\sim 1.2$}{\arcsec} pixel scale, the four IRAC point-spread functions are  undersampled.

\begin{table}
\caption[IRAC point-source sensitivity]{\label{table:IRAC:sensitivity}
IRAC point-source sensitivity ($1\sigma$ in units of \micro\Jy; see also \eqrefpage{eq:IRAC:sensitivity}) for one image frame of the stated integration time (frame time). The celestial background is assumed to be high, as is appropriate for most asteroid observations. Long exposures in channel~4 are background-limited; ``\unit{100}{\second}'' frames in channel~4 are automatically converted into 2 frames of \unit{50}{\second} each.
}
\centering
\begin{tabular}{r|llll}
\toprule
Frame time (sec) & \unit{3.6}{\micron} & \unit{4.5}{\micron} & \unit{5.8}{\micron} & \unit{8.0}{\micron} \\
\midrule
100 & 1.3  & 2.1 & 14  & 18 \\
30  & 2.5  & 4.1 & 27  & 32 \\
12  & 4.8  & 7.1 & 44  & 52 \\
2   & 34   & 41  & 180 & 156 \\
\bottomrule
\end{tabular}
\end{table}

\begin{table}
\caption[IRAC saturation limits]{\label{table:IRAC:saturation}
Largest non-saturated point-source flux for the four IRAC channels and the stated frame times. Units in this table are \milli\Jy, \emph{not} \micro\Jy\ as in \tableref{table:IRAC:sensitivity}! High celestial background is assumed.
}
\centering
\begin{tabular}{r|llll}
\toprule
Frame time (sec) & \unit{3.6}{\micron} & \unit{4.5}{\micron} & \unit{5.8}{\micron} & \unit{8.0}{\micron} \\
\midrule
100 & 3.8  & 3.9 & 27  & 28 \\
30  & 13   & 13  & 92  & 48 \\
12  & 32   & 33  & 230 &120 \\
2   & 190  & 200 & 1400&740 \\
\bottomrule
\end{tabular}
\end{table}

In standard observing mode, there are four selectable options for 
the  integration time per image frame (frame time) ranging from 2--\unit{100}{\second}, 
see \tablerefpage{table:IRAC:sensitivity} for the corresponding sensitivities and \tableref{table:IRAC:saturation} for the saturation limits. 
IRAC observations typically consist of a series of $n$ consecutive frames, which are later stacked; 
the total signal-to-noise ratio $S/N$  for a source with flux $f$ then equals
\begin{equation}
\label{eq:IRAC:sensitivity}
S/N (f) = \frac{f}{s} \sqrt{n}
\end{equation}
where $s$ denotes the single-frame sensitivity $s$ as in \tableref{table:IRAC:sensitivity}.
For deep IRAC observations, the dominant source of uncertainty is celestial background radiation due to the emission and scattering of sunlight by zodiacal dust.
The zodiacal dust is concentrated in the ecliptic plane, therefore IRAC's sensitivity is a strong function of ecliptic latitude. We planned all observations assuming high background, as is appropriate for most Solar-System objects.

IRAC is continuously exposed to the sky; there is a  shutter, which was used for laboratory measurements on ground, but it is not operated in flight in order to avoid any operational risk imposed by this moving part.
Therefore, all four detectors take science quality data simultaneously. Even if only one channel is requested, the same field is observed in a second channel, while an adjacent (``serendipitous'') field is observed in the two remaining channels; 
for some purposes, we found it useful to obtain observations in one FOV by requesting observations of a suitably offset field with the other FOV
\seesect{sect:IRAC:dithering}.
Data from all four channels are processed through the BCD pipeline and made available to the observer.

IRAC is offset from the optical axis of the Spitzer focal plane (see \figrefpage{fig:SST:focalplane}) leading to optical distortion, i.e.\ the  projected shape and size of pixels vary across the detector.
The distortion has been modeled by the Spitzer Science Center to an accuracy of \unit{0.1}{\arcsec} and 
must be corrected for in order to avoid significant photometric errors. 

For all four IRAC channels, a certain amount of stray light is scattered onto the detectors. This is only critical if bright background sources are in certain ``stray light avoidance zones'' close to the FOV  \seefigpage{fig:IRAC:FOVs}.

\subsection{Observing strategy for asteroid photometry}
\label{sect:IRAC:strategy}

IRAC observations have to be designed in a way quite different from ground-based mid-infrared observations:
\begin{itemize}
\item
Observation design must be very robust because
Spitzer executes all observations autonomously, without the possibility of real-time observer interaction.
Also, the exact time at which  observations are carried out is usually unknown at the planning stage---absolute timing constraints can be imposed by the observer but require a scientific justification. Once an observation is scheduled (roughly 2--5 weeks in advance), no more changes to the observation parameters can be made.
\item 
Observations of faint targets are plagued by background sources of comparable brightness, the abundance of which grows rapidly with observation depth.
While in the case of MIRSI, suitable targets are practically guaranteed to be the only observable source inside the FOV, 
IRAC observations must be designed in a way that mitigates against background contamination.
\item 
There is also a non-negligible number of heavily saturating sources which, if inside the FOV,
spoil the entire image frame and potentially  subsequent image frames through latent  effects. 
These must be prevented from entering the FOV.
\end{itemize}

IRAC observations are defined in the Spitzer Proposal Tool SPOT by specifying a target, the desired IRAC channels, 
the integration time per image frame, and the total number of image frames per channel to be taken. 
There are two options, which can be combined, for choosing the number of image frames: Dithering telescope boresight a user-defined number of times around the target location or identical in-place repeats. While dithering causes a small overhead for telescope slewing and settling, it enhances the accuracy and robustness of observations by mitigating against small-scale detector defects (e.g.\ permanently defective pixels or cosmic ray hits) and any residual pixel-to-pixel variations in sensitivity.

The timing of observations can be constrained by the observer in several ways, but it is stipulated in the Call for Proposals that 
they be minimized.
Timing constraints require a compelling scientific justification, otherwise a lower priority will be assigned to the program by both the Time Allocation Committee and, if the program is accepted, in the scheduling queue.

\subsubsection{Integration time}
\label{sect:IRAC:IT}

We determined the required total integration time from 
\eqrefpage{eq:IRAC:sensitivity},
 depending on expected target flux 
and on our predefined goal for the
total $S/N$ required to reach our scientific goals. 

The expected flux of our asteroid targets at IRAC wavelengths depends on several, generally unknown, asteroid parameters such as diameter and apparent color temperature.
This may induce a  flux uncertainty of up to a factor of 10.
The exact time of observations is not known at the planning stage, adding to the flux uncertainty.

For each target, we calculated upper and lower flux limits 
assuming appropriate end-members of the plausible range of physical properties and ephemerides. 
We requested conservatively deep integrations, such that we would
 barely reach our $S/N$-goal assuming the lower flux limit. We also checked against the appropriate IRAC saturation limits \seetablepage{table:IRAC:saturation}.

\subsubsection{Timing constraints}
\label{sect:IRAC:timing}

We generally had to prevent Spitzer from observing at undue times by
imposing absolute timing constraints, i.e.\ by explicitly specifying times during which the observations should take place.

\begin{figure}[bt]
\centering
   \includegraphics[width=0.6\textwidth]{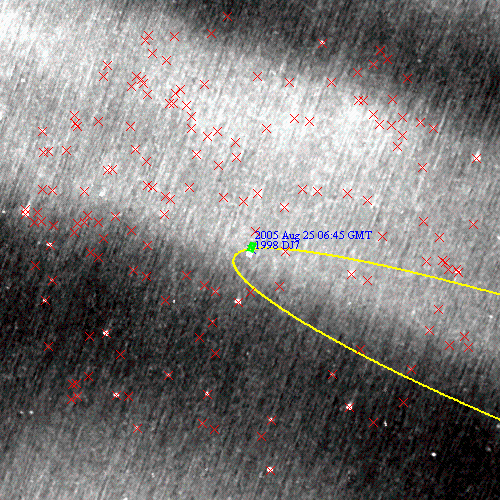}
\caption[SPOT-generated star chart overlaid with asteroid path, IRAS sources, and IRAC FOVs]{Star chart (\unit{12.5}{\degree} $\times$ \unit{12.5}{\degree}) derived from the IRAS \unit{12}{\micron} survey, overlaid with the Spitzer-centric path of asteroid (33143) 1998 DJ7 (yellow line), known IRAS sources (red crosses) and the two IRAC FOVs (green and white as in \figrefpage{fig:IRAC:FOVs}, but note the size difference!).}
\label{fig:IRAC:IRAS}
\end{figure}

For all asteroid targets,
we identified ``avoidance times'' during which the target path is close to bright known infrared sources, which would heavily saturate the detectors and  potentially lead to temporary detector damage if they enter a FOV or a straylight avoidance zone.
This was done by overlaying
star charts resulting from the IRAS \unit{12}{\micron} survey with the projection of the IRAC FOVs and the Spitzer-centric asteroid path (note the parallax between Spitzer and Earth) throughout the times of Spitzer observability; see  \figref{fig:IRAC:IRAS} for a typical example. To be sure, we considered every known IRAS source
to be potentially damaging IRAC, due to its much higher sensitivity.
Since heavy detector saturation may compromise subsequent observations through persistent latent images, we also kept the off-target FOV clear of IRAS sources.

Generally, this resulted in very mild timing constraints, typically barring a few hours per week, a notable exception being times during which the asteroid path intersects regions of high background source density, e.g.\ at low galactic latitudes.

For NEA targets, we additionally disabled observing at times
 when the target was too distant and thus too faint for successful observing.

\subsubsection{Dithering}
\label{sect:IRAC:dithering}

For most of our observations, telescope boresight is slightly altered between consecutive observations, such that the on-chip position of the target \emph{dithers} around the FOV center. While this causes a slight overhead for telescope reorientation and settling, it helps mitigate against the effects of defective pixels or residual pixel-to-pixel variations in sensitivity.

SPOT provides several standard dither patterns. As recommended by the SSC, we generally chose mid-scale dither patterns, which keep our  target on chip at all times.
We generally chose the longest possible frame time (\unit{100}{\second}) and the required number of dither positions  to reach the total integration time previously determined \seesect{sect:IRAC:IT}. We always used five or more   dither positions, however, and therefore reduced the frame time for shallow observations.

For most deep IRAC observations, we employed a non-standard dither pattern in order to mitigate more efficiently against contamination from celestial background sources.
For deep IRAC observations, practically no field in the sky is guaranteed to be clear of potentially interfering background sources, so it is impractical to avoid them by imposing timing constraints.
Furthermore, unresolved faint background sources (or diffuse but structured emission, e.g.\ from zodiacal emission or galactic cirrus) decrease the accuracy at which photometry can be performed, leading to  confusion noise which may limit the accuracy of deep IRAC observations.

This is mitigated against by the apparent movement of our targets: If the target moves by more than one PSF width between the first and last image in a series, then frames with background sources close to the asteroid path can be rejected \seesect{sect:IRAC:filtering}. Also, background structure
is reduced by coadding image frames with different asteroid positions relative to the inertial background \seesect{sect:IRAC:mosaicking}.

To this end, we have 
designed deep asteroid observations in a highly redundant way, consisting of up to 144 on-target image frames per observation. 
In order to maximize the time baseline and thus the apparent asteroid motion between first and last frames, we have used a non-standard dither pattern.
As opposed to the standard IRAC dither patterns, where the FOV shared by channels 2 and 4 is on target for the first half of the time, followed by the FOV of channels 1 and 3,
we had the FOVs ``take turns'' on target up to ten times. 
This  nearly doubled the temporal baseline  at the expense of a  small overhead for the additional telescope slews.

SPOT does not provide such dither patterns. We  implemented them in an indirect way 
taking advantage of the IRAC peculiarity that  even if only one channel is requested, all channels are exposed to the sky and read out, and all data are processed in the BCD pipeline.
We requested a ``moving cluster target''
 to be observed by channels 2 and 4, such that
odd-number cluster positions are on target (not offset) and  even-number cluster positions are offset by \unit{402}{\arcsec} along the row axis in IRAC array coordinates, centering the 2nd FOV  on target.
Typically, a standard dither pattern is performed on each cluster position.

\subsection{IRAC BCD pipeline}
\label{sect:IRAC:BCD_pipeline}

As for all Spitzer science data, IRAC data are partially reduced and calibrated at the SSC using an automated set of computer routines called the IRAC BCD pipeline \seesect{sect:SST:data}.
It outputs one BCD file per pointing and detector, i.e.\ an image file in FITS format.

In this section, those aspects of the  IRAC BCD pipeline which are relevant to our programs shall be sketched. The reference for most of the material presented here 
is the IRAC pipeline description document \citep{IRACpipelinedescription}.

\subsubsection{Correction for dark current}
\label{sect:IRAC:dark}

An integration-time dependent dark-current estimate is subtracted from all images in the BCD pipeline.
The detector dark-current levels for all possible integration times were determined on ground.
In orbit, the absolute dark-current level cannot be measured 
 since the shutter is not moved (see above).
However, drifts in dark-current level are monitored 
based on
sky dark frames, which are obtained from frequent, highly dithered observations of sky regions of very low emission (very low zodiacal background, only few bright sources such as stars).

\subsubsection{First-frame effect}
\label{sect:IRAC:firstframe}

For an unknown reason, the dark-current level is a function of the time elapsed since the end of the previous detector readout, the so-called first-frame effect.  This adds an unpredictable offset, chiefly to the first few frames in a series of observations. In order to mitigate against the first-frame effect, 
one or two redundant observing frames of short integration time are added by default at the beginning of each series of observations.
The remaining background-level uncertainty is not corrected for in the BCD pipeline and must be considered in the data reduction, cf.\ \sectref{sect:overlap}.

\subsubsection{Flat-fielding}
\label{sect:IRAC:flat}

To correct for the pixel-to-pixel variation in sensitivity, IRAC science data are divided by a gain map over the FOV normalized to one, called a flat field.
It is determined from 
frequent, highly dithered observations of predefined regions of high zodiacal emission (which is spatially relatively constant over the IRAC FOV; note again that IRAC's shutter is not used in flight) and only small numbers of stars or galaxies.
The images frames obtained are cleaned from localized sources and cosmic ray hits, then coadded and normalized to a median of one.

The flat-field frames thus obtained were seen to be constant within the observational uncertainty over the first two years of Spitzer operations. 

\subsubsection{Flux calibration}
\label{sect:IRAC:calibration}

Flux calibration factors were determined from observations of a set of calibration stars 
before the beginning of normal science operations
and are confirmed by regular reobservations \citep{Reach2005}. The absolute calibration accuracy is estimated to be within \unit{3}{\%}, limited by the uncertainty in calibration star flux. The relative accuracy (temporal stability) of IRAC is \unit{1.5}{\%} (rms) over a year. 

Calibration-star data-counts are obtained via synthetic aperture photometry using a particular set of aperture and sky-annulus radii (aperture radius: \unit{10}{\text{pixels}}, annulus radii: 10 and \unit{20}{\text{pixels}}); IRAC data should be reduced using the same parameters. 

For the reduction of most deep observations these parameters are impractical. 
Smaller aperture and sky annulus radii can be used, but flux values thus obtained require a 
so-called aperture correction to be applied \seesect{sect:IRAC:aperturecorrection}.
 
The calibration procedure assumes targets to be point-like; this is a safe assumption for our asteroid targets, since their angular size is well below the IRAC diffraction widths.

IRAC is a broad-band photometer \seetablepage{table:IRAC:channels}. In order to convert the passband-integrated detector response into  monochromatic flux values at the effective filter wavelengths,  assumptions must be made on the target spectrum.
BCD fluxes are corrected for a spectrum inversely proportional to wavelength.
Different target spectra require color corrections to be made, which can be significant in the case of asteroids \seesect{sect:IRAC:CC}. 

Calibrated images are in units of flux per solid angle. Note that this is \emph{not} equivalent to flux per pixel due to distortion (see above).

\subsubsection{Pointing refinement}
\label{sect:IRAC:pointing}
Spitzer's nominal pointing accuracy is \unit{0.5}{\arcsec}, less than half the width of an IRAC pixel. The astrometric accuracy of taken image frames is further refined  to an accuracy of \unit{0.15}{\arcsec} by matching the positions of detected sources with star catalogs (usually the 2 Micron All Sky Survey, 2MASS), more than sufficient for our needs.

\subsection{Data reduction steps taken by us}
\label{sect:IRAC:asteroids}

We start our data reduction efforts at the BCD stage.
First,  unusable or irrelevant (off-target) BCD frames are filtered out. This includes visual inspection of all image frames \seesect{sect:IRAC:filtering}.

Then, the remaining BCD files belonging to one observation are coadded to a  mosaic image in the asteroid rest frame. This includes several image manipulation tasks in order to mitigate against celestial background structure and to remove further detector effects and transient artefacts such as cosmic ray hits. These are described in sections \ref{sect:overlap} and \ref{sect:IRAC:mosaicking}.

Asteroid fluxes are determined from the  mosaic images.
This includes several corrections for known detector or source properties (see sections \ref{sect:IRAC:APEX} through \ref{sect:IRAC:CC}).

All IRAC BCD files considered in this thesis were produced by the  BCD pipeline version 14.
For mosaicking, the software package MOPEX was used. MOPEX  is provided by the SSC (\url{http://ssc.spitzer.caltech.edu/postbcd/download-mopex.html}; we used the version for Linux named 030106, released on 1 March 2006).

      \subsubsection{Filtering frames}
      \label{sect:IRAC:filtering}
Image frames that were irrelevant or unusable for our purposes were rejected 
from further analysis after visual inspection.

Due to IRAC's design with two non-overlapping fields-of-view \seefigpage{fig:IRAC:FOVs} constantly exposed to the sky, 
the fields-of-view  ``take turns'' on the target in a user-defined way when observing a point source in all four channels. Therefore, half of the downloaded image frames are off target and irrelevant for our purposes.
Those were deleted, which significantly saved  on storage space---and on computer time during sub-sequent analyses.%
\footnote{ Note that a single deep IRAC observation can easily result in several hundred megabytes of data.}%

Also, as introduced in \sectref{sect:IRAC:firstframe}, each series of IRAC observations starts with one or two
redundant integrations of reduced integration time
in order 
to mitigate against the first-frame effect.
Those images were deleted.

We visually inspected each remaining image frame before further analysis in order to reject unusable frames.
Those include frames with very bright on-chip sources (or particularly bad cosmic ray hits) compromising the image quality throughout the detector through blooming effects. Also frames with detector artefacts, cosmic ray hits, or background sources within a few pixels from the target location were rejected. Generally, only parts of the data were affected due to  dithering and the apparent movement of the asteroid target relative to the fixed star background.

	\subsubsection{Background matching}
        \label{sect:overlap}

The first-frame effect (\sectref{sect:IRAC:firstframe}) is a drift of the background levels in a series of observations, causing difficulties for outlier rejection during mosaicking (cf.\ \sectref{sect:IRAC:mosaicking}). First frames in a series are most affected.

Although the first frame is always rejected (see above), 
it is nevertheless required to match the background levels of image frames to be mosaicked. This is accomplished using the program
\textit{overlap.pl,} which is part of the MOPEX package.
It determines an offset value to each input image such that the pixel-by-pixel difference between overlapping areas of pairs of input images is minimized.

In a first step, the program  reprojects and interpolates the BCDs onto a common rectangular grid. 
In a second step, bright sources  above a user-specified detection threshold, particularly cosmic ray hits, which would bias the overlap correction, are masked and excluded from further processing. 
Then the additive overlap corrections are calculated and the corrected images are saved.

The detection threshold for bright object masking  must be adjusted on a case-by-case basis, after evaluation of the mosaicking process.

	\subsubsection{Mosaicking and outlier rejection}
        \label{sect:IRAC:mosaicking}

Generally, IRAC observations are not performed as a single frame of the required integration time,
but rather as a set of at least 5  correspondingly shorter integrations (our deepest IRAC observation consists of 144 individual BCD frames).
Usually, telescope boresight of the individual observations is slightly altered between consecutive observations, such that the on-chip position of the target dithers around the FOV center \seesect{sect:IRAC:strategy}.
Deriving target fluxes thus requires combining these exposures into a single high signal-to-noise image,  referred to as mosaic.

Mosaicking is the most time-consuming part of IRAC data reduction, both in terms of computer time and number of iterations generally required to achieve a satisfactory result.
We use the program \textit{mosaic.pl,}  part of the software package MOPEX, for this task.
Mosaicking is done in three consecutive steps:
\paragraph{Reprojection}
Image data are reprojected and interpolated
onto a common rectangular grid with a pixel scale of \unit{1.20}{\arcsec}.%
\footnote{ IRAC flux calibration is tied to a pixel scale of \unit{1.22}{\arcsec}. See  \sectref{sect:IRAC:validation} for a discussion.}
In particular, the effects of optical distortion are taken into account.

The asteroid rest frame is used as reference frame, i.e.\  the asteroid position is constant on all reprojected images
while the positions of inertial background sources vary. In some of our observations, background star trails are longer than \unit{10}{\arcsec}.

\paragraph{Outlier rejection}
Outlier pixels containing anomalously high or low signal are detected and rejected. These may be due to cosmic ray hits or detector artefacts, which may be permanently bad pixels or in response to a previous over-exposure, e.g.\ due to a cosmic ray hit \citep[see][]{IRACdatahandbook}. Also non-co-moving sources are partially rejected.

Outlier detection is based on pixel-wise image-to-image comparison. 
MOPEX provides three complementary algorithms for that; careful parameter tuning on a case-by-case basis is required in order to reliably reject outliers while keeping legitimate sources. It is crucial to inspect the maps of rejected pixels and to visually compare them with the input images in order to gage the quality of outlier rejection---the Airy rings around targets, in particular, are in danger of being erroneously rejected. Several reiterations may be required. If background matching was not well performed (see \sectref{sect:overlap}), large parts of one input image may be flagged as outliers, requiring a reiteration of that step, too.

\paragraph{Coaddition}
A mosaic is generated by pixel-wise averaging those parts of the input images flagged as good. 
Also a coverage map is generated, indicating the number of good input image pixels used for each output pixel.
Coverage maps are particularly useful for gaging the quality of  mosaics.


	\subsubsection{Automated source extraction and photometry using MOPEX}
        \label{sect:IRAC:APEX}

MOPEX contains a program to automatically determine the position and absolutely calibrated brightness of point sources called APEX. APEX fits segments of mosaic images with the point-spread function, which is constant over time to a high accuracy. If a satisfactory fit is reached, APEX assumes a detection and determines the source flux by scaling the point-spread function and (if requested by the user) fitting the background.

Using APEX for flux determination is not generally recommended by the SSC
(since flux calibration is performed using synthetic aperture photometry), but considered preferable inside fields containing many point sources: 
In the case of two or more sources whose point-spread functions overlap significantly, synthetic aperture is futile, whereas point-spread-function fitting still has a certain chance of success.

While we could determine flux values of calibration standard stars using APEX (c.f.\ \sectref{sect:IRAC:validation}), we got no meaningful results in the case of asteroids with near-by stellar sources.
We reckon the reason for this is the movement of the asteroid targets relative to the inertial background: While on mosaic images the point-spread function is probably a good match to the image of the asteroid itself, the trails of background stars are  only poorly matched.

	\subsubsection{Synthetic aperture photometry}
        \label{sect:IRAC:ATV}
Instead, we extracted asteroid flux densities from mosaicked images using the synthetic aperture photometry routines used for the analysis of IRTF data 
\citep[see \sectref{sect:IRTF:aperture}; the software has been developed by M.][]{Delbo2004}.
We generally used different combinations of aperture and sky annulus radii for which aperture correction factors were available  \seesect{sect:IRAC:aperturecorrection}.
Structure in the celestial background (e.g.\ background stars) often necessitated using small radii for both aperture and sky annulus. This effect is more severe in the short-wavelength channels 1 and 2 (the relative brightness of asteroids is larger in channels 3 and 4).

Before extracting fluxes, we converted the units of mosaic images from \mega\jansky\per\sr\ into  \micro\jansky\per$\text{pixel}$, such that the photometry output is in units of \micro\jansky.

	\subsubsection{Aperture correction}
        \label{sect:IRAC:aperturecorrection}

Let $f_{i,j,k}$ denote the flux contained in an aperture of radius $i$ minus the sky background content estimated from the flux inside an annulus of inner radius $j$ and outer radius $k$ (all radii in pixels).
In general, $f_{i,j,k}$ is a function of 
 $i$, $j$, and $k$: Varying $i$ changes the percentage of the point-spread function (PSF) energy content contained in the aperture, small values for $j$ and $k$ imply that parts of the source flux are inside the sky annulus, leading to sky-background overestimation and hence flux underestimation.

By virtue of the IRAC flux calibration scheme \seesect{sect:IRAC:calibration}, $f_{10,10,20}$ is generally the best available approximation to the ``true'' flux $f_\text{true}$.
Structure in the celestial background, however, often renders it impractical 
 to use such large radii (see above).
Fortunately, due to the high temporal stability of the IRAC PSFs, the ratio $f_{10,10,20}/f_{i,j,k}$ is very stable, such that smaller radii can safely be used if the resulting flux is later multiplied by the appropriate \textit{aperture correction factor} $AC_{i,j,k}$:
\begin{equation}
\label{eq:IRAC:aperturecorrection}
f_{10,10,20} \sim AC_{i,j,k} f_{i,j,k}.
\end{equation}

\begin{table}
\caption[IRAC aperture correction factors for a mosaic pixel scale of \unit{1.20}{\arcsec}]{Aperture correction factors
$AC_{i,j,k}$
 for the four IRAC channels and different aperture and sky annulus radii $i$, $j$, and $k$ (in mosaic pixels,  \unit{1.20}{\arcsec} each), see \eqref{eq:IRAC:aperturecorrection}.
The values in this table are assumed to be accurate to roughly one percent and differ slightly (but significantly) from those given in the IRAC data handbook (p.\ 53) due to the different pixel sizes used \seesect{sect:IRAC:validation}.}
\label{table:IRAC:aperturecorrection}
\centering
\begin{tabular}{rl|llll}
\toprule
Aperture & Sky annulus  & Channel 1 & Channel 2 & Channel 3 & Channel 4 \\
\midrule
2  & 2--6   & 1.253 & 1.278 & 1.414 & 1.615\\
2  & 10--20 & 1.236 & 1.259 & 1.393 & 1.582\\
3  & 3--6   & 1.128 & 1.133 & 1.154 & 1.255\\
3  & 10--20 & 1.110 & 1.114 & 1.131 & 1.235\\
5  & 5--10  & 1.057 & 1.067 & 1.065 & 1.081\\
5  & 10--20 & 1.043 & 1.049 & 1.050 & 1.060\\
10 & 10--20 & 1.000 & 1.000 & 1.000 & 1.000\\
\bottomrule
\end{tabular}
\end{table}

Aperture correction factors are obviously dependent on the pixel scale. Those quoted in the IRAC datahandbook (p.\ 53) are for a pixel scale around \unit{1.22}{\arcsec} and therefore not directly applicable to us, since we use a mosaic pixel scale of \unit{1.20}{\arcsec}.
We have therefore determined aperture correction factors to be used with our mosaic pixel scale of \unit{1.20}{\arcsec} from observations of the bright star HD~165459, which is also used as an IRAC photometric calibration standard (see also \sectref{sect:IRAC:validation}). 
See \tableref{table:IRAC:aperturecorrection} for our results, which we estimate to be accurate to roughly one percent, limited by the statistical uncertainty of the star flux determinations.

Note that this procedure can only be used in the case of a high temporal PSF stability---it cannot be used for, e.g.,  mid-infrared asteroid imaging  using the IRTF (where the PSF depends on variable parameters such as seeing or telescope focus).

	\subsubsection{Color correction}
        \label{sect:IRAC:CC}

IRAC is a broad-band photometer. Derivation of monochromatic flux values from the detector output therefore requires assumptions on the target spectrum to be made
and an appropriate choice of the effective filter wavelength $\lambda_0$.
Flux values obtained so far are correct for a nominal source spectrum $f_{nom}(\lambda) \propto \lambda^{-1}$ (with wavelength $\lambda$) \seesect{sect:IRAC:calibration}.
For other target spectra, this causes a systematic multiplicative offset, requiring a \emph{color correction} to be made.
Color correction factors for several target spectra are tabulated in the IRAC data handbook, p.\ 48 \citep{IRACdatahandbook}, but none of them is directly applicable to our asteroid targets.

When observing a source with flux distribution $f(\lambda)$, the detector response per integration time, $R$, equals
\begin{equation}
\label{eq:IRAC:CC:detectorresponse}
R = f\left(\lambda_0\right) \int \frac{f\left(\lambda\right)}{f\left(\lambda_0\right)} \frac{R\left(\lambda\right)}{hc/\lambda}  \text{d}\lambda
\end{equation}
with the  spectral detector response $R(\lambda)$ in units of electrons per photon (hence the division by the photon energy  $hc/\lambda$; all integrals in this section run from 0 to $\infty$).
The detector response to a target with spectrum $f_{tar}(\lambda)$, $R_{tar}$, and that to a calibration star with spectrum $f_{cal}(\lambda)$, $R_{cal}$, are then related to the monochromatic flux densities at the wavelength $\lambda_0$:
\begin{subequations}
\begin{align}
f_{tar}(\lambda_0) &= f_{cal}(\lambda_0) 
\frac{R_{tar}}{R_{cal}} \times K \\
K &= 
\frac{f_{tar}(\lambda_0)} {f_{cal}(\lambda_0)}
\frac{\int f_{cal}(\lambda) R(\lambda)/(hc/\lambda)\text{d}\lambda}{\int f_{tar}(\lambda) R(\lambda)/(hc/\lambda)\text{d}\lambda}
\end{align}
\end{subequations}
with the \emph{color-correction factor} $K$ taking account of the differences in spectral shape through the detector passband. Note in particular that $K$ is invariant under rescaling of both $f_{tar}$ and $f_{nom}$.

The IRAC BCD flux calibration includes a color correction assuming a nominal source spectrum $f_{nom}(\lambda) \propto \lambda^{-1}$.
This requires a second color correction to be made:
\begin{subequations}
\label{eq:IRAC:CC}
\begin{align}
f_{tar}(\lambda_0) &= \frac{f_{nom}(\lambda_0)}{K} \\
K    &= 
\frac{f_{nom}(\lambda_0)}{f_{tar}(\lambda_0)}
\frac{\int f_{tar}(\lambda) R(\lambda)/(hc/\lambda)\text{d}\lambda}{\int f_{nom}(\lambda) R(\lambda)/(hc/\lambda)\text{d}\lambda}.
\end{align}
\end{subequations}

\begin{table}[tb]
\caption[IRAC color-correction factors for NEATM spectra]{IRAC color-correction factors for NEATM asteroid spectra for several heliocentric distances $r$ (in \AU), solar phase angles $\alpha$ (in degrees), model parameters $\eta$, and geometric albedos \pv\ (the slope parameter $G$ is assumed to equal 0.15). The last column refers to the section in which the thus calculated color-correction factors are required.}
\label{table:IRAC:CC}
\centering
\begin{tabular}{rrrr|lllll}
\toprule
$r$ & $\alpha$ & $\eta$ & \pv & Channel 1 & Ch.\ 2 & Ch.\ 3 & Ch.\ 4 & \\ 
\midrule
1.00 & 60.0 & 2.0 & 0.2 & 1.138 & 1.070 & 1.033& 1.003& \Sectref{sect:PH5}\\
1.27 & 52.33 & 2.48 & 0.35 &  1.236 &  1.129 & 1.070 & 1.034 & \Sectref{sect:ML}\\
\bottomrule
\end{tabular}
\end{table}

We have developed a computer routine which calculates color correction factors for all IRAC channels and several model spectra, including  NEATM asteroid spectra with variable model parameters. 
It makes use of the measured spectral response curves of each IRAC detector tabulated on-line by the SSC (\url{http://ssc.spitzer.caltech.edu/irac/spectral_response.html}).
We have verified that our software reproduces all color correction factors quoted in the IRAC handbook (p.\ 48) to within one percent. See \tableref{table:IRAC:CC} for IRAC color correction factors for NEATM spectra with several model parameters.

The values of $\lambda_0$ used in the IRAC BCD pipeline 
were chosen in a way that minimizes the dependence of color-correction factors on the spectral slope.
They are defined the following way:
\begin{equation}
\label{eq:IRAC:effectivelambda}
\lambda_0 = \frac{\int R(\lambda)\text{d}\lambda}{\int R(\lambda)/\lambda \text{d}\lambda}.
\end{equation}
Evaluation of this integral for the four tabulated IRAC passbands yields
 effective wavelengths of 3.550, 4.493, 5.731, and \unit{7.872}{\micron}  \seetablepage{table:IRAC:channels}.

\subsubsection{Estimation of flux uncertainty}
        \label{sect:IRAC:uncertainty}

We estimated the statistical flux uncertainty in two mutually independent ways, which we added in quadrature:
\begin{itemize}
\item The standard statistical flux uncertainty derived from the scatter in the background flux level
\item The scatter in the aperture-corrected fluxes derived using different possible radii for the synthetic aperture and the sky annulus (see \sectref{sect:IRAC:aperturecorrection}). 
This scatter is estimated to be a measure of the PSF deviation from its nominal form due to background structure or noise.
\end{itemize}

The accuracy of observations in channels 1--3 was generally limited by statistical uncertainty, while our highest signal-to-noise channel-4 observations were occasionally limited by the \unit{3}{\%} \citep{Reach2005} calibration uncertainty.

On the basis of our data reduction pipeline validation experiments described in \sectref{sect:IRAC:validation}
we estimate that our data reduction pipeline does not introduce major systematic uncertainties. 
We therefore neglect any systematic uncertainties
resulting from, e.g.,
aperture correction (\sectref{sect:IRAC:aperturecorrection}) or color correction (\sectref{sect:IRAC:CC}).

\subsection{Validation of our data reduction pipeline}
        \label{sect:IRAC:validation}

\begin{table}
\caption[Determined and expected fluxes of IRAC calibration standard star HD~165459]{Fluxes of the photometric calibration standard star HD~165459 from IRAC observations on 13 February 2006 (upper row) and expected values \citep[lower row, quoted from][Table 6]{Reach2005}. Values in the upper row are \emph{not} color-corrected to enable comparison with the \citeauthor{Reach2005}\ values. Uncertainties in the first row are statistical (from the synthetic aperture procedure), those in the second row are systematic (uncertainties in stellar flux modeling).}
\label{table:IRAC:validation}
\centering
\begin{tabular}{r|rrrr}
\toprule
 & Channel 1 & Channel 2 & Channel 3 & Channel 4 \\
 & (\milli\Jy)  & (\milli\Jy)  & (\milli\Jy)  & (\milli\Jy) \\
\midrule
Determined & $643.9\pm 4.1$ & $410.2 \pm 7.3 $ & $264.3 \pm 8.2$ & $147.0 \pm 2.6$ \\
Expected & $647\pm17$ & $421\pm 11$ & $268.7 \pm 7.1$ & $148.1 \pm 3.9$ \\
\bottomrule
\end{tabular}
\end{table}

\paragraph{Check against a calibration star}
We have validated  our data reduction pipeline by 
determining flux values which were known \textit{a priori:}
The brightness of a star used in the photometric calibration of IRAC.
A list of such stars and their fluxes at IRAC wavelengths has been published by \citet{Reach2005}.
We have searched the database of publicly available Spitzer data and retrieved  IRAC observations of the 
calibration standard star HD~165459 taken on 13 Feb 2006; the on-target integration time per FOV was 5 dither positions times \unit{2}{\second}. 
These observations were performed as part of a regular IRAC calibration program; the downloaded data were processed through the standard IRAC BCD pipeline.

We have determined flux values for the four IRAC channels using the method described in this section without color correction. As can be seen in \tableref{table:IRAC:validation}, fluxes agree with those given by \citet{Reach2005} within the error bars. 

\paragraph{Check pixel-scale effects}
We use a mosaic pixel scale of \unit{1.20}{\arcsec}, while a pixel scale of {$\sim 1.22$}{\arcsec} is used for IRAC calibration.
Note that the photometric calibration is tied to synthetic aperture photometry with an aperture of 10~pixels, so the pixel scale would be expected to matter.

In order to verify that  this  does not introduce an appreciable calibration uncertainty, 
we have mosaicked the above-mentioned HD~165459 data twice, using pixel scales of 1.20 and \unit{1.22}{\arcsec}, respectively.
Fluxes resulting from synthetic aperture photometry (with ``default'' radii of 10, 10, and 20 pixels) agree to within a few permille.

We have also determined aperture correction factors on the \unit{1.22}{\arcsec} pixel scale mosaics using the method described in \sectref{sect:IRAC:aperturecorrection}. 
As expected, the thus determined aperture correction factors differ significantly  from those quoted in \tablerefpage{table:IRAC:aperturecorrection}, but  match those tabulated in the IRAC datahandbook, p.\ 53, to within \unit{1}{\%}.


%% file: IRS_general.tex
The InfraRed Spectrograph \citep[IRS; see][]{IRS} 
covers the mid-wavelength range of Spitzer's spectral coverage
with spectroscopic capabilities over the wavelength range from 5.3 to \unit{38}{\micron} at two different spectral resolutions (referred to as low and high resolution in the following); and additional small-FOV imaging capabilities (peak-up imaging, PUI) at central wavelengths around 16 and \unit{22}{\micron}.
At IRS wavelengths,  asteroid flux is practically purely thermal.

After a short overview of the instrument over-all layout and its capabilities as used in our programs (\sectref{sect:IRS:layout}), we detail our observing strategies and data reduction techniques in section \ref{sect:IRS:PUI} and \ref{sect:IRS:lowres} for peak-up imaging and low-resolution spectroscopy, respectively. 
The high-resolution spectroscopy mode  is not used in our programs and is generally disregarded in the following.

More information on the instrument  can be found in \citet{IRS}, the \citet[chapter 7]{SOM}, and in the IRS data handbook provided on-line \citep{IRSdatahandbook} (we made use of version 2.0 from 1 Apr.\ 2006, \url{http://ssc.spitzer.caltech.edu/irs/dh/}). A technical reference for the IRS BCD pipeline (see \sectref{sect:SST:data}) is the IRS pipeline description \citep{IRSpipelinedescription}.

		\subsection{Instrument layout}
                \label{sect:IRS:layout}

Similar to IRAC, IRS is composed of four detectors, located in the Cryogenic Telescope Assembly at a helium bath temperature of \unit{1.4}{\kelvin}, and a set of  control units  located in the Spacecraft Bus at a higher temperature (see \figrefpage{fig:SST:design}).
The four blocked-impurity-band detectors  have $128\times128$ pixels each,
two detectors cover the IRS wavelength range at low spectral resolution ($\lambda/\Delta\lambda$  between 64 and 128)
and two  at high resolution around 600. 
The two short-wavelength detectors are Si:As arrays and referred to as Short-Low (SL) and Short-High (SH) for low and high spectral resolution, respectively.
Analogously, the long-wavelength detectors, Si:Sb arrays, are referred to as Long-Low (LL) and Long-High (LH). 
The low-resolution modules use single diffraction gratings,
the high-resolution modules employ a cross-dispersed echelle layout.
Light from the two imaging apertures is projected onto two designated fields on the SL chip, with an effective FOV of $\unit{54}{\arcsec}\times \unit{81}{\arcsec}$ each at a pixel scale around \unit{1.8}{\arcsec} ($30\times45$ pixels).
The focal plane apertures of the IRS modules have no overlap on the sky; 
bringing a target from one aperture to another requires slewing the spacecraft since IRS does not have any moving parts.

The spectroscopy modules employ narrow slit designs, where the slit width equals the point-spread function (PSF) width at the largest wavelength---due to diffraction-limited imaging the PSF width increases with wavelength.
The projected slit widths are \unit{3.6}{\arcsec}~(SL), \unit{4.7}{\arcsec}~(SH), \unit{10.5}{\arcsec}~(LL), and \unit{11.1}{\arcsec}~(LH).

Ground-based tests showed the photometric response of the detectors to be temporally stable within \unit{1}{\%}.
The absolute photometric accuracy of spectroscopic observations is typically limited by spill-over losses due to source mis-centering, which are reduced by a procedure called peak up (see below).

For science observations, the IRS detectors use a read-out system named ``sample-up-the-ramp'' whereby each pixel is read out non-destructively a number of times at the beginning of each measurement, and an equal number of times at the end of each measurement before the detector is reset.
This reduces read-out noise and  helps mitigate cosmic ray hits.
It also enables meaningful data to be extracted from moderately saturated pixels, although at reduced signal-to-noise.

                     \paragraph{Peak up}
                     \label{sect:IRS:peakup}
The $1\sigma$ accuracy of blind Spitzer pointings is  \unit{0.5}{\arcsec} (see \sectrefpage{sect:SST:telescope}), therefore significant parts of the PSF are at risk of being placed outside the slit, in particular for the narrow short-wavelength slits. Mis-centering of spectroscopy targets results in systematic underestimation of the absolute flux level and furthermore introduces a spurious spectral slope over the sub-module's spectral range because the PSF broadens with wavelength, leading to larger losses.

IRS offers a mechanism, called \emph{peak up,} to center 
spectroscopy targets in the slit more accurately, using more accurate small  slews after refining spacecraft pointing on one of two dedicated imaging arrays called \emph{peak-up arrays.}%
\footnote{ There is a third peak-up option, which is however not available for moving targets \citep[see][]{SOM}.}
The latter are also used for photometric imaging in the PUI mode which is discussed in \sectref{sect:IRS:PUI}.
Peaking up also enables spectroscopy of targets with relatively large positional uncertainty, which would otherwise be impossible to acquire.

Peak up zeroes in on the brightest source within a $24\times24$-pixel ($\unit{43.2}{\arcsec}\times \unit{43.2}{\arcsec}$) square centered at the nominal position of the peak-up target.
The peak-up target can be the spectroscopy target itself or a near-by source whose position relative to the spectroscopy target is accurately known \citep[see][for details]{SOM}. 
The brightness of peak-up targets should be within 5--\unit{150}{\milli\jansky} at \unit{16}{\micron} or 15--\unit{340}{\milli\jy} at \unit{22}{\micron}, depending on the peak-up array selected.

		\subsection{Peak-Up Imaging (PUI)}
                \label{sect:IRS:PUI}
Although originally designed for peak up only, the two peak-up FOVs can be used to provide  science-quality photometry. Since Spitzer cycle II (summer 2005), Peak-Up Imaging (PUI) is an IRS operation mode fully supported by the Spitzer Science Center, which includes regular calibration observations and a dedicated  BCD pipeline.

Light entering the two peak-up apertures is passed through  suitable bandpass filters and then projected onto two designated parts of the SL detector, which are not used for spectroscopy.
Both FOVs have
a total size of $41\times56$ pixels with significant vignetting at the boundaries; the un-vignetted part in either FOV is $30\times45$ pixels corresponding to roughly $\unit{54}{\arcsec}\times \unit{81}{\arcsec}$ at a pixel scale around \unit{1.8}{\arcsec}. 
After a PUI observation, the entire SL chip is read out, providing low-resolution spectra in two orders and imaging in both peak-up fields. Only one of the four fields-of-view, however, is on a given point-source target, the others  observe neighboring ``serendipitous'' fields.

\begin{figure}
\centering
  \includegraphics[width=0.6\linewidth, angle=-90]{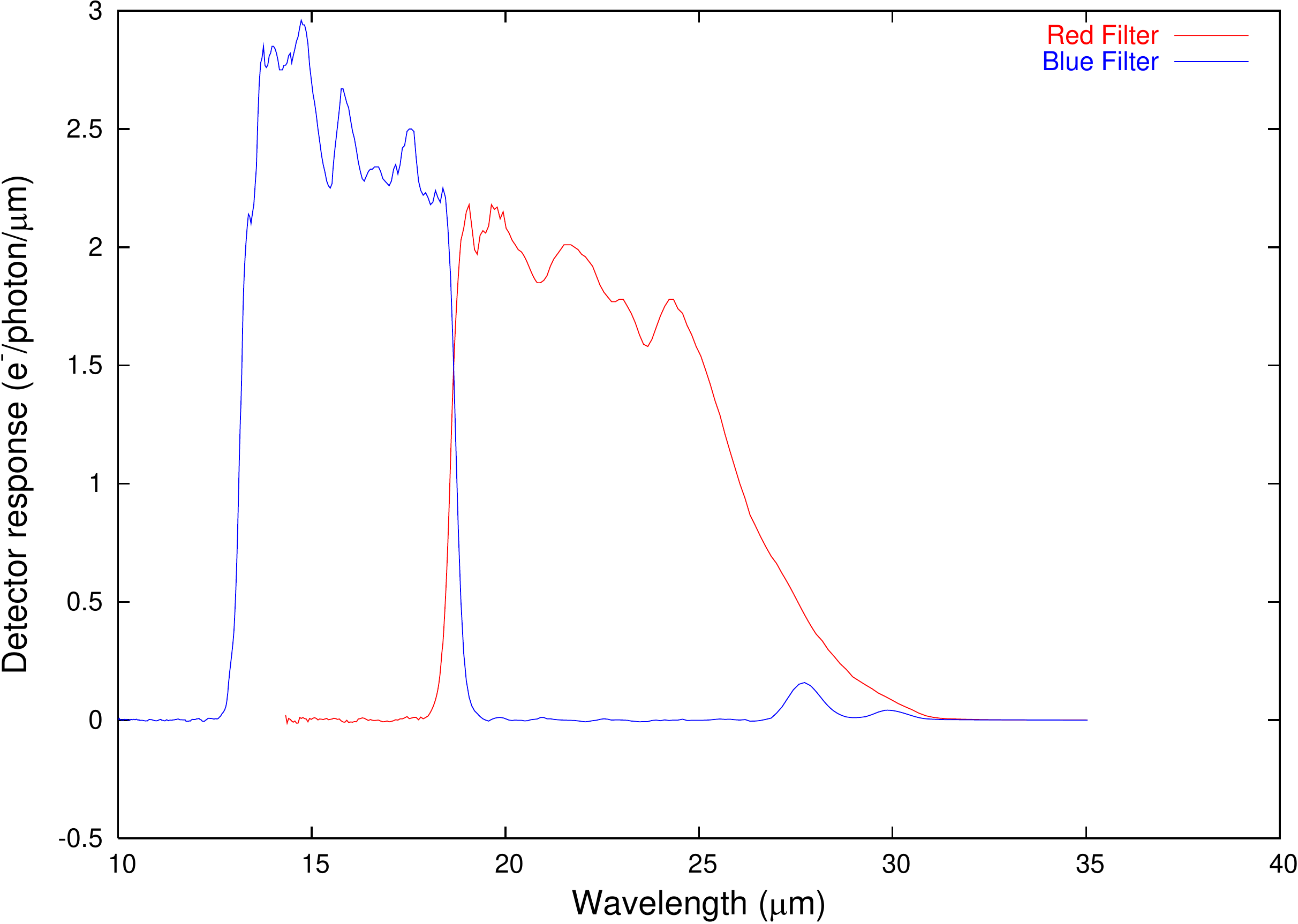}
\caption[Detector response curves for the two IRS peak-up arrays]{Detector response curves for the two IRS peak-up arrays. Depicted is the product of the  total filter transmission and the  spectral detector quantum efficiency, from the files linked at \url{http://ssc.spitzer.caltech.edu/irs/spectral_response.html}. Note the slight leak in the blue filter longward of \unit{25}{\micron}.}
\label{fig:IRS:PUIfilters}
\end{figure}

See \figref{fig:IRS:PUIfilters} for  detector response curves for the two PU arrays. The effective wavelength of the ``blue'' filter is around \unit{16}{\micron}, that of the ``red'' filter is about \unit{22}{\micron}, the FWHM spectral breadth of either is some \unit{30}{\%}.

\subsubsection{Instrument performance}
\label{sect:PUI:performance}

\begin{table}
\caption[PUI point-source sensitivity]{\label{table:PUI:sensitivity}
PUI point-source sensitivity ($1\sigma$ in units of \micro\Jy; see also \eqrefpage{eq:IRAC:sensitivity}) for one image frame of the stated integration time (frame time). The celestial background is assumed to be high, as is appropriate for most asteroid observations. 
}
\centering
\begin{tabular}{r|ll}
\toprule
Frame time (sec) & ``Blue'' filter (\unit{16}{\micron}) & ``Red'' filter (\unit{22}{\micron}) \\
\midrule
30 & 80 & 120\\
14 & 120 & 180\\
6  & 190 & 280\\
\bottomrule
\end{tabular}
\end{table}

\begin{table}
\caption[PUI saturation limits]{\label{table:PUI:saturation}
Largest non-saturated point-source flux (in units of \milli\Jy) for the two PUI filters and high celestial background. Units in this table are \milli\Jy, \emph{not} \micro\Jy\ as in \tableref{table:PUI:sensitivity}! Note that imaging of somewhat brighter sources is possible regardless of saturation, although at reduced signal-to-noise, due to the sample-up-the-ramp read-out mode \seesect{sect:IRS:layout}.
}
\centering
\begin{tabular}{r|ll}
\toprule
Frame time (sec) & ``Blue'' filter (\unit{16}{\micron}) & ``Red'' filter (\unit{22}{\micron}) \\
\midrule
30 & 35 & 80\\
14 & 80 & 190 \\
6  & 180 & 410 \\
\bottomrule
\end{tabular}
\end{table}

Three values for the frame time, i.e.\ the integration time per frame, are selectable: 6, 14, and \unit{30}{\second}. See \tableref{table:PUI:sensitivity} for the corresponding point-source sensitivities and \tableref{table:PUI:saturation} for the saturation limits.
As in the case of IRAC \seesect{sect:IRAC:strategy}, longer integrations can be achieved by repeating frames in-place or by dithering; this will be discussed in \sectref{sect:PUI:strategy}. 
For deep PUI observations, the dominant source of uncertainty is celestial background radiation due to the thermal emission of zodiacal dust which is concentrated in the ecliptic plane. Most of our asteroid observations took place at low ecliptic latitudes, hence we assumed the most conservative sensitivity values (``high'' background) for planning of observations.
In the IRS datahandbook \citep[p.\ 50]{IRSdatahandbook}, systematic uncertainties in the PUI flux calibration are advertised to be within \unit{6}{\%} (see also \sectref{sect:PUI:validation}).

Similar to IRAC, PUI imaging suffers from a slight amount of optical distortion due to the instrument's offset from the telescope optical axis. While less severe than in the case of IRAC \seesect{sect:IRAC:layout} due to the smaller field-of-view, distortion must be accounted for in the data reduction to avoid significant photometric errors.

\subsubsection{Asteroid observation strategies}
\label{sect:PUI:strategy}

The design of our PUI observations was similar to that of our IRAC  observations \seesect{sect:IRAC:strategy}---note that all our PUI targets were also observed with IRAC.

Required integration times were estimated on the basis of the sensitivities given in \tableref{table:PUI:sensitivity} and lower limits on expected target flux (see also \sectref{sect:IRAC:IT}); a typical flux uncertainty at PUI wavelengths is a factor 3.
Identical timing constraints were imposed on IRAC and PUI observations of the respective asteroid targets, thus preventing bright IRAS sources from  temporarily damaging the detector through latent effects \seesect{sect:IRAC:timing}.
We typically dithered \seesect{sect:IRAC:dithering} around the nominal target positions (4 or more dither positions) to mitigate cosmic ray hits and defective pixels; note that due to the limited FOV size, PUI dithering distances are much smaller than for IRAC.

\subsubsection{BCD pipeline}
\label{sect:PUI:BCD}

As for all Spitzer science data, PUI data are partially reduced and calibrated at the Spitzer Science Center (SSC) using an automated set of computer routines called the PUI BCD pipeline
\seesect{sect:SST:data}.
The PUI BCD pipeline is \emph{mutatis mutandis} nearly identical to the IRAC BCD pipeline \seesect{sect:IRAC:BCD_pipeline}, we  here  sketch the former, focusing on  differences among the two.
One BCD file per filter and pointing is produced, i.e.\ a FITS file with most known instrument artefacts removed and in physical flux units (\mega\Jy\per\sterad).
Each BCD file is associated with several ancillary files including maps of estimated flux uncertainty and masks where potentially unreliable pixels are flagged (e.g.\ suspected cosmic ray hits, vignetted pixels, and permanently damaged pixels); those ancillary files are useful for further data reduction.
Technical references for the PUI BCD pipeline are the IRS pipeline description \citep{IRSpipelinedescription}
and the IRS datahandbook \citep{IRSdatahandbook}.

Corrections for dark current are determined as in the case of IRAC \seesect{sect:IRAC:dark}, but no time-dependent drift has been found, i.e.\ there is no first-frame effect \seesect{sect:IRAC:firstframe}. PUI data are flat-fielded similar to IRAC data \seesect{sect:IRAC:flat} and flux calibrated against a set of stellar calibrators \citep[see \sectref{sect:IRAC:calibration} and][p.\ 178; flux calibration will be further discussed in \sectref{sect:PUI:validation}]{SOM}. As in the case of IRAC, PUI data are flux calibrated assuming a target spectrum inversely proportional to wavelength, necessitating color corrections to be made \seesect{sect:PUI:reduction}. Pointing refinement, as is done for IRAC data \seesect{sect:IRAC:pointing}, is not performed on PUI data; it would not be feasible in general due to the low density of bright stars  at PUI wavelengths.

All IRS detectors including the SL chip used for PUI are subject to the damaging effects of solar flares, resulting in an increasing number of permanently or temporarily defective pixels. As a part of calibration activities, the SSC monitors the pixel responsivity and flags permanently damaged pixels as ``dead'' in the BCD pipeline. Some pixels, called ``rogue pixels,'' are overly responsive at times, but operate normally most of the time. The SSC issues regularly updated lists of rogue pixels, but rogue pixels are not flagged in BCD files.

\subsubsection{Data-reduction steps taken by us}
\label{sect:PUI:reduction}

Reduction of PUI data is largely analogous to IRAC data reduction \seesect{sect:IRAC:asteroids}.
The most significant differences are the following:

\paragraph{Visual inspection} While  all PUI data frames were visually inspected, no unusable PUI frames were found. This is  due to the much lower density of bright stars at PUI wavelengths relative to IRAC wavelengths, but also due to the sample-up-the-ramp read-out mode \seesect{sect:IRS:layout} mitigating the effects of cosmic-ray hits.
Rogue pixels \seesect{sect:PUI:BCD}, however, are clearly recognizable as such in visual inspection. It turned out to be easier to manually add them to the bad-pixel map provided by the BCD pipeline than to tweak the mosaic routines (see below) to reliably recognize rogue pixels without falsely flagging good pixels. To lighten this task, an IDL routine was developed which flags a rogue pixel for all BCDs belonging to a series of consecutive observations.
\paragraph{Background matching} Since no first-frame effect is present, background matching \seesect{sect:overlap} is not needed.
\paragraph{Mosaicking} A mosaic pixel scale of \unit{1.8}{\arcsec} is used. Generally, no cosmic-ray hits or background sources are present in BCD images. 
Rogue pixels were  rejected manually (see above). It is important to have \emph{mosaic.pl} disregard pixels in the vignetted region close to the FOV boundary, where no reliable flat fielding is possible. They are flagged as vignetted in the BCD-provided bad-pixel map but not rejected by default. 

\paragraph{Flux extraction, aperture correction, color correction}
The PUI flux calibration is based on synthetic aperture photometry using a sky annulus with inner and outer radii of 8 and 14 pixels, respectively, and aperture radii of 3 (blue) and 4 (red) pixels \citep{IRSdatahandbook}.%
\footnote{ These values appear to have changed with the new BCD pipeline version 15---see \sectref{sect:PUI:validation}.}
These aperture radii are small compared to the FWHM width of the PSF, which is around 2 (blue) and 2.5 (red) pixels, hence fluxes are aperture corrected by default.
We generally used larger aperture radii and the quoted sky annulus radii; see \tableref{table:PUI:aperturecorrection} for aperture correction factors.
See \tableref{table:PUI:CC} for color-correction factors which were determined like their IRAC counterparts \seetablepage{table:IRAC:CC} but using the PUI detector response \seefigpage{fig:IRS:PUIfilters}.

\begin{table}
\caption[IRS PUI aperture correction factors]{Aperture correction factors $AC_{i,j,k}$ (see \eqrefpage{eq:IRAC:aperturecorrection})
 for the two PUI filters and different aperture radii (in mosaic pixels of \unit{1.8}{\arcsec} each). Inner and outer sky annulus radii are 8 and 14 pixels, respectively.
\citep[from][p.\ 50]{IRSdatahandbook}
}
\label{table:PUI:aperturecorrection}
\centering
\begin{tabular}{r|ll}
\toprule
Aperture & Blue filter & Red filter \\
\midrule
2 & 1.69 & 2.13 \\
3 & 1.38 & 1.57 \\
4 & 1.16 & 1.36 \\
5 & 1.07 & 1.18 \\
6 & 1.05 & 1.07 \\
7 & 1.04 & 1.03 \\
8 & 1.03 & 1.02 \\
9 & 1.02 & 1.02 \\
10& 1.01 & 1.01 \\
11& 1.00 & 1.01 \\
12& 1.00 & 1.01 \\
\bottomrule
\end{tabular}
\end{table}

\begin{table}
\caption[PUI color-correction factors for NEATM spectra]{IRS PUI color-correction factors for NEATM asteroid spectra for several heliocentric distances $r$ (in \AU), solar phase angles $\alpha$ (in degrees), model parameters $\eta$, and geometric albedos \pv\ (the slope parameter $G$ is assumed to equal 0.15). The last column refers to the section in which the thus calculated color-correction factors are required. The effective wavelengths were calculated from \eqrefpage{eq:IRAC:effectivelambda}.}
\label{table:PUI:CC}
\centering
\begin{tabular}{rrrr|lll}
\toprule
$r$ & $\alpha$ & $\eta$ & \pv & Blue (\unit{15.8}{\micron}) & Red (\unit{22.3}{\micron}) & \\ 
\midrule
1.00 & 60.0 & 2.0 & 0.2 & 0.982 & 0.994 & \Sectref{sect:PH5}\\
\bottomrule
\end{tabular}
\end{table}

\subsubsection{Validation}
\label{sect:PUI:validation}


We aimed at validating our PUI data reduction pipeline, like we have validated our IRAC pipeline  \seesect{sect:IRAC:validation}.
However, as of April 2007, 
the Spitzer Science Center has not announced 
which flux standards were used for PUI flux calibration.
Upon request,  the Spitzer helpdesk informed us (private email communication from 24 Feb 2006) that the primary flux  standard used for PUI calibration was the star HD~163466, with PUI fluxes of \unit{54.29}{\milli\Jy} (blue) and \unit{27.44}{\milli\Jy} (red) before color correction (similar to the IRAC calibration standard quoted in \sectref{sect:IRAC:validation}).

PUI observations of that star were performed on 20 Apr, 26 Apr, and 26 May 2006 during regular IRS calibration observations, data are public. BCD files from these three observations were reduced by us using the data reduction pipeline described herein.
The resulting flux values and their deviations from the nominal values are given in \tableref{table:PUI:validation}.

\begin{table}
  \caption[Determined PUI fluxes of the star HD~163466 which was used for PUI flux calibration]{Determined fluxes of the star HD~163466, which was used for PUI flux calibration, from PUI observations on 20 Apr, 26 Apr, and 26 May 2006. Fluxes should be (Spitzer helpdesk, private email communication from 24 Feb 2006): \unit{54.29}{\milli\Jy} (blue) and \unit{27.44}{\milli\Jy} (red) before color correction.}
  \label{table:PUI:validation}
  \centering
  \begin{tabular}{r|llll}
    \toprule
    Date & ``blue'' flux  & off by & ``red'' flux & off by \\
         & (\milli\Jy)  & (\%)   & (\milli\Jy)& (\%)  \\
    \midrule
    2006/04/20 & $51.06\pm0.30$ & -5.9 & $28.61 \pm 0.16$ & +4.3\\
    2006/04/26 & $50.98\pm0.41$ & -6.1 & $28.86 \pm 0.13$ & +5.2\\
    2006/05/26 & $51.57\pm0.47$ & -5.0 & $28.95 \pm 0.17$ & +5.5\\
    \bottomrule
  \end{tabular}
\end{table}

We see significant deviations from the nominal values, roughly \unit{-5.5}{\%} in the blue filter and roughly \unit{+5}{\%} in the red filter. We have reported this discrepancy to the Spitzer helpdesk over summer and fall 2006. The helpdesk found no flaw in our data reduction pipeline but stated that our deviation in photometric zeropoints was well within the quoted uncertainty of \unit{$\pm 6$}{\%}. We feel, however, that our results reveal a rather serious calibration issue with the BCD pipeline 14, possibly related to the fact that flux calibration is tied to synthetic aperture photometry with very small radii, leading to relatively large uncertainties due to the needed aperture correction.

It should also be noted that in PUI observations of an exoplanet, repeated over a timespan of \unit{6}{\hour}, \citet{Deming2006} found a time dependent drift in PUI photometric gain, which cannot be explained to date.

\paragraph{Update: pipeline version 15}
The validation attempt presented in this section is based on the IRS BCD pipeline version 14, issued in Jul 2006. 
 The Spitzer Science Center has issued an updated IRS BCD pipeline, version 15, on 28 Feb 2007
 ``which results in changes in measured photometric values of $<\unit{15}{\%}$.''%
\footnote{ C.f.\ \url{http://ssc.spitzer.caltech.edu/archanaly/plhistory/irs.html}.}
 All PUI data in the archive (including ours) have been reprocessed; however, at the time of writing, no updated pipeline documentation is available.
We hope that this new pipeline version fixes the issues found herein and we will reprocess all our PUI data based on the reprocessed BCDs. \emph{All PUI fluxes quoted in this thesis are therefore preliminary---updated PUI fluxes based on the IRS BCD pipeline version 15 will be used for journal publication.}

		\subsection{Low-resolution spectroscopy}
                \label{sect:IRS:lowres}

Two modules (named SL and LL) provide low-resolution spectroscopy at a relative spectral resolution between 64 and 128 at wavelengths between 5.2 and \unit{38.0}{\micron}.
Each module has two sub-apertures in its slit, bringing source light in 1st and 2nd diffraction order of its diffraction grating onto the respective chip.
The two diffraction orders do not overlap on the chip (this is prevented using suitable bandpass filters in the optical path) nor on the sky.
The telescope must be slewed in order to bring an observational target from one sub-slit into another.
In total, four telescope pointings are required to observe a spectrum over the full low-resolution wavelength range, one per sub-slit. 
The projected slit widths of the two modules match the FWHM PSF widths at the largest wavelengths (Spitzer imaging is diffraction limited, hence the PSF width grows with wavelength). Their length is much larger, 
such that one-dimensional spatial resolution is acquired along the slit in addition to the spectral dimension. 
There is some spectral overlap among spectrally adjacent IRS modules and spectral orders, allowing  spectra to be cross-checked.
See \tableref{table:IRS:slits} for an overview of slit dimensions and spectral ranges.

\begin{table}
\caption[IRS low-resolution modules: Overview of spectral coverage, detector pixel scale, and projected slit dimensions.]{IRS low-resolution modules: Overview of spectral coverage, detector pixel scale, and projected slit dimensions. Note the slight spectral overlap between the sub modules. PUI is performed on the SL detector.}
\label{table:IRS:slits}
\centering
\begin{tabular}{l|llll}
\toprule
& Wavelength range &Pixel scale & Slit width & Slit length \\
\midrule
SL 2nd order & \unit{5.2--8.7}{\micron}   & \unit{1.8}{\arcsec} & \unit{3.6}{\arcsec} & \unit{57}{\arcsec} \\
SL 1st order & \unit{7.4--14.5}{\micron}  & \unit{1.8}{\arcsec} & \unit{3.7}{\arcsec} & \unit{57}{\arcsec} \\
LL 2nd order & \unit{14.0--21.7}{\micron} & \unit{5.1}{\arcsec} & \unit{10.5}{\arcsec} & \unit{168}{\arcsec} \\
LL 1st order & \unit{19.5--38.0}{\micron} & \unit{5.1}{\arcsec} & \unit{10.7}{\arcsec} & \unit{168}{\arcsec} \\
\bottomrule
\end{tabular}
\end{table}

IRS spectroscopy observations are specified in SPOT. Adjustable parameters are: 
\begin{itemize}
\item Peak-up method and parameters \seesect{sect:IRS:peakup}
\item Sub-modules to be used; if more than one module is selected, observations are ordered  by wavelength with short wavelengths first, i.e.\ SL2, SL1, LL2, LL1 if all four low-resolution modules are selected%
\footnote{ Pointing and tracking accuracy decays with time; this has a more profound influence on short-wavelength observations using relatively narrow slits.}
\item Integration time per frame and sub-module, where possible choices are 6, 14, 60, and \unit{240}{\second} for the SL modules;  6, 14, 30, and \unit{120}{\second} for the LL modules
\item Number of frames (``cycles'') to be taken at each position to increase the total integration time
\end{itemize}

There are two different spectroscopy modes:
\emph{staring} and \emph{mapping}. Staring mode is primarily designed for point sources after successful peak up;  in each sub-module the source is first placed at \unit{33}{\%} of the slit length, then at \unit{66}{\%}, at each position the specified number of frames is taken---the total integration time is thus twice the number of cycles times the frame time. The purpose of this ``nodding'' strategy is to enable correction for diffuse background emission.
In mapping mode, the source is placed at a selectable number of offsets parallel and/or perpendicular to the slit direction, where the angular offset distances are adjustable. The mapping mode was designed for spectral maps of extended sources, but is also useful for spectroscopy of sources which cannot be peaked up on \seesect{sect:Patroclus}.

\paragraph{Instrument performance}

\begin{figure}[tb]
\centering
\parbox{0.45\linewidth}{
  \includegraphics[width=\linewidth]{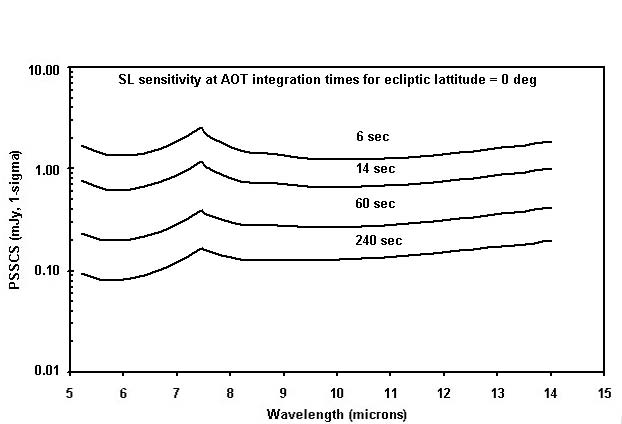}    
}
\hfill
\parbox{0.45\linewidth}{
  \includegraphics[width=\linewidth]{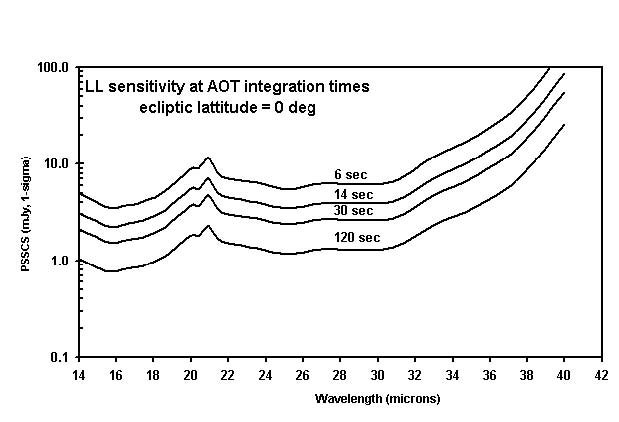}  
}
  \caption[IRS low-resolution sensitivities]{$1\sigma$ point-source sensitivity per frame for the SL (left) and LL (right) modules for the four possible frame times  assuming low ecliptic latitude, i.e.\ high background due to zodiacal emission, as is appropriate for most asteroid observations. \citep[Figs.\ from][p.\ 160 and 161]{SOM}}
  \label{fig:IRS:sensitivity}
\end{figure}

\begin{table}
  \caption[Absolute photometric uncertainties due to spill-over losses for the IRS SL and LL modules]{Absolute photometric uncertainties due to spill-over losses for the SL and LL modules and the three selectable peak-up options. Note that the position uncertainty after a ``low''-accuracy peak-up  is actually worse than Spitzer's nominal pointing accuracy; it should only be used on sources of poorly known position. \citep[see][p. 187]{SOM}}
  \label{table:IRS:PUlosses}
\centering
  \begin{tabular}{llll}
\toprule
 & High & Moderate & Low \\
\midrule
$1\sigma$ radial position uncertainty & \unit{0.4}{\arcsec} & \unit{1.0}{\arcsec}  & \unit{2.0}{\arcsec} \\
Photometric uncertainty (SL) & \unit{15--20}{\%} & \unit{37--42}{\%} & \unit{$\sim100$}{\%} \\
Photometric uncertainty (LL) & \unit{2--5}{\%} & \unit{5--10}{\%} & \unit{15--20}{\%} \\
\bottomrule
  \end{tabular}
\end{table}

See \figref{fig:IRS:sensitivity} for an overview of the sensitivities of the IRS low-resolution modules;  the total signal-to-noise of an observation consisting of a series of $n$ frames is $\sqrt{n}\times f/s$ with flux $f$ and sensitivity $s$, provided other noise sources (e.g.\ shot noise) can be neglected. 
The maximum unsaturated flux is typically a few thousand times larger.

Laboratory tests on ground  showed the temporal stability of the photometric response to an accurately centered source to be better than \unit{1}{\%}; in flight, the photometric stability is limited by source mis-centering. The overall absolute photometric uncertainty of IRS is around \unit{10}{\%}, dominated by uncertainty in stellar flux models \citep[p.\ 210]{SOM}.
See \tableref{table:IRS:PUlosses} for an overview of photometric uncertainties due to spill-over losses.





%% file: resultsintro.tex
This chapter contains results of thermal-infrared studies of 8 asteroids. 
Each of the eight sections is devoted to a self-contained study of an individual asteroid. Object-specific aspects of our results are discussed in the appropriate section, implications on asteroid science in general will be discussed in \chaptref{chapt:discussion}.

The first five sections contain our main results: the thermal inertia, size, and albedo of 5 near-Earth asteroids (NEAs) with diameters spanning the range 0.1--\unit{17}{\km}. 
These results are based on  application of our thermophysical model (TPM; see \chaptref{chapt:TPM}) to extensive sets of thermal-infrared data, most of which have been obtained using the IRTF \seechapt{chapt:IRTF} or the Spitzer Space Telescope \seechapt{chapt:SST}.
Section \ref{sect:Eros}, in particular, contains a study of (433) Eros, which had been scrutinized using the NEAR-Shoemaker spacecraft. The primary aim of our Eros study is to validate the TPM for application to NEA data.

Sections \ref{sect:Lutetia} and \ref{sect:ML} contain studies of asteroid targets of future spacecraft encounters: (21) Lutetia, target of a flyby of the ESA spacecraft Rosetta in 2010, and (10302) 1989~ML, a nominal target of the planned ESA mission Don Quijote.
Based on new thermal-infrared observations, these objects' size and albedo have been determined, their surface mineralogy and thermal inertia have been constrained.

In section \ref{sect:Patroclus}, 
the first  thermal-infrared observations of an eclipsing binary asteroid system, (617) Patroclus,  are reported.
These allowed us to determine the system's thermal inertia in a fascinatingly direct way. 



%% file: Eros.tex
In order to study the applicability of our
thermophysical model (TPM) to NEAs,
it was used to analyze published high-quality thermal-infrared data \citep[][]{HarrisDavies} of an extremely well studied object:  (433) Eros, the NEA target of the rendezvous mission NEAR-Shoemaker.

We obtain a best-fit diameter of
\unit{17.8}{\km}, within \unit{$\sim$5}{\%} of the NEAR-Shoemaker result of \unit{16.9}{\km} \citep{Thomas2002}.
Depending on model assumptions,
the best-fit thermal inertia is
100--\unit{200}{\TIunit}, 
a very plausible value given the boulder-strewn yet regolith-dominated surface structure determined from spacecraft imaging, and furthermore
in excellent agreement with  previous estimates.

We conclude that the TPM is well suited to be applied to NEA data.

\subsection{Introduction}
        \label{sect:Eros:intro}

Eros is a well studied object.
E.g., the  NEAR Shoemaker spacecraft scrutinized it for over one year while  in orbit around Eros \citep[see][and references therein]{Veverka2000}.
\citet{Thomas2002} report  a regolith-dominated surface with a moderate amount of craters and boulders, and 
a volume of \unit{2535}{\km\cubed}, corresponding to an effective diameter of \unit{16.9}{\km} (see \eqrefpage{eq:Deff}).
Eros' disk-integrated albedo is  $\pv=0.29\pm0.02$ (at $\lambda=\unit{550}{\nano\metre}$) and $A=0.12\pm0.02$ \citep{Domingue2002}.
The exact spin state was determined by \citet{Thomas2002} along with an accurate model of its shape; the latter is available in a computer-readable format on-line at the Planetary Data System (\url{http://pdssbn.astro.umd.edu/volume/nieros_4001/data/}).

Eros' thermal inertia has not been directly determined from spacecraft measurements. 
However, Eros was observed extensively during an exceptionally favorable close approach in 1974/75 enabling thermal studies of unprecedented detail \citep{Morrison1976,LebofskyRieke1979}. 
Although Eros' shape had not been known in detail at that time, the authors could make use of detailed optical lightcurves available to them.
Correcting the thermal lightcurve to the optical lightcurve (and allowing for a relative phase shift), \citet{Morrison1976} obtained an upper limit on thermal inertia%
\footnote{ As was usual at that time, thermal inertia is defined by \citeauthor{Morrison1976}\ as the reciprocal of our definition (\eqrefpage{eq:def_TI}). Values are given in units of $\text{cal}^{-1}$\usk\centi\metre\squared\usk $\second^{1/2}$\usk\kelvin.}
around \unit{100}{\TIunit}. \citet{LebofskyRieke1979}, on the other hand,  fitted a crude shape model to
available optical light curves and based a thermophysical model on that shape model. In their thermophysical model, thermal conduction is explicitly modeled whereas beaming is approximated in a rather simple way.
They inferred a thermal inertia between 140 and \unit{280}{\TIunit}, significantly above the lunar value around \unit{50}{\TIunit}. The database available to \citeauthor{LebofskyRieke1979}  is a superset of that available to \citeauthor{Morrison1976}, they also used a more realistic thermophysical model. We therefore take their thermal-inertia estimate to be more reliable.

Eros has been reobserved in the thermal infrared by \citet{HarrisDavies}.
From a NEATM analysis, the authors indirectly concluded that Eros' thermal inertia was around \unit{170}{\TIunit}, well within the range determined by \citet{LebofskyRieke1979}.

\subsection{Modeling}
\label{sect:Eros:modeling}

\begin{figure}
  \centering
\includegraphics[width=0.6\linewidth]{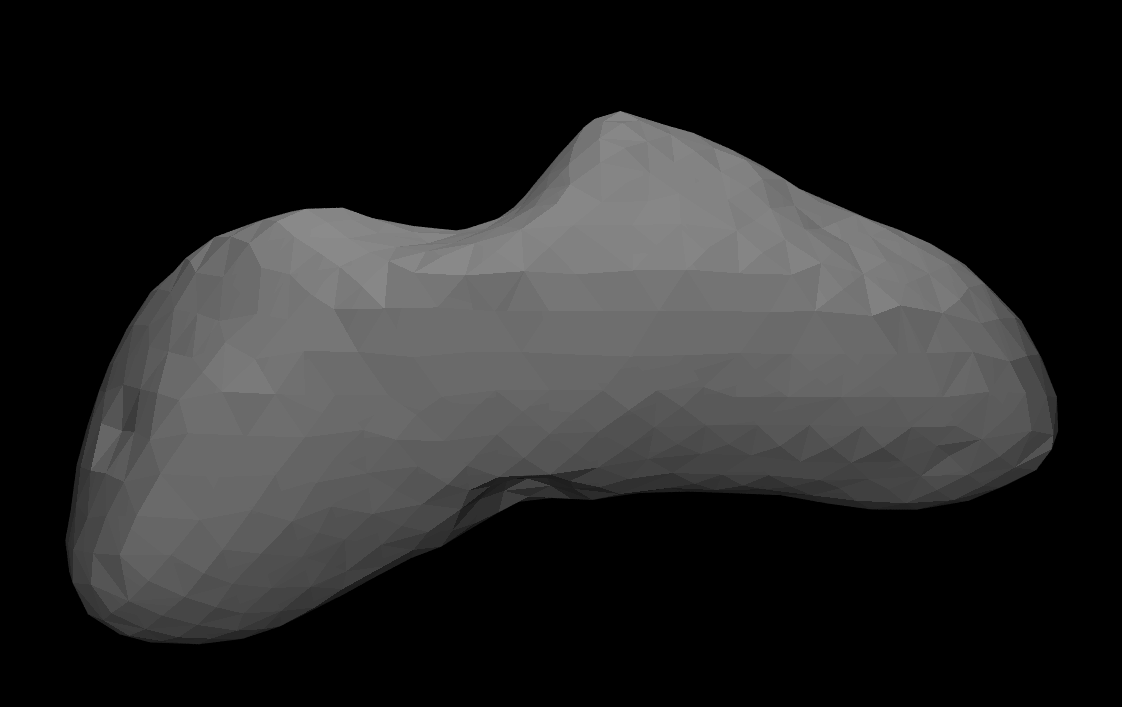}  
  \caption[Depiction of the Eros shape model by \citet{Thomas2002} used by us.]{Depiction of the Eros shape model by \citet{Thomas2002} used by us. The figure was created using the freeware ray-tracing software POV-ray.}
  \label{fig:Eros:shape}
\end{figure}
Our thermophysical modeling is based on the model of Eros' shape and spin state by \citet{Thomas2002},  the version with 1708 triangular facets is used (see \figref{fig:Eros:shape}).
To validate our numerical treatment of the model geometry, we have in a first step attempted to reproduce published optical lightcurves of Eros,
finding  excellent agreement between synthetic optical lightcurves and published observational data by \citet{Hicks1999} and \citet{Erikson2000}.

N-band data are analyzed which were obtained by \citeauthor{HarrisDavies}  on June 27--30 1998 using the United Kingdom Infrared Telescope UKIRT on Mauna Kea~/ \Hawaii\ with the CSG3 spectrometer \citep[see][for details]{HarrisDavies}.
There is a total of 7  spectra, each providing simultaneous photometry at 25 wavelengths between 8.06 and \unit{13.04}{\micron}.
%
%
While \citet{HarrisDavies} referred all data to a common observing geometry and to the flux level of lightcurve average, we rather aim at reproducing each spectrum at the time of observation with the corresponding observing geometry.
To this end, our TPM \seechapt{chapt:TPM} is used.

Most required input parameters have been determined from NEAR-Shoemaker results \citep{Domingue2002,Thomas2002}:
$H=10.82$ as implied by $D=\unit{16.9}{\km}$ and $\pv=0.29$, while the \pv-value and the reported Bond albedo imply $G=0.181$ (see \eqrefpage{eq:q}).
The  J2000 ecliptic coordinates of the spin axis are $\beta=\unit{+11.35}{\degree}$, $\lambda=\unit{17.22}{\degree}$, implying a subsolar latitude at the time of the UKIRT observations of \unit{+34}{\degree}, a sub-Earth latitude of \unit{$\sim+64$}{\degree}, and an hour angle of the Sun at the sub-Earth point  of \unit{+0.9}{\hour} corresponding to a ``local time'' of \unit{12.9}{\hour}, i.e.\ the afternoon side was observed.

\subsection{Results}
\label{sect:Eros:results}

\begin{figure}
  \centering
\includegraphics[angle=-90,width=0.6\linewidth]{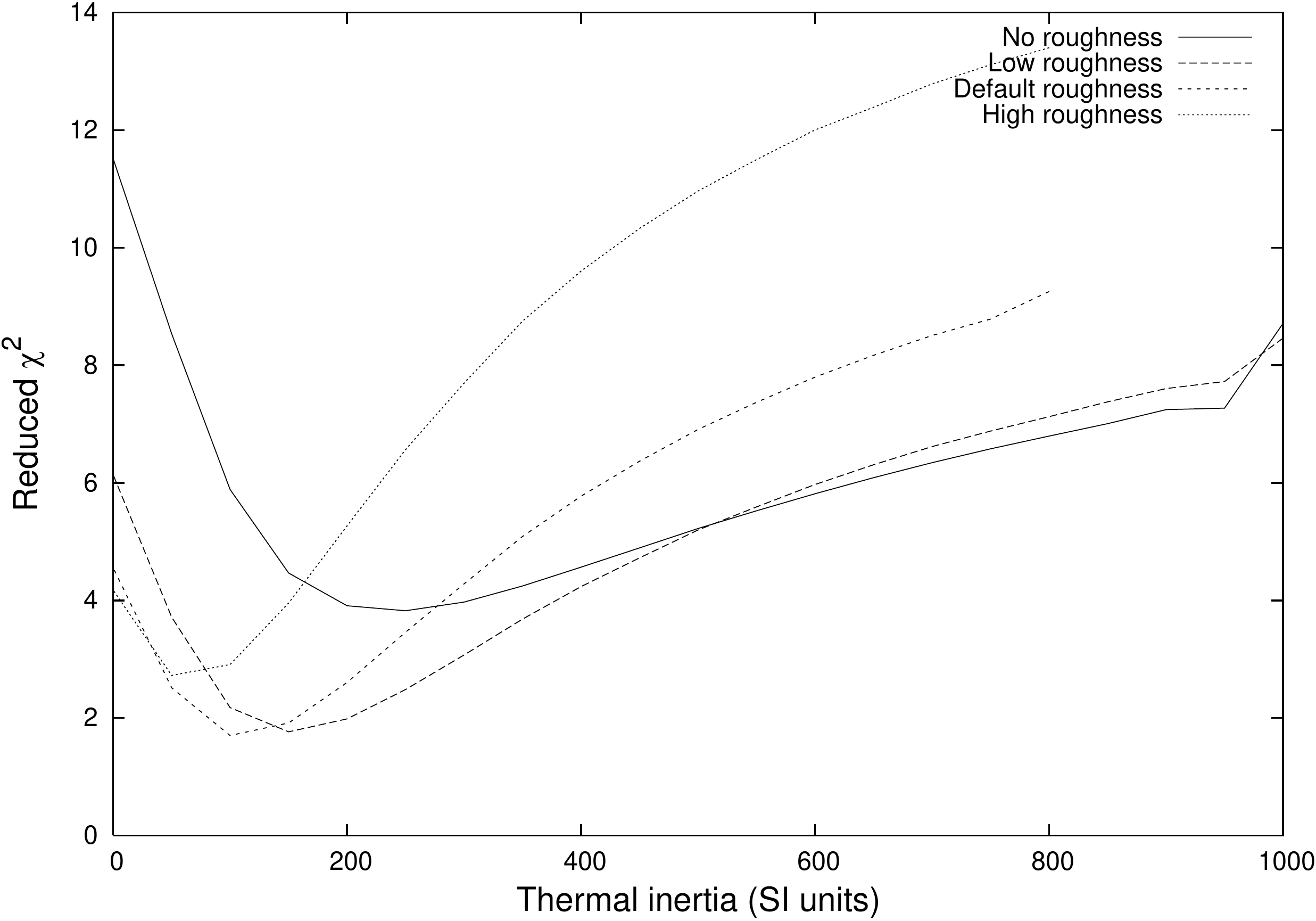}
  \caption[Eros: Reduced $\chi^2$ as a function of thermal inertia for different roughness parameters.]{Eros: Reduced $\chi^2$ as a function of thermal inertia for different roughness parameters. With $7\times25$ data points and two fit parameters (diameter and thermal inertia), the reduced $\chi^2$ equals $\chi^2/173$.}
  \label{fig:Eros:chi2}
\end{figure}

\begin{figure}
  \centering
\includegraphics[angle=-90,width=0.6\linewidth]{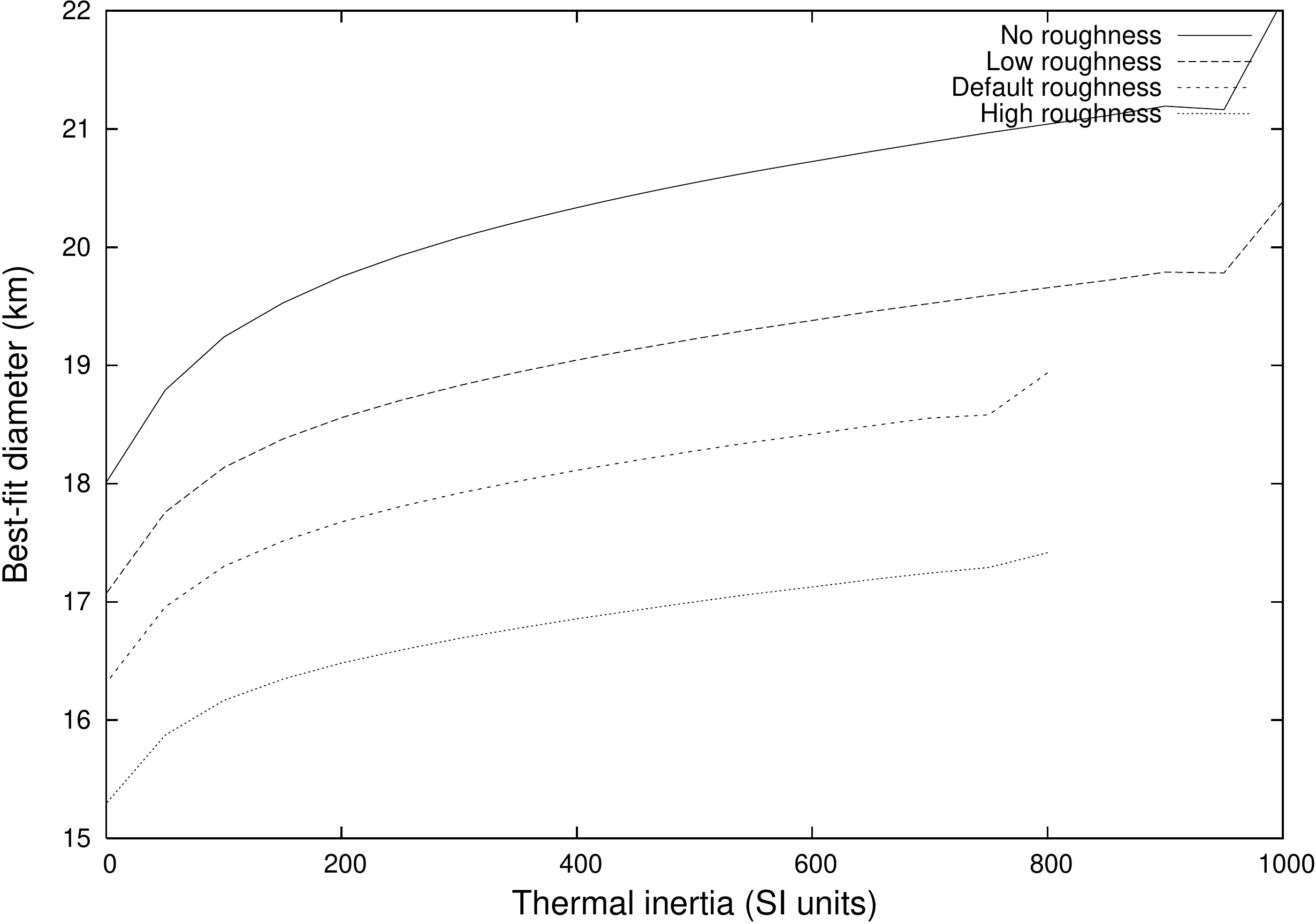}
  \caption{Eros: Best-fit diameter as a function of thermal inertia for different roughness parameters.}
  \label{fig:Eros:D}
\end{figure}

See Figures \ref{fig:Eros:chi2} and \ref{fig:Eros:D} for plots of the best-fit $\chi^2$ and diameters as a function of thermal inertia, using the methods detailed in \sectref{sect:TPM:fitting}.
Both figures contain slight ``wiggly'' artefacts for large thermal-inertia values due to the approximation mentioned therein; these artefacts quickly become unnoticeable with decreasing thermal inertia.
The respective scale factor, $\kappa$, is closely related to the derivative of the curves given in \figref{fig:Eros:D}, which are seen to be relatively flat; it has been verified that $\kappa$ is within \unit{1}{\%} of unity for all but the lowest and largest thermal-inertia values considered.
To increase the numerical accuracy at low thermal inertia, one may decrease the step size in thermal inertia.

As apparent from \figref{fig:Eros:chi2}, 
the best-fit thermal inertia is dependent on the assumed roughness parameters, where increasing roughness decreases the best-fit thermal inertia.
The data are best fit for intermediate roughness parameters, 
roughly in the range
of 100--\unit{200}{\TIunit} in thermal inertia.

As can be seen in 
\figref{fig:Eros:D}, 
the best-fit diameter for low roughness and a thermal inertia of \unit{150}{\TIunit} (the best-fit value for these roughness parameters) is around \unit{18.3}{\km}, while for default roughness the best-fit diameter is \unit{$\sim17.3$}{\km}.

\subsection{Discussion}
\label{sect:Eros:discussion}

The best-fit diameter is
slightly model dependent, but within the range $17.8\pm\unit{0.5}{\km}$, where the quoted uncertainty reflects solely  the statistical scatter inherent in the data and the systematic uncertainty from the unknown roughness parameters. 
Other sources of uncertainty, which are hard to estimate \emph{a priori} but are probably dominated by systematic modeling uncertainties, add to the error budget.
The NEAR-Shoemaker result is \unit{16.9}{\km} \citep{Thomas2002}, some \unit{5}{\%} below our result.

The best-fit thermal inertia is 100--\unit{200}{\TIunit},  in excellent agreement with previous estimates by \citeauthor{LebofskyRieke1979} (140--\unit{280}{\TIunit})
and \citeauthor{HarrisDavies} (\unit{$\sim170$}{\TIunit}).
We reckon that our result supersedes previous estimates since it makes explicit use of Eros' spin state and shape model determined from the NEAR-Shoemaker spacecraft.

Eros' thermal inertia is more than an order of magnitude below that of bare rock and roughly three  times the lunar value \seetablepage{table:thermalproperties}, a very plausible value given spacecraft imaging of Eros' surface, which reveal a regolith-dominated surface with a moderate amount of craters and boulders \citep{Thomas2002}; the latter being consistent with the higher-than-lunar thermal inertia.

We conclude that applying our TPM, based on a detailed shape model, to published thermal-infrared data of Eros results in estimates for diameter and thermal inertia which are in excellent agreement with previous estimates including ``ground truth'' from spacecraft observations. 
In particular, the diameter uncertainty due to systematic modeling uncertainties does not appear to vastly exceed \unit{5}{\%} for the favorable circumstances (large data set, accurate model of shape and spin state, moderate solar phase angle of \unit{29}{\degree}) of this study.
No such studies for an NEA have been published before.

Eros' shape model by \citet{Thomas2002} is not convex but contains several prominent concavities \seefigpage{fig:Eros:shape}. 
Nevertheless, the convex-shape TPM produces results in excellent agreement with independently obtained results.
This is somewhat unexpected because the indentations present on Eros' surface would be expected to lead to shadowing effects and to mutual heating of surface elements
as discussed in \chaptref{chapt:TPMconcave} in the appendix.
We conclude that neglecting the effects of non-convex global shape does apparently not lead to a critical systematic diameter offset for objects as irregularly shaped as Eros.

In general, neglecting shadowing leads to an overestimation of thermal flux and consequently to a diameter underestimation; the converse applies to neglecting mutual heating.
The fact that our derived diameter is slightly larger than the 
spacecraft result
may indicate that at the moderate phase angle of the considered observations (\unit{$\sim29$}{\degree}) the effect of mutual heating dominates  the cooling effect of shadowing, which may become more relevant at larger phase angles. Further analysis using a more general TPM accounting for shadowing and mutual heating are required to study this speculation.

We would expect more accurate constraints on Eros' physical properties to result from a TPM fit to all thermal-infrared data available in the literature, including the flux values quoted in \citet{LebofskyRieke1979}, such that the database would stretch a broader range in solar phase angle and aspect geometry.
Note that, in contrast to simple models, our model can be used to consistently fit data obtained at different epochs. 
However,  combining data obtained by different observers using different instruments requires a very careful assessment of 
potential cross-calibration issues and of 
the relative accuracies of the data sets, which is beyond our current scope.

\subsection{Summary}
\label{sect:Eros:summary}

In order to validate our TPM for application to NEAs,  thermal-infrared observations \citep{HarrisDavies} of the NEA (433) Eros
have been reanalyzed.
Eros had been
 scrutinized from ground and through the NEAR-Shoemaker spacecraft.

Eros' diameter is reproduced to within \unit{$\sim5$}{\%} of the NEAR-Shoemaker result \citep{Thomas2002}.
The best-fit thermal inertia is
100--\unit{200}{\TIunit} depending on the assumed model roughness parameters,
in excellent agreement with earlier estimates. 
Eros' thermal inertia is
more than an order of magnitude below that of bare rock and roughly three  times the lunar value, a very plausible value given Eros' surface structure as revealed by spacecraft imaging.

We conclude that  our convex-shape TPM 
is well suited to analyze
 thermal-infrared observations of NEAs and to derive quantitative estimates of their diameter, albedo, and thermal inertia.


%% file: itokawa.tex
In 2005, the Japanese spacecraft Hayabusa has rendezvoused with the NEA (25143) Itokawa.
Prior to that, in 2004, we had performed thermal-infrared observations of Itokawa with the IRTF and MIRSI, which are presented here. 
Preliminary results
have been accepted for publication in 
the proceedings of the 1st Hayabusa symposium \citep{Ito1}, which has not yet been printed, however.

Additional thermal-infrared observations were published by
\citet{MuellerItokawa},%
\footnote{ In an attempt to reduce confusion, Thomas \Mueller, MPG Garching, is spelled \Mueller\ (with umlaut) throughout this thesis, while my last name is spelled  Mueller (no umlaut). \label{allesmuelleroderwas}}
also before the arrival of Hayabusa.
There included is a recalibration of other data points found in the literature, which he had used in our preliminary analysis.

We here present TPM analyses of different combinations of thermal-infrared data sets, including the first analysis of the complete set of reliable thermal data of Itokawa.
The resulting best-fit diameter is
0.32--\unit{0.33}{\km},
in excellent agreement with Hayabusa results.
Together with the results of our Eros study
\seesect{sect:Eros}, this suggests that the systematic diameter uncertainty inherent in our thermophysical modeling does not exceed \unit{10}{\%}.

Itokawa's thermal inertia is found to be
$700\pm\unit{100}{\TIunit}$, refining the estimate by
\citet{MuellerItokawa}, $750\pm\unit{250}{\TIunit}$.

\subsection{Introduction}
        \label{sect:ito:intro}

(25143) Itokawa (previously known as 1998~SF36) is an S-type NEA which is a particularly favorable spacecraft target due to the relatively moderate amount of energy and time required to reach it.
It has been studied in detail from the spacecraft Hayabusa during a rendezvous in 2005 \citep[see][and references therein]{Fujiwara2006}.
Apart from a detailed study of the asteroid's physical properties using remote sensing techniques while hovering within a few kilometers of the asteroid surface, 
Hayabusa was scheduled to take small samples of asteroid material and to return them to Earth.
Due to technical problems it is currently unclear whether Hayabusa has succeeded in taking samples and whether it will manage to return to Earth; the planned Earth-arrival date is in 2010.

We have observed Itokawa  at the IRTF in July 2004, before the arrival of Hayabusa at its target. Apogee data obtained by us have been used by \citet{KaasalainenIto} in the derivation of Itokawa's shape and spin state.
Our MIRSI data, combined with other data found in the literature, have been used by us to determine Itokawa's size and thermal inertia.

A vast body of information on Itokawa has been collected both before and after the arrival of Hayabusa using different techniques, 
we here focus on those which are most relevant for our purposes, i.e.\ 
estimates of size, shape, and thermal inertia.

\subsubsection{Pre-Hayabusa}

\citet{Sekiguchi2003} report single-wavelength N-band photometry and estimate the diameter of Itokawa to be \unit{$0.35\pm0.03$}{\km}, using the STM \seesect{sect:STM}.
Combining the \citeauthor{Sekiguchi2003}\ flux measurement with newly obtained $\text{M}^\prime$-band photometry at \unit{4.68}{\micron}, \citet{Ishiguro2003} derive dimensions for a triaxial ellipsoid of $(620\pm140)\times(280\pm60)\times(160\pm30)$\usk\metre, corresponding to an effective diameter of roughly 
\unit{$300\pm30$}{\metre}.
\citeauthor{Ishiguro2003}\ also report a thermal inertia
of \unit{290}{\TIunit}, which they derive from the two thermal-infrared data points at their disposal.
Their thermal model assumes a spherical shape.

From radar observations, 
\citet{Ostro2004} determine a shape model of Itokawa and a volume-equivalent diameter of \unit{$358\pm36$}{\metre}. After obtaining new radar observations in June 2004, \citet{Ostro2005} publish an updated shape model and size estimate, 
with ellipsoidal axes within \unit{10}{\%} of $594 \times 320 \times 288$~\metre, corresponding to a volume-equivalent diameter of \unit{$380\pm38$}{\metre}.

\citet{KaasalainenIto} determined 
the spin state of Itokawa and a convex-definite model of its shape
from the inversion of optical lightcurve data.

From a preliminary analysis of 
the thermal-infrared data published by \citet{Sekiguchi2003} and \citet{Ishiguro2003}, and new data which had been obtained at the ESO 3.6-m telescope \citep[later published in][]{Delbo2004} and at the IRTF (see below), we \citep{Ito1}
determined a diameter around \unit{280}{\metre} and a thermal inertia of some \unit{350}{\TIunit}.
Our analysis was based on the \citet{KaasalainenIto} shape model and the TPM described in \chaptref{chapt:TPM}.
The thermal-inertia result is intermediate between lunar regolith and bare rock (but closer to the former) and was interpreted as ``incompatible with a surface dominated by bare rock.'' An Eros-like surface was proposed but with a coarser regolith and/or a larger number of boulders.

\citet{MuellerItokawa} obtained further thermal-infrared flux measurements at the ESO 3.6-m telescope.
Combining their data with those of \citet{Sekiguchi2003} and \citet{Delbo2004} but with neither  our MIRSI data (which were not available to them)
nor the \citet{Ishiguro2003} M-band value, they determined a diameter of \unit{$0.32\pm0.03$}{\km}
and a thermal inertia of \unit{750}{\TIunit}, which they interpret as ``an indication for a bare rock dominated surface.''
Their results are based on the thermophysical model by \citet{LagerrosI,LagerrosIII,LagerrosIV} and the shape model by 
\citet{KaasalainenIto} which had also been used by us.

\begin{table}
  \caption{Itokawa: Overview of published diameter estimates, $D$, and reported uncertainties, $\sigma$.}
  \label{table:ito:D}
  \centering
  \begin{tabular}{lll}
\toprule
$D(\km)$ & $\sigma (\km)$ & Source \\
\midrule
0.35 & 0.03 & \citet{Sekiguchi2003} \\
0.30 & 0.03 & \citet{Ishiguro2003} \\
0.36 & 0.04 & \citet{Ostro2004} \\
0.38 & 0.04 & \citet{Ostro2005} \\
0.28 & --- & \citet{Ito1} \\
0.32 & 0.03 & \citet{MuellerItokawa} \\
\midrule
0.327 & 0.006 & \citet{Demura2006} \\
\bottomrule
  \end{tabular}
\end{table}

\subsubsection{Hayabusa results}
From Hayabusa observations,
\citet{Demura2006} determined a volume within \unit{5}{\%} of \unit{0.018378}{\km\cubed}, corresponding to a volume-equivalent diameter of \unit{$327\pm6$}{\metre} (see \eqref{eq:Deff}).
See \tableref{table:ito:D} for an overview of published diameter estimates.

\citeauthor{Demura2006}\ also report an updated shape model which appears to be in broad agreement with that by \citet{KaasalainenIto},
however at the time of writing it has not yet been released to the scientific community.%
\footnote{ The Hayabusa data archive including an update on the shape model by \citet{Demura2006} has been made public on 24 April 2007 (see \url{http://hayabusa.sci.isas.jaxa.jp/}).}
A spin axis is reported within \unit{3.9}{\degree}  of $\lambda=\unit{128.5}{\degree}$ and $\beta=\unit{-89.66}{\degree}$ (J2000 ecliptic coordinates), in very good agreement with the estimate by \citet{KaasalainenIto}.

From global imaging of Itokawa (see, e.g., \figrefpage{fig:intro:ito}) one recognizes a dichotomy between rough and smooth terrains  on Itokawa's surface, which cover some 80 and \unit{20}{\%} of the surface area, respectively \citep{Saito2006}.
Rough terrains are densely covered in boulders of various sizes, in contrast to smooth areas which display very low boulder abundance.

During operations in preparation of the spacecraft  touchdowns for sample taking and during the two touchdowns themselves, 
high-resolution
imaging was obtained of the vicinity of the touchdown sites, i.e.\ of parts of the Muses-C region and of adjacent rough terrain \citep{Yano2006}.
The obtained images of smooth terrain display a dense cover of coarse regolith consisting of gravel in the \milli\metre--\cm\ range with no apparent sign of finer dust on the surface.
Itokawa's regolith is  coarser than the regolith found in Eros' ponds \citep{Yano2006}.
The smooth regolith-covered regions were found to coincide with the minima of the gravitational potential \citep[see][]{Fujiwara2006}.
Based on an analysis of high-resolution imaging, 
\citet{Miyamoto2007} report evidence for ``landslide-like granular migration'' on Itokawa's surface, i.e.\ for regolith migration aligned with the local gravity slope.
They conclude that Itokawa's surface is unconsolidated and suggest that vibrations due to, e.g., impact-induced seismic shaking cause gravel fluidization and subsequent down-slope movement, where the resulting grain mobility increases with decreasing grain size.

A very low abundance of crater-like features was found on Itokawa's surface, 
totaling  no more than 100 for crater diameters exceeding \unit{70}{\centi\metre} \citep[see][]{Saito2006}. 
Craters of small and intermediate size may be buried in migrating regolith; 
\citet{Miyamoto2007} present images of
regolith concentrations on the floors of crater-like features.

One estimate of a local value of thermal inertia  is available, obtained from an indirect temperature measurement during touchdown.
The temperature close to the sampling site in the Muses-C terrain is estimated 
by \citet{Yano2006} to be \unit{$310\pm10$}{\kelvin}, which they conclude is indicative of a thermal inertia in the range between 100 and \unit{1000}{\TIunit},
although there appear to be significant systematic uncertainties in their thermal modeling.

\subsection{Thermal-infrared observations}
        \label{sect:ito:data}

We have observed Itokawa with the NASA IRTF \seechapt{chapt:IRTF}
on Mauna Kea~/ \Hawaii\ using MIRSI and Apogee  on 10 July 2004 (UT). 
The Apogee data are not needed here, hence they shall be disregarded in the following.
Additional observations were scheduled for 28 and 29 July but failed due to bad weather.
The night of 10 July was photometric but had a relatively high (constant) level of atmospheric humidity.

During our observations, Itokawa's heliocentric distance  was \unit{1.06}{\AU}, its topocentric distance from Mauna Kea was \unit{0.05}{\AU}, the solar phase angle was \unit{28.3}{\degree}.
Itokawa's  declination was \unit{$-40$}{\degree}, the airmass at meridian transit ($\sim$ 12:15 UT) was 2.0.
Unfortunately, technical problems%
\footnote{ This was the first occurrence of the MIRSI ``striping'' artefact visible in \figrefpage{fig:IRTF:MIRSIgarbled}, which hit us unexpectedly at the time of the Itokawa observations.}
disabled MIRSI observations for some 90 minutes around transit, 
hence the reported MIRSI observations have been obtained at larger airmasses.
The \citet{CohenVII,CohenX} calibration standard stars $\beta$~And and $\alpha$~Lyr have been observed at similarly large airmasses before and after the Itokawa observations.
MIRSI data have been reduced using the methods described in \sectref{sect:IRTF:reduction}; see \tableref{table:ito:MIRSI} for the resulting flux values.

\begin{table}
          \caption[MIRSI observations of Itokawa: fluxes values]{Flux values resulting from MIRSI observations of Itokawa on 10 July 2004 (UT).
Times given refer to the middle of the exposure and are corrected for 1-way light-time.}
          \label{table:ito:MIRSI}
          \centering
          \begin{tabular}{llll}
\toprule 
Time  & Wavelength  & Flux & $\sigma$ flux \\
(\hour)  & (\micron) & $\left[\frac{10^{-15}\watt}{\metre^{2}\micron}\right]$ & $\left[\frac{10^{-15}\watt}{\metre^{2}\micron}\right]$ \\
\midrule 
11.73888 & 11.7 &16.7 &2.2 \\
11.79576 & 11.7 &15.8 &2.0 \\
13.52352 & 11.7 &20.0 &2.5 \\
13.67784 & 9.8 & 24.7 &3.9 \\
13.84128 & 9.8 & 17.8 &3.8 \\
\bottomrule
\end{tabular}
\end{table}



In \citet{Ito1}, preliminary TPM fits are reported based on
  the MIRSI data quoted in \tableref{table:ito:MIRSI}, thermal fluxes obtained at the ESO 3.6-m telescope with TIMMI2 on 8 April 2001 \citep[measurements from 9 April are also reported but were not used  due to the reported unfavorable atmospheric conditions in that night]{Delbo2004}, and respectively one thermal flux value from  \citet[\unit{11.9}{\micron} photometry obtained with TIMMI2]{Sekiguchi2003} and \citet[\unit{4.68}{\micron} photometry obtained at Subaru]{Ishiguro2003}.
Later, we were informed that the  \unit{4.68}{\micron} flux value reported by  \citeauthor{Ishiguro2003}\ was likely to be compromised
(Hasegawa, 2004, private communication; Hasegawa is a co-author of that paper).

The database of \citet{MuellerItokawa} contains new TIMMI2 measurements in addition to those used by us (\citeauthor{Delbo2004} and \citeauthor{Sekiguchi2003}).
Our MIRSI data were not available to them. They chose  not to use the \unit{4.68}{\micron} flux by \citeauthor{Ishiguro2003}
Flux values originally reported by \citeauthor{Delbo2004}\ and \citeauthor{Sekiguchi2003}\ have been recalibrated by \citeauthor{MuellerItokawa}

We have used our TPM described in \chaptref{chapt:TPM} to fit different combinations of data sets.
Itokawa's absolute optical magnitude is assumed to be $H=19.5$ throughout our analysis, which appears to be consistent with our Apogee data. The exact choice of $H$ would be expected to have negligible influence on diameter and thermal-inertia results.
The following data sets are analyzed:
\begin{enumerate}
\item 
That considered in \citet{Ito1}, i.e.\ MIRSI data, TIMMI2 data by \Delbo\ (8 April 2001) + \citeauthor{Sekiguchi2003}, and M-band data by \citeauthor{Ishiguro2003}
Unlike in \citet{Ito1}, recalibrated TIMMI2 fluxes are used. (total: 12 data points)
\item 
As above, but without the \citeauthor{Ishiguro2003}\ M-band data: MIRSI + recalibrated TIMMI2 data by \Delbo\ and \citeauthor{Sekiguchi2003} (total: 11 data points)
\item 
The  ``high-quality'' data points reported in \citet{MuellerItokawa}, including  \citeauthor{Delbo2004}\ and \citeauthor{Sekiguchi2003}\ data as above. (total: 15 data points)
\item 
All reliable data, i.e.\ the 15 data points as above in addition to our MIRSI data. (total: 20 data points)
\end{enumerate}

\subsection{Results}
        \label{sect:ito:results}

\begin{figure}
          \centering
  \begin{minipage}[t]{0.48\linewidth}
    \includegraphics[angle=-90, width=\linewidth]{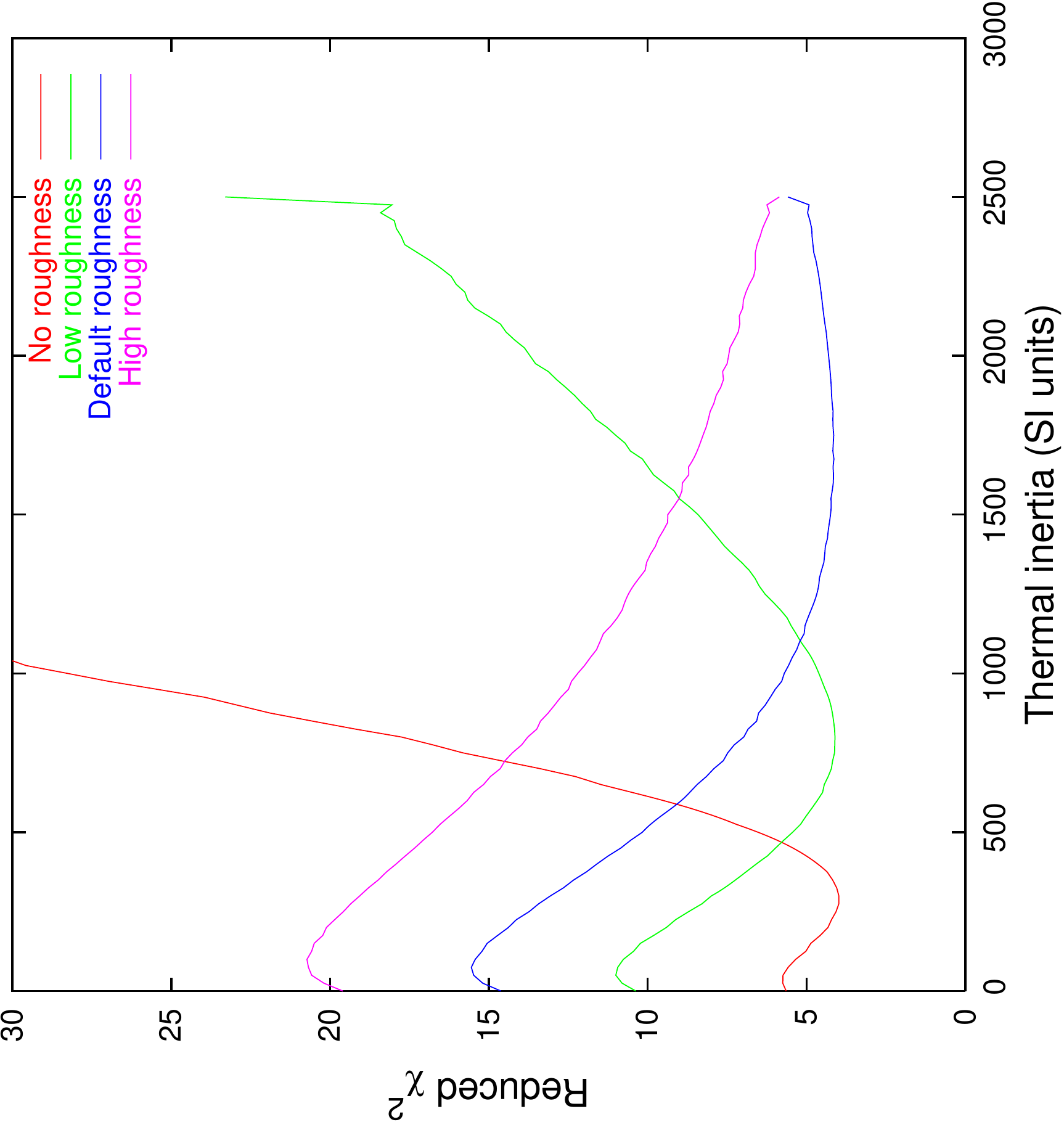}
  \end{minipage}
  \begin{minipage}[t]{0.48\linewidth}
    \includegraphics[angle=-90, width=\linewidth]{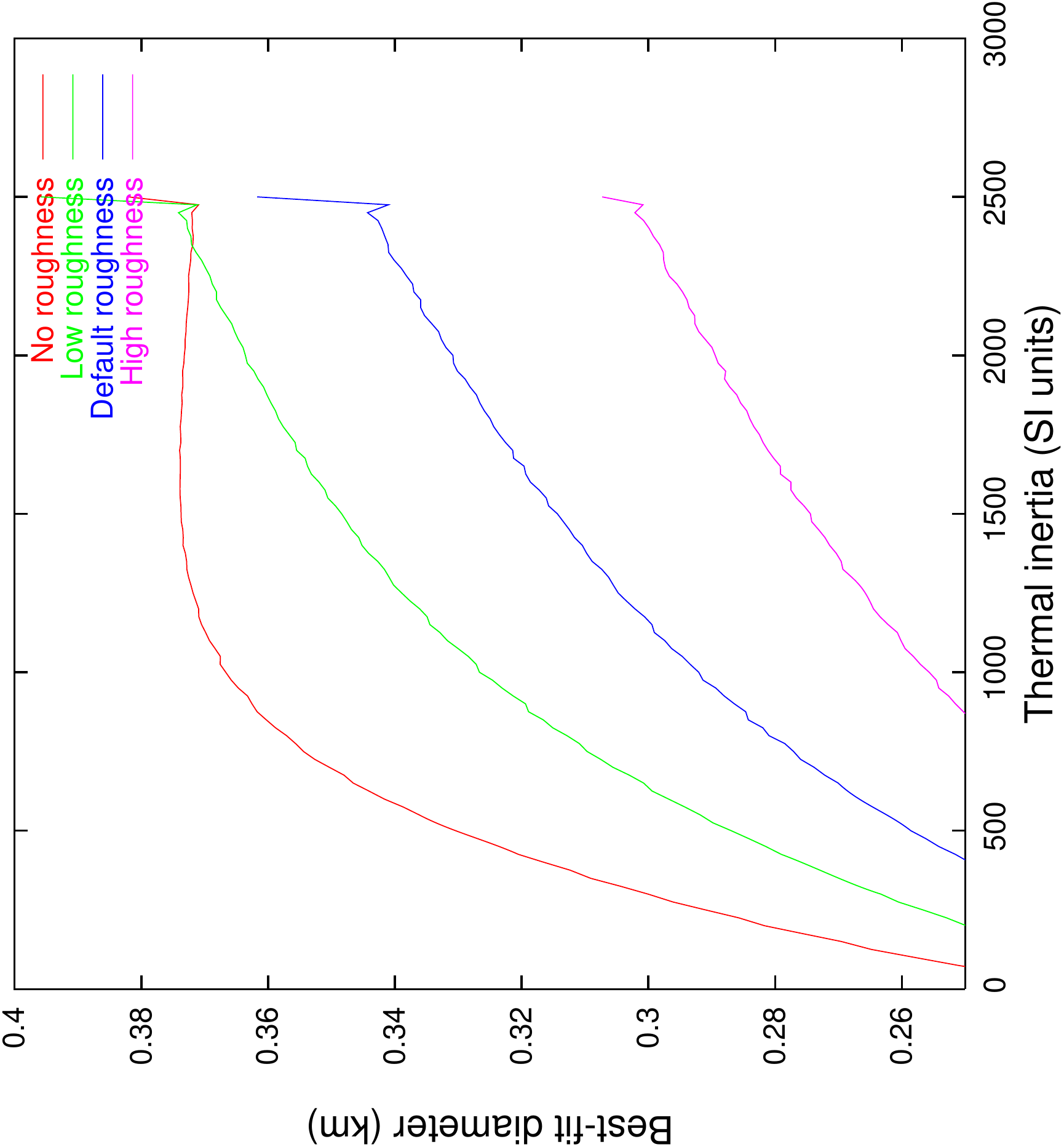}
  \end{minipage}
\caption{Itokawa: TPM fits to data set 1, i.e.\ the flux values considered in \citet{Ito1}. Left: Reduced $\chi^2$ as a function of thermal inertia for four sets of roughness parameters \seesect{sect:TPM:fitting}. Right: Best-fit diameter as a function of thermal inertia.}
          \label{fig:ito:mueller04}
\end{figure}

\begin{figure}
          \centering
  \begin{minipage}[t]{0.48\linewidth}
    \includegraphics[angle=-90, width=\linewidth]{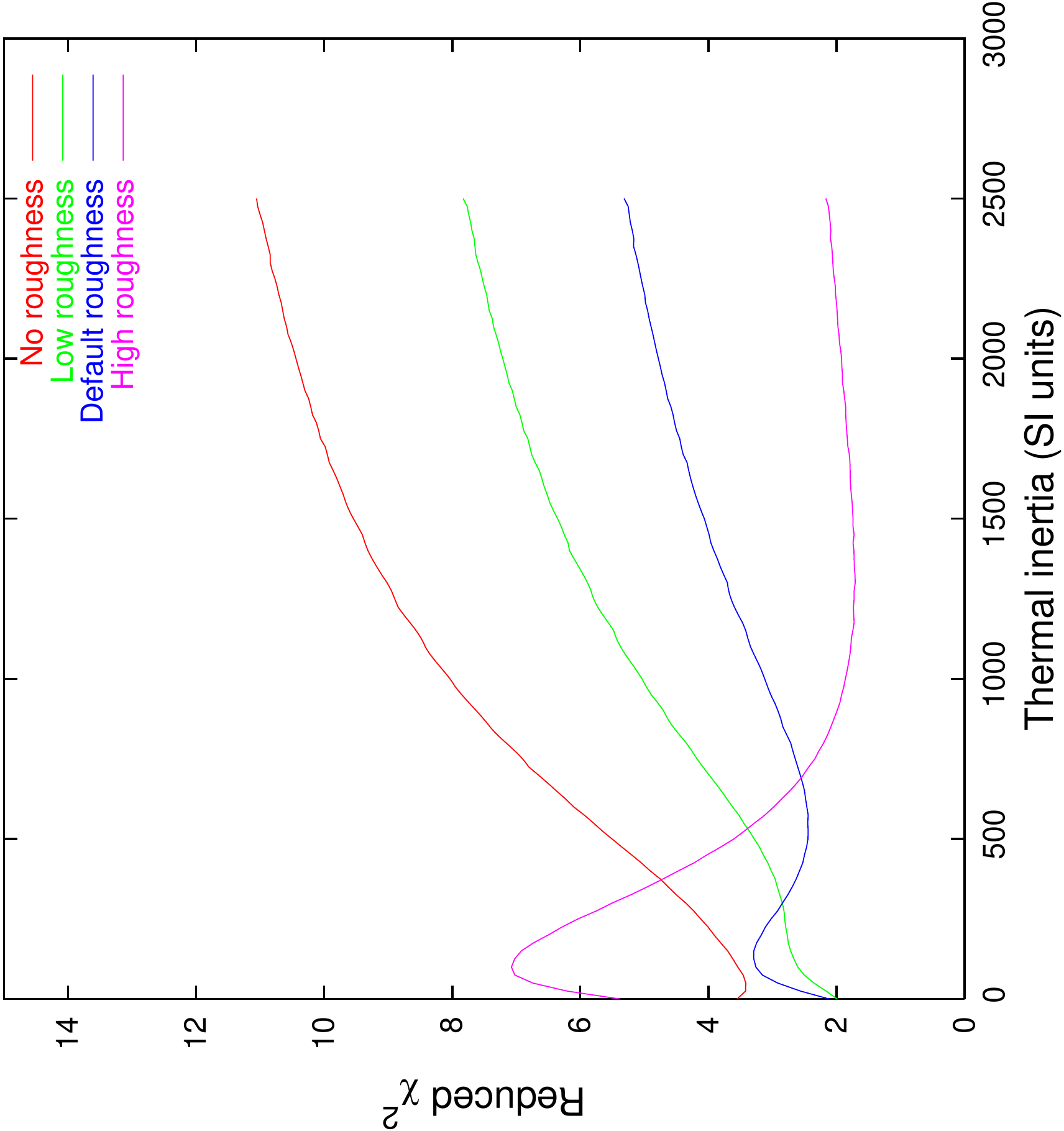}
  \end{minipage}
  \begin{minipage}[t]{0.48\linewidth}
    \includegraphics[angle=-90, width=\linewidth]{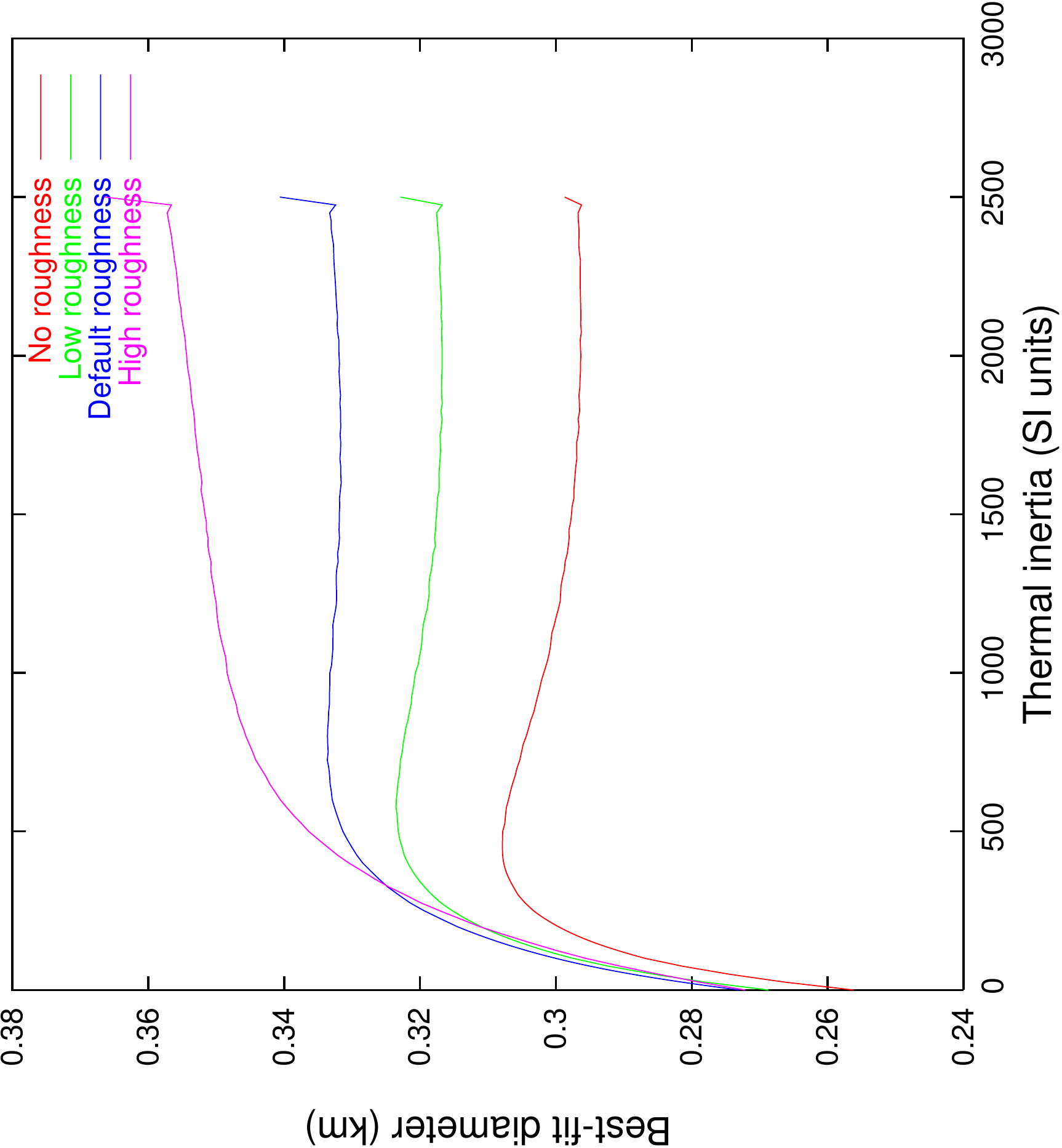}
  \end{minipage}
\caption{As \figref{fig:ito:mueller04}, but without the \unit{4.68}{\micron} flux value reported by \citet{Ishiguro2003} (data set 2).}
          \label{fig:ito:mueller04-M}
\end{figure}

\begin{figure}
          \centering
  \begin{minipage}[t]{0.48\linewidth}
    \includegraphics[angle=-90, width=\linewidth]{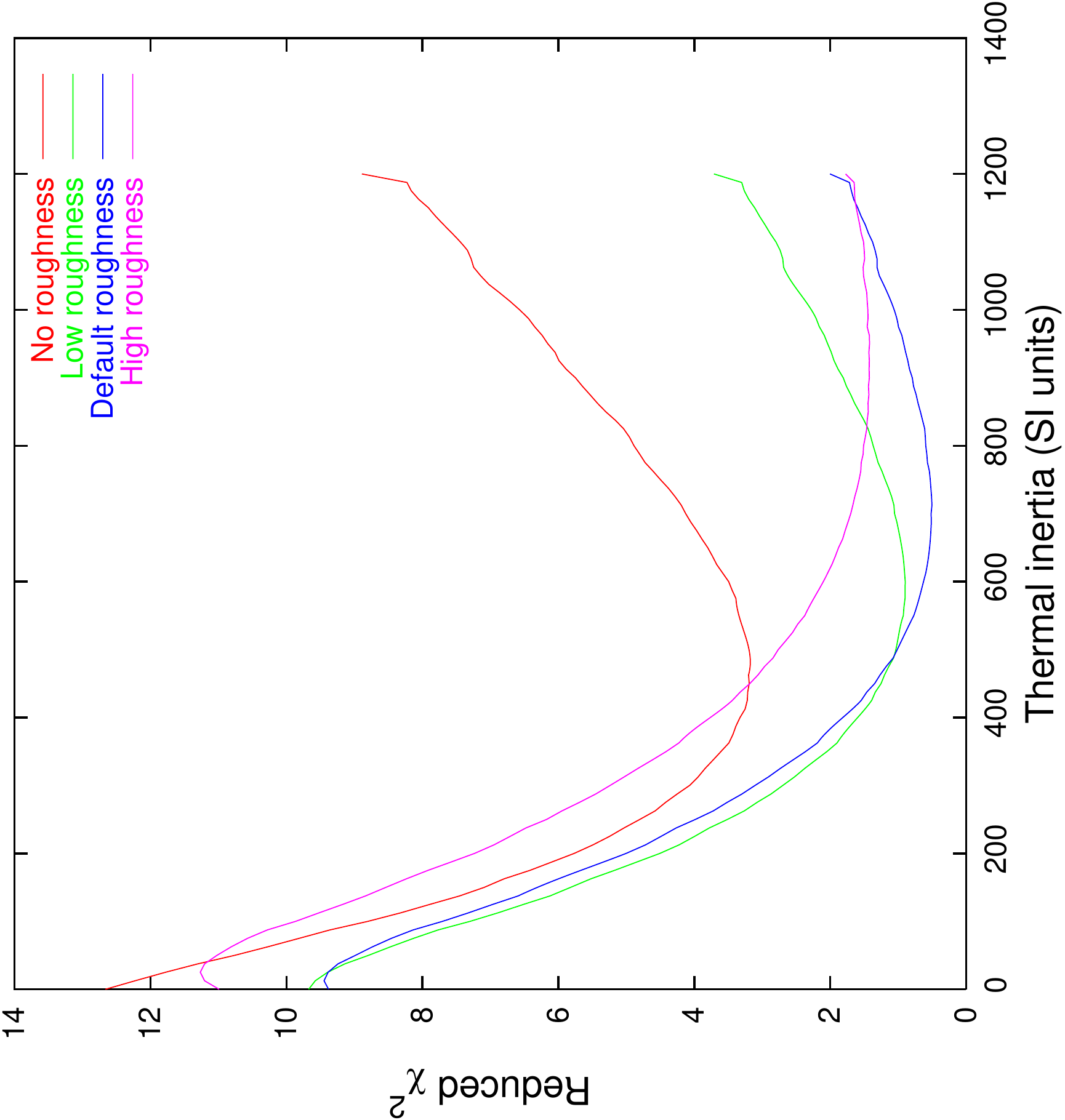}
  \end{minipage}
  \begin{minipage}[t]{0.48\linewidth}
    \includegraphics[angle=-90, width=\linewidth]{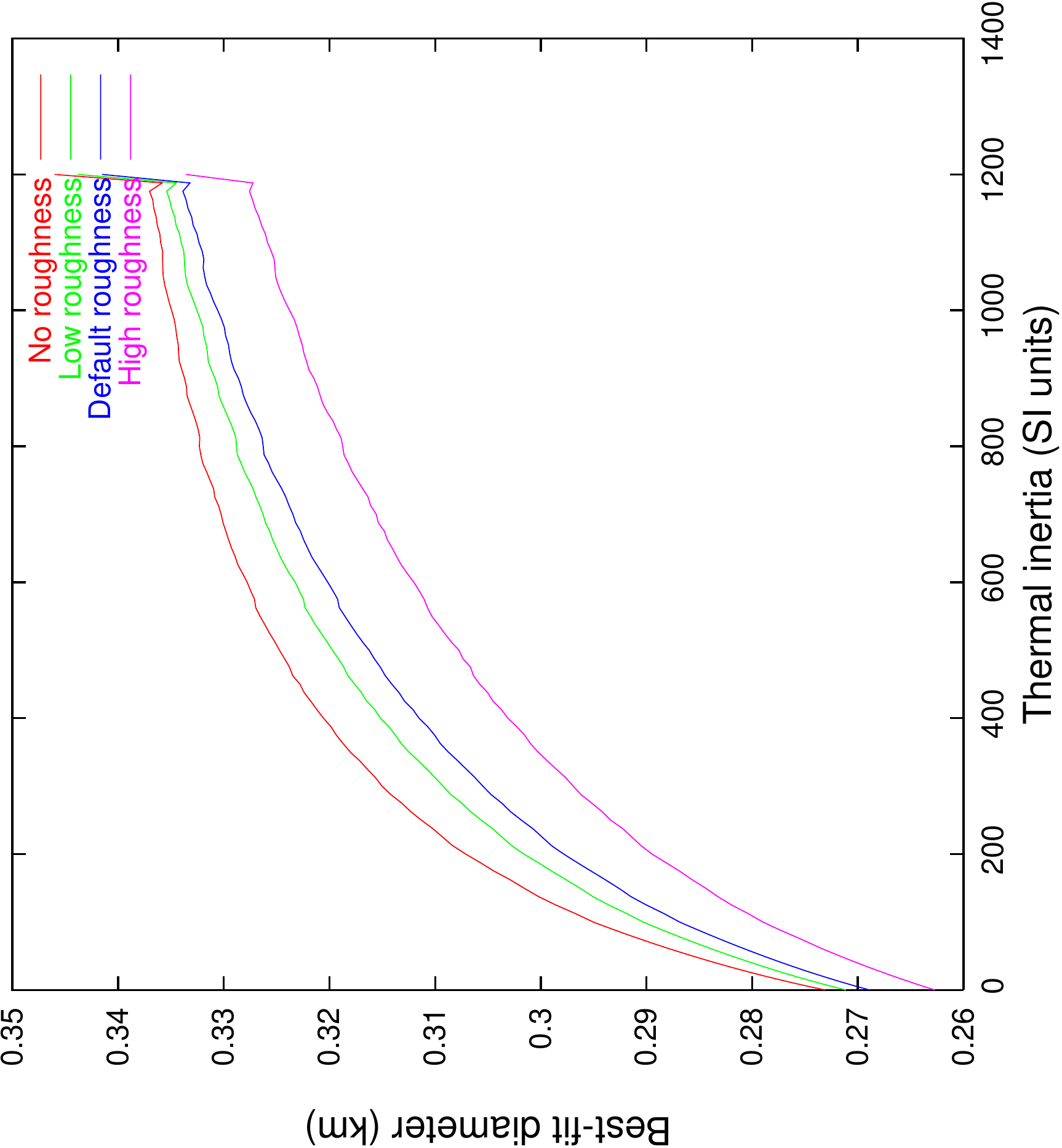}
  \end{minipage}
\caption{As \figref{fig:ito:mueller04}, but for the data set quoted by \citet{MuellerItokawa} (data set 3).}
          \label{fig:ito:ThM15}
\end{figure}

\begin{figure}
          \centering
  \begin{minipage}[t]{0.48\linewidth}
    \includegraphics[angle=-90, width=\linewidth]{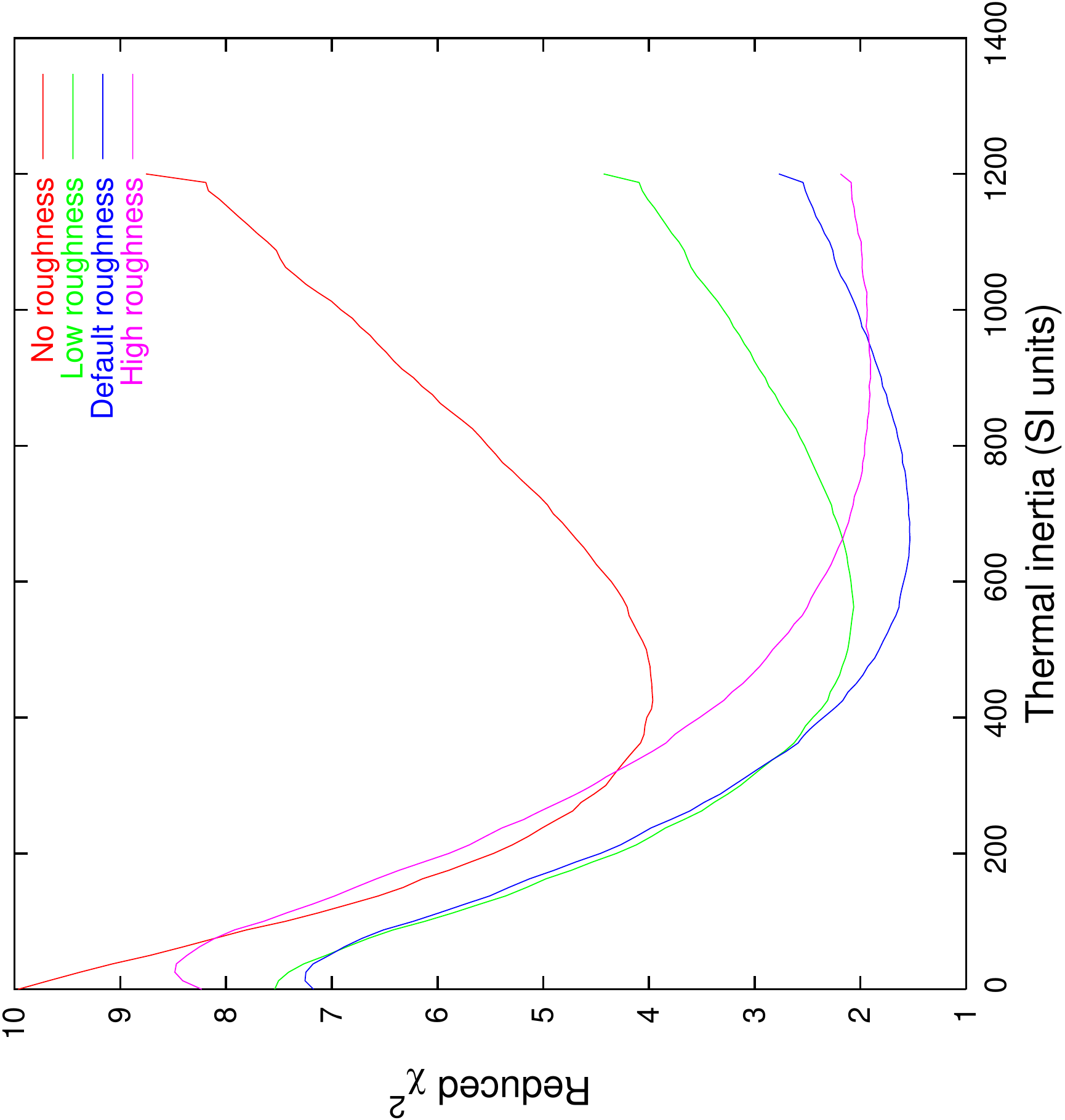}
  \end{minipage}
  \begin{minipage}[t]{0.48\linewidth}
    \includegraphics[angle=-90, width=\linewidth]{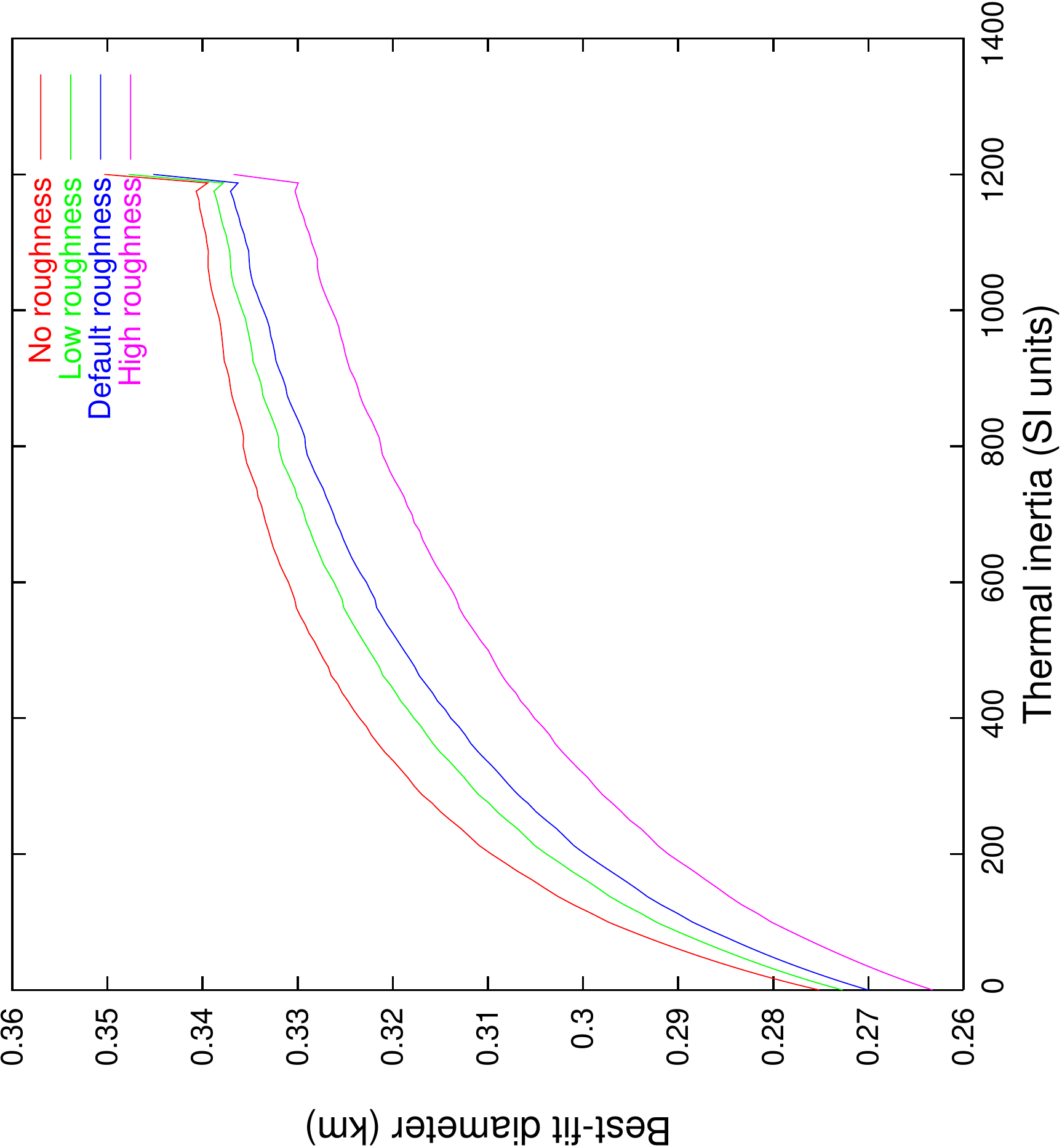}
  \end{minipage}
\caption{As \figref{fig:ito:mueller04}, but for the complete set of reliable thermal-infrared data, i.e.\ those quoted by \citet{MuellerItokawa} in addition to the MIRSI data given in \tableref{table:ito:MIRSI} (data set 4).}
          \label{fig:ito:ThM15+MIRSI}
\end{figure}

See Figures \ref{fig:ito:mueller04}, \ref{fig:ito:mueller04-M}, \ref{fig:ito:ThM15}, and \ref{fig:ito:ThM15+MIRSI} for  our model analyses of the four data sets described above.
See \sectref{sect:Eros} for a detailed discussion of analogous plots and their interpretation.
Our results for the individual data sets are:
\begin{enumerate}
\item 
Depending on roughness parameters,  $\chi^2$ is minimized at very different  thermal-inertia values. Thermal inertia is not well constrained by the data, although values much below \unit{250}{\TIunit} would appear to be inconsistent.
The corresponding best-fit diameters 
vary between \unit{296}{\metre} (no roughness, $\Gamma=\unit{275}{\TIunit}$) and \unit{319}{\metre} (default roughness, $\Gamma=\unit{1650}{\TIunit}$).
The difference in best-fit diameter when compared to \citet{Ito1} is due to the recalibration of the ESO fluxes discussed above.
Fitting the data for individual sets of roughness parameters clearly shows that there is no well constrained global $\chi^2$ minimum. This was somewhat obscured by the approach used in \citet{Ito1}, where crater density was used as a variable fit parameter.
\item
Thermal inertia is unconstrained within the range 0--\unit{2500}{\TIunit},  the largest value considered in our analysis (thermal inertia of bare rock). Zero roughness appears to be a poor fit to the data, but none of the remaining three roughness models is excluded. 
The corresponding best-fit diameters vary between 
0.27 and \unit{0.35}{\km}. Note that 
by disregarding a single data point we could significantly improve the quality of the fit: 
the reduced $\chi^2$ now reaches much lower values than for data set 1.
\item
There is a clear global minimum for ``default'' roughness at thermal-inertia values between 600 and \unit{800}{\TIunit}, although slightly lower values ($\Gamma$ between 500 and \unit{700}{\TIunit}) and ``low'' roughness are not  ruled out. 
Zero and ``high'' roughness would appear to be less consistent with the data. 
For the quoted ranges in thermal inertia, the corresponding best-fit diameters vary between 0.320 and \unit{0.326}{\kilo\metre}. 
\item
The global minimum for ``default'' roughness and thermal inertia between 600 and \unit{800}{\TIunit} remains, ``low'' roughness is less well consistent with the data than before. The corresponding best-fit diameters vary between 0.323 and \unit{0.329}{\km}.
\end{enumerate}
These results are summarized in \tableref{table:ito:results}.

\begin{table}
  \centering
\caption{Itokawa: Results of TPM fits to the four data sets considered (see Figs.\ \ref{fig:ito:mueller04}--\ref{fig:ito:ThM15+MIRSI}).}
\label{table:ito:results}
  \begin{tabular}{rrr}
\toprule
Data set & Diameter  & Thermal inertia \\
         & (\metre)& (\TIunit) \\
\midrule 
1 & 296--319 & $\geq250$ \\
2 & 270--350 &  0--2500\\
3 & 320--326& 500--800\\
4 & 323--329& 600--800\\
\bottomrule
  \end{tabular}
\end{table}

\subsection{Discussion}
        \label{sect:ito:discussion}

\subsubsection{Diameter}

Taken at face value, the  best-fit diameter resulting from our TPM analysis of all available reliable thermal-infrared data is $326\pm\unit{3}{\metre}$, in amazingly good agreement with the Hayabusa result of $327\pm\unit{6}{\metre}$ \citep{Demura2006}. We caution, however, that 
solely the scatter in the flux values is reflected in the quoted range of uncertainty.
Systematic flux calibration uncertainties and systematic uncertainties inherent in our TPM and in the shape model  add to the error budget.
Nevertheless,
this close match emphasizes the potentially very high accuracy of diameter estimates derived from TPM analysis of a suitable thermal-infrared database.
Combining this with 
results from an  analysis of Eros data (see also \sectref{sect:Eros}), where we were able to reproduce the diameter estimate derived from spacecraft measurements to within \unit{$\sim5$}{\%}, we conclude that \unit{10}{\%} is a conservative upper limit on the systematic diameter uncertainty inherent in our thermophysical modeling \seesect{sect:NEA:D}, resulting in a final diameter result of $0.32\pm\unit{0.03}{\km}$.

We  note that our preliminary results reported in \citet{Ito1} were plagued by inaccuracies in  flux values quoted from other sources.
Using recalibrated TIMMI2 flux values 
resolves the diameter discrepancy between our preliminary analysis ($D\sim\unit{0.28}{\km}$) and the Hayabusa result, $0.327\pm\unit{0.006}{\km}$ rejecting the unreliable (Hasegawa, 2004, private communication) M-band flux by \citeauthor{Ishiguro2003}\ significantly increased the goodness of fit.

\subsubsection{Thermal inertia}

The best-fit thermal inertia is  $700\pm\unit{100}{\TIunit}$. This refines the estimate by \citet[$750\pm\unit{250}{\TIunit}$]{MuellerItokawa}, which was based on a subset of the database available to us.
A reanalysis of their data set confirms their result.

The TPM-derived  thermal-inertia result is significantly larger than the estimate by \citet{Ishiguro2003} of \unit{$\sim290$}{\TIunit}.
This is not entirely surprising given the small data set of only two thermal-infrared data points available to them.
Furthermore, \citeauthor{Ishiguro2003}\ analyzed their data using a simplistic thermal model based on spherical geometry which we would expect to be inappropriate for a quantitative determination of thermal inertia.

\subsubsection{Geological interpretation of thermal inertia in the light of Hayabusa results}

A thermal inertia of $700\pm\unit{100}{\TIunit}$ corresponds to some 14 times the lunar value, so Itokawa's surface would not be expected to be dominated by fine Moon-like regolith.
On the other hand, \unit{700}{\TIunit} is less than a third of the thermal inertia of solid  rock (see \tablerefpage{table:thermalproperties}).
As highlighted by \citet{MuellerItokawa},
Itokawa's corresponding thermal conductivity is  $\kappa\sim\unit{0.3}{\kappaunit}$ which they conclude is a plausible value for ``porous stony material''.%
\footnote{ Thermal inertia is  proportional to the square root of the mass density, which further decreases the thermal inertia of porous material. The effect of thermal conductivity dominates, however.}
As can be seen from \tableref{table:thermalproperties}, 
an extremely large porosity would be required to explain this conductivity: 0.3 is smaller than the value for solid rock by a factor of ten, and only twice as large as the value for the very vesicular material pumice, which is an implausible model for asteroidal ``bare rock''.

Among the materials listed in \tableref{table:thermalproperties}, the best matches to Itokawa's thermal inertia are compact snow, sandy soil, and coal. Given Itokawa's taxonomic classification as S type, sandy soil would be expected to be the best analogue, although we caution that the  atmosphere (and furthermore humidity inside the soil) skew thermal-inertia values measured on Earth towards larger values relative to asteroid surfaces.

We note that the thermal skin depth (\eqrefpage{eq:skindepth}), i.e.\ the length scale for the penetration of the diurnal heat wave, is of the order of several centimeters given our thermal-inertia result. Gravel with grain sizes up to some \unit{10}{\centi\metre}  would therefore be expected to display a reduced thermal inertia relative to boulders made of the same material.
Furthermore, a dust coating (thin compared to the skin depth) would reduce the thermal inertia of a boulder. This will be further discussed in \sectref{sect:NEA:barerock}.

Based on Hayabusa measurements obtained while descending to a sampling site in smooth terrain, a local thermal-inertia value between 100 and \unit{1000}{\TIunit} was obtained \citep{Yano2006}.
The sampling site was seen from close-up imaging to be covered with coarse regolith, with typical grain sizes in the \milli\metre--\centi\metre\ range. 
It is reasonable to expect the sampling site to be representative of Itokawa's smooth terrains, which 
cover some \unit{20}{\%} of the total area \citep[see][]{Saito2006}.
For the rough terrains, which cover some \unit{80}{\%} of the surface area, 
no local thermal-inertia measurement is available. A small fraction of the rough terrain, adjacent to the touchdown site in the smooth Muses-C region, was imaged in high resolution; the images reported by \citet{Miyamoto2007} show much larger grain sizes than in smooth terrain. While particle sizes for the rough terrain are not explicitly reported by \citeauthor{Miyamoto2007}, their Fig.\ 2c appears to display a very low abundance of pebble-sized gravel but rather appears to be dominated by larger cobbles which are
comparable to the thermal skin depth (centimeters) in size.

From our global thermal-inertia result it appears that the typical thermal inertia on Itokawa's surface does not greatly exceed that of the sampling site, although we caution that the latter is only poorly constrained so far.
We speculate that the large boulders apparent in Hayabusa imaging of the rough terrain \citep[see, e.g.,][Figs.\ 3 and 4]{Saito2006}
are covered with a thin dust coating. 
Alternatively, 
the ``bare rock'' of which Itokawa's boulders are composed may be a poor thermal conductor relative to granite found  on Earth; this could be caused by, e.g., a high amount of porosity at length scales smaller than the skin depth, i.e.\ at the \milli\metre-scale or smaller. Note that \citet{Christensen2003} report a thermal inertia up to \unit{2200}{\TIunit} for  outcrops of exposed bedrock on Mars (the floor of the Nili Patera caldera).

We wish to stress that the dependence of thermal inertia on grain size, which is well established for the atmospheric pressures prevalent on Earth \citep[see][]{Clauser1995} and Mars \citep[see][]{Presley1997}, is significantly less well studied in a vacuum. Laboratory measurements of the thermal inertia of particulate materials in a vacuum chamber would be very valuable for the interpretation of our results. The same applies to the effect of porosity on thermal inertia---note that the residual atmospheric gas inside pores enhances the thermal conductivity and hence the thermal inertia.

\subsubsection{Mutual consistency of thermophysical models}

The data set of 15 flux values
considered by
\citet{MuellerItokawa} to be of ``high quality'' has been reanalyzed.
While \citeauthor{MuellerItokawa}\ used the TPM by \citet{LagerrosI,LagerrosIII,LagerrosIV} in their analysis, we use that described in \chaptref{chapt:TPM} which is largely based on Lagerros' but has been implemented independently.

They report $D=0.32\pm\unit{0.03}{\km}$  and $\Gamma=\unit{750}{\TIunit}$. No formal thermal-inertia uncertainty is given, but 
values between 500 and \unit{1000}{\TIunit} are reported to be consistent with their data. Both results are largely reproduced in our reanalysis.
Comparing their Fig.\ 4 with the ``default roughness'' curve in  \figrefpage{fig:ito:ThM15}, it appears that our result is slightly skewed towards lower thermal inertia.
This may be due to the different fitting techniques used 
(\citeauthor{MuellerItokawa}\ do not attempt to minimize $\chi^2$ but rather the fractional formal diameter uncertainty, which is not entirely equivalent) and does not lead to significantly different results.

We conclude that the TPM by 
\citet{LagerrosI,LagerrosIII,LagerrosIV}
and that described herein are mutually consistent.

\subsection{Summary}
        \label{sect:ito:summary}

A TPM analysis of all available thermal-infrared data of the NEA (25143) Itokawa, the target of the Japanese mission Hayabusa, 
results in a diameter estimate between 320 and \unit{330}{\metre}, in excellent agreement with the Hayabusa result of $327\pm\unit{6}{\metre}$. Combining this with the results of our Eros study \seesect{sect:Eros} we conclude that \unit{10}{\%} is a conservative upper limit on the systematic diameter uncertainty inherent in our thermophysical modeling.

We find Itokawa's thermal inertia to be $700\pm\unit{100}{\TIunit}$, thus refining the estimate by \citet{MuellerItokawa}, which is based on a subset of the data available to us.
A reanalysis of their database reproduces their results, thus demonstrating the mutual consistency of our TPM and that by \citet{LagerrosI,LagerrosIII,LagerrosIV} which was used by \citeauthor{MuellerItokawa}

Itokawa's thermal inertia is intermediate between that of bare rock and dusty regolith. This is not straightforward to reconcile with Hayabusa close-up imaging results, which reveal a surface dominated by relatively large boulders.
We speculate that boulders on Itokawa are covered with a thin coating of dust. This will be further discussed in \sectref{sect:NEA:barerock}.
We call for a reanalysis of the thermal-infrared database using the accurate Hayabusa-derived shape model, which has been released to the scientific community on 24 April 2007. 


%% file: Betulia.tex
The C-type NEA (1580) Betulia is a highly unusual object for which earlier radiometric observations, interpreted on the basis of simple thermal models, indicated a surface of unusually high thermal inertia \seesect{sect:thermal:overview}.

We report results from extensive multi-wavelength thermal-infrared observations of Betulia obtained in 2002 with the NASA Infrared Telescope Facility IRTF.
From a TPM analysis we determine that Betulia's thermal inertia is only moderate, around \unit{180}{\TIunit}, comparable to our other NEA results.
We determine an effective diameter of \unit{$4.57\pm0.46$}{\kilo\metre} and an albedo of $\pv=0.077\pm0.015$, consistent with expectations based on the taxonomic type.
Our results are in broad agreement with an independent NEATM analysis of the same data set but are hard to reconcile with  previous results with indicate a much larger diameter and, indirectly, a high thermal inertia indicative of bare rock.

After publication, our diameter and albedo estimates have been confirmed on the basis of a reanalysis of available polarimetric data and, independently, on the basis of new radar observations.

\subsection{Introduction}
\label{sect:Betulia:intro}

The C-type NEA (1580) Betulia is well known for its unusual
lightcurve, the amplitude and form of which changes
dramatically with changing solar phase angle; 
particularly, at large phase angles it 
exhibits a triply-peaked lightcurve as opposed to the doubly-peaked lightcurves of virtually all other asteroids
 \citep{TedescoBetulia}.
These and other lightcurve data were used by \citet{KaasalainenBetulia} to determine Betulia's spin state and a convex-definite model of its shape. They found the  shape to be ``very peculiar with   a large planar area on one side'' which they note may conceal a considerable concavity (their method of shape determination is designed to produce convex models).
It is important to bear Betulia's unusual shape 
 in mind when interpreting observational
data with the aid of standard photometric, polarimetric,
and radiometric techniques. 

Betulia is also significant in being the first NEA for which thermal-infrared observations indicated a surface of high thermal inertia \seesect{sect:thermal:overview}. 
\citet{LebofskyBetulia} obtained thermal-infrared data at a single thermal wavelength (\unit{10.6}{\micron}).
Using the STM \seesect{sect:STM}, which neglects thermal inertia, they obtained a diameter of
\unit{$4.20\pm0.80$}{\km}, inconsistent with a lower bound on the diameter  of \unit{$D>5.8\pm0.4$}{\km} derived from radar observations \citep{PettengillBetulia}
and furthermore with  polarimetric observations, 
which appeared to indicate a diameter around \unit{7}{\km} \citep{TedescoBetulia}.
Because of this inconsistency, \citeauthor{LebofskyBetulia}\ rejected the STM diameter and used instead the FRM \seesect{sect:FRM}, obtaining
\unit{$D=7.5\pm0.34$}{\km},  concluding that Betulia had a very high thermal inertia, comparable to that of bare rock.

Betulia was observed extensively during an observing run with
the NASA Infrared Telescope Facility (IRTF) in June 2002.
Good quality data over a significant part of the rotation period
were obtained at five wavelengths in the range 7--\unit{20}{\micron}.
We have fitted the resulting flux data with the 
TPM for convex shapes \seechapt{chapt:TPM}
based on the shape model by \citet{KaasalainenBetulia}.

\subsection{Observations and data reduction}
\label{sect:Betulia:observations}

\begin{table}
\caption[Betulia: Observing geometry]{Betulia: Observing geometry. Values for the subsolar and sub-Earth latitude are based on the spin axis by \citet{KaasalainenBetulia}.}
\label{table:Betulia:geometry}
  \centering
  \begin{tabular}{lll}
\toprule
 & 2002 June 02 & 2002 June 05 \\
\midrule
Heliocentric distance, $r$ (\AU) & 1.143 & 1.150 \\
Geocentric distance, $\Delta$ (\AU) & 0.246 & 0.263 \\
Solar phase angle, $\alpha$ (\degree) & 52.9 & 53.3 \\
Subsolar latitude (\degree) & 13 & 15 \\
Sub-Earth latitude (\degree) & -39 & -34 \\
\bottomrule    
  \end{tabular}
\end{table}

The observations were performed on 2002 June 2 and
June 5 UT with the IRTF and JPL's $128\times128$~pixel, 7--\unit{25}{\micron}
infrared astronomical camera, MIRLIN. For details of MIRLIN
see \citet{MIRLIN} and \url{http://cougar.jpl.nasa.gov/mirlin.html}. Measurements were made in N-band filters centered at (bandwidth in brackets): 7.91 (0.76), 10.27 (1.01),
and 11.7 (1.11)~\micron, and in Q-band filters at: 17.93 (0.45) and
20.81 (1.65)~\micron. 
See \citet{Betulia} for details on the observations.

Standard synthetic aperture
procedures 
\seesect{sect:IRTF:reduction}
were used for the derivation of raw signal counts
from the MIRLIN images. Absolutely calibrated fluxes for
the target asteroids were obtained by multiplying the integrated
absolute fluxes of the calibration stars by the ratios
of the target/calibration star raw counts in each filter. Absolutely
calibrated infrared spectra for the calibration stars,
$\alpha$~Boo and $\alpha$~Hya, were taken from the database of \citet{CohenX}.
Color corrections for the different flux distributions
of the calibration stars and asteroids in the narrow filter
pass bands were found to be no more than a few percent \citep{Delbo2004} and were not applied.

\begin{table}
\caption{Betulia: Measured flux values}
\label{table:Betulia:fluxes}
  \centering
\footnotesize{
  \begin{tabular}{llllll}
\toprule
Date (UT) & Time & Julian date & Wavelength & Flux & Error \\
          &(UT) & (days-2,452,420)& (\micron)& (\milli\Jy) & (\milli\Jansky)  \\
\midrule
2002-06-02 & 05:59 & 7.7493& 7.91 &1400& 57 \\
 &07:03 & 7.7937 & 7.91 & 1789  &61  \\
 &08:09 & 7.8396 & 7.91  &1803  &67  \\
 &06:07 & 7.7549 & 10.27 & 2270 & 36  \\
 &07:11 & 7.7993 & 10.27 & 3334 & 32  \\
 &08:17 & 7.8451 & 10.27 & 3315 & 45  \\
 &06:16 & 7.7611 & 11.70 & 2347 & 38  \\
 &07:19 & 7.8049 & 11.70 & 3900 & 36  \\
 &08:24 & 7.8500 & 11.70 & 3166 & 37  \\
 &06:29 & 7.7701 & 17.93 & 4445 & 568 \\
 &07:33 & 7.8146 & 17.93 & 4745 & 311 \\
 &08:37 & 7.8590 & 17.93 & 4670 & 343 \\
2002-06-05 &  05:50 &  10.7431 &  7.91 &  1150 &  67  \\
 & 07:04 &  10.7944 &  7.91  & 1277  & 74   \\
 & 08:13 &  10.8424 &  7.91  & 1302  & 84   \\
 & 05:57 &  10.7479 &  10.27 &  2171 &  60   \\
 & 07:11 &  10.7993 &  10.27 &  2364 &  74   \\
 & 08:20 &  10.8472 &  10.27 &  1998 &  71   \\
 & 06:02 &  10.7514 &  11.70 &  2606 &  38   \\
 & 07:16 &  10.8028 &  11.70 &  2312 &  51   \\
 & 08:25 &  10.8507 &  11.70 &  2458 &  39   \\
 & 06:13 &  10.7590 &  17.93 &  2635 &  247  \\
 & 07:25 &  10.8090 &  17.93 &  3406 &  247  \\
 & 08:35 &  10.8576 &  17.93 &  2528 &  290  \\
 & 06:28 &  10.7694 &  20.81 &  3235 &  177  \\
 & 07:39 &  10.8187 &  20.81 &  2663 &  189  \\
 & 08:47 &  10.8660 &  20.81 &  3172 &  176  \\
\bottomrule
  \end{tabular}
} 
\end{table}

\Tableref{table:Betulia:geometry}
lists the observing geometry. The resulting fluxes
are listed in \tableref{table:Betulia:fluxes}. The quoted uncertainties in the flux
measurements refer to the formal statistical uncertainties in
the synthetic aperture procedure, only. Errors in the absolute
calibration and fluctuations in atmospheric conditions during
the observations increase the scatter in the flux data.
On 2002 June 2, an R-band (visible) lightcurve was obtained simultaneously with our IRTF observations
on the University of Hawaii's 88-inch telescope and
kindly made available  by Yan \Fernandez\ (2002, private communication).

\subsection{Results}
\label{sect:Betulia:results}

\begin{figure}
  \centering
\includegraphics[angle=-90,width=0.6\linewidth]{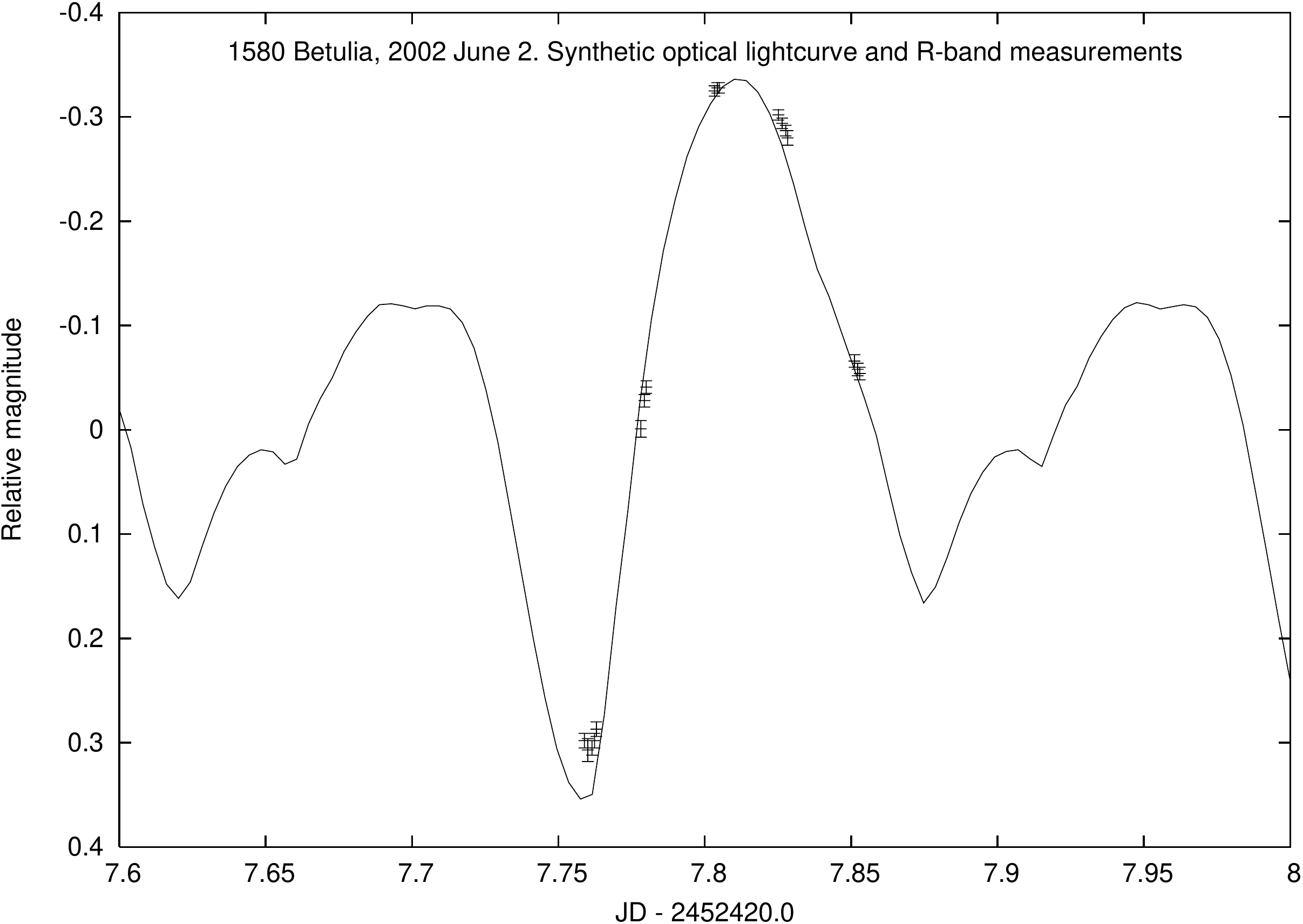}
  \caption[Betulia: Synthetic optical lightcurve and R-band measurements]{Betulia: Synthetic optical lightcurve using the shape model from \citet{KaasalainenBetulia} for the observing geometry of 2002 June 2. There is very good agreement in shape and amplitude between the synthetic curve and the corresponding R-band measurements. The positioning of the synthetic lightcurve relative to the R-band measurements was adjusted manually on both axes to obtain a good fit, leading to a slight correction to the rotational period within the uncertainties quoted by \citeauthor{KaasalainenBetulia}\ (see text for details).}
  \label{fig:Betulia:optical}
\end{figure}

Our TPM analysis is based on the 
convex shape model by \citet{KaasalainenBetulia} which consists of 3192 triangular facets. The corresponding spin axis is
 \unit{$\beta=21.92211922$}{\degree} and \unit{$\lambda=133.275794$}{\degree} (J2000 ecliptic latitude and longitude, respectively; from M.\ Kaasalainen, 2003, private communication).
\citet{KaasalainenBetulia} give a rotation period of \unit{$6.13836\pm0.00001$}{\hour}
 on the basis of observations between the years 1976 and 1989.
Using these parameters, we have generated a synthetic optical lightcurve for the observing geometry of 2002 June 2 and compared it to the  R-band data by \Fernandez\ \seefig{fig:Betulia:optical}. We found a slight offset in rotational phase of around \unit{29}{\minute}
which was corrected for by adjusting the rotation period to a value of 
\unit{6.1383602}{\hour}, well within the quoted range of uncertainty.

Applying our TPM \seechapt{chapt:TPM} to the IRTF measurements
listed in \tableref{table:Betulia:fluxes}, assuming $H = 15.1$ and $G=0.15$ \citep[see][and references therein]{Betulia}, resulted
in an effective diameter, $D_\text{eff}$, of \unit{$4.57 \pm 0.15$}{\km}, a geometric albedo, \pv, of $0.077 \pm 0.005$, and a thermal inertia of
\unit{$180\pm50$}{\TIunit}.
The quoted errors reflect only the
statistical scatter about the model curve; modeling uncertainties
are much larger than the formal errors. 
Comparison
with independent results for $D_\text{eff}$ in the cases of (433) Eros and (25143) Itokawa suggests the uncertainty in diameter
determinations with the TPM does not exceed
\unit{10}{\%} \seesect{sect:NEA:D}.

\begin{figure}
  \centering
\includegraphics[width=0.5\linewidth,angle=-90]{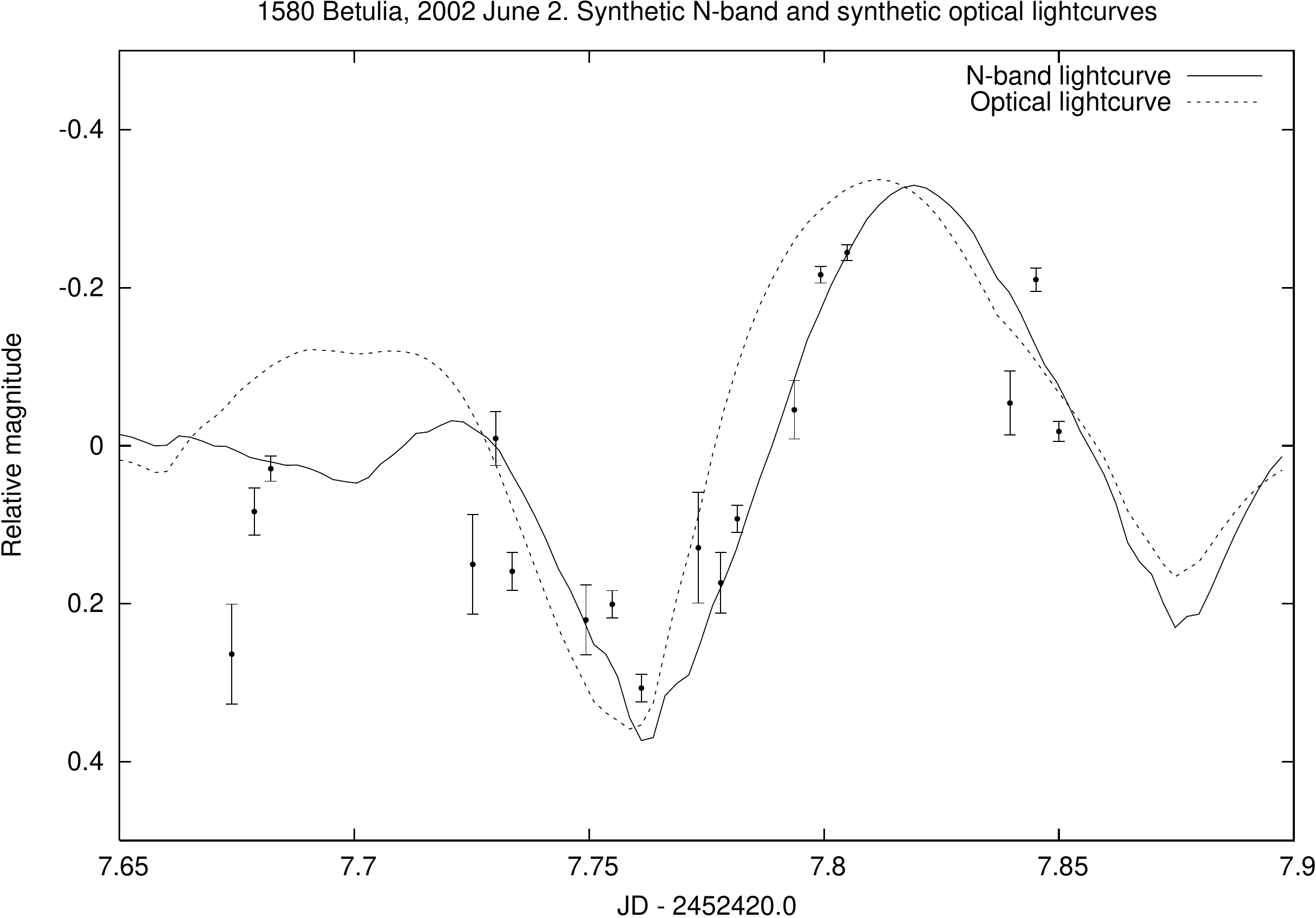}  
  \caption[Betulia: Comparison of synthetic N-band and synthetic optical lightcurves, points corresponding to the measured N-band fluxes are superimposed.]{Betulia: Comparison of synthetic N-band and synthetic optical lightcurves 
for the observing geometry of 2002 June 2. Points corresponding to the measured N-band fluxes are superimposed. For  this plot,  data obtained on June 5 are referred to June 2 by subtracting 12 rotational periods and by adjusting the flux values for the changes in observing geometry (whereby the dominant effect is due to the change in geocentric distance, see \tableref{table:Betulia:geometry}).
No  manual adjustments to the N-band synthetic lightcurve or measurement points were made.}
  \label{fig:Betulia:lightcurveN}
\end{figure}

\begin{figure}
  \centering
\includegraphics[width=0.5\linewidth,angle=-90]{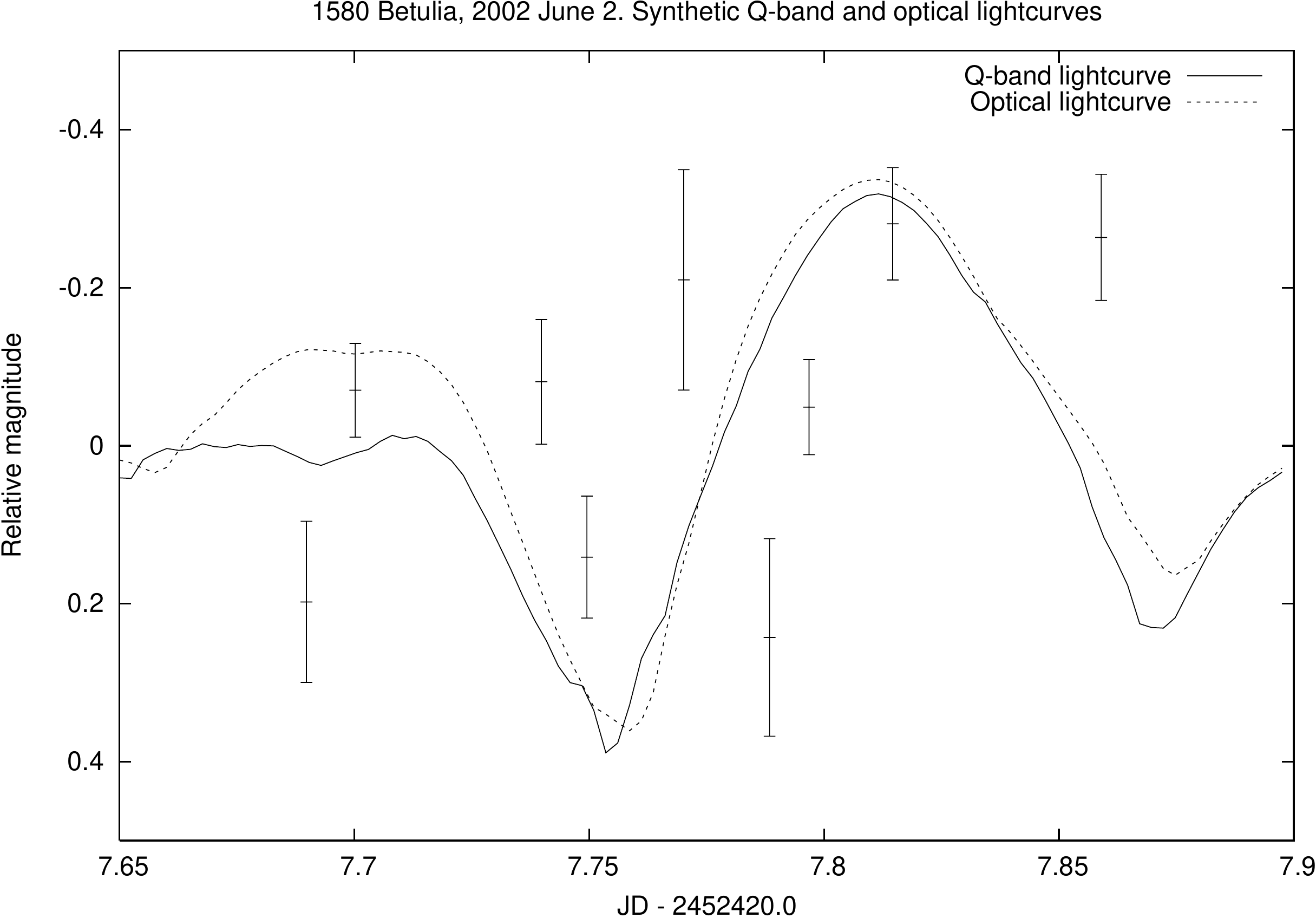}  
  \caption[Betulia: Comparison of synthetic Q-band and synthetic optical lightcurves, points corresponding to the measured Q-band fluxes are superimposed.]{Betulia: Comparison of synthetic Q-band and synthetic optical lightcurves, points corresponding to the measured Q-band fluxes are superimposed as in \figref{fig:Betulia:lightcurveN}. The plotted Q-band lightcurve results from parameters which   best fit  the total data set including N-band data.}
  \label{fig:Betulia:lightcurveQ}
\end{figure}

There is evidence in \figref{fig:Betulia:lightcurveN} for a phase lag between the observed optical and thermal-infrared lightcurves which has the same sense and appears to be roughly the same fraction of the rotation period as that observed by \citet{LebofskyRieke1979} in the case of (433) Eros,
which they explained in terms of surface thermal inertia (see also \sectref{sect:Eros}).
Given the differences in the structure of the optical and N-band synthetic lightcurves evident in a comparison of the two lightcurves over a full rotation period, the significance of the apparent small phase lag is not clear.
The synthetic Q-band lightcurve (\figref{fig:Betulia:lightcurveQ}) does not display a clear phase lag.
The resolution of this discrepancy
may lie in shortcomings of the shape model and the TPM: for example, possible variations in albedo
and thermal inertia over the surface are ignored, and treatment
of the surface structure in the TPM is
limited to an idealistic distribution of hemispherical craters.
We have no satisfactory quantitative explanation for the discrepancy
at present and further development of the model
may be required to more accurately reproduce the details of
the observations.

\subsection{Discussion}
\label{sect:Betulia:discussion}

\begin{table}
\caption[Betulia: Summary of diameter and albedo determinations]{Betulia: Summary of diameter and albedo determinations.
Values given in brackets were determined by us, assuming $H=15.1$, in order to facilitate comparison.}
\label{table:Betulia:diameters}
  \centering
  \begin{tabular}{lll}
\toprule
Source & $D_\text{eff}$ (\km) & \pv \\
\midrule
This work & $4.57\pm0.46$ & $0.077\pm0.015$ \\
\citet{TedescoBetulia}, polarimetry & $\sim 7$ & $\sim0.033$ \\
\citet{LebofskyBetulia}, STM & $4.20\pm0.80$ & ($0.091\pm0.035$) \\
", FRM & $7.48\pm0.34$ & ($0.029\pm0.003$) \\
\citet{PettengillBetulia}, radar & $>5.4$ & ($<0.055$) \\
\citet{Betulia}, NEATM & $3.82\pm0.58$ & $0.11\pm0.04$ \\
Tedesco (2005, private communication) & $4.76\pm0.74$ & $0.071\pm0.005$ \\
\citet{Magri2007}, radar & $5.39\pm0.54$ & ($0.055\pm0.011$) \\
\bottomrule
  \end{tabular}
\end{table}

\paragraph{Diameter and albedo}
See \tableref{table:Betulia:diameters} for a comparison of our diameter and albedo result with previous determinations. 
It is instructive to first compare our results with the NEATM results by \citet{Betulia}, which are based on the same data set.
While  the TPM gives a somewhat
larger diameter and lower albedo (the latter being more in
line with expectations for a C-type asteroid) the two sets of
results are in reasonable agreement given the uncertainties
associated with both models, especially at the large solar
phase angle at which our observations were made.

Our  results are, however, inconsistent with the radar-derived lower limit on diameter of \unit{$5.8\pm0.4$}{\km} \citep{PettengillBetulia} and with the polarimetric albedo result by \citet{TedescoBetulia}.
Based on thermal-infrared observations at a single thermal wavelength (\unit{10.6}{\micron}), 
\citet{LebofskyBetulia} obtained two model-dependent diameter values, based on the STM (which neglects thermal inertia, see \sectref{sect:STM}) and the FRM (which effectively assumes infinite thermal inertia, see \sectref{sect:FRM}), respectively.
Our results are in
good agreement with his STM result and inconsistent with his FRM result.
\citet{LebofskyBetulia}, however, rejected the STM diameter in favor of the FRM result due to the better consistency of the latter with the results of \citet{TedescoBetulia} and \citet{PettengillBetulia}.
Betulia was the first asteroid for which the STM appeared to produce inconsistent results.

After the publication of our results, 
E.\ Tedesco (2005, private communication) has reanalyzed his published polarimetric data  using a more recent calibration scheme, leading to a revised albedo estimate of $\pv=0.071\pm0.005$. Together with the $H$ value given in \citet{Betulia}, $H=15.1\pm0.3$, this implies \unit{$D=4.76\pm0.74$}{\km} in excellent agreement with our result.

\citet{Magri2007} report new radar observations of Betulia from which they determine an effective diameter of \unit{$5.39\pm0.54$}{\km}, some \unit{15}{\%} larger than our result but within the combined ranges of uncertainty (see also \sectref{sect:NEA:D}), and significantly below \unit{7}{\km}.

\paragraph{Shape}
From their radar observations combined with optical lightcurve data available in the literature, \citet{Magri2007} determined a new model of Betulia's shape.
They  confirm the spin-state estimate by  \citet{KaasalainenBetulia} 
and largely confirm their convex-definite shape model. The radar shape model does, however, contain a few  concavities including a very large concavity with a diameter comparable to Betulia's radius. 
The location of the latter is quoted to coincide with the large planar area in the \citeauthor{KaasalainenBetulia}\ model, where a concavity had been suggested by \citeauthor{KaasalainenBetulia}

We call for a reanalysis of
our data based on the new radar shape model. At the phase angles at which our observations took place (52.9 and \unit{53.3}{\degree}), shadowing inside the large concavity would be expected to lead to lower thermal fluxes relative to model fluxes for a convex shape, and might thus have caused us to  underestimate the diameter.
Depending on rotational phase, the concavity may or may not be oriented towards the observer, so it should produce an observable difference in the shape of the thermal lightcurve.
Indeed, judging from Figures  \ref{fig:Betulia:lightcurveN} (on p.\ \pageref{fig:Betulia:lightcurveN}) and \ref{fig:Betulia:lightcurveQ}, there appears to be a flux drop in both the N and the Q band at times shortward of 7.7, which is not reproduced by our TPM based on the convex shape model. 
Unfortunately, no optical lightcurve data are available for that time, so observational artefacts cannot be ruled out at the present time. 
The  concavity is situated ``on the southern hemisphere'' \citep{Magri2007} which was visible during our observations \seetablepage{table:Betulia:geometry}.

\paragraph{Thermal inertia}
We obtain the first quantitative  estimate of Betulia's thermal inertia, \unit{$180\pm50$}{\TIunit}, some three times larger than lunar regolith but an order of magnitude below bare rock (see \tablerefpage{table:thermalproperties}).
We therefore find no evidence for a bare-rock surface of Betulia, in contrast with the suggestion by \citet{LebofskyBetulia}. 
We wish to emphasize that the thermal data
at their disposal were taken at a single thermal wavelength and hence did not allow direct conclusions to be drawn on thermal properties (see also \figrefpage{fig:thermal:singlewavelength}).
In the light of the revised diameter estimates (see above) their reason to reject the STM diameter in favor of the FRM result no longer applies.
In fact, their STM diameter is in good agreement with our result, providing indirect support for our moderate-thermal-inertia result.

Our thermal-inertia result is consistent
with NEATM fits to our data set by \citet{Betulia}, which resulted in 
$\eta = 1.09$, a value consistent with the
presence of some thermally insulating regolith \citep[see][]{Delbo2003}.

In particular, our thermal-inertia result for Betulia is in line with our other thermal-inertia results for NEAs as discussed in \sectref{sect:NEA:TI}, while a very large thermal inertia as suggested by \citeauthor{LebofskyBetulia}\ would be unusual.

\subsection{Summary}
\label{sect:Betulia:summary}

From a TPM analysis of thermal-infrared
flux measurements of the NEA (1580) Betulia
obtained at the NASA IRTF, 
we obtain a diameter of
\unit{$4.57 \pm 0.46$}{\km}, an albedo of $\pv = 0.077 \pm 0.015$,
and a surface thermal inertia of
\unit{$180\pm50$}{\TIunit}, or some three times the lunar value.
This value of thermal inertia is less than \unit{10}{\%} of that expected
for a bare-rock surface and implies that the surface
of Betulia has a significant thermally insulating regolith, in
contrast to the conclusions of earlier work but 
consistent with our other thermal-inertia results for NEAs \seesect{sect:NEA:TI}.

Our results for diameter and albedo are in broad agreement with a NEATM analysis of the same data set \citep{Betulia}. The corresponding value of the model parameter, $\eta=1.09$, is consistent with the presence of thermally insulating regolith. The difference between the results
from the TPM and the NEATM is probably due
to the complex shape and nature of Betulia and the more sophisticated
treatment of the TPM, although
we caution that neither model has been thoroughly tested at
the high solar phase angles at which the IRTF observations
were made.

Our diameter and albedo results are inconsistent with previous estimates based on polarimetric and radar data.
After publication, however, our results  were confirmed through a reanalysis of available polarimetric data based on up-to-date calibration schemes (Tedesco, 2005, private communication) and, independently, by \citet{Magri2007} based on new radar observations.


%% file: PH5.tex
We present preliminary results of Spitzer observations of the small NEA (54509) YORP (known until 2 April 2007 as 2000~PH5), which has  an ultra-fast rotation period of only 12~minutes.
Its rotation period has recently \citep[published on-line on 8 March]{Lowry2007,Taylor2007}
been demonstrated to be steadily decreasing
due to the YORP effect \seesect{sect:intro:Yarko} after which it has been named; together with the work by \citet{Kaasalainen2007} this constitutes the first direct detection of the YORP effect.

Further interest in (54509) YORP stems from the fact that 
it would be expected to be regolith-free due to its fast rotation: For any plausible mass density, gravity is overwhelmed by the centrifugal force on most of its surface.

Using the Spitzer Space Telescope we have obtained thermal lightcurves of (54509) YORP at wavelengths of 8, 16, and \unit{22}{\micron} with a time resolution of \unit{14}{\second} or better.
From these, we obtain the diameter and thermal inertia of the asteroid.

The Spitzer Science Center has recently issued an updated calibration scheme which applies to two of our three data sets and is expected to lead to significant flux changes \seesect{sect:PUI:validation}. All results presented herein are based on the superseded calibration scheme and are therefore preliminary.


\subsection{Introduction}
\label{sect:PH5:intro}

The NEA (54509) YORP was discovered on 3 August 2000 by the MIT Lincoln Laboratory's near-Earth asteroid search program \citep[LINEAR;][]{Stokes2000}.
It is approximately co-orbital with Earth on  a horseshoe orbit \citep[see, e.g.,][Fig.\ S1 in the supporting on-line material]{Taylor2007}
and has undergone 
annual very close approaches with Earth in the years 2000--2005.
It was soon found to be an ultra-fast rotator, with a period of only \unit{12.17}{\min}.

From extensive optical photometric observations in the years 2001--2005, \citet{Lowry2007} determined a steady linear increase in 
the angular velocity $\omega$ of 
$\textd\omega/\textd t = 2.0 (\pm0.2)\times \unit{$10^{-4}$}{\deg\usk\power{day}{-2}}$.
From radar observations obtained in the years 2001--2005 combined with the optical data by \citeauthor{Lowry2007}, \citet{Taylor2007} determined the spin state, shape, and diameter of YORP. 
They used their results as input parameters for model calculations of the  change in spin rate expected from the YORP effect (see also \sectref{sect:intro:Yarko}) and concluded that they have detected the latter.
Together with \citet{Kaasalainen2007}, who independently detected the YORP effect on another NEA, this constitutes the first direct detection of the YORP effect.
\citeauthor{Taylor2007}\ determined the J2000 ecliptic coordinates of the spin axis of YORP to be $\lambda=\unit{180}{\degree}$, $\beta=\unit{-85}{\degree}$; a highly irregular shape; and a volume-equivalent diameter around \unit{114}{\metre}. No formal diameter uncertainty is reported, in the following we assume a fractional uncertainty of \unit{10}{\%} as is usually reported for radar-derived diameters.
Together with the absolute optical magnitude of $H=22.562$ (NeoDys as of 11 May 2007),
this diameter implies an albedo of $\pv=0.13$. 

\citet{Gietzen2007} 
report results of near-infrared spectroscopic observations of YORP.
While a large amount of noise is apparent in their YORP data
they report a clear detection of silicate features at \unit{1}{\micron} and \unit{2}{\micron}.
They conclude that YORP belongs to either of the silicaceous taxonomic classes S or V. An S-type classification would suggest a moderate albedo around $\pv=0.2$ (see, e.g., \sectref{sect:intro:mineralogy}), 
a much larger albedo would be associated with a V-type classification.

So far, no asteroid has been unambiguously shown to display a high thermal inertia indicative of a bare-rock surface.
Fast rotators such as YORP are widely expected to be regolith-free bare rock
 \citep[see, e.g.][]{Whiteley2002}
mainly because
their surface gravity cannot match the centrifugal force on most of their surface for reasonable values of mass density.

We have observed YORP with the Spitzer Space Telescope \seechapt{chapt:SST} at three thermal wavelengths (8, 16, and \unit{22}{\micron}) in order to determine its size and thermal inertia.
At each wavelength we have sampled the thermal lightcurve at a time resolution of \unit{14}{\second} or better.
We have used our TPM described in \chaptref{chapt:TPM} to fit the data, based on a preliminary version of the shape model described in \citet{Taylor2007}, which was kindly made available to us in computer-readable form by P.\ Taylor in 2006.

\subsection{Spitzer observations}
\label{sect:PH5:data}

Our Spitzer observations have used the InfraRed Array Camera (IRAC, see \sectref{sect:IRAC:general}) for the \unit{8}{\micron} data and the InfraRed Spectrograph IRS in peak-up-imaging mode (PUI, see \sectref{sect:IRS:PUI}) for observations at central wavelengths of $\sim16$ and \unit{$\sim22$}{\micron}.
Due to the design of IRAC, the asteroid was simultaneously observed at an effective wavelength of \unit{4.5}{\micron}; as expected, the asteroid signal in those ``serendipitous'' data 
is too noisy to be used.
Our IRAC observations started on 
18 August 2005,  17:50 UT,
our PUI observations on 
17 August 2005, 12:33  UT.
The heliocentric distance at that epoch was \unit{1.09}{\AU}, the distance to Spitzer was \unit{0.146}{\AU} for the IRAC observations and \unit{0.147}{\AU} for the PUI observations, the solar phase angle was \unit{59.3}{\degree} for both. Assuming the spin axis given by \citet{Taylor2007}, the aspect was nearly equatorial, with a subsolar latitude of \unit{5}{\degree} and sub-observer latitude of \unit{13}{\degree} at the epoch of the IRAC observations. 
The local time at the sub-observer point was \unit{8.2}{\hour}, i.e.\ the cold ``morning'' side was observed (determined from the projections of the vectors towards the Sun and Spitzer onto the asteroid equator).

In order to obtain the finest possible time resolution, our observations were not dithered but used ``in-place repeats'' (see \sectref{sect:SST:solarsystem} for a discussion of these observing modi).
IRAC 
was pointed onto the asteroid 
(channels 2 and 4, at wavelengths of 4.5 and \unit{8}{\micron}, were simultaneously on target)
for 60 consecutive exposures with 
 frame times of \unit{12}{\second} each.
Including short ``dead'' times between consecutive frames for, e.g., detector readout, the last exposure ended \unit{13.6}{\minute} after the beginning of the first exposure, 
corresponding to 1.12 rotation periods and a time resolution of \unit{13.6}{\second}.
With PUI, the asteroid was observed 55 times with PUI ``blue'' (effective filter wavelength \unit{$\sim22$}{\micron}), then 55 times with PUI ``red'' (effective filter wavelength \unit{$\sim16$}{\micron}). Each PUI frame had an integration time of \unit{6}{\second}; 
including ``dead'' times between consecutive frames, each series of 55 frames lasted \unit{12.78}{\minute} corresponding to  $1.05$ rotation periods and a time resolution of \unit{13.9}{\second}.
We expected the first few IRAC frames to be unusable due to the first-frame effect \seesect{sect:IRAC:firstframe}, hence the slightly larger overlap between the two consecutive periods.

As recommended by the Spitzer Science Center (SSC), we did not perform dedicated calibration observations but relied on the calibration provided by the SSC, i.e.\ we started our data-reduction efforts at the level of Basic Calibrated Data (BCD) frames provided by the SSC (see also \sectref{sect:SST:data}).
BCD frames are flux-calibrated images with most detector-specific artefacts removed.
Note that the BCD pipeline for the PUI data has been updated on 28 February 2007; the PUI fluxes reported herein are based on the superseded pipeline version, 
significant calibration changes are to be expected from the new pipeline version
 (see also \sectref{sect:PUI:validation}).

Because our observations of YORP were not dithered we used a non-standard version of our data reduction pipeline.
In particular, MOPEX could not be used for the automated rejection of 
outlier pixels such as cosmic ray hits.
Instead, all BCD frames were visually inspected 
and rejected if  obvious outliers were found within the vicinity of the target. 
The target was found to be clearly visible and well centered in all frames for wavelengths of 8, 16, and \unit{22}{\micron}. 
The serendipitous IRAC \unit{4.5}{\micron} observations of an identical field
showed stellar background sources much more clearly than the simultaneous \unit{8}{\micron} frames. We additionally rejected IRAC \unit{8}{\micron} frames 
when the corresponding \unit{4.5}{\micron} frame displayed significant background structure close to the center, i.e.\ the asteroid position (which was discernible in some \unit{4.5}{\micron} frames).
No background sources were found in the PUI frames.

The remaining IRAC frames were in a first step reprojected onto a rectangular grid to correct them for  optical distortion, using MOPEX 
(see \sectref{sect:IRAC:mosaicking}).
Then, they were converted from flux units of \mega\jansky\per\sterad\ into units of \milli\jansky\per pixel using a dedicated IDL routine.
On the resulting frames, synthetic aperture photometry was performed 
\seesect{sect:IRAC:ATV}
for different aperture radii. The resulting flux values were aperture corrected to account for flux losses due to the limited radius of the synthetic apertures \seesect{sect:IRAC:aperturecorrection}.
Flux uncertainties were estimated from the statistical scatter of the individual flux vales (for any particular aperture radius) and the scatter in the flux set for all aperture radii; the uncertainty contributions were added in quadrature \seesect{sect:IRAC:uncertainty}.
Analogous procedures were used for PUI frames.

Final fluxes for each wavelength were averaged and fitted using the NEATM. Color-correction factors 
(see \sectref{sect:IRAC:CC} and \ref{sect:PUI:reduction})
were determined for the resulting NEATM model parameters. This procedure was reiterated using the color-corrected fluxes as NEATM input until stable color-correction factors have been obtained (after the second iteration).

For each observation, the time of mid-exposure was determined and one-way lighttime was subtracted. Given the Spitzer-centric distance of \unit{$\sim0.05$}{\AU}, lighttime was \unit{$\sim0.4$}{\minute}, or \unit{$\sim12$}{\degree} in rotational phase.

\begin{figure}
  \centering
\includegraphics[angle=-90,width=0.6\linewidth]{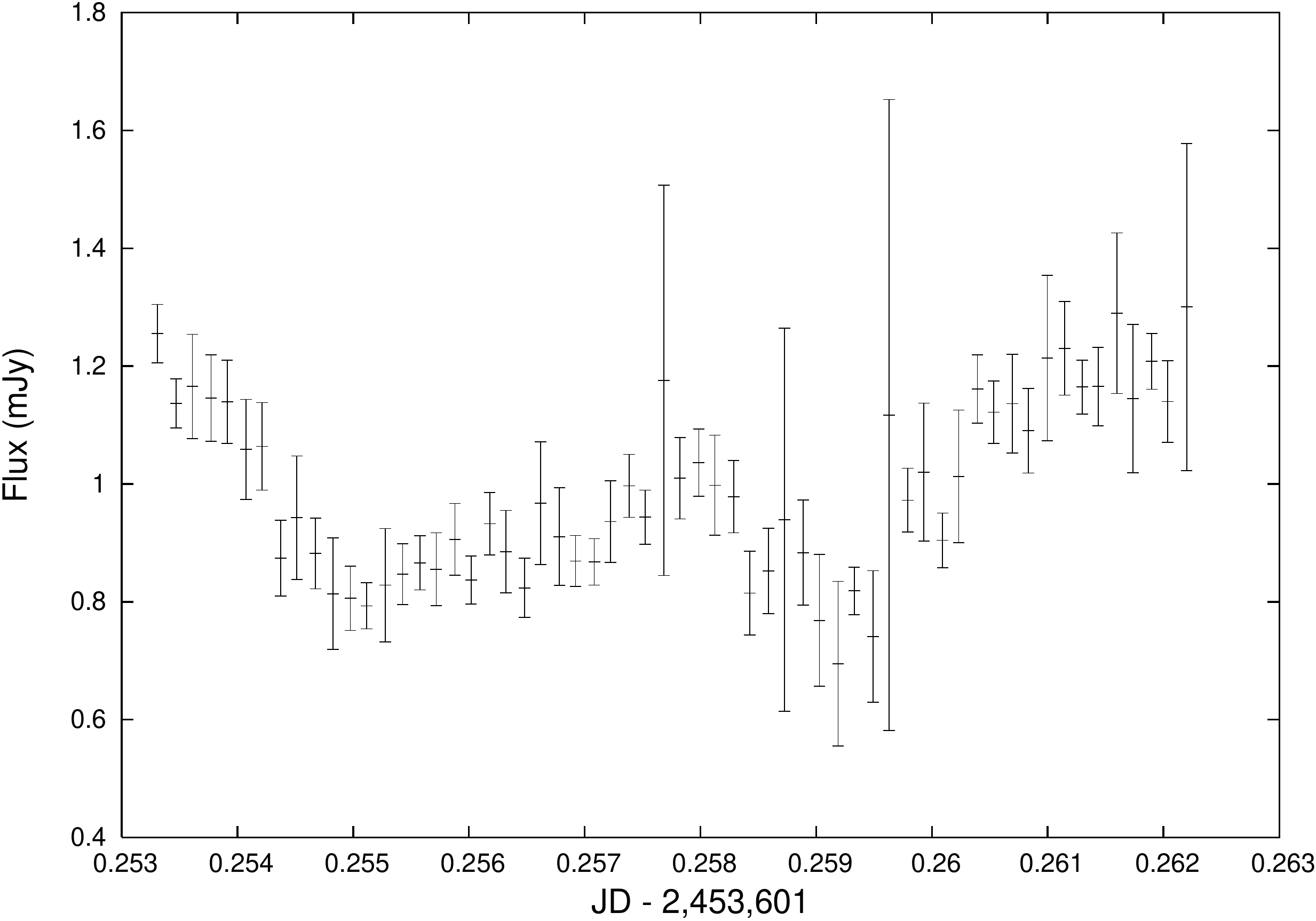}
  \caption{Plot of IRAC-channel-4 fluxes 
obtained for YORP.}
  \label{fig:PH5:IRAC}
\end{figure}

\begin{figure}
  \centering
\includegraphics[angle=-90,width=0.6\linewidth]{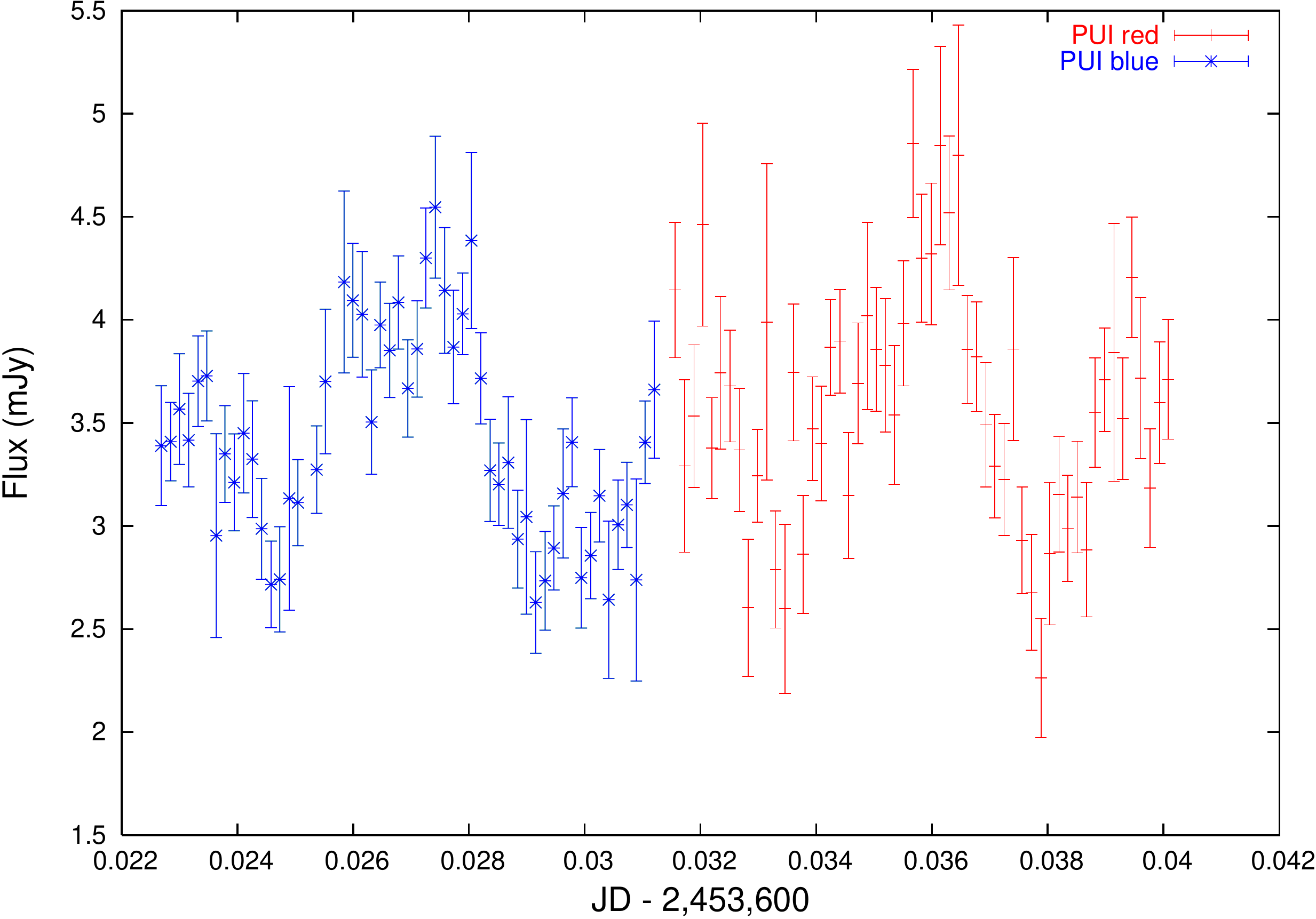}
  \caption{Plot of PUI fluxes obtained for YORP.}
  \label{fig:PH5:PUI}
\end{figure}

See \figref{fig:PH5:IRAC} for the resulting IRAC fluxes and \figref{fig:PH5:PUI} for the resulting PUI fluxes. 
There are some data points with much larger flux uncertainties than the others. It has been verified for a few of them that there are slight image artefacts close to the source which have not been caught 
during first visual inspection.
They were kept in the database to avoid the introduction of subjective bias; due to their large error bars those fluxes are given duly low weight in the fitting process.
As detailed in \chaptref{chapt:SST}, the effective wavelengths are 7.872, 15.7929, and \unit{22.3272}{\micron}.

\subsection{Results}
\label{sect:PH5:results}

\begin{figure}
  \centering
  \begin{minipage}[t]{0.48\linewidth}
    \includegraphics[angle=-90, width=\linewidth]{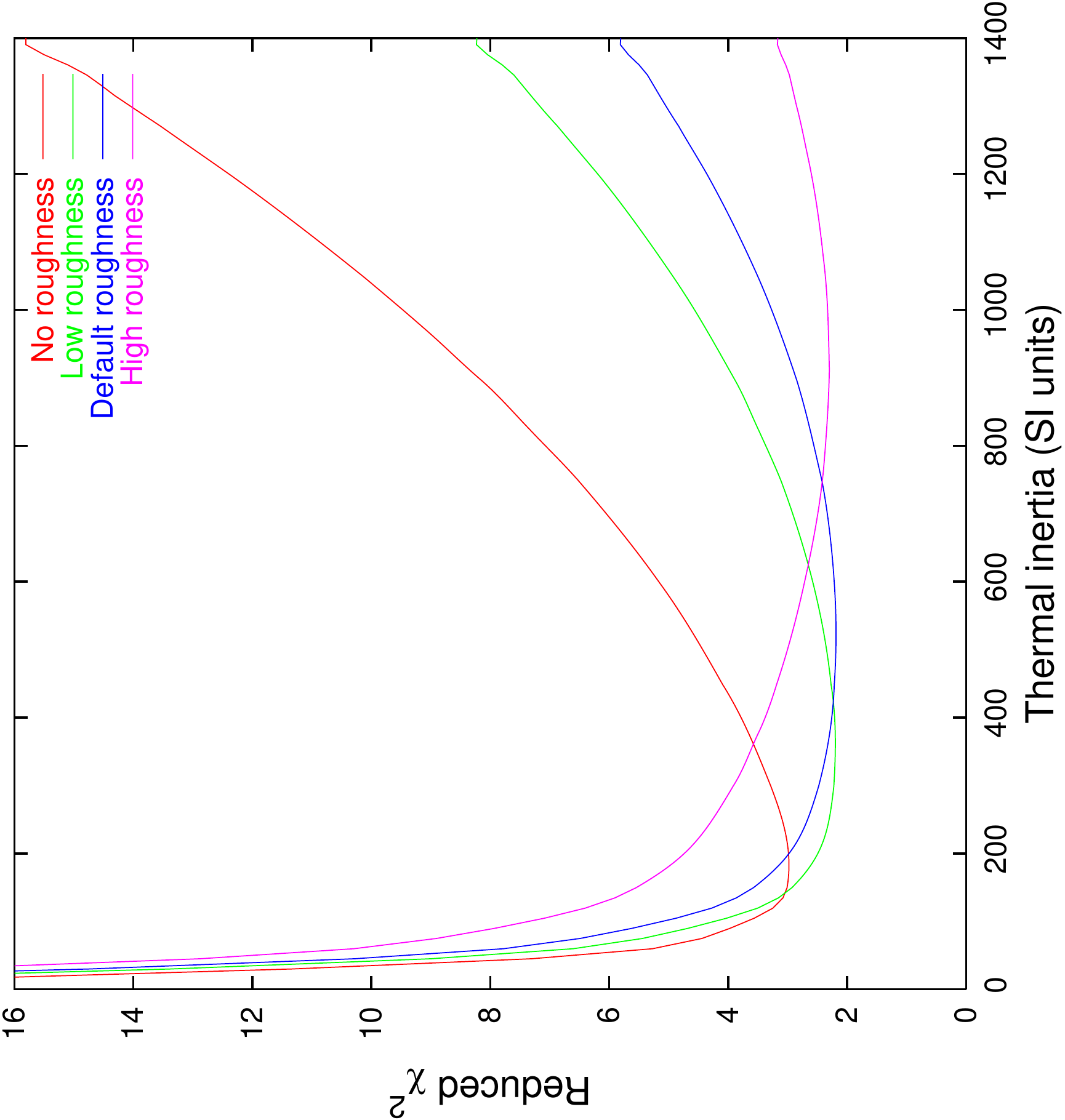}
  \end{minipage}
  \begin{minipage}[t]{0.48\linewidth}
    \includegraphics[angle=-90, width=\linewidth]{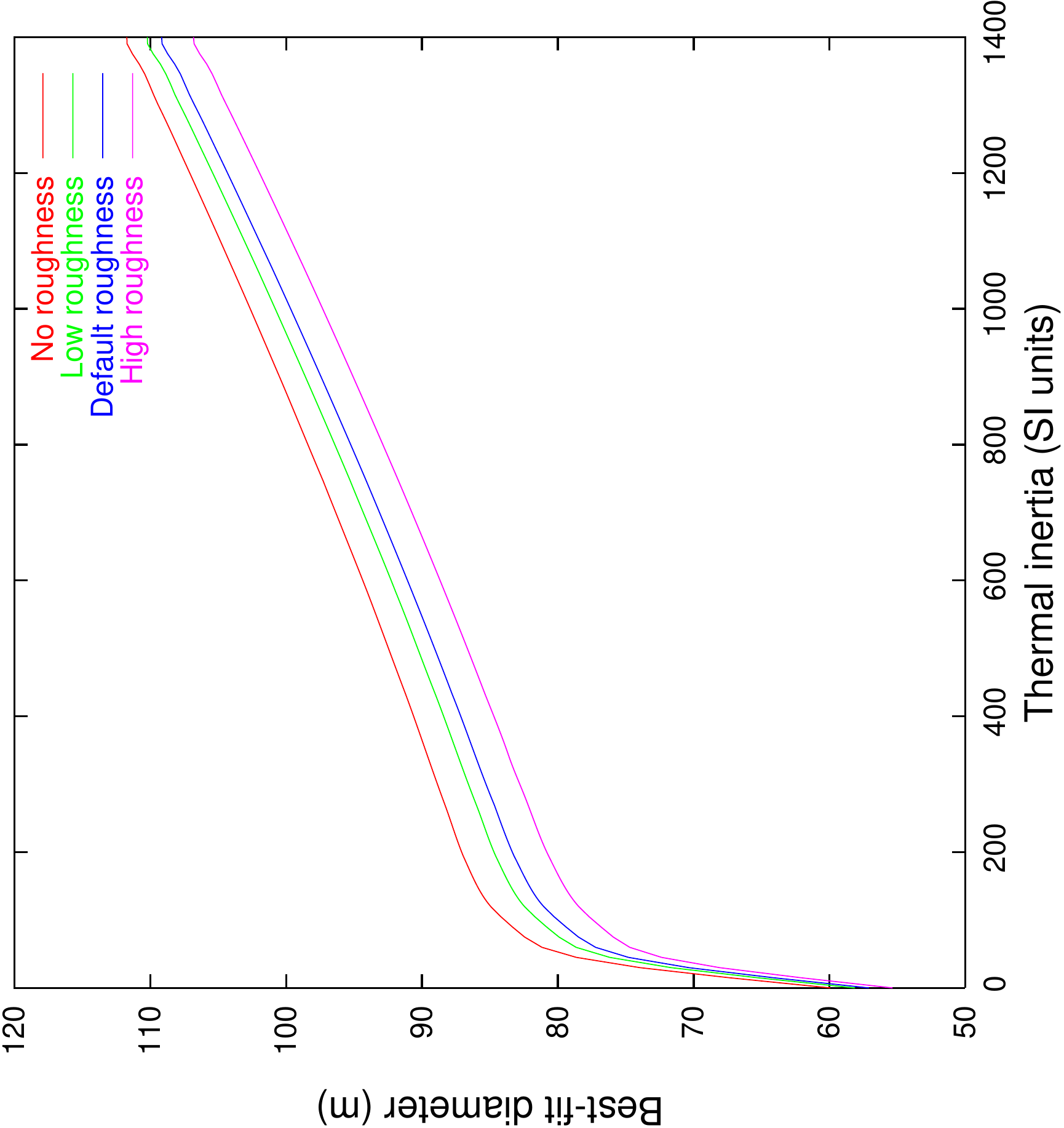}
  \end{minipage}
  \caption{TPM fits to our Spitzer data of YORP: reduced $\chi^2$ (left) and best-fit diameter (right) as a function of thermal inertia for different sets of roughness parameters.}
  \label{fig:PH5:fits}
\end{figure}

We used our  TPM 
 described in \chaptref{chapt:TPM}
together with a preliminary version of the the shape model reported by \citet{Taylor2007} for YORP
to fit the resulting flux values.
The reported value for the angular acceleration caused by the YORP effect was used in the determination of rotational phases for a given epoch, no manual adjustment of rotational phase has been made.
See \figref{fig:PH5:fits} for the results
(see \sectref{sect:Eros} for a detailed discussion of analogous plots).
As apparent in the left figure, there are different $\chi^2$ minima depending on the roughness parameters used. While zero roughness appears to be less consistent with the data than non-zero roughness, the remaining three roughness models appear to fit the data equally well.
Thermal-inertia values in the range 
200--\unit{1200}{\TIunit} are consistent with the data.
The corresponding range in best-fit diameter is
88--\unit{96}{\metre}, 
corresponding to a geometric albedo \pv\ between 0.18 and 0.22.

\begin{figure}
  \centering
\includegraphics[angle=-90,width=0.6\linewidth]{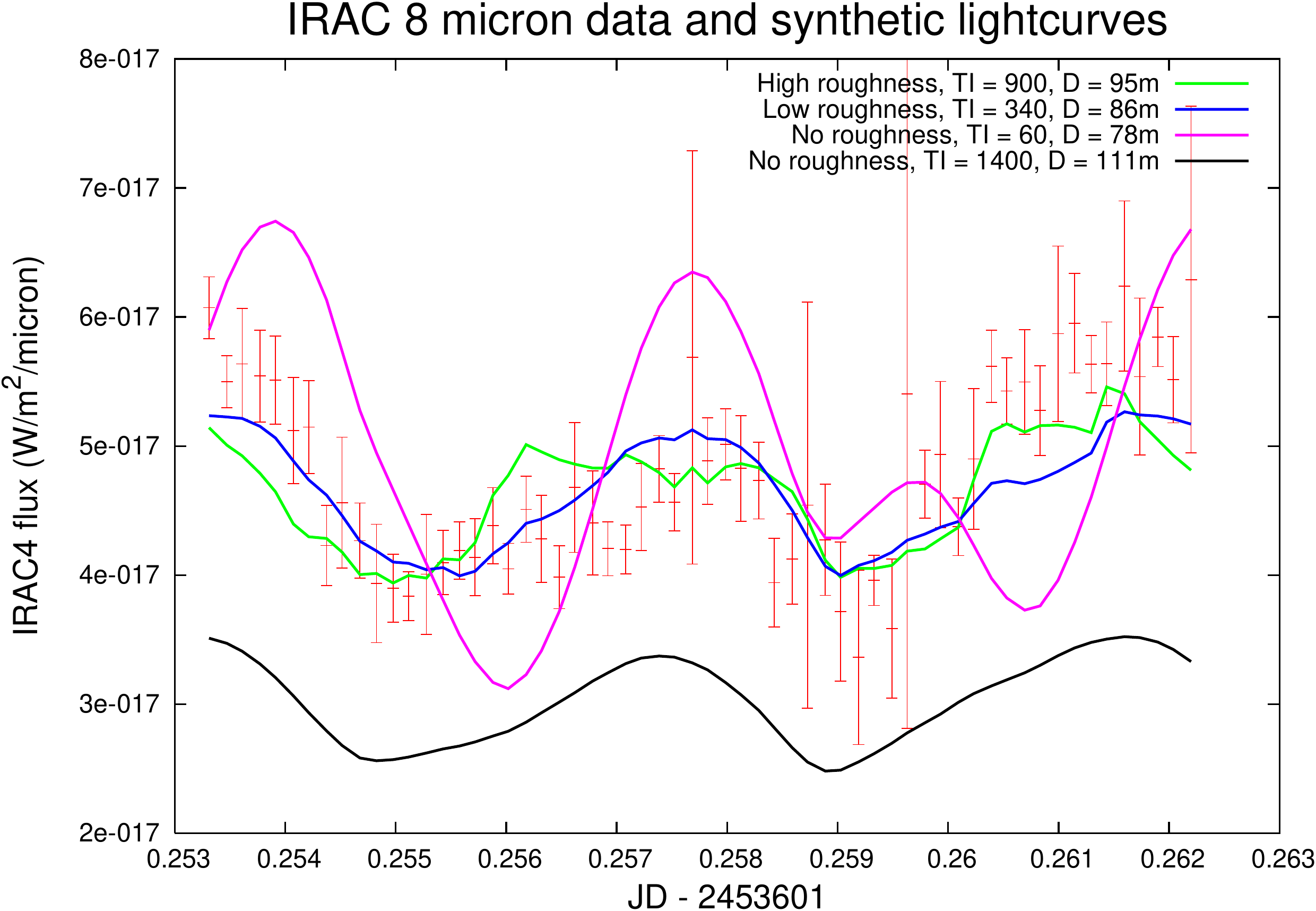}

\medskip

\includegraphics[angle=-90,width=0.6\linewidth]{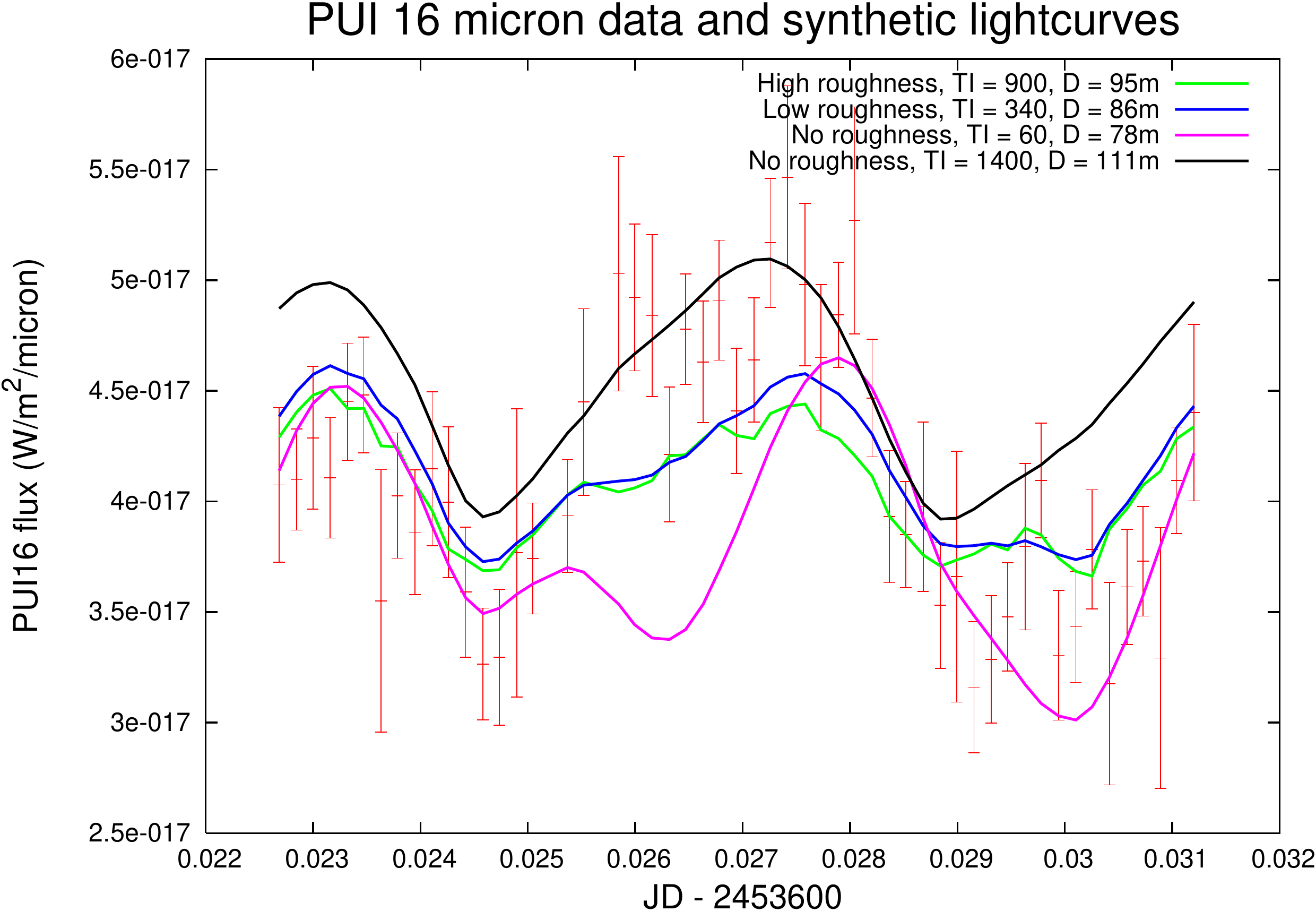}
  \caption{IRAC data as shown in \figref{fig:PH5:IRAC} and ``blue'' PUI data as shown in \figref{fig:PH5:PUI}
superimposed on four synthetic lightcurves. In both plots, the green and blue lines correspond to best-fitting parameters for different roughness models (cf.\ \figref{fig:PH5:fits}). 
The lines in magenta and black illustrate the effect of very low and very high thermal inertia: 
Low thermal inertia leads to large lightcurve amplitudes; high thermal inertia smoothes the lightcurve shape and amplitude. Furthermore, thermal inertia influences the ratio of flux levels at different wavelengths (color temperature); clearly, the black lines poorly fit the color temperature.
}
  \label{fig:PH5:lc}
\end{figure}

Synthetic model lightcurves for different parameters have been superimposed on two of our three data sets \seefig{fig:PH5:lc}. A similar plot for the third data set would look qualitatively identical.

\subsection{Discussion}
\label{sect:PH5:discussion}

\subsubsection{Fit quality}
As apparent in \figref{fig:PH5:lc}, our  TPM 
based on the preliminary shape model by \citet{Taylor2007}
provides a reasonably good fit to our Spitzer data.
It is clearly seen that very large and very low values of thermal inertia can be ruled out on the basis of the resulting lightcurve shape.

We wish to emphasize that no manual adjustment was made to the rotational phase  determined from the shape model and the Spitzer-centric observing geometry at the time of observation.
The good phase agreement
 between our Spitzer data and the synthetic lightcurves provides further evidence for the validity of the  spin up of YORP due to the YORP effect \citep[see][note 23]{Taylor2007}.

We also note that 
some lightcurve features are not adequately reproduced (e.g.\ that around a time of  0.026 in the \unit{16}{\micron} lightcurve).
This may be due to the fact that we use
the convex-shape TPM  although the shape model by \citet{Taylor2007} is not convex.
At the time of writing, the concave-shape TPM described in \chaptref{chapt:TPMconcave} is not yet sufficiently well tested and stable to be used for this purpose.
We also note that we do not yet have the final shape model by 
\citeauthor{Taylor2007}\ at our disposal. 
Our results are therefore preliminary.

\subsubsection{Diameter and albedo}
Assuming a systematic modeling uncertainty in diameter of \unit{10}{\%} \seesect{sect:NEA:D}, the range in diameter that best fits our data is $92\pm\unit{10}{\metre}$, corresponding to $\pv=0.20\pm0.04$.
The true modeling uncertainty in this case may be larger due to the concavities present in the \citeauthor{Taylor2007}\ shape model, which are currently neglected in our TPM.
At the phase angle of our observations (\unit{59.3}{\degree}), shadowing effects would reduce the amount of observable flux relative to our model calculations, hence we would expect our model to somewhat underestimate the diameter and, correspondingly, to   overestimate \pv.

An additional source of uncertainty is the flux calibration of the PUI data sets which account for two thirds of our database; the new calibration pipeline recently issued by the Spitzer Science Center is reported to cause calibration changes of up to \unit{15}{\%}.
If our diameter result were entirely based on PUI fluxes, the corresponding change in diameter would be 
up to \unit{7.5}{\%}.

Taken together, our preliminary diameter result appears not to be inconsistent with the radar-derived diameter of $114\pm\unit{12}{\metre}$ by \citet{Taylor2007}, where we assume a fractional diameter uncertainty of \unit{10}{\%} as is usually reported for radar-derived diameters
(no diameter uncertainty is reported  by \citeauthor{Taylor2007}).

From spectroscopic observations, \citet{Gietzen2007} conclude that YORP belongs to either of the taxonomic classes S or V. 
S-type classification would indicate an albedo of $\pv\sim0.2$, a much larger value would be expected for V types.
An S-type classification would be in excellent agreement with our albedo result and would not be inconsistent with that implied by the radar diameter ($\pv=0.13\pm0.03$).
A V-type classification would appear to be inconsistent with both diameter determinations.

\subsubsection{Thermal inertia}
Unfortunately, the thermal inertia of YORP is not well constrained by our preliminary data analysis, 
although a low thermal inertia comparable to the lunar value is clearly inconsistent with the data \seefig{fig:PH5:fits}.
Interestingly, 
also large thermal-inertia values close to that expected for bare rock (\unit{$\sim2500$}{\TIunit}) appear to be inconsistent, 
with \unit{1200}{\TIunit} being a tentative upper limit.
We wish to emphasize that these are only preliminary results and that a final analysis may well result in a higher thermal inertia.


\subsection{Summary}
\label{sect:PH5:summary}

Using the Spitzer Space Telescope, we have obtained thermal lightcurves of the ultra-fast rotating NEA (54509) YORP at wavelengths of 8, 16, and \unit{22}{\micron}.
A preliminary data analysis demonstrates the potential of our TPM to constrain the thermal properties of this  $D\sim\unit{100}{\metre}$ object.
As expected, a very low thermal inertia indicative of a thick dusty regolith is clearly excluded by our data, while we cannot  confirm our expectation, namely a large thermal inertia indicative of bare rock.
Rather, an intermediate range between some 200 and \unit{1200}{\TIunit} appears to best fit the data. See \sectref{sect:NEA:barerock} for a discussion.
We caution that our result is preliminary due to uncertainties in the flux calibration. 


%% file: WT24.tex
The potentially hazardous asteroid (33342) 1998~WT24 approached the Earth within \unit{0.0125}{\AU} on 16  December 2001 and was the target of a number of optical, infrared, and radar observing campaigns. 
Interest in 1998~WT24 stems from its having an orbit with an unusually low perihelion distance, which causes it to cross the orbits of the Earth, Venus, and Mercury, and its possibly being a member of the E spectral class, which is rare amongst NEAs. We present 
a TPM analysis of an extensive database resulting from 
thermal-infrared observations of 1998~WT24 obtained in December 2001 with the 3-m NASA Infrared Telescope Facility (IRTF) on Mauna Kea, \Hawaii\ and the ESO 3.6-m telescope in Chile at large solar phase angles.

We have obtained best-fit values of \unit{$0.35 \pm 0.04$}{\km} for the effective diameter, $0.56 \pm 0.2$ for the geometric albedo, \pv, and 100--\unit{300}{\TIunit} for the thermal inertia. Our results are in good agreement with   independent analyses of the same data set. Our values for the diameter and albedo are consistent with results derived from radar and polarimetric observations. The albedo is one of the highest values obtained for any asteroid and, since no other taxonomic type is associated with albedos above 0.5, supports the suggested rare E-type classification for 1998~WT24. The thermal inertia is an order of magnitude higher than values derived for large main-belt asteroids but consistent with the relatively high values found for other near-Earth asteroids.

\subsection{Introduction}
\label{sect:WT24:intro}

(33342) 1998 WT24 was discovered on 1998 November 25 by the LINEAR \citep{Stokes2000} search program. Observations of physical characteristics published to date suggest that 1998 WT24 is a rare E-type Aten NEA with a high polarimetric albedo of around 0.43 \citep{Lazzarin2004b,Kiselev2002}.
The E class is the taxonomic class which is associated with the highest albedo values, typically in the range $\pv=$ 0.3--0.6 (see also \sectref{sect:intro:mineralogy}).
Examples of main-belt E-types are (44) Nysa \citep[IRAS $\pv = 0.55$,][]{SIMPS} and (64) Angelina \citep[$\pv = 0.40$,][]{Tedesco2002b}.
Only four other E-type NEAs are known so far:
(3103) Eger \citep{Clark2004a,Gaffey1992},
(10302) 1989~ML \seesect{sect:ML},
and (4660) Nereus and (5751) Zao \citep{Delbo2003}.

For 1998~WT24, \citet{Krugly2002} report a rather short rotation period of \unit{3.697}{\hour} and a lightcurve amplitude measured at solar phase angles between \unit{50}{\degree} and \unit{60}{\degree} of \unit{0.26}{\text{mag}}. The absolute magnitude at maximum brightness derived by \citet{Kiselev2002} is \unit{$18.69 \pm 0.08$}{\text{mag}}, or \unit{$18.39 \pm 0.08$}{\text{mag}} if a brightness opposition effect is assumed similar to that observed for the E-type asteroids (44) Nysa and (64) Angelina. Given the reported \unit{0.53}{\text{mag}} lightcurve amplitude at the time of the \citeauthor{Kiselev2002}\ photometric observations (2001 December 2.9), the \citeauthor{Kiselev2002}\ result of $H_\text{max} = \unit{$18.39 \pm 0.08$}{\text{mag}}$ is compatible with $H$~(lightcurve mean)~=\unit{$18.54 \pm 0.1$}{\text{mag}} derived independently by \citet{Delbo2004} using the method of \citet{HG} on observations made at ESO on 2001 December 2 and 4, taking $G = 0.4$. Taking $H = 18.5 \pm 0.3$, the diameter of (33342) 1998~WT24 inferred from an albedo of $\pv = 0.43$ is \unit{$D = 0.40 \pm 0.06$}{\km}. It should be noted, however, that the derivation of albedos from polarimetric observations is based on a method that depends on empirically derived relations between the polarization parameters and the solar phase angle and albedo and that different calibrations of this method have been published, most recently by 
\citet{Cellino1999}. \citeauthor{Kiselev2002}\
 used the relations of
\citet{LupishkoMohamed1996}. If
we take the \citeauthor{Kiselev2002}\ value of $h_v = \unit{0.039}{\%\ \text{deg}^{-1}}$ for the slope of the linear part of the polarization-phase curve and apply the appropriate relation of \citet{Cellino1999}, we find a higher albedo of $\pv = 0.62$. Assuming $H = 18.5\pm0.3$, the implied diameter is reduced to $D = \unit{$0.34\pm0.05$}{\km}$.
Goldstone radar images of 1998~WT24 made available on the
web by S.\ Ostro and colleagues (\url{http://neo.jpl.nasa.gov/images/1998wt24.html}) suggest a slightly elongated body with a possible concavity. However, no formal report of the Goldstone observations is available at the time of writing. \citet{Zaitsev2002} made radar observations of 1998~WT24 on 2001 December 16 and 17 and derived lower limits for the maximum pole-on breadth of the asteroid of $D_\text{max} = 0.42$ and \unit{0.40}{\km}, respectively, from observations of the Doppler-broadened echo bandwidth, $B$, on the two dates. $B$ is proportional to $D_\text{max}\times\sin\psi$, where $\psi$ is the angle between the spin vector and the radar line-of-sight. Since the pole orientation of 1998~WT24 is unknown, only a lower limit for the maximum pole-on dimension can be determined from the radar data.

No pole solution for 1998~WT24 has been published to date. However, clues concerning the pole direction can be derived by comparing diameter results from different sources.
In particular, a comparison of the available polarimetric and radar results suggests that the sub-Earth latitude at the time of the \citeauthor{Zaitsev2002}\ radar observations was small (or, equivalently, that the angle $\psi$ defined above was close to \unit{90}{\degree}).
Based on a simple-model analysis of the thermal-infrared 
data discussed herein, \citet{WT24} suggest that  the subsolar latitude was  close to zero during the time of their observations.
Taken together, these two constraints lead to a crude estimate on the spin axis position with ecliptic latitude $\beta$ and longitude $\lambda$ close to 
$\beta=\unit{52}{\degree}$, $\lambda=\unit{175}{\degree}$ or $\beta=\unit{-52}{\degree}$, $\lambda=\unit{355}{\degree}$, depending on whether the rotation is prograde or retrograde, respectively, where ``the quoted solutions may be in error by several tens of degrees.'' \citep{WT24}

Thermal-infrared flux measurements were obtained at wavelengths in the range 7--\unit{21}{\micron}.
The observing nights were partially compromised by poor weather and instrument problems but the availability of independent data sets from four nights in total with a broad range of observing geometries \seetable{table:WT24:geometry} has nevertheless enabled substantial, self-consistent results to be derived.

\begin{table}
\caption[1998 WT24: Observing geometry.]{1998 WT24: observing geometry. Note: The sense of the solar phase angle changed on December 15. For example, if rotation is retrograde, the cooler morning side of the asteroid was observed prior to December 15 and the warmer afternoon side after December 15.}
  \label{table:WT24:geometry}
  \centering
\footnotesize{
  \begin{tabular}{lllll}
\toprule
  &  2001 Dec.\ 4 & 2001 Dec.\ 18& 2001 Dec.\ 19 &2001 Dec.\ 21 \\
\midrule
Telescope &ESO& IRTF& IRTF& IRTF\\
Heliocentric distance, $r$ (\AU) & 1.0148& 0.9901& 0.9874& 0.9817\\
Geocentric
distance, $\Delta$ (\AU) & 0.0621& 0.0162& 0.0198& 0.0284\\
Solar phase angle,
$\alpha$ (\degree)& $-60.4$ &67.5& 79.3& 93.4\\
\bottomrule
  \end{tabular}
} 
\end{table}

\subsection{Data}
\label{sect:WT24:data}

(33342) 1998 WT24 was observed with the ESO 3.6-m telescope and the IRTF in December 2001.
I have not contributed to obtaining these data, the flux values are quoted here for completeness; see \citet{WT24} for details on the observations and the data reduction. 

\begin{table}
  \caption{1998 WT24: measured fluxes, quoted after \citet{WT24}.}
  \label{table:WT24:fluxes}
  \centering
\footnotesize{
  \begin{tabular}{llllll}
\toprule
Date (UT) & Time & Julian date & Wavelength & Flux & Error \\
          &(UT) & (days-2,452,240)& (\micron)& (\milli\Jy) & (\milli\Jansky)  \\
\midrule
2001-12-04 & 09:10 & 7.8819 & 8.73 & 144 & 31 \\
& 08:53 & 7.8701 & 10.38 & 239 & 15 \\
& 08:50 & 7.8681 & 11.66 & 262 & 37 \\
2001-12-18 & 05:32 & 21.7306 & 7.91 & 2859 & 246 \\
& 05:33 & 21.7312 & 7.91 3& 109 & 265 \\
& 05:57 & 21.7479 & 10.27 & 5017 & 189 \\
& 05:58 & 21.7486 & 10.27 & 4226 & 161 \\
& 05:30 & 21.7292 & 11.70 & 4887 & 83 \\
& 05:38 & 21.7347 & 11.70 & 4592 & 121 \\
& 05:39 & 21.7354 & 11.70 & 5177 & 92 \\
& 05:55 & 21.7465 & 11.70 & 4880 & 91 \\
& 05:34 & 21.7319 & 17.93 & 5849 & 1219 \\
& 05:45 & 21.7396 & 17.93 & 7005 & 887 \\
2001-12-19 & 05:48 & 22.7417 & 7.91 & 1758 & 62 \\
& 06:17 & 22.7618 & 7.91 &  1416 & 54 \\
& 06:49 & 22.7840 & 7.91 &  1444 & 66 \\
& 05:05 & 22.7118 & 11.70 & 3180 & 68 \\
& 05:51 & 22.7437 & 11.70 & 3220 & 48 \\
& 06:20 & 22.7639 & 11.70 & 3294 & 51 \\
& 06:52 & 22.7861 & 11.70 & 2658 & 63 \\
& 05:12 & 22.7167 & 17.93 & 3847 & 217 \\
& 05:57 & 22.7479 & 17.93 & 3402 & 230 \\
& 06:25 & 22.7674 & 17.93 & 3079 & 293 \\
& 06:59 & 22.7910 & 17.93 & 3805 & 382 \\
& 05:23 & 22.7243 & 20.81 & 4959 & 736 \\
2001-12-21 & 04:53 & 24.7035 & 7.91 & 489 & 50 \\
& 05:48 & 24.7417 & 7.91 & 546 & 57 \\
& 05:01 & 24.7090 & 10.27 & 915 & 31 \\
& 05:55 & 24.7465 & 10.27 & 1058 & 45 \\
& 05:07 & 24.7132 & 11.70 & 1146 & 33 \\
& 06:01 & 24.7507 & 11.70 & 1384 & 49 \\
& 05:20 & 24.7222 & 17.93 & 1915 & 103 \\
& 06:14 & 24.7597 & 17.93 & 2119 & 202 \\
\bottomrule    
  \end{tabular}
} 
\end{table}


Thermal fluxes are
listed in \tablerefpage{table:WT24:fluxes}.
The quoted uncertainties in the flux measurements refer to the formal statistical uncertainties in the synthetic aperture procedure, only. The data quality  is variable due to fluctuating atmospheric and instrumental circumstances.
In particular, the Dec.\ 18 observations were affected by partially non-photometric conditions and problems with the filter wheel. The scatter of multiple measurements made in the same filter 
\citep[see, e.g.,][Fig.\ 1]{WT24}
reveals the presence of non-statistical variability in the data, which is probably due mainly to rotation of the asteroid and variable atmospheric conditions. However, our model fitting routines effectively take account of the scatter in multiple measurements, in addition to the statistical uncertainties, and weight the data accordingly, which mitigates against serious errors due to non-statistical variability.

\subsection{Results}
\label{sect:WT24:results}

We have applied our  TPM described in \chaptref{chapt:TPM} to the thermal-infrared data of 1998~WT24 given in \tableref{table:WT24:fluxes}.
Since no shape model has yet been published for 1998~WT24, we first tried modeling the asteroid as a biaxial ellipsoid spinning about one of the two shorter axes.
The ellipsoid was modeled as a mesh of 6144 triangular facets. Several types of ellipsoid were tried, all of which were consistent with published optical lightcurves, but the results turned out to be largely independent of the shape of the ellipsoid. In fact, it was found that no ellipsoid fitted the thermal data significantly better than a sphere. Therefore, we pursued our analysis assuming a spherical shape.

For both possible spin directions \seesect{sect:WT24:intro} surface roughness, thermal inertia, and size were adjusted until the best agreement was obtained with the full set of observational results listed in \tableref{table:WT24:fluxes}, i.e.\ $\chi^2$ was minimized. It was found that the fit significantly improved when the data from December 18 were excluded, which is consistent with these data being of inferior quality, as discussed in \citet{WT24}. They are disregarded in the following.

\begin{figure}
  \centering
\includegraphics[angle=-90,width=0.6\linewidth]{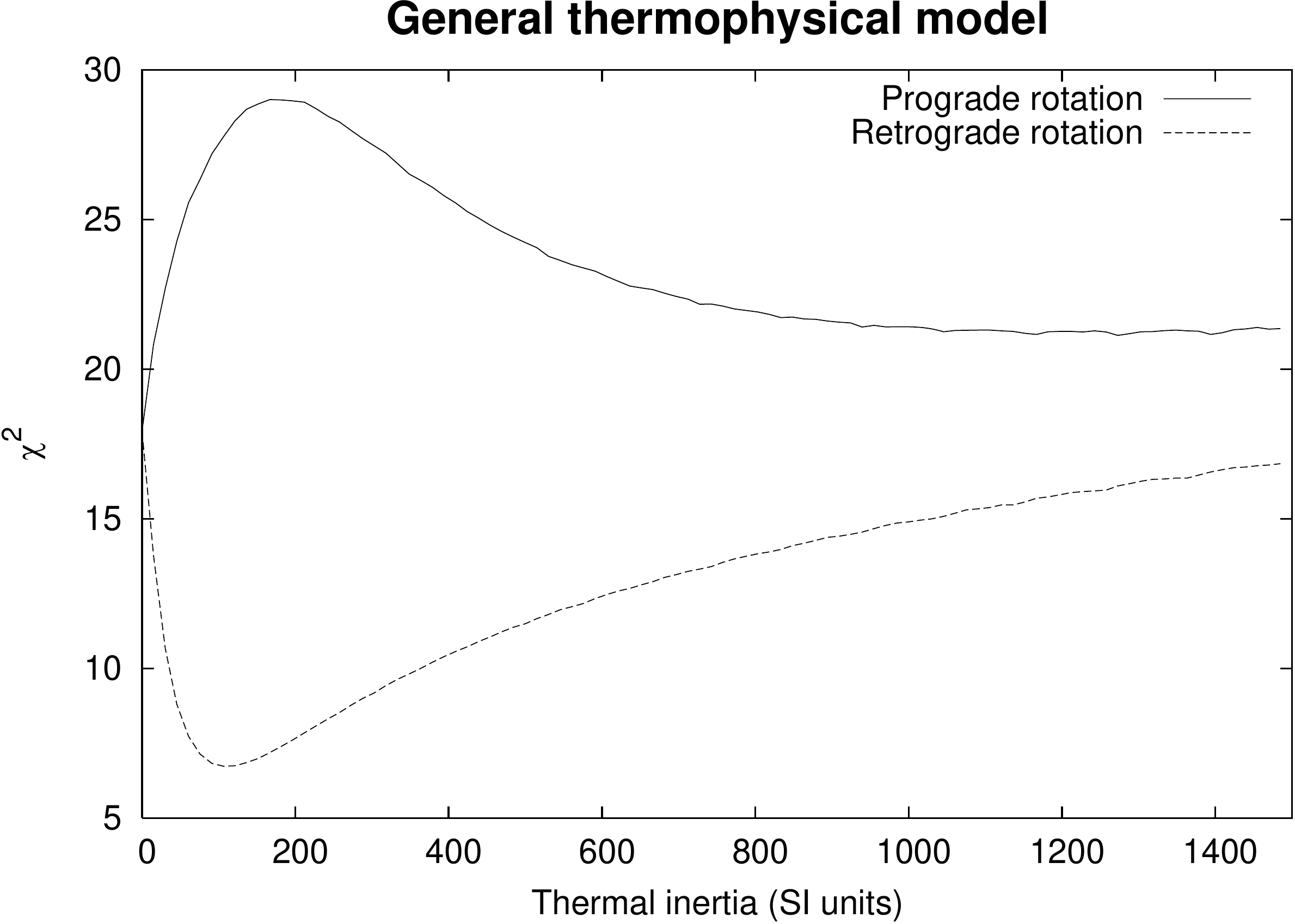}
  \caption[1998 WT24: Reduced $\chi^2$ as a function of thermal inertia for the two possible spin axis orientations.]{1998 WT24: Goodness of fit (reduced $\chi^2$) as a function of thermal inertia for the two possible spin axis orientations and a smooth surface. Adding surface roughness does not alter the clear preference for retrograde rotation. Note the convergence of the two curves as thermal inertia approaches zero,
where the temperature distribution is no longer sensitive to the spin state.}
  \label{fig:WT24:retro}
\end{figure}

\begin{figure}
  \centering
\includegraphics[angle=-90, width=0.6\linewidth]{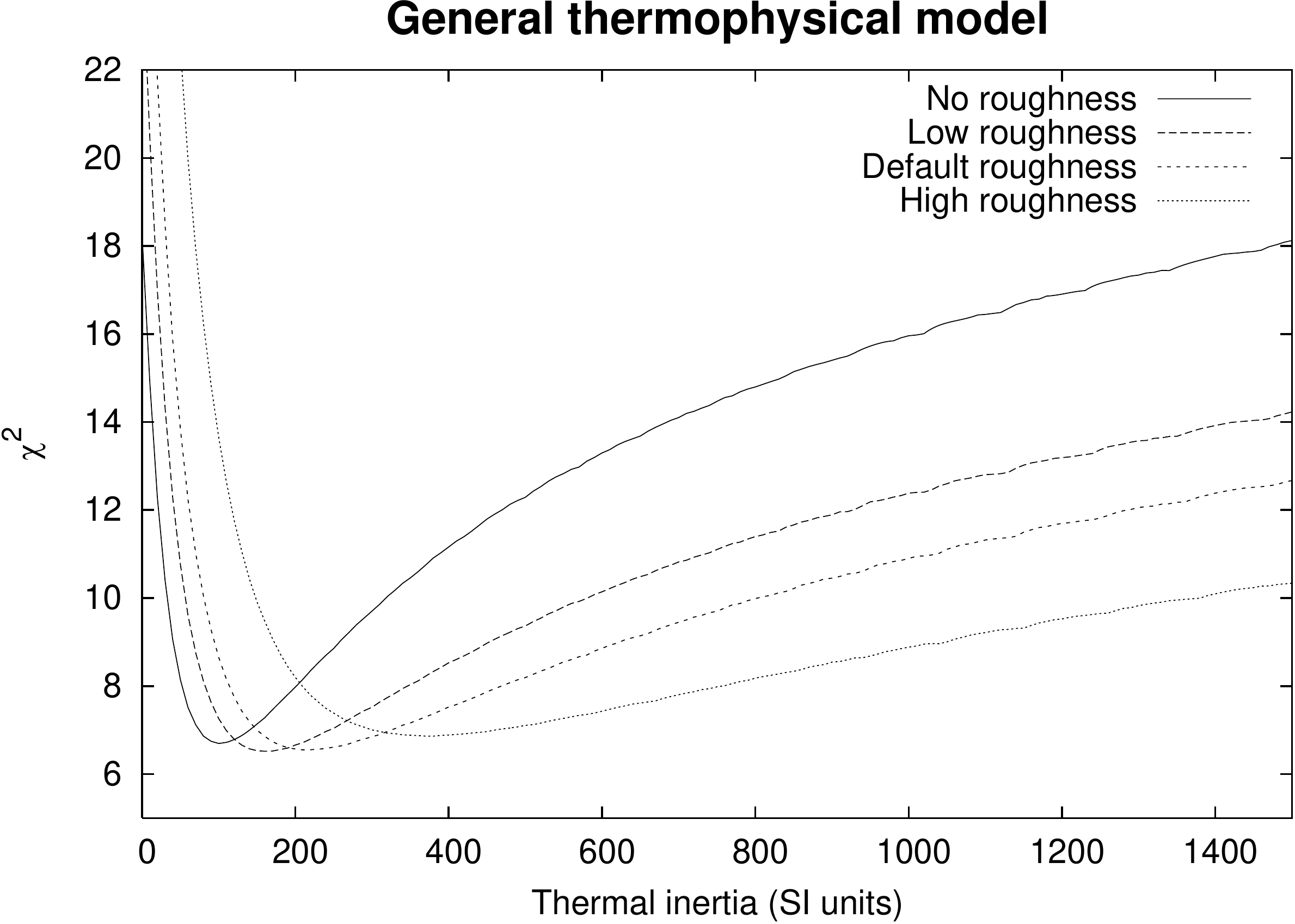}
  \caption[1998 WT24: Goodness of fit (reduced $\chi^2$) as a function of thermal inertia for retrograde rotation and different degrees of surface roughness.]{1998 WT24: Goodness of fit (reduced $\chi^2$) as a function of thermal inertia for retrograde rotation and different degrees of surface roughness.}
  \label{fig:WT24:chi2}
\end{figure}

\begin{figure}
  \centering
\includegraphics[angle=-90,width=0.6\linewidth]{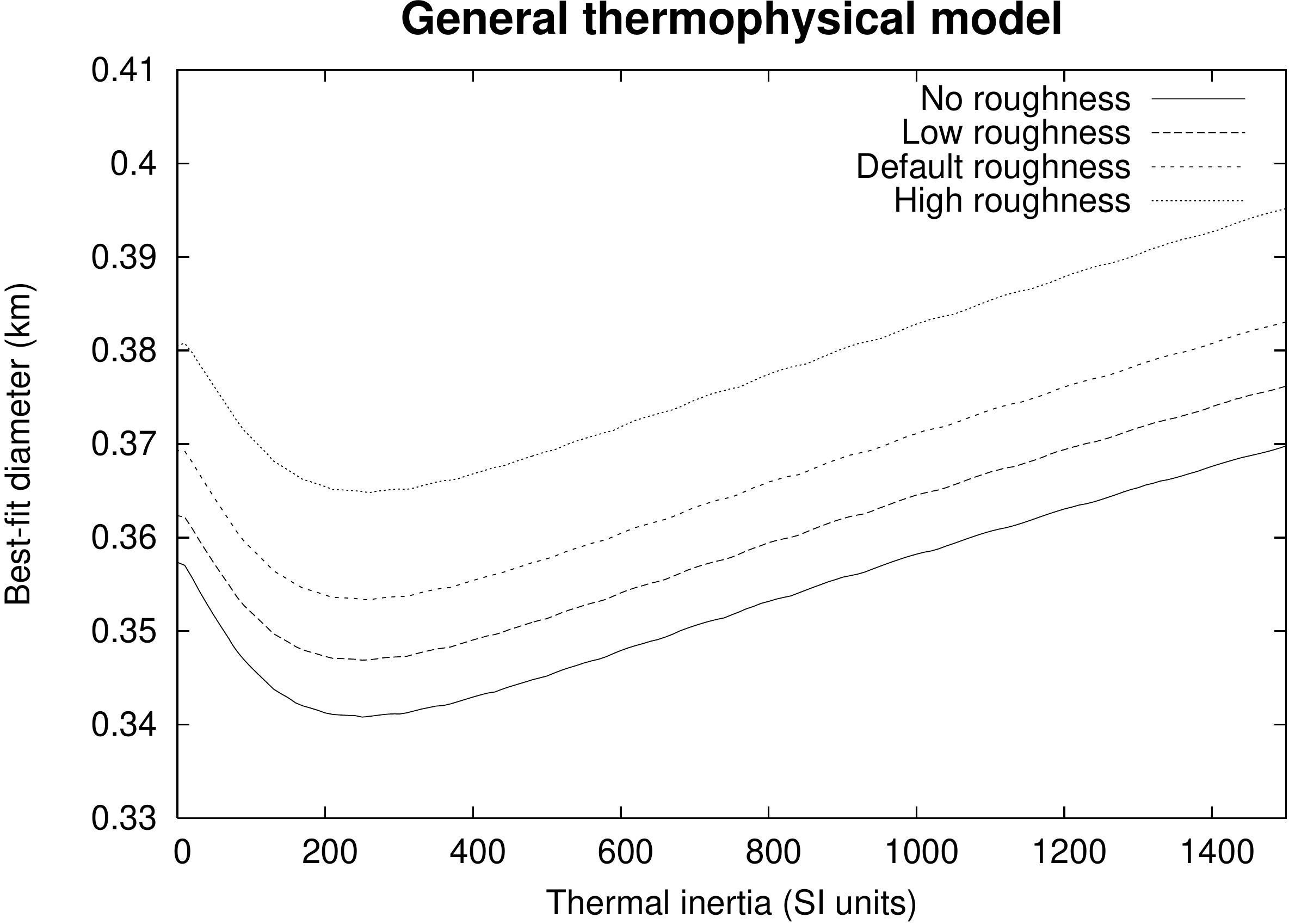}
  \caption{1998 WT24: Dependence of best-fit diameter on thermal inertia for different degrees of surface roughness.}
  \label{fig:WT24:D}
\end{figure}

We found the retrograde pole solution  to fit the data significantly better than the prograde solution \seefigpage{fig:WT24:retro}.
For the retrograde pole solution, and different degrees of surface roughness (see \sectref{sect:TPM:fitting} for a discussion of the roughness parameters used), the best-fit thermal inertia is about \unit{200}{\TIunit}, with an uncertainty of some \unit{100}{\TIunit} \seefig{fig:WT24:chi2}.
Zero roughness and the highest degree of roughness (saturated crater coverage) give slightly worse fits to the data. In the case of zero roughness we note that our TPM gives very similar results to those of the independently developed smooth-sphere model presented by \citet{WT24}. 
The corresponding best-fit values of diameter and albedo are \unit{$0.35\pm0.02$}{\km} (where the uncertainty solely reflects the flux scatter) and $\pv = 0.56\pm0.2$ (assuming $H = 18.5\pm0.3$). 
\Figref{fig:WT24:D}
illustrates the dependence of diameter on roughness and thermal inertia.

\begin{figure}
  \centering
\includegraphics[angle=-90, width=0.6\linewidth]{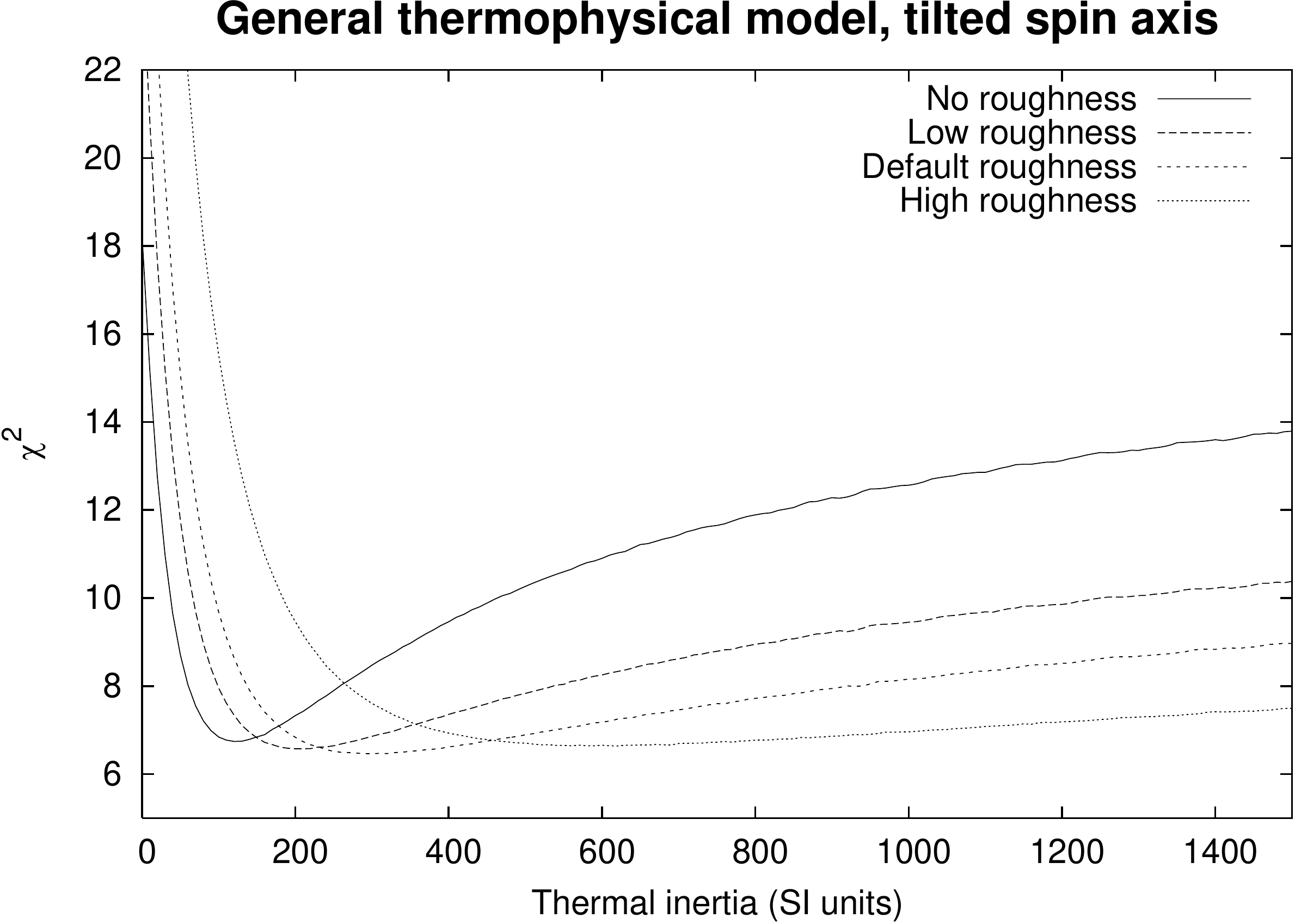}
  \caption{Like \figref{fig:WT24:chi2} but with a spin axis tilted  towards the Sun by \unit{30}{\degree} relative to the nominal retrograde solution (for the ephemeris of the radar observations).}
  \label{fig:WT24:chi2tilted}
\end{figure}

\begin{figure}
  \centering
\includegraphics[angle=-90,width=0.6\linewidth]{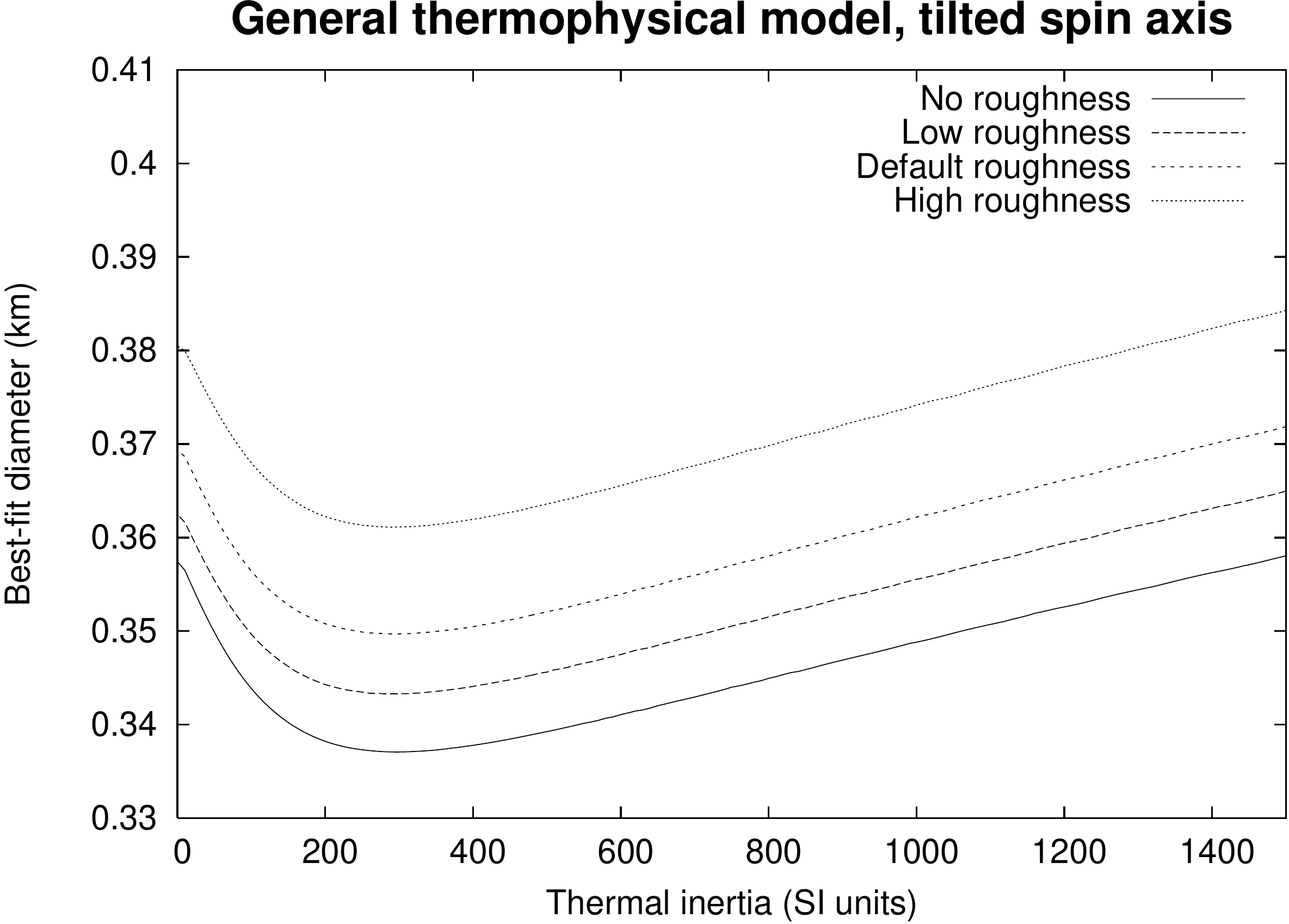}
  \caption{Like \figref{fig:WT24:D} but with a tilted spin axis as in \figref{fig:WT24:chi2tilted}.}
  \label{fig:WT24:Dtilted}
\end{figure}

We note that 
a crucial TPM input parameter, namely
 the spin axis determined by \citet{WT24},  is uncertain to within a few tens of degrees.
To study 
the maximum possible effect thereof on our results, 
 we have repeated our TPM analysis assuming a hypothetical spin axis position which is tilted 
towards the Sun \citep[for the ephemeris of the radar observations by][]{Zaitsev2002} by \unit{30}{\degree} 
relative to the nominal retrograde spin solution.
We estimate that a tilt angle largely exceeding \unit{30}{\degree} would be inconsistent with the arguments presented by \citet{WT24}.
Tilting the axis towards the Sun rather than to the observer or into some random direction
maximizes the effect on the model temperature distribution and hence on the resulting thermal-inertia estimate. In particular, the effect of thermal inertia on diurnal temperature curves is maximized for a subsolar latitude of zero and vanishes for a subsolar latitude of \unit{$\pm90$}{\degree}.
At an increased subsolar latitude, therefore, the model would be expected to require a larger thermal-inertia value to match the observed data.
Indeed, 
\figref{fig:WT24:chi2tilted} displays $\chi^2$-minima which are, compared to \figref{fig:WT24:chi2}, slightly shifted towards larger thermal-inertia values for all choices of roughness parameters.
Keeping in mind that this simulation is  close to a worst-case scenario, we feel that in this case we can neglect the spin-axis-induced uncertainty in thermal inertia.
The same applies to the resulting best-fit diameter (\figref{fig:WT24:Dtilted}).

\subsection{Discussion}
\label{sect:WT24:discussion}

\begin{table}
  \centering
\caption{Summary of diameter, albedo, and thermal inertia determinations for (33342) 1998~WT24. Values in brackets are calculated by us assuming $H=18.5\pm0.3$.}
  \label{table:WT24:D}
\scriptsize{
  \begin{tabular}{llll}
\toprule
$D_\text{eff}$ (\km) & \pv & Thermal inertia & Source \\
 & & (\TIunit)&\\
\midrule
($0.40\pm0.13$) & $0.43\pm0.15$& --- & \citet{Kiselev2002}, \\
& & & \citep[][calibration]{LupishkoMohamed1996} \\
($0.34\pm0.10$) & $0.62\pm0.17$ & --- & \citet{Kiselev2002},\\
 & & & \citep[][calibration]{Cellino1999} \\
$D_\text{max}>0.40$ & --- & --- & \citet{Zaitsev2002} \\
$0.34\pm0.02$ & $0.60\pm0.2$ & $\sim100$ & \citet{WT24}, ``smooth-sphere TPM'' \\
$\sim0.36$ & $\sim0.56$ & --- & \citet{WT24}, data from Dec.\ 21, FRM\\
$<0.41$ & $>0.42$ & --- & \citet{WT24}, data from Dec.\ 4, 19, NEATM\\
$0.35\pm0.02$ & $0.56\pm0.2$ & 100--300 & This work \\
\bottomrule
  \end{tabular}
}
\end{table}

The thermal-infrared data of 1998~WT24 considered here have been obtained at large phase angles exceeding \unit{60}{\degree}.
At such large phase angles, unresolved topographic structure, such as large concavities or boulders, could significantly influence the temperature distribution on the surface, so we caution that the modeling uncertainties in our results are considerable, in particular since we do not have a physical shape model of 1998~WT24 at our disposal.
Nevertheless, despite the fact that our TPM and all the thermal models used in analyses of the same data set \citep{WT24} have their shortcomings and have not been thoroughly tested at the high phase angles of our 1998~WT24 observations,
the overall agreement of the results from the various models is good \seetable{table:WT24:D}.
We assume that the uncertainty in our diameter estimate is dominated by systematic modeling uncertainties, which we estimate to be \unit{10}{\%} at most \seesect{sect:NEA:D}, so our conservative estimate of the diameter of 1998~WT24 is $0.35\pm\unit{0.04}{\km}$.

Furthermore, our results agree with those by \citet{Kiselev2002}, which are based on polarimetric observations.
\citet{Zaitsev2002} derived lower limits for the maximum
pole-on breadth of 0.42 and \unit{0.40}{\km}, respectively, from observations of the Doppler-broadened radar echo bandwidth on two consecutive dates. \citeauthor{Zaitsev2002}\ interpret the constancy of their radar echo bandwidths as indicating a roughly spherical shape.
The lightcurve amplitude of \unit{0.26}{\text{mag}} peak to peak reported by \citet{Krugly2002} at phase angles between \unit{50}{\degree} and \unit{60}{\degree} suggests an axial ratio of roughly $a/b = 1.15$, after reducing the lightcurve amplitude by a factor of 1.7 to crudely correct for the phase-angle dependence \citep{Zappala1990}.
Given the effective diameter of \unit{0.35}{\km} derived in this work, the corresponding dimensions of a biaxial ellipsoid would be $a = \unit{0.38}{\km}$ and $b = \unit{0.33}{\km}$. The larger dimension is consistent with the results of \citet{Zaitsev2002}, given the uncertainties. 

We conclude that our results are consistent with published values for the diameter and albedo derived using independent techniques (\tableref{table:WT24:D}),
which increases our confidence in the thermal inertia  derived in this work.
Our thermal-inertia result is well in line with our other thermal-inertia results for NEAs.

Our albedo for 1998~WT24 of $\pv = 0.56 \pm 0.2$ is at the high end of the range associated with E types in general and means that this object has one of the highest albedos measured for any asteroid. The uncertainty in our derived albedo is, however, relatively large, due to modeling uncertainties and the uncertainty in the $H$ value. A fainter $H$ value would lead to a lower albedo but would have little influence on the derived diameter or thermal inertia.


\subsection{Summary}
\label{sect:WT24:summary}

Using our TPM to fit thermal-infrared flux measurements of the NEA 1998~WT24 results in an effective  diameter of \unit{$0.35\pm0.04$}{\km} and an albedo of $\pv=0.56\pm0.2$ (the conservative uncertainties allow for modeling errors and, in the case of \pv, the uncertainty in $H$), and indicates that the surface thermal inertia is around 100--\unit{300}{\TIunit}, or a few times the lunar value. The high albedo is consistent with the suggestion that 1998~WT24 is a member of the E spectral class.
The thermal inertia is much lower than that expected for a bare-rock surface and implies that 1998~WT24 has significant areas of thermally insulating regolith, consistent with our other thermal-inertia results for NEAs. Our results suggest that 1998~WT24 is a retrograde rotator.
It has been verified that the uncertainty in the spin-axis position reported by \citet{WT24} induces a negligible uncertainty in thermal inertia, size, and albedo, relative to other sources of uncertainty.

Given the large solar phase angles, in excess of \unit{60}{\degree}, at which the  thermal-infrared observations took place, 
our diameter and albedo results are in remarkably good agreement with results of several other analyses of the same data set and furthermore with results from polarimetric and radar observations.


%% file: Lutetia.tex
The ESA spacecraft Rosetta, which was launched in 2004 and is currently on its way to 
comet 67P/Tschurjumow-Gerasimenko,
will fly by the main-belt asteroid (MBA) (21) Lutetia in 2010,
at a planned flyby distance of about \unit{3000}{\km} and a relative velocity of \unit{15}{\km\per\second}.
Lutetia is classified as an M type, but recent spectroscopic observations indicate a primitive, carbonaceous-chondrite-like (C-type) surface composition for which a low geometric albedo would be expected; this is incompatible with the IRAS albedo of $0.221\pm 0.020$.
To assist the flyby planning, we have observed Lutetia using the IRTF.
We infer that
 Lutetia has a diameter of \unit{$98.3 \pm 5.9$}{\km} and a geometric albedo of $0.208 \pm 0.025$, in excellent agreement with the IRAS value and consistent with an M-type classification.
We can thus rule out a low albedo typical of a C-type taxonomic classification.
Furthermore, we find that Lutetia's thermal properties are well within the range expected for an asteroid of its size.

\subsection{Introduction}
During its journey to the comet 67P/Tschurjumow-Gerasimenko,
the ESA spacecraft Rosetta is scheduled to fly by two MBAs  \citep{Barucci2005}: 2867 \v{S}teins in September 2008 and 21 Lutetia in July 2010.
Due to its small size of just a few kilometers, little is known about \v{S}teins.
On the other hand, Lutetia with a diameter of about \unit{100}{\km} is rather well observed at various wavelengths. However, the emerging picture of its surface composition is ambiguous:
Based on color measurements and the IRAS albedo of $0.221 \pm 0.020$,  \citet{Tholen1989} classified Lutetia as M type, 
therefore it was generally believed to have a metallic surface composition. 
\citet{Howell1994} confirm this classification using a larger data set (they noted that Lutetia has an affinity to the C-class but ruled it out on the basis of the IRAS albedo).
Based on CCD spectroscopy, \citet{BusBinzel} classified Lutetia as $X_k$-type, which is 
compatible
with a metallic surface composition.
However, in recent work by \citet{Birlan2004, Lazzarin2004, Barucci2005,Birlan2006},
spectral features
were found which are similar to those of carbonaceous chondrites. This seems to hint at a more ``primitive'' surface composition, which is usually associated with a C-type classification and a geometric albedo  below 0.1, incompatible with the IRAS value. Interestingly, \citet{LupishkoMohamed1996} derived a geometric albedo of 0.100 from polarimetry.
Also, \citet{Magri1999} found Lutetia's radar albedo to be the lowest measured for any M-type MBA---metallic objects are expected to display a high radar albedo.
Furthermore, results of \citet{Rivkin2000} and \citet{Lazzarin2004} indicate the presence of hydrated material on Lutetia's surface.

In short, the IRAS albedo seems incompatible with recent results.
\citet{Barucci2005} thus called for a new determination of Lutetia's albedo.



We have observed Lutetia at visible and thermal-infrared wavelengths using the IRTF \seechapt{chapt:IRTF}. Data were analyzed  using the NEATM \seesect{sect:NEATM} and our TPM \seechapt{chapt:TPM}.

\subsection{Observations}

The observations were performed on June 24 2004, between roughly 14:15 and 15:00~UT, at the  IRTF on Mauna Kea  using the mid-infrared spectrometer and imager  MIRSI \citep[see also \sectref{sect:MIRSI}]{MIRSI} in imaging mode and the optical CCD Apogee (see \sectref{sect:apogee}). 
Observing conditions were good with low humidity and no discernible clouds.
See \tableref{table:21:geometry} for the observing geometry.

\begin{table}
\caption{Observing geometry and modeling assumptions \citep[$H$ and $G$ are from][]{IRAS}.
\label{table:21:geometry}}
\centering
\begin{tabular}{l l}
\toprule
Heliocentric distance $r$ & \unit{2.065}{\AU} \\
Geocentric distance $\Delta$ & \unit{2.199}{\AU} \\
Solar phase angle $\alpha$ & \unit{27.4}{\degree} \\
Absolute magnitude $H$ & 7.35 \\
Slope parameter $G$ & 0.11  \\
Thermal emissivity $\epsilon$ & 0.9
\\ \bottomrule
\end{tabular}
\end{table}

We used three of MIRSI's narrow-bandwidth filters centered at 8.7, 11.6, and \unit{18.4}{\micron} (see also \tablerefpage{table:MIRSI:filters}).
The photometry was calibrated against subsequent observations of the standard stars $\gamma$~Aql and $\beta$~Peg, for which absolutely calibrated spectra have been published by \citet{CohenVII,CohenX}.
We also obtained absolutely calibrated V-magnitudes of Lutetia 
just before and just after the MIRSI observations, 
using Apogee.
These were calibrated against the \citet{Landolt1973} standard star 93-101.

\begin{table}
\caption[IRTF observations of 21 Lutetia]{IRTF observations of 21 Lutetia. Times are those of mid  exposure on June 24 2004 and  are not light-time corrected. Filter ``V'' denotes an Apogee measurement; numbers denote the central wavelength (in \micron) of the MIRSI filter used.
 \label{table:21:1}}
\centering
\begin{tabular}{l l l l}
\toprule
UT & Filter & Flux & $\sigma_{\textrm{Flux}}$ \\
\midrule
14:17 &  V   & 11.87 mag & 0.01 mag \\
14:19 & 11.6 & \unit{12.88}{\Jy} & \unit{1.29}{\Jy} \\
14:26 & 8.7  &  \unit{5.81}{\Jy} & \unit{0.58}{\Jy} \\
14:34 & 18.4 &  \unit{17.1}{\Jy} & \unit{4.1}{\Jy} \\
14:37 &  V   & 11.81 mag & 0.01 mag \\
\bottomrule
\end{tabular}
\end{table}

We used our IRTF data reduction techniques as described in \sectref{sect:IRTF:reduction}. 
See  \tableref{table:21:1} for a list of the optical and thermal-IR fluxes. 
For MIRSI N-band data (8.7 and \unit{11.6}{\micron}), the errors  are dominated by the calibration uncertainty including possible atmospheric variability between observations; this uncertainty was estimated to be \unit{10}{\%}.
For the Q-band filter centered at \unit{18.4}{\micron}, uncertainties caused by the airmass correction and the statistical scatter resulting from the synthetic aperture procedure contributed an additional \unit{14}{\%} to the error budget.

\subsection{Data analysis}

\subsubsection{Observing geometry and rotational phase}
\label{sect:21:LC}

From the inversion of multi-epoch optical lightcurves observed from 1962 through 1998, \citet{Torppa2003} derived a physical model of Lutetia's shape and spin state. 
Two pole directions are given, one of which seems to be superior to the other one (M.\ Kaasalainen, private communication, 2005). 
The J2000 ecliptic latitude of both solutions is \unit{+3}{\degree},  the longitude is \unit{39}{\degree} for the preferred first solution and \unit{220}{\degree} for the secondary. This implies that during our observations, on 2004 June 24, the sub-Earth and subsolar latitudes were \unit{-75}{\degree} and \unit{-48}{\degree}, respectively. Using the second pole orientation results in the same sub-Earth and subsolar latitudes, but with their signs changed.

\begin{figure}[bt]
\centering
\includegraphics[angle=-90, width=0.7\textwidth]{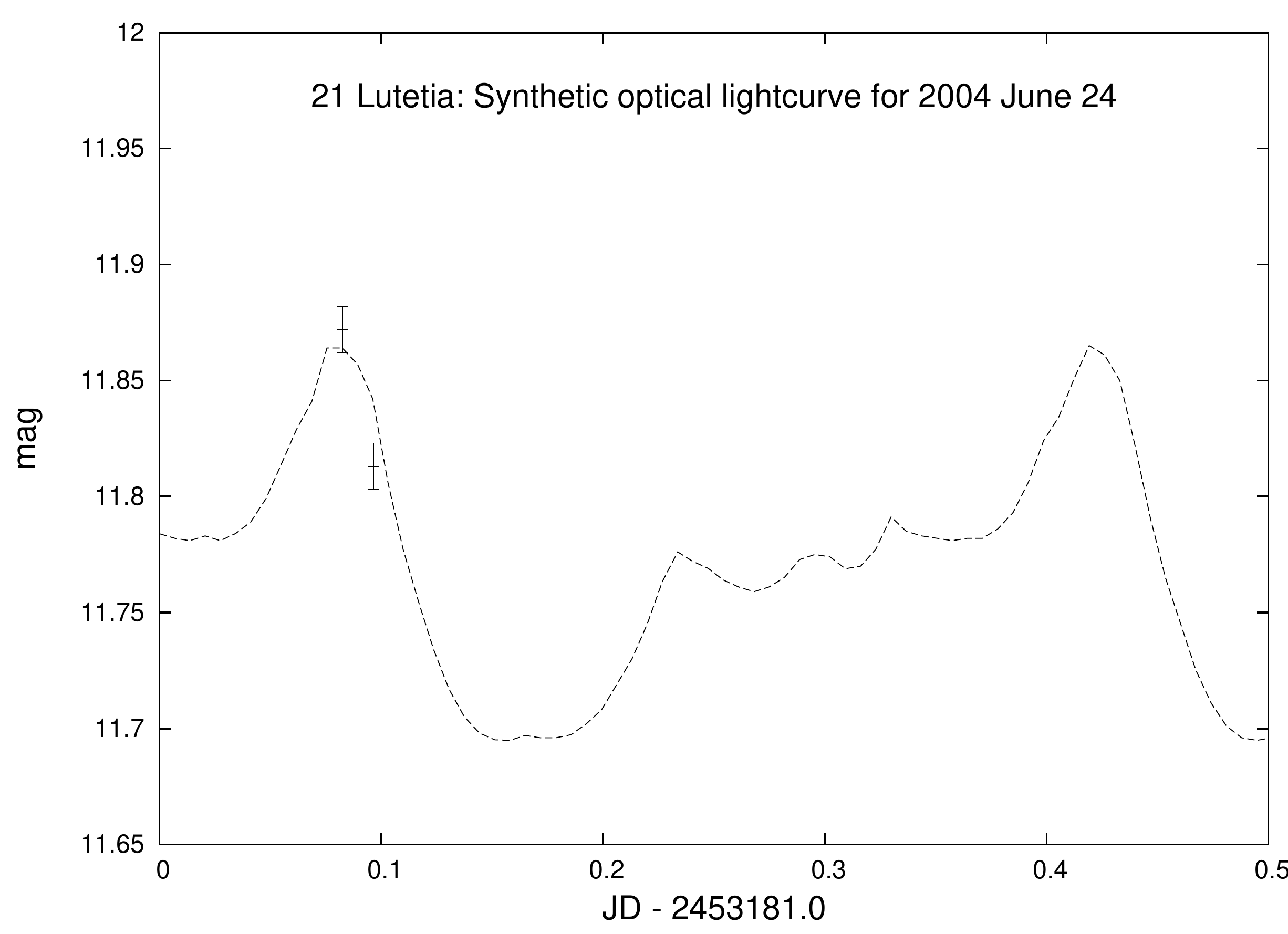}
 \caption[Measured V-band data and synthetic optical lightcurve]{
Measured V-band data (see \tableref{table:21:1}) and synthetic optical lightcurve generated using the shape model by \citet{Torppa2003}.
The rotational period is \unit{8.165455}{\hour}, so some 1.5 cycles are displayed.
}
 \label{fig:21:optical}
\end{figure}

The shape model of \citet{Torppa2003} was used to generate a synthetic optical lightcurve. In \figref{fig:21:optical} the resulting lightcurve is plotted, together with our measured V-band data.
Judging from \figref{fig:21:optical} 
our observations took
place near lightcurve minimum; 
the lightcurve amplitude is roughly 0.17 magnitudes from minimum to maximum.

Between 1983 Apr 25 and May 4, Lutetia was sighted five times by IRAS \citep{IRAS}; the observing geometry was practically constant, with a sub-Earth latitude of about \unit{18}{\degree} and a subsolar latitude of about \unit{-2}{\degree}.
Again, choosing the second pole solution changes the signs.

\subsubsection{NEATM}
\label{sect:21:NEATM}

The NEATM \seesect{sect:NEATM} was used to fit the
 MIRSI data (see  \tablerefpage{table:21:1}) and the IRAS flux values published by \citet{IRAS}.
IRAS was equipped with four broad-band filters centered at 12, 25, 60, and \unit{100}{\micron}. The filter breadth requires color corrections of the fluxes \citep{IRAS_CC};%
\footnote{ See also \url{http://irsa.ipac.caltech.edu/IRASdocs/exp.sup/ch6/tabsupC6.html}.}
we assumed a black-body temperature of \unit{230}{\kelvin}. We considered only the 12, 25, and \unit{60}{\micron} filters since there are significant uncertainties concerning both the calibration of the \unit{100}{\micron} data and the applicability of the model at these wavelengths. 
All five IRAS sightings of Lutetia were used.

Throughout our data analysis we assumed Lutetia's thermal emissivity to be 0.9, and its absolute (optical) magnitude in the HG-system \citep{HG} to be $H = 7.35$ with $G  = 0.11$ \citep{IRAS}. 

\begin{table}
\caption[NEATM fits to the available thermal-infrared data]{
NEATM fits to the available thermal-infrared data: The MIRSI data as given in \tableref{table:21:1}, and the five IRAS sightings \citep{IRAS}. No attempt at lightcurve correction was made. 
 \label{table:21:NEATM}}
\centering
\begin{tabular}{c c c r}
\toprule
 & $\eta$ & \pV & D (\km) \\
\midrule
MIRSI & 0.93
& 0.188 & 103.8 \\
IRAS 1 & 0.93
& 0.178 & 106.6 \\
IRAS 2 & 0.94
 & 0.226 & 94.7 \\
IRAS 3 & 1.06
 & 0.161 & 112.2 \\
IRAS 4 & 0.96
& 0.191 & 103.0 \\
IRAS 5 & 0.82
 & 0.231 & 93.6 \\
\bottomrule
\end{tabular}
\end{table}

\begin{table}
\caption{NEATM diameters and albedos of Lutetia for the two  data sets
\label{table:21:NEATM:final}}
\centering
\begin{tabular}{c c c}
\toprule
 & D (\km) & \pV  \\
\midrule
MIRSI & $104\pm 16$ & $0.188\pm 0.057$ \\
IRAS (mean of five)  & $102\pm 16$ & $0.197\pm 0.059$ \\
\bottomrule
\end{tabular}
\end{table}

See \tableref{table:21:NEATM} for an overview of the NEATM results. 
The mean results from the IRAS data are $\pV = 0.197\pm 0.027,$
$D = \unit{$102.0 \pm 7.1$}{\km},$
and $\eta = 0.94 \pm 0.08$
(errors are from the internal scatter only), in excellent agreement with our MIRSI results.
We estimate that
NEATM diameters and albedos are generally accurate to within \unit{15}{\%} and \unit{30}{\%}, respectively \seesect{sect:NEATM:accuracy}.
Our final results with conservative uncertainties are given in table \ref{table:21:NEATM:final}.

We did not attempt to correct our data to the flux level of the lightcurve average;
if the shape model by \citet{Torppa2003} adequately describes the epoch of our observations \seesect{sect:21:LC},
this should lead to a slight underestimation of the diameter of up to a few percent, well inside the range of uncertainty quoted.

\citet{Walker2003} calculated a mean value for $\eta$  of $1.067 \pm 0.087$ from IRAS observations of 694 asteroids. Our result for Lutetia, $\eta = 0.94 \pm 0.08$, is comparable to that of \citeauthor{Walker2003}, which indicates that Lutetia has thermal properties rather typical for a large MBA, i.e. low thermal inertia and some surface roughness.

\subsubsection{Thermophysical model (TPM)}
\label{sect:21:TPM}

We have analyzed our data using the detailed TPM \seechapt{chapt:TPM}.
We use the shape model by \citet{Torppa2003} discussed in \sectref{sect:21:LC}, which consists of a convex mesh of 2040 triangular facets.


\begin{figure}
\centering
\includegraphics[angle=-90, width=0.7\textwidth]{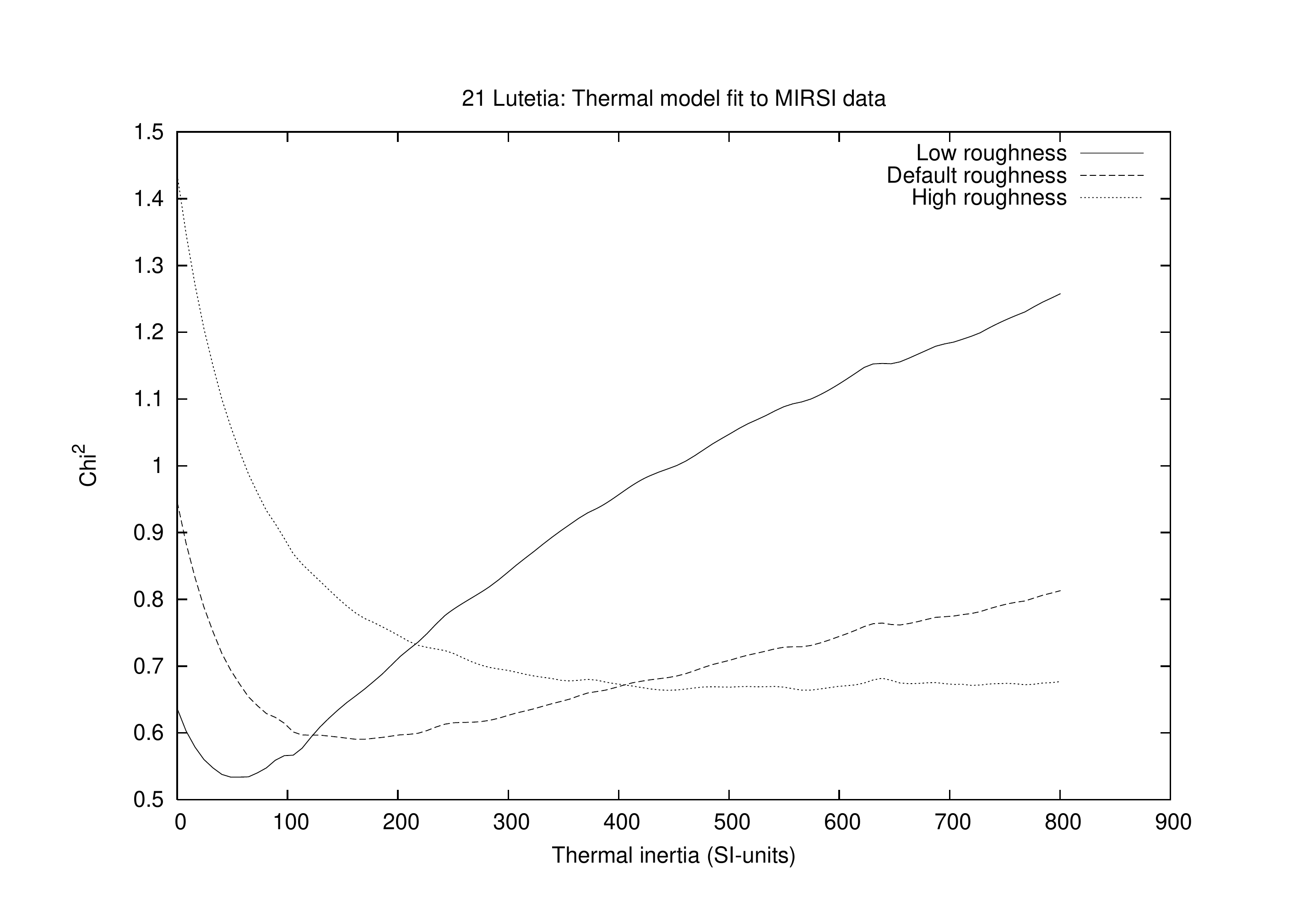}
 \caption[Goodness of fit $\chi^2$ vs.\ thermal inertia for the MIRSI data]{
Goodness of fit $\chi^2$ vs.\ thermal inertia for the MIRSI data. For each value of thermal inertia, the best-fitting diameter is found. The SI-unit of thermal inertia is \TIunit.}
 \label{fig:21:Chi2_IRTF}
\end{figure}

\begin{figure}
\centering
\includegraphics[angle=-90, width=0.7\textwidth]{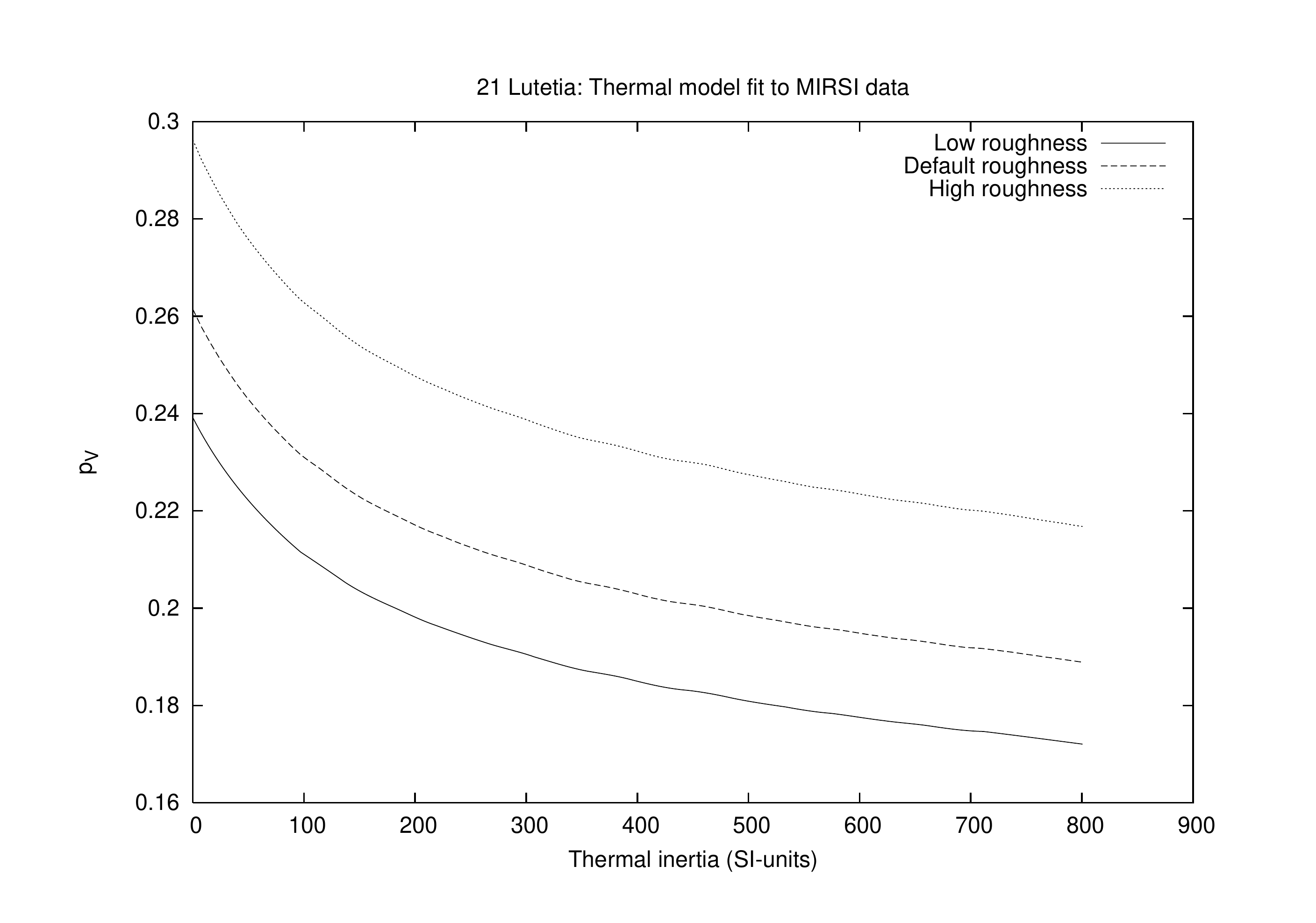}
\caption{
Best-fit geometric albedo \pV\ vs.\ thermal inertia for the MIRSI data; cf. \figref{fig:21:Chi2_IRTF}.}
 \label{fig:21:pV_IRTF}
\end{figure}

It may not be appropriate to fit the MIRSI data with the TPM
since its usage generally requires more than just three data points.
As can be seen in \figref{fig:21:Chi2_IRTF}, 
the best fit suggests low roughness and low thermal inertia of about \unit{50}{\TIunit}, about the thermal inertia of lunar soil. However, considerably higher thermal inertias can also be fitted by adding more surface roughness (note the scale on the $\chi^2$-axis!).
According to \citet{MuellerLagerros1998}, typical MBA thermal inertias range between 5 and \unit{25}{\TIunit}.

Nevertheless, as can be seen in \figref{fig:21:pV_IRTF}, the best-fit geometric albedo \pV\ for the respective best-fit thermal inertia is largely independent of surface roughness:
$\pV =  0.225 \pm 0.020$ and
\unit{$D = 94.9\pm 4.3$}{\km},
in excellent agreement with our NEATM findings  (\sectref{sect:21:NEATM}). The uncertainties quoted here reflect only the scatter of the fitting procedure; we estimate the systematic uncertainties inherent in the modeling and calibration to be significantly higher:  \unit{10}{\%} for the diameter and \unit{20}{\%} for albedo.
Our MIRSI data alone do not significantly constrain Lutetia's thermal inertia or surface roughness.

\begin{figure}
\centering
 \includegraphics[angle=-90, width=0.7\textwidth]{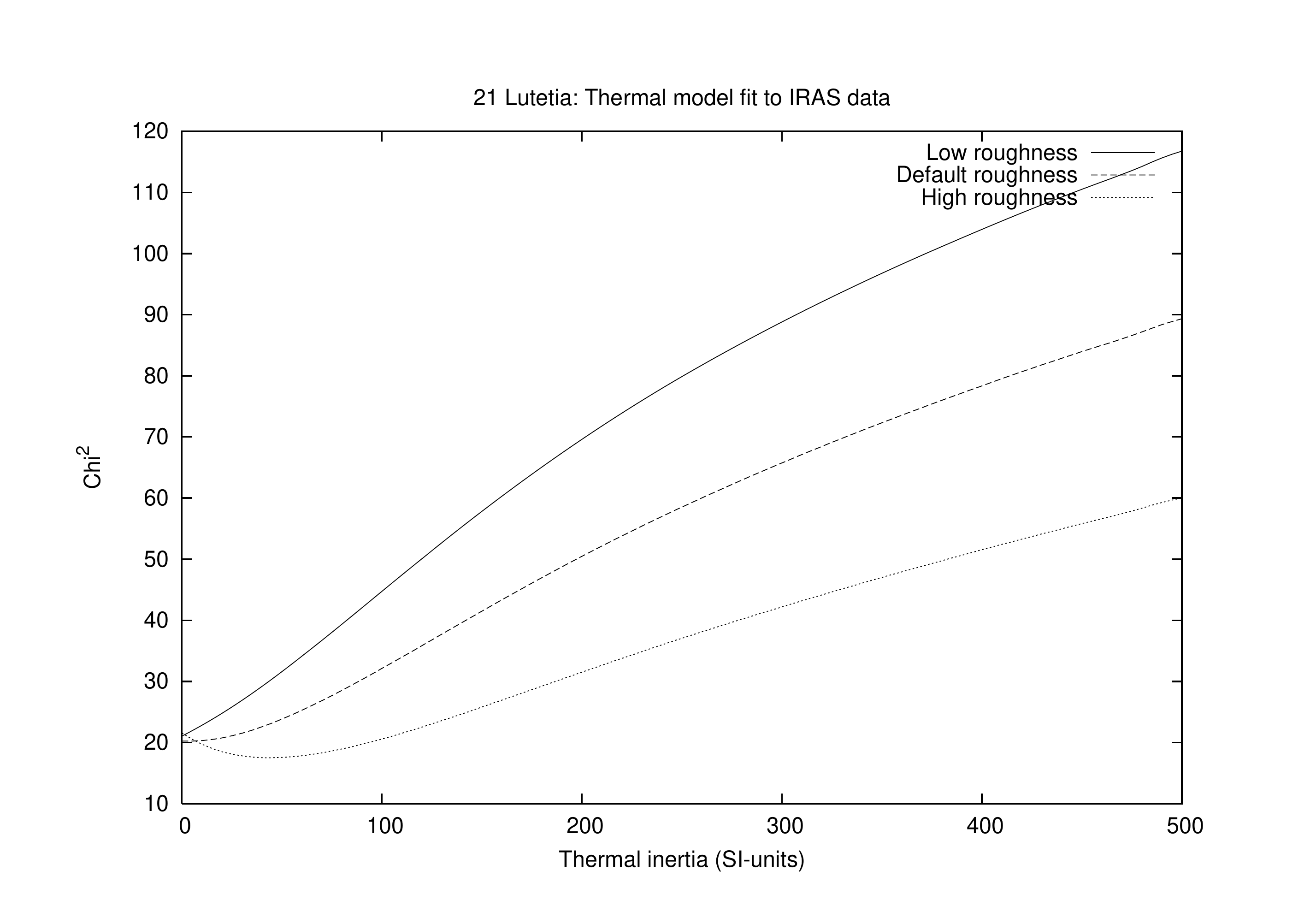}
 \caption{
As \figref{fig:21:Chi2_IRTF}, but for IRAS data.}
 \label{fig:21:Chi2_IRAS}
\end{figure}

We have also used the TPM to fit the flux values measured by IRAS in 1983 \seesect{sect:21:NEATM}.
As can be seen in \figref{fig:21:Chi2_IRAS}, the IRAS data are also best fitted with a thermal inertia of about \unit{50}{\TIunit},
however, this time the best fit is reached assuming high surface roughness rather than low surface roughness.
The best-fit diameter and albedo are
$\pV = 0.21 \pm 0.04$ and
\unit{$D = 98.3 \pm 9.4$}{\km}
(statistical errors only, see above), in excellent agreement with our previous findings.


We have also attempted to simultaneously fit both data sets, MIRSI + IRAS. This leads to a best-fit thermal inertia of zero, 
whichever roughness is assumed. In particular, changing the roughness does not alter the goodness of the fit, so roughness is not constrained at all.
The corresponding diameter and albedo are
$\pV =  0.235 \pm 0.027$ and
\unit{$D   =  92.9 \pm 5.4$}{\km}
(statistical errors only),
in agreement with our previous findings.

While the TPM fits to the MIRSI, IRAS, and MIRSI + IRAS data sets give very similar values of $D$ and \pV, results for thermal inertia and, in particular, surface roughness are not well constrained.
However, our results strongly suggest that the thermal inertia does not greatly exceed \unit{100}{\TIunit}.

Linking multi-epoch data taken some 20 years apart requires very accurate knowledge of the rotational period. 
The one used in this work \citep[\unit{8.165455}{\hour},][]{Torppa2003} seems to be sufficiently accurate: it is based on a large temporal baseline of some 36 years (lightcurves from 1962--1998, including some from 1983), so it should safely bridge the six years  1998--2004. Judging from \figrefpage{fig:21:optical} this is indeed the case, although we caution that we have no more than two  data points.

Another potential source of error is the sub-Earth latitude, which varies considerably between apparitions: in 1983, the sub-Earth latitude was \unit{+18}{\degree}, as opposed to \unit{-75}{\degree} in 2004 \seesect{sect:21:LC}.
This could cause the shape model to be biased towards one hemisphere,
although we note that lightcurves at various sub-Earth latitudes were used in the construction of the shape model
(for example, the sub-Earth latitude in Oct. 1962 was \unit{-54}{\degree}).
Furthermore, the surface composition or roughness could vary across the surface; our model assumes it to be homogeneous.

\subsection{Discussion}

\subsubsection{Diameter and albedo}

\begin{table}
\caption[Summary of diameter and albedo determinations for Lutetia]{
Summary of diameter and albedo determinations for Lutetia. Values in parentheses are calculated by us on the basis of \eqrefpage{eq:FowlerChillemi} and $H=7.35$. NEATM results are quoted with 15 and \unit{30}{\%} uncertainty in diameter and  albedo, respectively; TPM results  with \unit{10}{\%} uncertainty in diameter and \unit{20}{\%} uncertainty in albedo (see \sectref{sect:21:NEATM} and \ref{sect:21:TPM}).
 \label{table:21:diameters}}

\centering
\begin{tabular}{l l l}
\toprule
$D$ (\km) & \pV & Notes\\
\midrule
$95.8\pm 4.1$ & $0.221\pm 0.020$ & \citet[using the STM]{IRAS} \\
(142.4) & 0.100 & \citet[from polarimetry]{LupishkoMohamed1996}  \\
$116\pm 17$ & ($0.151\pm 0.045$) & \citet[from radar observations]{Magri1999}  \\
$104\pm 16$   & $0.188\pm 0.057$ & NEATM fit to MIRSI data \\
$102\pm 16$   & $0.197\pm 0.059$ & NEATM fit to IRAS data \\
$94.9\pm 9.5$ & $0.225\pm 0.045$ & TPM fit to MIRSI data \\
$98.3\pm 9.9$ & $0.210\pm 0.042$ & TPM fit to IRAS data\\
$92.9\pm 9.3$ & $0.235\pm 0.047$ & TPM fit to MIRSI+IRAS\\
\hline
$98.3 \pm 5.9$ & $0.208 \pm 0.025$ & Weighted average of rows 4--7\\
\bottomrule
\end{tabular}
\end{table}

We have determined the radiometric size and albedo of 21 Lutetia from two data sets (IRAS and new IRTF measurements) and two  independent thermal models with different levels of sophistication, see \sectref{sect:21:NEATM} and \ref{sect:21:TPM}.
Given the progress since the IRAS results in both thermal  modeling \citep{NEATM, LagerrosIV} and  mid-IR calibration 
\citep{CohenVII,CohenX}, our radiometric results should be more reliable than those of \citet{IRAS}.
Our results and those taken from the literature are summarized in \tableref{table:21:diameters}. 
Given the uncertainties, there is good mutual agreement among the radar results by \citet{Magri1999} and Tedesco's and our radiometric results,
while the polarimetric albedo of \citet{LupishkoMohamed1996}  is inconsistent with all other albedo determinations.
In particular, our results are incompatible with typical C-type albedos of $\leq 0.1$, which may be indicated by recent spectroscopic findings \citep{Birlan2004, Lazzarin2004, Barucci2005}.

\subsubsection{Thermal properties}

From our data we have determined not only  the radiometric diameter and albedo of Lutetia, but also the apparent color temperature, from which conclusions on the surface thermal properties, such as thermal inertia and roughness, can be drawn.

Using the NEATM, we found the fit parameter $\eta$, which describes the apparent color temperature, to be 
0.93 / 0.94
for the MIRSI / IRAS-observations, respectively. 
This implies that Lutetia's thermal properties are rather typical for a MBA,
i.e.\ low thermal inertia and some surface roughness.

Our results from the TPM confirm this picture in general: for both data sets, the best-fit thermal inertia is around \unit{50}{\TIunit}, about the thermal inertia of lunar regolith and somewhat higher than the typical thermal inertia of large MBAs \citep{MuellerLagerros1998}.
Values in the range 0--\unit{100}{\TIunit} are consistent with our data. A larger data set would be required to conclusively determine Lutetia's thermal inertia.


\subsection{Summary}

From new thermal-infrared spectrophotometric measurements and detailed thermophysical modeling we infer that
 Lutetia has a diameter of \unit{$98.3 \pm 5.9$}{\km} and a geometric albedo of $0.208 \pm 0.025$, in good agreement with the results from 
IRAS radiometry \citep{IRAS} and radar observations \citep{Magri1999}.
We can rule out a low albedo typical of a C-type taxonomic classification as indicated by recent spectroscopic findings.
Further spectroscopic observations should be made to check for variegation of  spectral features with rotational phase and sub-Earth latitude.

Furthermore we  confirm that Lutetia's surface must be covered with 
thermally insulating  regolith. 
A lunar-like thermal inertia of $50\pm \unit{50}{\TIunit}$
is compatible with both our MIRSI data and the IRAS flux values.


%% file: ML.tex

The NEA (10302) 1989~ML is a nominal target of the planned ESA mission \emph{Don Quijote}. To assist the target selection process, we were awarded 
Director's Discretionary Time with the Spitzer Space Telescope, which allowed us to determine the asteroid's size and albedo, critical parameters for mission planning.
Combining our Spitzer results with optical and near-infrared data, we could furthermore classify 1989~ML as an E-type asteroid, thereby severely constraining its surface mineralogy.

\subsection{Introduction}
\label{sect:ML:intro}

The most accessible asteroids for rendezvous missions are those with orbits similar to that of the Earth. Indeed, some NEAs are easier to reach than the Moon.
The energy ($\Delta v$) and flight time required to reach the Amor-type NEA (10302) 1989 ML (period \unit{1.44}{\yr},  eccentricity 0.14,  inclination \unit{4.4}{\degree}) are relatively small, comparable to those for the Hayabusa target (25143) Itokawa
\citep{Perozzi2001, Christou2003, Binzel2004}, making it a very attractive spacecraft target for rendezvous missions.
1989~ML has been considered as a possible target for both the Japanese   Hayabusa  and
the European Don Quijote missions
\citep[see][]{Binzel2001,NEOMAP};
at the time of writing further missions to NEAs are being planned in Japan, Europe, and the USA \seesect{sect:intro:spacecraft}, for which 1989~ML may be considered as a target. 
\textit{However, a serious and urgent problem for mission planning is the lack of information on the physical properties of this asteroid.}

Preliminary optical lightcurve measurements by \citet{Weissman1999} and \citet{Abe2000} suggest the peak-to-peak amplitude is about \unit{1}{$\text{mag}$}, corresponding to a very elongated shape. 
\citeauthor{Weissman1999}\ report a rotation period near \unit{19}{\hour}. However, according to \citeauthor{Abe2000}, \unit{$\sim 32$}{\hour} is also possible.
This long-period high-amplitude lightcurve has limited the accuracy of determinations of the absolute optical magnitude, $H$: \citet{Abe2000} report $H=19.7$ (assuming $G=0.15$); NEODys (as of 18 April 2007) reports $H=19.39$ ($G=0.15$); \citet{Weissman1999} report an absolute magnitude in the R-band of $H_R = 19.14$, which implies $H=19.47$ using $V-R = 0.37 \pm 0.03$ \citep{ML}. Since published $H$ values for NEAs are notoriously unreliable, we adopt the average value of $H=19.5$ with a conservative uncertainty of $\pm 0.3$. 

\citet{Binzel2001} report a neutral $X_c$-type spectrum at optical wavelengths, implying that 1989~ML belongs to one of the E, M, or P spectrally degenerate classes \seesect{sect:intro:mineralogy}. E-type asteroids have a high geometric albedo $\pv$ ($0.3 < \pv < 0.6$) and may be related to enstatite achondrite meteorites; M-type asteroids have moderate albedos around 0.1--0.2 and some are probably related to metallic meteorites; P-type asteroids have very low albedos ($\pv \leq 0.1$) and appear to be organic-rich, similar to carbonaceous chondrites \citep[see, e.g.,][]{Clark2004a,Clark2004b}. Determination of the albedo of an X-type asteroid is therefore very important for constraining its composition.

We  observed 1989~ML in the thermal infrared using the Infrared Array Camera IRAC 
on board the Spitzer Space Telescope 
at a phase angle of \unit{52.3}{\degree}. 
We employed the 
NEATM \seesect{sect:NEATM}
to derive the effective diameter from the Spitzer data. 

\subsection{Observations and data reduction}
\label{sect:ML:observations}

1989~ML was observed on 2006 June 2 and 3 with IRAC on board of Spitzer  \seesect{sect:IRAC:general}. We were awarded a total of \unit{1.2}{\hour} Director's Discretionary Time for this project. 

The asteroid was observed six times, each observation providing nearly simultaneous photometry in all 4 IRAC channels with five dither positions, of \unit{30}{\second} integration time each, per field-of-view. 
The observation time in each case was about 12~minutes, including dead times for telescope slewing and settling, which is significantly shorter than the asteroid rotation period.
In order to trace the rotational flux variability (about 1~mag at visible wavelengths) we requested time gaps (by imposing follow-on constraints on the individual observations)
of $\sim\unit{3.2}{\hour}$  between consecutive observations, corresponding to $\sim\unit{60}{\degree}$ in rotational phase assuming the nominal rotation period of \unit{19}{\hour} \citep{Weissman1999}. The observing geometry did not change significantly during our observations: the heliocentric distance was \unit{1.270}{\AU}, the distance to Spitzer was \unit{0.891}{\AU}, and the solar phase angle, $\alpha$, was \unit{52.3}{\degree} (source: JPL Horizons System; all values are constant during our Spitzer observations to $\pm 2$ in the last quoted digit or better). 

We analyzed the obtained BCD images \seesect{sect:IRAC:BCD_pipeline} using the data reduction techniques presented in \sectref{sect:IRAC:asteroids}.
A discussion of the color corrections to the  flux densities derived in this case is given in \sectref{sect:ML:results}.

The mosaics for observations 1--5 display clear asteroid signals at the predicted positions, but observation \#6 failed, because the target asteroid was within \unit{2}{\arcsec} of a stellar background source of comparable brightness. Also data from observation \#4 are compromised by the presence of a faint background source, which can be neglected for IRAC channels 3 and 4 but is comparable in flux to the asteroid at shorter wavelengths. 

\begin{table}
\caption[Time-resolved flux densities of 1989~ML from IRAC channel 4]{Time-resolved flux densities of 1989~ML from IRAC channel 4 (7.872 \micron) and observations 1--5. Flux values in parentheses are color corrected. Times refer to the beginning of the observation and are not light-time corrected.}
\label{table:Spitzerfluxes:ML1}
\centering
\begin{tabular}{lll}
\toprule
JD - 2453889.0 & Flux (\micro Jy) & $\sigma$ flux (\micro Jy)\\
\midrule
0.47827 & 126 (122) & 19 (18)\\
0.60011 & 49 (47)   & 16 (15)\\
0.75043 & 115 (112) & 23 (22)\\
0.86787 & 188 (182) & 17 (16)\\
1.01833 & 91  (89)  & 21 (20)\\
\bottomrule
\end{tabular}
\end{table}

\begin{table}
\caption[Flux densities of 1989~ML for all four IRAC channels, derived from stacking observations 1--5]{Flux densities for all four IRAC channels derived from stacking observations 1--5. As in \tableref{table:Spitzerfluxes:ML1}, flux values in parentheses are color corrected.}
\label{table:Spitzerfluxes:ML2}
\centering
\begin{tabular}{lll}
\toprule
Central wavelength (\micron) &  Flux (\micro Jy) & $\sigma$ flux (\micro Jy)\\
\midrule
3.550 & 4.14 (4.14) & 0.77 (0.77) \\
4.493 & 4.1 (3.8) & 1.3 (1.2) \\
5.731 & 21.7 (20.3) & 7.3 (6.8) \\
7.872 & 111.3 (108)  &  17.7 (18) \\
\bottomrule
\end{tabular}
\end{table}

The highest signal-to-noise ratio is obtained in IRAC channel 4, from which photometry can be extracted for each observation yielding a coarse thermal lightcurve (see \tableref{table:Spitzerfluxes:ML1}). For all channels, we stacked images from observations 1--5 and derived an average flux value from the resulting mosaic images \seetable{table:Spitzerfluxes:ML2}. 

\subsection{Results}
\label{sect:ML:results}

\begin{figure}[tb]
\centering
\includegraphics[angle=-90, width=0.7\textwidth]{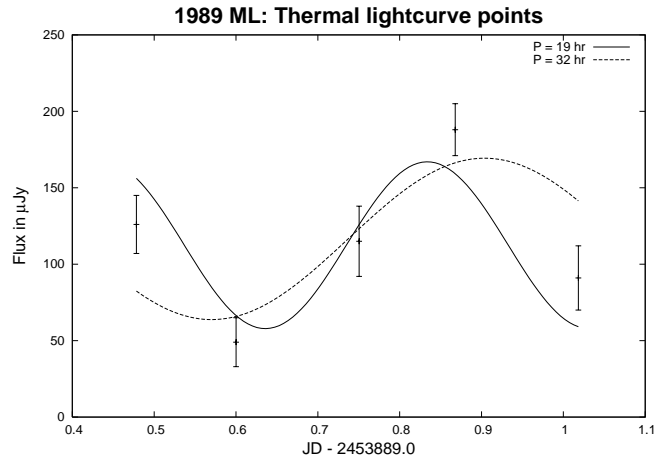}
\caption[Time-resolved channel-4 data as given in \tableref{table:Spitzerfluxes:ML1} overlaid with two sinusoidal lightcurves corresponding to the rotation periods proposed in the literature]{Time-resolved channel-4 data (not color corrected) as given in \tableref{table:Spitzerfluxes:ML1} overlaid with two sinusoidal lightcurves corresponding to the rotation periods proposed in the literature. As is usual, the lightcurves are assumed to be double-peaked, i.e.\ the photometric periods equal half the rotation periods (9.5 and \unit{16}{\hour}).
\label{fig:ML:lc}
}
\end{figure}

Our time-resolved channel-4 data given in \tableref{table:Spitzerfluxes:ML1} (\unit{7.872}{\micron}) are consistent with a high-amplitude long-period lightcurve as proposed in the literature. Sinusoidal double-peaked lightcurves corresponding to the two possible rotation periods 
of 19 and \unit{32}{\hour} 
fit the data well, cf.\ \figref{fig:ML:lc}.
The average flux levels of the two fitted lightcurves are consistent with one another and with the flux value obtained from stacking images from observations 1--5 (\tableref{table:Spitzerfluxes:ML2}).
Agreement is to within a few percent, a negligible difference compared to the statistical flux uncertainty. We cannot further constrain the rotation properties from our Spitzer data, but conclude that for all IRAC channels the flux values obtained from stacking observations 1--5 are good proxies to the lightcurve average flux level.

IRAC is a broad-band photometer, so the derived flux values must be color corrected \seesect{sect:IRAC:CC}. Due to their different spectral shapes, different color corrections apply to the thermally emitted flux component and to reflected sunlight; color corrections to the latter are negligible. 
We estimated the amount of reflected sunlight in channels 1--4 assuming a solar black body temperature of \unit{5800}{\kelvin} and a relative reflectance of $\sim 1.2$ between \unit{3.6}{\micron} and the V band \citep[Fig.\ 2]{ML}.
Reflected sunlight was found to contribute virtually all the measured flux in channel 1, only negligible amounts in channel 4, and \unit{$\sim 1$}{\micro\jansky} in channel 3 (\unit{$\sim 5$}{\%} of the measured flux). The relative contributions to the channel 2 flux cannot be easily determined.

\begin{figure}[tb]
\centering
\includegraphics[angle=-90, width=0.7\textwidth]{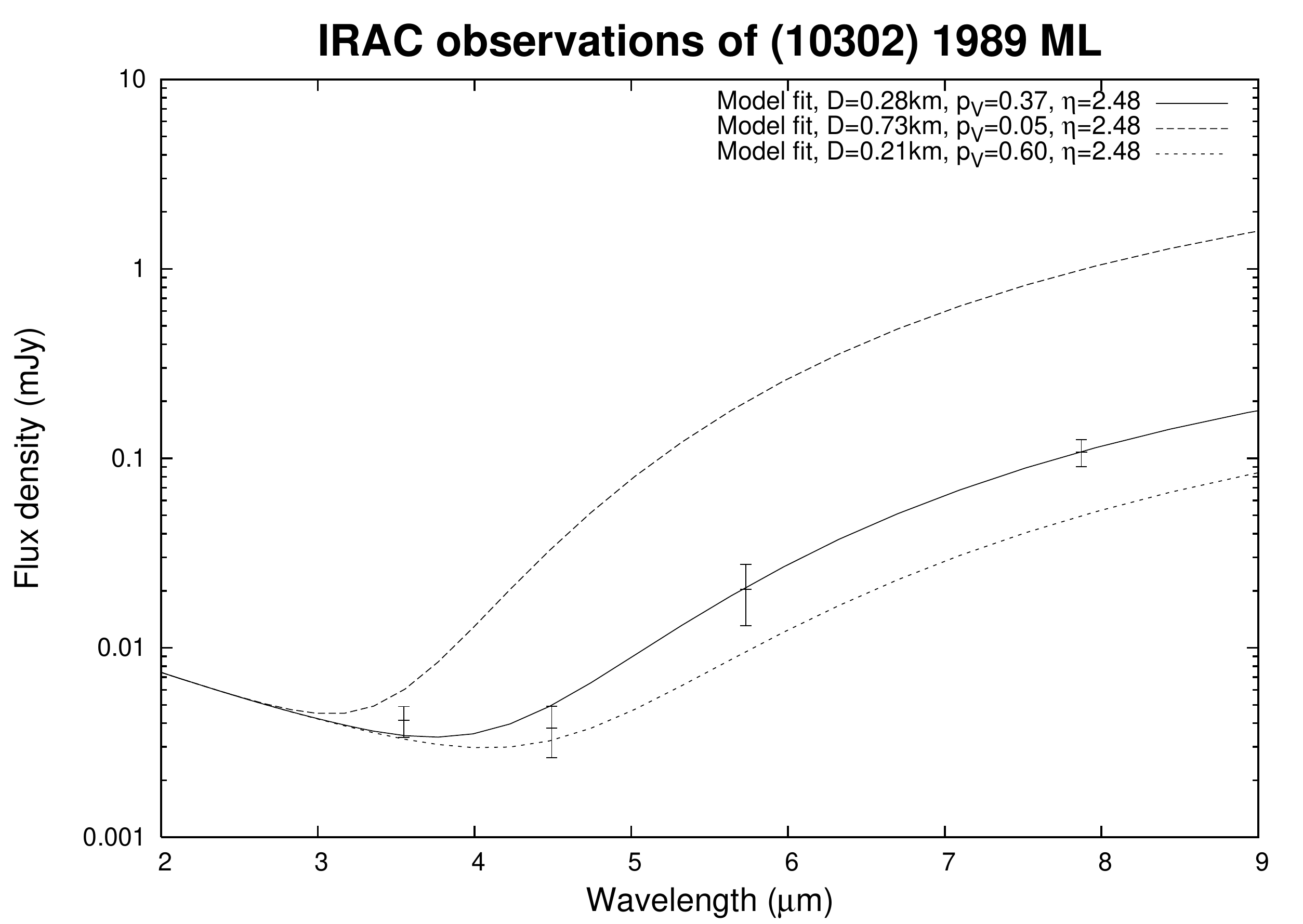}
\caption[Model fits to the photometric data given in \tableref{table:Spitzerfluxes:ML2}]{Model fits to the photometric data given in \tableref{table:Spitzerfluxes:ML2}. 
Both reflected sunlight and thermal emission are modeled (cf.\ text). The thermal contributions to channels 3 and 4 (wavelengths 5.7 and \unit{7.9}{\micron}) are fitted using the NEATM; the reflected sunlight is extrapolated from the predicted V magnitude assuming a relative reflectance of 1.2 \citep[see text and][Fig.\ 2]{ML}. An albedo of $\pV = 0.05$, which would be typical for P-type asteroids, is clearly incompatible with the data (dashed line). 
\label{fig:ML:NEATM}
}
\end{figure}

We fitted the NEATM to the thermal flux values from channels 3 and 4 and calculated color-correction factors using the model parameters (diameter, \pv, and $\eta$) derived. 
We repeated this procedure with color-corrected thermal fluxes, after which the procedure was seen to have converged; a second iteration brought about changes significantly below the \unit{1}{\%} level. The resulting model spectrum with the observational data overlaid is shown in \figref{fig:ML:NEATM}.

The best-fit parameters are: Diameter $D = \unit{0.276}{\kilo\metre}$, $\pv = 0.37$, and $\eta = 2.48$. Color-corrected fluxes were obtained by dividing the thermal flux contributions  by 1.129 (channel 2), 1.070 (channel 4) and 1.034 (channel 4).  Uncertainties in $D$, \pV, and $\eta$  were estimated using a Monte-Carlo analysis.
To this end, we generated a random set of synthetic thermal flux values at the channel-3 and channel-4 wavelengths, normally distributed around the measured values, and fitted them using  the NEATM. We rejected unrealistic results with $\eta > 3$ or $\pv > 0.7$.
The remaining sample of 15,000 results gave $D = 0.246 \pm \unit{0.037}{\km}$, $\pV = 0.46 \pm 0.13$, and $\eta = 2.23 \pm 0.44$ ($1 \sigma$  standard deviations---note, however, that the distribution of resulting model parameters is highly non-Gaussian). We adopt the fractional uncertainties from this simulation for our best-fit results stated above, yielding: $D = 0.276 \pm \unit{0.041}{\km}$, $\pV = 0.37 \pm 0.11$, and $\eta = 2.48 \pm 0.49$.
The uncertainty in our adopted value for $H$ ($19.5 \pm 0.3$) contributes an additional \unit{30}{\%} to the albedo error (added in quadrature) so $\pv = 0.37 \pm 0.15$. This albedo is suggestive of an E classification. P types, for which \pv\ should not exceed 0.1, would appear to be ruled out (see also \figref{fig:ML:NEATM}). 
We note that in addition to the statistical errors there is a systematic modeling uncertainty that increases with solar phase angle and thermal inertia \citep[see][and \sectref{sect:NEATM:accuracy}]{Harris2006}, but in the sense of underestimating \pV. So the systematic uncertainty in the NEATM results in this case would tend to increase \pV\ above the value of 0.37 derived here.  

Our albedo result is consistent with the
available photometric and spectroscopic data in the optical and near-infrared wavelength ranges, which also favor an E-type classification \citep[see][and references therein; that part of the paper is by co-author Prof.\ Fitzsimmons]{ML}.

The derived $\eta$ value of $2.5 \pm 0.5$ is rather high for a solar phase angle of \unit{52}{\degree} \citep[][see also \figrefpage{fig:NEATM:etas}]{Delbo2003} and is consistent with a high surface thermal inertia, corresponding to a lack of thermally insulating dust or regolith.
Better spectral coverage and a thermophysical model would be required to derive conclusive statements about surface thermal properties.

\subsection{Discussion}
\label{sect:ML:discussion}

1989~ML is an attractive spacecraft target due to its low-$\Delta v$ orbit. However, virtually no conclusive information on its physical properties has been available so far.
Our determination of the object's size, albedo, and surface mineralogy is therefore helpful in the process of selecting suitable targets for NEA missions.

Knowledge on the target diameter is particularly mission relevant.
Our result of $D = 0.276 \pm \unit{0.041}{\km}$ is much lower than estimates based on the $H$ value and common default values of \pv; e.g., assuming $\pv=0.2$ results in a  \unit{$\sim35$}{\%} larger diameter, whereas the diameter of a P-type asteroid ($\pv\sim0.05$) would be 2.7 times larger than our result.
This corresponds to  volume ratios of 2.5 or 20, respectively, and  correspondingly large differences in mass. 
Constraints on the target mass are particularly important for the design of orbiting spacecraft.

For phase-A studies of the Don-Quijote mission, 1989~ML was chosen as one of two nominal targets \citep[see, e.g.,][]{NEOMAP} assuming a diameter of \unit{0.5}{\kilo\meter},  above our estimate by a factor of $\sim1.8$ and corresponding to a $\sim6$ times larger volume.
Don Quijote consists of a kinetic impactor and an orbiter, 
its primary aim is to measure the impact-induced orbital change of the target NEA using the orbiter.

In the target definition process, it was found that a suitable target for Don Quijote should have a diameter up to 
\unit{0.5}{\kilo\metre} in order to guarantee that the momentum transfer leads to an accurately observable change in orbit; we have found 1989~ML to clearly satisfy this constraint. 
On the other hand, 
too small asteroids do not allow stable spacecraft orbits around them and are harder to target with a kinetic impactor; note that a non-central impact would reduce the momentum transfered by ejecta.
While no numerical value for a minimum diameter has been specified by \citet{NEOMAP}, our results may imply that 1989~ML is too small to be considered a suitable target for Don Quijote.

Another target selection criterion for the Don Quijote mission is  taxonomic type: 
Dark, C-type-like objects would be preferred \citep{NEOMAP}.
With an albedo of $\pv=0.37\pm0.15$, 1989~ML is clearly not a dark object.

\subsection{Summary}
\label{sect:ML:summary}

On the basis of thermal-infrared photometric observations of the NEA (10302) 1989~ML using the IRAC camera on board the Spitzer Space Telescope we have determined its effective diameter to be $0.28 \pm \unit{0.05}{\kilo\metre}$ and its geometric albedo $\pV$ to be $0.37 \pm 0.15$. This high albedo is incompatible with a P-type classification, is only marginally consistent with an M-type classification, but is fully consistent with an E-type classification. 
The available optical and near-infrared data also favor E type. Taken together, we conclude that the available data suggest an E classification for (10302) 1989~ML---note that only 4 E-type NEAs were known beforehand \citep{Clark2004a}.

1989~ML is an attractive spacecraft target due to its Earth-like orbit.
Virtually nothing was previously  known about the physical properties of this easy-to-reach asteroid; our results will inform the target selection process of presently planned, and potentially of future  NEA missions.
In particular, we have found that the diameter of 1989~ML is much smaller than assumed in phase-A studies of the ESA mission Don Quijote. 
This, together with our result for its taxonomic type, most probably implies that 1989~ML is not a suitable target for Don Quijote.


%% file: Patroclus.tex
A particularly direct way of measuring thermal inertia is from time-resolved observations of the thermal response to eclipse events,
effectively allowing one to see  shadowed surface elements cool down and heat back up in real time.
Recent progress in the orbital modeling of binary asteroid systems now allows the reliable prediction of eclipse events in such systems.

We here present the first thermal-infrared observations of an eclipsing binary asteroid system. 
Our target is the binary Trojan  (617) Patroclus, which consists of two components of roughly equal size.
The mutual orbit has been determined by \citet{Marchis2006}, 
allowing them to determine 
the system's mass. Combining their result with a diameter estimate by \citet{Fernandez2003}, they determine the average 
mass density to be \unit{$0.8^{+0.2}_{-0.1}$}{\gramm\usk\power{\cm}{-3}}.

Using IRS on board the Spitzer Space Telescope, we have obtained a total of 18 thermal-infrared spectra ($\sim8$--\unit{33}{\micron}) of the system, providing good temporal coverage of two mutual events in June 2006. 

A preliminary analysis of the Spitzer data
results in a thermal inertia of
$\sim\unit{90}{\TIunit}$, slightly larger than that of Galilean satellites and significantly above published upper limits on the thermal inertia of two other Trojans. The latter are found  to be methodologically unreliable.
This is consistent with a relatively coarse regolith, coarser than on main-belt asteroids or our Moon.

The diameters of the two components are found to be
$100\pm10$ and $108\pm\unit{11}{\km}$, respectively,
implying an average mass density of $1.15\pm\unit{0.37}{\gramm\usk\power{\cm}{-3}}$.
This is consistent with the \citeauthor{Marchis2006}\ estimate at the $1\sigma$ level but allows for a larger mass fraction consisting of rock and a lower porosity than estimated by \citeauthor{Marchis2006}
We wish to highlight the importance of accurate diameter estimation in the case of binary systems of known mass.

We caution that the results presented herein are based on a variant of the thermophysical model which is not yet fully tested. Furthermore, the \citeauthor{Marchis2006}\ model of the mutual orbit is currently being refined on the basis of newly obtained observational data (Marchis, 2007, private communication); a final analysis of our data must await the availability of the latter.

\subsection{Introduction}
\label{sect:Patroclus:intro}

Thermal-infrared observations during eclipse events are a well-established technique to determine the thermal inertia of planetary satellites.
E.g., \citet{PettitNicholson1930} determined the thermal inertia of our Moon from observations during a total lunar eclipse; 
\citet{MorrisonCruikshank1973} report observations of the Galilean satellites while the latter were eclipsed by Jupiter; \citet{Neugebauer2005} report observations of  Iapetus during eclipses by Saturn's rings.
Asteroids are not frequently shadowed by planets. Binary asteroid systems, however, are well known to undergo eclipses, when respectively one component shadows the other. 
Targeted observations of an eclipse event require the ability to predict its timing, 
which is reliably doable for only a handful of binary asteroid systems to date.

Our target, (617) Patroclus, is the only currently known binary system in the population of  Trojans co-orbital with Jupiter \seesect{sect:intro:populations}.
Its two components have a diameter ratio around 1.1 and diameters around 100--\unit{120}{\km} (see below).
Trojan orbits are stable over most of the age of the Solar System \citep[see][and references therein]{Emery2006}. The origin of the Trojans is currently under debate, they may have originated at their present location or may have been captured during the epoch of the Late Heavy Bombardment \seesect{sect:intro:populations}.
Trojans have generally low albedos of $\pv\sim0.04$ and virtually featureless, highly reddened reflection spectra in the visible and near-IR wavelength ranges; in both respects, they resemble cometary nuclei \citep[e.g.][and references therein]{JewittLuu1990,EmeryBrown2003,Fornasier2004,Emery2006}.
Trojans smaller than some \unit{70}{\kilo\metre} in diameter appear to be collisional fragments, while larger bodies  appear to be primordial ``accretion survivors,'' i.e.\ bodies whose current form and internal structure have remained unchanged since the time of their formation \citep{Jewitt2000}.

Patroclus was found to be binary by \citet{Merline2001}.
They  determined the components' difference in  near-IR brightness to be only \unit{0.2}{\text{mag}}, implying a diameter ratio close to unity.
The effective diameter of the (spatially not resolved) system was determined by \citet{SIMPS} to be $140.9\pm\unit{4.7}{\km}$
(based on IRAS observations analyzed using the STM---see \sectref{sect:STM}), corresponding to a geometric albedo of $\pv=0.047\pm0.003$ with an absolute optical magnitude of $H=8.19$.
\citet{Fernandez2003} obtained
new \unit{12.5}{\micron} observations of Patroclus at the Keck II telescope. Using the STM, they reproduced the IRAS result, but rejected it in favor of
 a diameter of $166.0\pm\unit{4.8}{\km}$ ($\pv=0.036\pm0.004$) based effectively on the NEATM \seesect{sect:NEATM} assuming $\eta=0.94$, which they argue is more representative of Trojans than the STM value of $\eta=0.756$.
It is worth pointing out that the quoted uncertainties solely reflect the statistical uncertainty although there are considerable systematic uncertainties,
due mostly to the uncertainty in $\eta$. This is acknowledged by \citeauthor{Fernandez2003}\ to limit the accuracy of their results, but no quantitative discussion of the systematic uncertainty is given.
Note that the quoted diameters are area-equivalent diameters $D_\Area$ of the system as a whole. $D_\Area$ is related to the components' diameters, $D_1$ and $D_2$, through
$D_\Area{}^2 = D_1{}^2 + D_2{}^2$.

The system's mutual orbit  was determined by \citet{Marchis2006} based on spatially resolved adaptive-optics observations.
They report a purely Keplerian orbit (without precession) to fit their data well.
The best-fit orbit is roughly circular (eccentricity $0.02\pm0.02$) with a center-to-center separation of the two components of $680\pm\unit{20}{\km}$, corresponding to a maximum angular separation around \unit{0.2}{\arcsecond}.
The J2000 ecliptic coordinates of the spin pole are $\lambda = 234\pm\unit{5}{\degree}$, $\beta=-62\pm\unit{1}{\degree}$,
the orbital period equals $102.8\pm\unit{0.1}{\hour}$. 
Combining their result with the size estimate by \citeauthor{Fernandez2003}, \citeauthor{Marchis2006}\ infer a mass density of only \unit{$0.8^{+0.2}_{-0.1}$}{\gramm\usk\power{\cm}{-3}}, 
compatible with a composition dominated by water ice and moderate porosity.
The size uncertainty accounts for most of the uncertainty in the mass density.
The brightness ratio of the two components was found by \citeauthor{Marchis2006}\ to be roughly identical for two different near-IR filters, indicative of an identical surface mineralogy.
Assuming identical albedo, they infer a diameter ratio of 1.082 and, using the \citeauthor{Fernandez2003}\ size estimate, diameters of 112.6 and \unit{121.8}{\km} for the two components, respectively.

Optical lightcurve observations \citep[Mottola, unpublished work]{Angeli1999} reveal a low amplitude indicative of roughly spherical components.
The spin properties of the individual components are not well constrained, but there is no indication for multi-periodicity in the available lightcurve data which were taken at epochs when no mutual events occurred.
This is consistent with a  synchronization of the individual spin periods with the orbital period, possibly through tidal damping.

The orbit model by \citet{Marchis2006} allowed a series of 
mutual events in 2006 to be predicted.
A large campaign has been launched for targeted observations of these events using optical telescopes.
The orbit model is currently being refined on the basis of data obtained in the course of that campaign (Marchis, 2007, private communication).

We  used the InfraRed Spectrograph (IRS; see \sectref{sect:IRS:general}) on board the Spitzer Space Telescope to observe two eclipse-occultation events in June 2006,
where one component obstructed the line of the sight from the other component towards the Sun and the observer, respectively.
A total of 18 spectra ($\sim8$--\unit{33}{\micron}) was obtained, nine per event, providing good temporal coverage.

Our thermophysical model (TPM) was generalized to allow for the effects of eclipses and occultations (see \chaptref{chapt:TPMconcave} in the appendix).
Applying the generalized TPM to our Spitzer data, we could determine the thermal inertia of the Patroclus system in a way which is nearly unaffected by systematic uncertainties due to thermal-infrared beaming \seesect{sect:TPM:beaming}.
This represents the first determination of an asteroid thermal inertia from eclipse observations and the first reliable determination of the thermal inertia of a Trojan.
We furthermore determine the size and albedo of the object,
allowing us to refine the mass-density estimate by \citet{Marchis2006}.

We caution that the generalized TPM is not yet fully tested; moreover, the orbit model of the Patroclus system, on which our modeling is based, is currently being refined (see above). All results presented in this section are therefore preliminary.

\subsection{Observations}
\label{sect:Patroclus:obs}

Patroclus was observed using the InfraRed Spectrograph (IRS) on board the Spitzer Space Telescope. IRS was used in ``low-resolution'' spectroscopy mode \seesect{sect:IRS:lowres} using the 
modules SL1, LL1, and LL2 \seetablepage{table:IRS:slits}
to obtain spectra in the nominal wavelength range  7.4--\unit{38}{\micron} 
at a relative spectral resolution $\lambda/\Delta\lambda$ between 64 and 128.
The observed flux is practically purely thermal.
The angular separation of the two components of the Patroclus system is never larger than \unit{0.2}{\arcsecond} (and minimized during occultations) while the IRS pixel scale is \unit{1.8}{\arcsecond} or coarser \seetablepage{table:IRS:slits}, hence the system was not spatially resolved.

Time-resolved observations were obtained 
during two consecutive mutual events in June 2006,  referred to as events 1 and 2 in the following. 
In event 1, the  larger component shadowed the smaller, and vice-versa in event 2.
The diameter ratio is only $\sim1.1$ and the mutual orbit is roughly circular, hence the two events 
produced very similar observable effects.
Both events were combined eclipse-occultation events and lasted about \unit{6.5}{\hr}; the predicted chronology of event 2 is:
\begin{description}
\item June 26, 23:15 UT:  Start of the eclipse
\item June 27, 02:45 UT: Start of the occultation (eclipse ongoing)
\item June 27, 05:45 UT:  End of the event
\end{description}
The $1\sigma$ timing uncertainty is $\sim\unit{30}{\minute}$, much more accurate predictions are expected from the refinement of the orbit model, which is currently under development.

\begin{table}
\caption[(617) Patroclus: Times of our Spitzer observations.]{(617) Patroclus: Times of our Spitzer observations. There were nine observations per event, referred to as 1.0--1.8 and 2.0--2.8, respectively. 
Times refer to the beginning of the observations measured at Spitzer.
Each observation ended after about \unit{6}{\minute}.
}
\label{table:Patroclus:timing}
\centering
\begin{tabular}{rrl|rrl}
  \toprule
    &  Day & Time & & Day & Time \\
    & (June 2006) & (UT) & &(June 2006) & (UT) \\
\midrule
1.0 & 24 & 18:40 & 2.0&  26& 10:42 \\
1.1 & 24 & 21:54 & 2.1&  26& 23:22 \\
1.2 & 24 & 22:47 & 2.2&  27& 00:24\\
1.3 & 24 & 23:54 & 2.3&  27& 01:31\\
1.4 & 25 & 00:41 & 2.4&  27& 02:19\\
1.5 & 25 & 01:47 & 2.5&  27& 03:29\\
1.6 & 25 & 02:49 & 2.6&  27& 04:24\\
1.7 & 25 & 04:12 & 2.7&  27& 05:55\\
1.8 & 25 & 05:24 & 2.8&  27& 06:52\\
\bottomrule
\end{tabular}
\end{table}

\begin{table}
  \caption[(617) Patroclus: Observing geometry of our IRS observations]{(617) Patroclus: Observing geometry at the epoch of our observations:
heliocentric distance $r$, Spitzer-centric distance $\Delta$,  solar phase angle $\alpha$, and J2000 ecliptic coordinates (longitude and latitude) in a heliocentric and Spitzer-centric frame, respectively.
All values are constant during our observations to $\pm1$ in the last quoted digit or better.
The absolute optical magnitude equals $H=8.19$ \citep[quoted after][]{SIMPS}, the slope parameter $G=0.15$.
}
\label{table:Patroclus:obsgeometry}
\centering
\begin{tabular}{r|ll}
\toprule
Event:& 1 &2 \\
\midrule
  $r$        & \unit{5.947}{\AU}  & \unit{5.947}{\AU} \\
  $\Delta $   & \unit{5.95}{\AU}  & \unit{5.98}{\AU}\\
  $\alpha$& \unit{9.80}{\degree}  & \unit{9.77}{\degree}\\
Heliocentric & \unit{170.8}{\degree}, \unit{+18.03}{\degree} & \unit{170.9}{\degree}, \unit{+18.00}{\degree}\\
Spitzer-centric& \unit{160.5}{\degree}, \unit{+18.2}{\degree}& \unit{160.7}{\degree}, \unit{+18.1}{\degree}\\
\bottomrule
\end{tabular}  
\end{table}

A total of 18 thermal-infrared spectra of Patroclus has been obtained, nine per event, providing good temporal coverage of both events and their aftermath (see \tableref{table:Patroclus:timing} for the observation times and \tableref{table:Patroclus:obsgeometry} for the observing geometry). 
To enable comparison, two observations were performed before the predicted start of the events (observations 1.0 and 2.0), while the purpose of the observations after the end of the events (1.7--1.8, 2.7--2.8) was to observe the heating up of the system.

To prevent wavelength-dependent spill-over losses, IRS spectroscopy targets must be accurately centered into the slit.
The projected width of the LL slits is above \unit{10}{\arcsec}, while that of the SL1 slit is \unit{3.7}{\arcsec}
\seetablepage{table:IRS:slits}, comparable to the width of the point-spread function (PSF) in the respective wavelength range.
 ``Blind'' telescope pointings have a $1\sigma$-accuracy of \unit{0.5}{\arcsec} and are therefore adequate to center sources of well known position (such as Patroclus) in the wide LL slits but risk placing large portions of the PSF outside the SL1 slit.
IRS has an automated peak-up mechanism \seesect{sect:IRS:peakup} to refine the pointing based on imaging with dedicated peak-up detectors.
This mechanism could not be used for our observations, however, because Patroclus would have saturated the peak-up detectors.
Instead, small spectral maps were created for the SL1 observations, with three small steps perpendicular to the slit, each offset by \unit{2}{\arcsec} (roughly half the slit width). 
This enables one to estimate the target offset from the slit center and to correct for the effects thereof in the data analysis
\citep[this strategy was developed and previously used by our colleague J.\ Emery,  see][]{Emery2006}.

The apertures of the used IRS modules have a projected length of \unit{57}{\arcsec} or more, 
hence one-dimensional spatial information can be obtained in addition to the spectral dimension. 
The usual ``nod'' strategy was used, i.e.\ the target was placed  at about 33 and \unit{66}{\%} of the slit length in each module in order to enable subtractive correction for diffuse background emission (``sky background'').
The spectral maps for the LL observations consist of two pointings for the two nod positions, that for the SL1 observations consists of a total of 6 positions (two nod positions times three  pointings perpendicular to the slit).

The obtained data have been reduced by  Emery using the methods described in \citet[sect.\ 3]{Emery2006}.
The detectors contain permanently and temporarily damaged pixels, which required measurements for some wavelengths to be excluded from further consideration, e.g.\ all wavelengths above \unit{33}{\micron}.
For each data point, the statistical flux uncertainty is estimated based on photon count statistics and the observed difference between the two nod positions. Data from the different IRS submodules (which have some spectral overlap) were multiplicatively matched to one another.
%
Systematic flux uncertainties due to, e.g., the residual effect of source mis-centering or uncertainties in the absolute flux calibration 
are estimated to be \unit{3}{\%} (see, e.g., \tablerefpage{table:IRS:PUlosses})
and added in quadrature.
The total flux uncertainty is dominated by the systematic uncertainty except for the Wien slope at the shortest wavelengths.

\begin{figure}
  \centering
\includegraphics[angle=-90,width=0.6\linewidth]{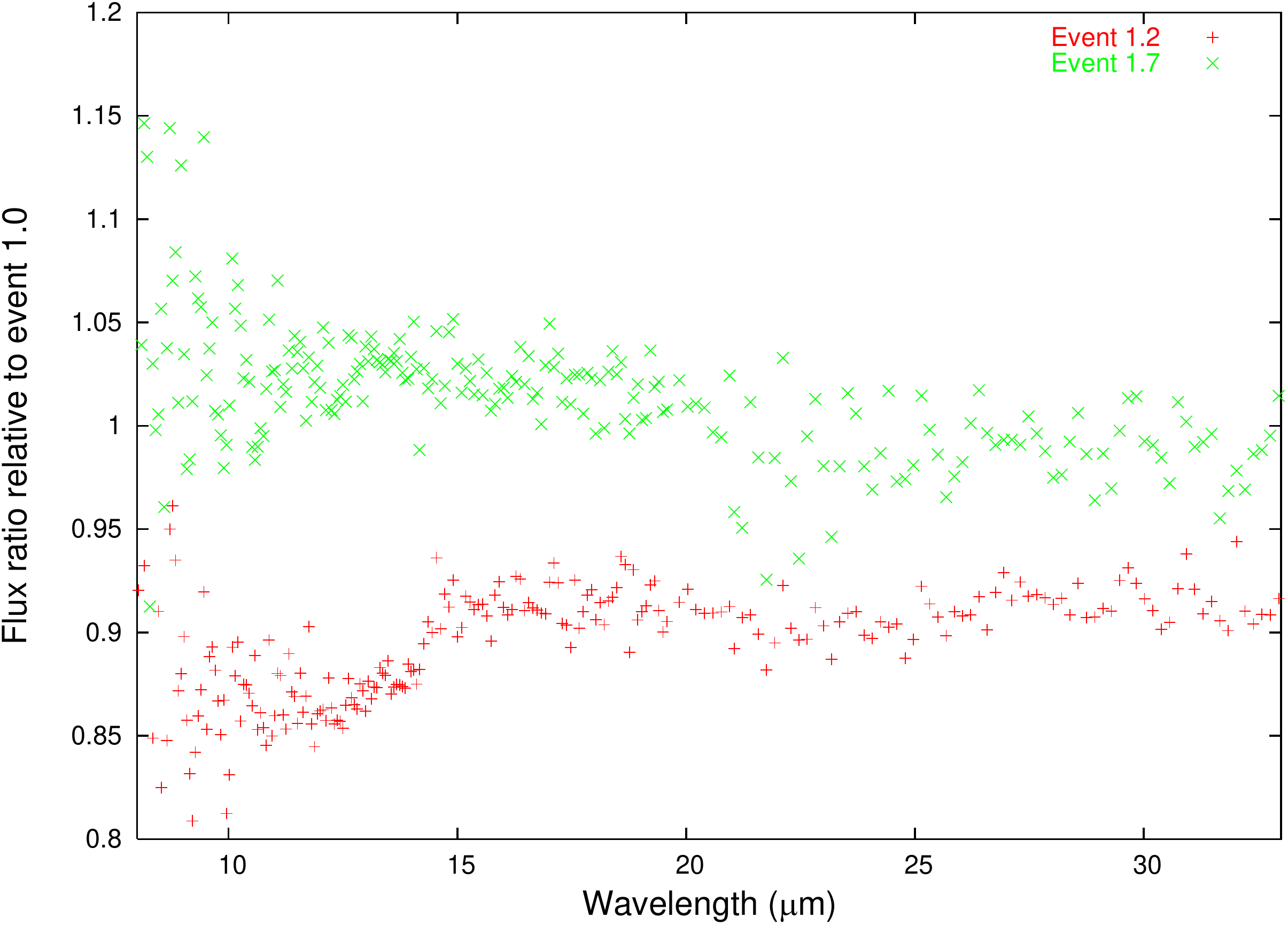}
  \caption[(617) Patroclus: Ratio of measured thermal spectra.]{(617) Patroclus: Ratio of two measured thermal spectra relative to pre-event measurement 1.0. Observation 1.2 was taken during the eclipse, observation 1.7 in the aftermath. Note the difference in flux level and spectral slope, which is consistent with the eclipse-induced temperature drop!}
  \label{fig:Patroclus:fluxratios}
\end{figure}

See \figref{fig:Patroclus:fluxratios} for a representative plot of flux ratios during event 1.

The observed spectra contain slight spectral features due to silicates within the wavelength ranges  10--\unit{12}{\micron} and 18--\unit{22}{\micron} (Emery, 2006, private communication).
In order to avoid biases, those wavelengths were disregarded in the following, leaving 179 data points per observation. 

\subsection{NEATM analysis}
\label{sect:Patroclus:NEATM}

\begin{figure}
  \centering
\includegraphics[angle=-90, width=0.6\linewidth]{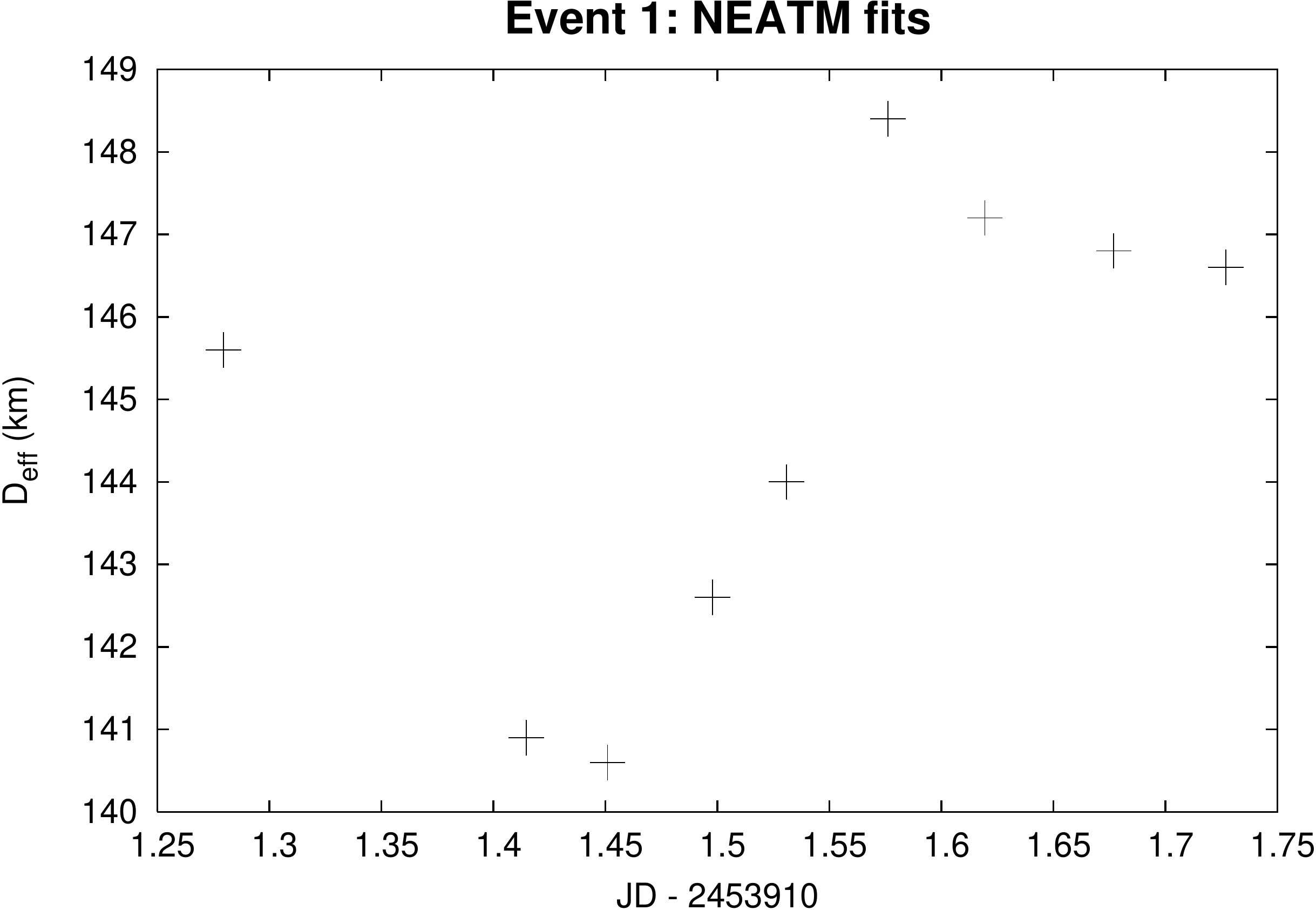}

\bigskip

\includegraphics[angle=-90, width=0.6\linewidth]{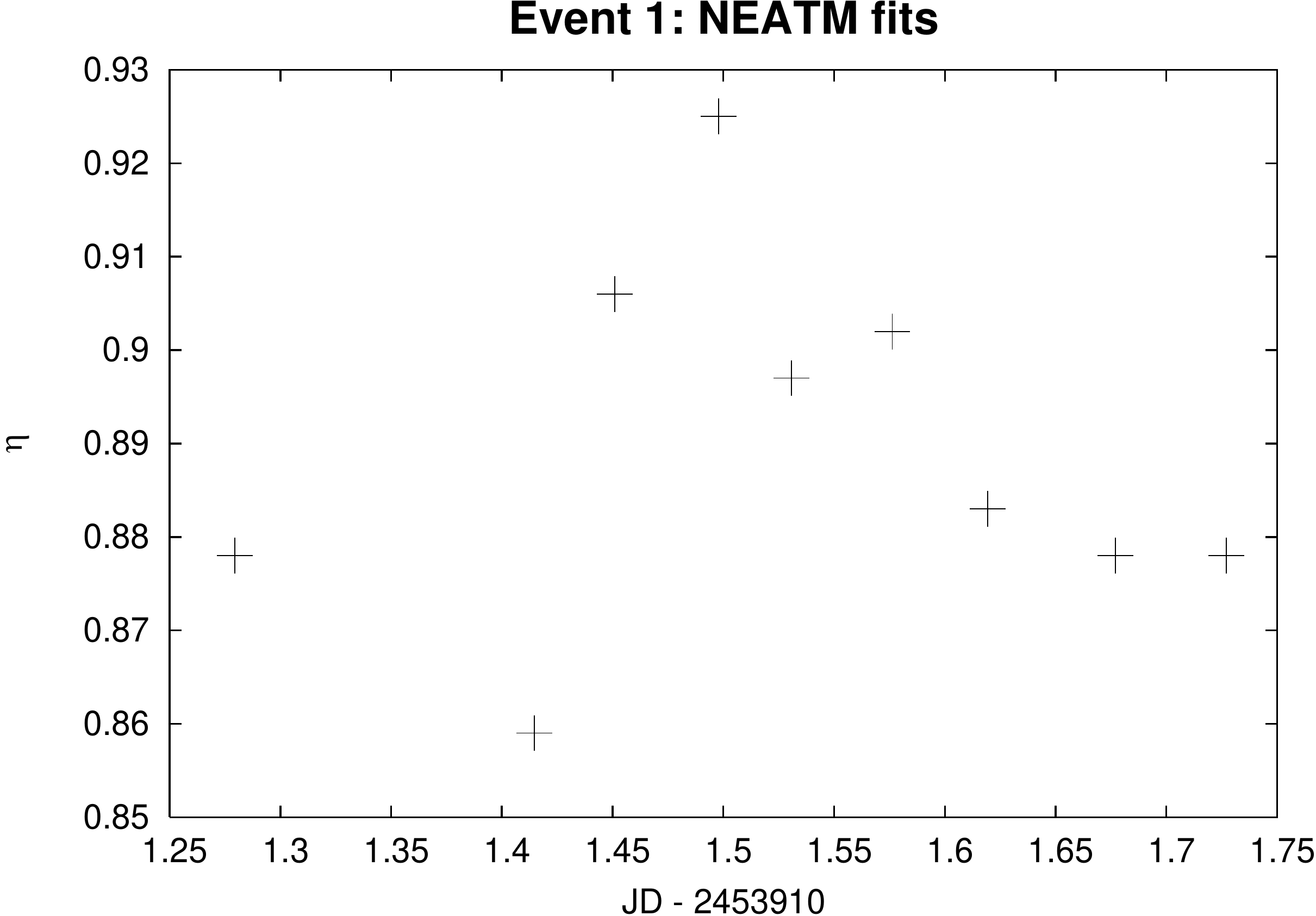}
  \caption[(617) Patroclus: NEATM fits to spectra for event 1.]{(617) Patroclus: NEATM fits to spectra for event 1. Above: Best-fit effective diameter against observing time. Below: Best-fit $\eta$ against observing time.}
  \label{fig:Patroclus:NEATM}
\end{figure}

As a first step, the obtained spectra were analyzed using the NEATM \seesect{sect:NEATM}; see \figref{fig:Patroclus:NEATM} for results from the nine observations of event 1, results for event 2 are qualitatively identical.
Observations 1.7 and 1.8, which were obtained after the  end of the event, imply an 
effective diameter around \unit{147}{\km} and $\eta\sim0.88$, intermediate between previous estimates by \citet{SIMPS} and \citet{Fernandez2003}.
During the events, there is a  clearly recognizable dip in best-fit diameter which reflects the event-induced flux drop.
Simultaneously, the best-fit $\eta$ rises (with the exception of observation 1.1), corresponding to a lower apparent color temperature
due to the eclipse-induced cooling of the shadowed parts.

While the NEATM takes no direct account of shadowing or occultation, 
the NEATM results clearly indicate that we have indeed observed the thermal response to mutual events.

\subsection{Thermophysical modeling}
\label{sect:Patroclus:modeling}

In order to obtain a reliable estimate of the thermal inertia and diameter, the data were analyzed using a generalized thermophysical model (TPM)
(see \chaptref{chapt:TPMconcave} in the appendix).
In the modeling of shadowing effects and thermal inertia, the spin axis position \citep{Marchis2006} and the observing geometry \seetablepage{table:Patroclus:obsgeometry} are explicitly taken into account.
Thermal-infrared beaming is modeled in terms of a ``beaming factor'' $\eta$ \seesect{sect:TPMconcave:beaming} which is not to be confused with the NEATM $\eta$ (the latter accounts for the \emph{combined} effect of thermal inertia and beaming, rather than for beaming alone).
The Patroclus system is assumed to be tidally locked and on a circular mutual orbit, such that the system is at rest in a co-rotating frame \seesect{sect:TPMconcave:binary}; this is consistent with all available data (see \sectref{sect:Patroclus:intro}---the orbital eccentricity of $0.02\pm0.02$ is negligible).
Both components are assumed to be spherical and homogeneous in terms of mass density, albedo, roughness, and thermal inertia.

The timing uncertainty, which was estimated to be $\pm\unit{30}{\minute}$ before the Spitzer observations, 
is non-negligible compared to the event duration of $\sim\unit{6.5}{\hr}$, 
requiring a non-standard fit technique to be used:
Variable model parameters are the time offset $\Delta t$,  thermal inertia $\Gamma$, beaming parameter $\eta$, and area-equivalent diameter $D_\Area$ (see \sectref{sect:thermal:size-albedo} for a definition and \eqrefpage{eq:TPMconcave:rescale}).
Patroclus' albedo is very low, $\pv\sim0.04$, so its absorptivity (on which temperatures depend) is close to unity.
Consequently,  temperatures are virtually independent of the precise value of \pv\ \seefigpage{fig:beaming:Bond}. Model fluxes were calculated for $\pv=0.041$ (corresponding to $D_\Area=\unit{151.054}{\km}$ assuming $H=8.19$) and later rescaled with a spectrally constant factor $\kappa$ (see below) to vary the diameter.

Synthetic lightcurves were generated for the wavelengths at which data had been obtained (disregarding the position of spectral features; see above) and for different values of thermal inertia and $\eta$.
Fluxes were calculated for 1000 time points per revolution, corresponding to a time resolution of $\sim\unit{6}{\minute}$ or roughly the length of one Spitzer observation.
After initial runs over wider and coarser grids of $\Gamma$ and $\eta$ values, 
an equidistant grid was considered with $\Gamma$ varying between 14 and \unit{110}{\TIunit} (step width  \unit{3}{\TIunit}) and  $\eta$ between 0.56 and 0.95 (step width  0.03), totaling to 462 $\Gamma$--$\eta$ combinations.

Data were analyzed separately for each event.
To this end, 
 model fluxes were interpolated for
the times of the respectively nine observations plus a variable time offset $\Delta t$. 
Initial runs showed the 
required range in $\Delta t$ to be $-1.5$---\unit{0}{\hr} (i.e.\ events occured somewhat earlier than predicted), the resolution of the $\Delta t$ grid was \unit{2}{\minute}.
For each combination of $\Gamma$, $\eta$, and $\Delta t$, a flux-scale-factor $\kappa$ was found which  minimizes $\chi^2$
\seeeqpage{eq:NEATM:fitting}; $\kappa$ and the corresponding $\chi^2$ were stored.
After searching the grid, that combination of $\Gamma$, $\eta$, and $\Delta t$ was determined which lead to the global minimum in $\chi^2$. 
The corresponding best-fit diameter equals $\unit{151.054}{\km} \times \sqrt{\kappa}$  (fluxes are proportional to $D^2$).

In order to study the accuracy of the results,  a Monte-Carlo technique was employed: 
For each event, $9\times5000$ random spectra were generated, 
normally distributed around the measured data.
Best-fit parameters were determined for each  set of nine random spectra;
see \tableref{table:Patroclus:results} for their mean values and standard deviations.

%
\begin{table}
\caption[(617) Patroclus: Best-fit TPM parameters for events 1 and 2.]{(617) Patroclus: Best-fit TPM parameters for events 1 and 2 (average and standard deviation of 5000 Monte-Carlo runs; see text). $\Delta t$ is given in units of \hr, thermal inertia in units of \TIunit, $D_\Area$ in \km. The total number of data points per event is $9\times179=1611$, hence the reduced \chitwo\ equals   $\chi^2/1607$ (there are four fit parameters).}
\label{table:Patroclus:results}
  \centering
  \begin{tabular}{r|rrrrr}
\toprule 
 & Reduced \chitwo & $\Delta t$ & $\Gamma$ & $\eta$ & $D_\Area$ \\
\midrule
Event 1 & $14.2\pm0.2$& $-0.47\pm0.02$ & $89\pm3$ & $0.62\pm0.00$ & $149.5\pm0.4$ \\
Event 2 & $5.6\pm0.2$ & $-1.02\pm0.02$ & $90\pm4$ & $0.62\pm0.01$ & $144.6\pm0.4$ \\
\bottomrule
  \end{tabular}
\end{table}

We note that the standard deviations of the best-fit values are much lower than realistic expectations on the accuracy of our results, particularly so for $\eta$. 
This may indicate an underestimation of the flux uncertainty, which would also be consistent with the fact that our values for the reduced \chitwo\ significantly exceed unity.
Alternatively, 
we may be unable to resolve the true uncertainties 
if they are comparable to the resolution of the search grid used by us.
A detailed analysis of the uncertainties is 
deferred to a later stage, when the improved orbit model will be available.
Keeping this in mind, we conclude that the best-fit $\Gamma$, $\eta$, and $D_\Area$ for the two events are in excellent agreement with one another.
The two values for the best-fit diameter average to \unit{147}{\km}, in excellent agreement with the NEATM analysis \seesect{sect:Patroclus:NEATM}.
The best-fit thermal inertia  is  $90\pm\unit{4}{\TIunit}$, where the uncertainty is likely to be underestimated.
The best-fit TPM $\eta$ is 0.62, significantly below the best-fit NEATM $\eta$. This is consistent, given the fact that the TPM $\eta$ only reflects the effect of beaming, rather than the combined effect of thermal inertia and beaming as its NEATM counterpart.
The found $\eta$ value suggests a very rough surface, probably even rougher than the lunar surface \citep[for which $\eta\sim0.72$;][]{Spencer1989}.

Results for the best-fit time offset $\Delta t$ for the two events  differ at a statistically significant level:
Apparently, both events took place earlier than predicted, by $\sim\unit{0.5}{\hr}$ in the case of event 1, $\sim\unit{1.0}{\hr}$ in the case of event 2.
The discrepancy of \unit{0.5}{\hr} corresponds to a rotational phase $<\unit{2}{\degree}$.
We note that the orbit model by \citet{Marchis2006} has a slight eccentricity of $0.02\pm0.02$ which is neglected in our thermophysical modeling.
We speculate that this caused the slight discrepancy in timing offsets.

Additional simulation runs were performed in which the components were assumed to be biaxial ellipsoids rather than spheres.
Their longest axes were assumed to be aligned with one another and with the line connecting the two components; this is the stable attractor state of such a system.
Axis ratios (identical for the two components) of 1.03, 1.06, and 1.09 were tried, leading to very similar results compared to the spherical case. Larger axis ratios would appear to be inconsistent with the observed low lightcurve amplitude.

\subsection{Discussion}
\label{sect:Patroclus:discussion}


\paragraph{Diameter and mass density}

Two independent analyses of the data (using the NEATM and the TPM) resulted in a best-fit diameter of
$D_\Area\sim\unit{147}{\km}$; the corresponding albedo is $\pv=0.0433$.
This is intermediate between previous estimates by 
\citet{SIMPS} and \citet{Fernandez2003}. 
Note that in the derivation of both, the apparent color temperature (effectively: $\eta$) was not derived but rather assumed. 
Our result, on the other hand, is based on data with broad wavelength coverage, allowing us to determine the color temperature reliably.
Consequently, our  results would be expected to be more reliable.

The systematic diameter uncertainty inherent in the TPM is hard to estimate in a quantitative way.
In the case of near-Earth asteroids, it was found not to exceed \unit{10}{\%} \seesect{sect:NEA:D}.
Thermophysical modeling of Patroclus, with roughly spherical  components and  observed at low phase angles $<\unit{10}{\degree}$,
would appear to be much less challenging (apart from the effect of mutual events) and potentially more accurate.
We conclude that \unit{10}{\%} is a conservative upper limit on the diameter uncertainty.

The diameters of the components are $100\pm10$ and $108\pm\unit{11}{\km}$, respectively.
This corresponds to a total volume of $(1.18\pm0.36)\times\unit{$10^{6}$}{\power{\km}{3}}$, implying an average mass density of $1.15\pm\unit{0.37}{\gramm\usk\power{\cm}{-3}}$.
This is consistent at the $1\sigma$ level with the previous estimate by \citet{Marchis2006}. 
In particular, it seems that the accuracy of the \citet{Fernandez2003} diameter estimate was overestimated.

We wish to stress the importance of  accurate 
diameter measurements of binaries:
Since the mass density is linear in mass but inversely cubic in diameter, mass densities are particularly susceptible to diameter uncertainties; 
e.g., a diameter uncertainty of \unit{15}{\%} translates into a mass-density uncertainty around \unit{45}{\%}.
For this reason, the uncertainty in mass density is typically limited by diameter inaccuracies \citep[see][]{Merline2002,Richardson2006}.

\paragraph{Thermal inertia}
We derive a thermal inertia of $\sim\unit{90}{\TIunit}$ for the Patroclus system. 
The uncertainty is hard to estimate in a quantitative way, but the good mutual agreement between results for the two events and the very low scatter found in a Monte-Carlo analysis \seetable{table:Patroclus:results} are reassuring.

As expected, the measured thermal inertia is much below that of bare rock (\unit{2500}{\TIunit}) implying the existence of a thermally insulating layer of regolith.
However, our result is significantly larger than the value for lunar regolith (\unit{50}{\TIunit}; see, e.g., \tablerefpage{table:thermalproperties}) or for large main-belt asteroids, which average around \unit{15}{\TIunit} \citet{MuellerLagerros1998}.
This suggests a relatively coarse regolith.

It must be taken into account that the heliocentric distance of Patroclus is $r\sim\unit{6}{\AU}$, hence its surface is much colder than, e.g., that of the Moon. In general, thermal inertia is temperature dependent; for purely radiative heat conduction (as expected for fine regolith) thermal inertia would be expected to scale with $r^{-3/4}$ \seesect{sect:NEA:TI:trend}.
Under this assumption, the thermal inertia of Patroclus at a heliocentric distance of \unit{1}{\AU} would be around $\unit{350}{\TIunit}$, comparable to our findings for the typical thermal inertia of near-Earth asteroids \seesect{sect:NEA:props}.
While this appears to be somewhat surprising,
an analysis of spectral features found in our data also indicates the presence of a coarse regolith
(Emery, 2007, private communication).

\begin{table}
\caption{Small Solar-System bodies beyond the asteroid main belt: Overview of previously published estimates of thermal inertia. KBO is for Kuiper belt object.}
\label{table:Patroclus:TI}
  \centering
  \begin{tabular}{r|ll}
\bottomrule
      & Thermal inertia & Reference \\
      & (\TIunit)       & \\
\midrule 
(617) Patroclus (Trojan) & $\sim 90$ & This work \\
(2363) Cebriones (Trojan) & $<14$ & \citet{Fernandez2003}\\
(3063) Makhaon  (Trojan)  & $<30$ & \citet{Fernandez2003}\\
Ganymede (Jovian satellite) & $\sim70$ & \citet{Spencer1987} \\
Callisto (Jovian satellite) & $\sim50$ & \citet{Spencer1987} \\
Europa (Jovian satellite) &45--70 & \citet{Spencer1999} \\
(8405) Asbolus  (Centaur) & $<11$ & \citet{Fernandez2002} \\
(2060) Chiron  (Centaur)& $3^{+5}_{-3}$    & \citet{Groussin2004} \\
(10119) Chariklo (Centaur)& $0^{+2}_{-0}$    & \citet{Groussin2004} \\
(55565) 2002 AW197 (KBO)&$<20$  & \citet{Cruikshank2005}\\
(134340) Pluto (KBO) & 30--50 &  \citet{Lellouch2006} \\
\bottomrule    
  \end{tabular}
\end{table}

Our thermal-inertia result is significantly above published upper limits on the thermal inertia of two other Trojans \citep[see also \tableref{table:Patroclus:TI}]{Fernandez2003}.
We note that the latter are based on a significantly less extensive database and were obtained using an indirect method:
\citeauthor{Fernandez2003}\  found their data to be more consistent with the STM than with the FRM \seesect{sect:thermal:STM-FRM}.
From this they conclude that the thermal parameter \seeeqpage{eq:def_thermalparameter} should be $\leq 1$ (no quantitative discussion is given) which they transform into an upper limit on thermal inertia.
In their analysis, the effect of thermal-infrared beaming is neglected; at the small phase angles at which Trojans are observed, beaming increases the apparent color temperature while thermal inertia  lowers it.
Moreover, the spin axis orientation of their targets is unknown, they appear to assume equatorial aspect.
Both beaming and non-equatorial aspect reduce the observable effect of thermal inertia, and would therefore increase thermal-inertia estimates derived from any given data set.
The upper limit published by \citet{Fernandez2003} is therefore methodologically unreliable.
While it remains unclear how representative Patroclus is of the Trojan population as a whole, 
we conclude that the typical thermal inertia of Trojans appears to be much larger than previously thought.

It is instructive to compare our result to thermal-inertia estimates for other atmosphereless bodies in the outer Solar System, although we caution that their surface composition may be very different from that of Trojans.

Since thermal parameters are temperature dependent (see above), 
the probably best analogues are the satellites of Jupiter, with which the Trojans are  co-orbital. 
Using close-up spacecraft observations in the thermal infrared, 
\citet[using Voyager data]{Spencer1987} and \citet[using Galileo data]{Spencer1999} determined the thermal inertia of three Galilean satellites, superseding previous estimates by \citet{MorrisonCruikshank1973}.
Their results \seetable{table:Patroclus:TI} are only slightly lower than our result, hence the surface of Patroclus may be expected to resemble those of Galilean satellites. We caution that the dynamics of ejecta, which are likely to be crucial for the formation of regolith, are very different on planetary satellites compared to asteroids (see also \sectref{sect:NEA:Mars}). Furthermore, Jovian satellites as opposed to Trojans are significantly influenced by absorption of Jovian emission.

There are  published estimates of the thermal inertia of three Centaurs \seetable{table:Patroclus:TI}. Centaurs are icy planetoids which orbit the Sun between the orbits of Jupiter and Neptune, at typical heliocentric distances between 5 and \unit{30}{\AU}. Some Centaurs display cometary activity near perihelion \citep[e.g.][]{Tholen1988}.
The upper limit on the thermal inertia of Asbolus reported by \citet{Fernandez2002} is based on an analysis analogous to that by \citet{Fernandez2003}  (see above) and would therefore seem equally unreliable.
Assuming a spin axis perpendicular to the respective orbital plane, 
\citet{Groussin2004} estimate the thermal inertia of the nuclei of the active Centaurs (2060) Chiron and (10119) Chariklo.
Some observational evidence is provided for the assumed equatorial aspect in the case of Chiron, but not for Chariklo.
We conclude that \citeauthor{Groussin2004}'s thermal-inertia estimate for Chiron appears to be reliable 
while that for Chariklo seems unreliable.
It is intriguing that the thermal inertia of a Centaur should be so much lower than Patroclus'.

\citet{Cruikshank2005} report 
Spitzer observations of the 
Kuiper belt object (55565) 2002~AW197 at a heliocentric distance of \unit{47.15}{\AU}.
They used the \citet{Spencer1990} TPM to derive an upper limit of \unit{8.7}{\TIunit} on this object's thermal inertia assuming equatorial aspect and an upper limit on the rotation period of \unit{154}{\hr}. 
Unfortunately, nothing is known about the pole orientation of this object, not even its rotation period is known. Nevertheless, \citeauthor{Cruikshank2005}\ conclude that the thermal inertia of 2002~AW197 is likely to be ``well below \unit{20''}{\TIunit}.
Assuming thermal conduction to be dominated by radiative heat transfer, the corresponding thermal inertia at a heliocentric distance of \unit{1}{\AU} would be $<\unit{360}{\TIunit}$, again comparable to our results for near-Earth asteroid and Patroclus.


The thermal inertia of Pluto has been measured by \citet{Lellouch2006} based on extensive Spitzer observations \citep[refining a previous estimate by][which was based on ISO observations]{Lellouch2000}.
Pluto's spin state and shape are well known, hence we would consider the results of \citeauthor{Lellouch2006}\ reliable.
Note, however, that the thermal conduction on Pluto would be expected to be enhanced by its thin atmosphere, hence its thermal inertia cannot be readily compared to that of 
 atmosphereless bodies.

\paragraph{Model assumptions}

The thermophysical modeling is based on the  orbit model by \citet{Marchis2006} 
which is currently being refined \seesect{sect:Patroclus:intro}.
The slight orbital eccentricity of $0.02\pm0.02$ is neglected, i.e.\ the mutual orbit is assumed to be circular.
Furthermore, the system is assumed to be tidally locked, which is consistent with all observational data known to us.

We caution that the spin states of the components are presently not well constrained.
Furthermore, our results indicate that the orbital eccentricity has an influence on the event timing and may have an influence on the event geometry (although results from the mutually independent analyses of the two events are in otherwise excellent agreement).
Orbital eccentricity induces a slight time variability in the orbital distance of the two components.
An improved modeling approach, in which orbital eccentricity would be included to first order, would be to use two different models of the systems for the two events: In each, the mutual orbit would remain circular (as required by our current TPM, see \sectref{sect:TPMconcave:binary}) but the orbital distance would be different for the two events; those distances would need to be determined from the future refined orbit model.

Both components are assumed to have a spherical shape.
This is expected to be a fair approximation given the reportedly low lightcurve amplitude at epochs when no mutual events occur. The TPM allows for ellipsoidal shapes, first model simulations with different axis ratios were seen to result in virtually identical results. 

Mass density, albedo, surface roughness, and thermal inertia are assumed to be homogeneous over the Patroclus system.
While this is a nontrivial assumption, spatially resolved observations of the two components through different filters reveal a roughly constant brightness ratio \citep{Merline2001,Marchis2006}, consistent with our assumption.
It is also supported by the good mutual agreement of the independent analyses of the two observed events: Although the system was not spatially resolved, the observed eclipse-induced system would be expected to depend primarily on the eclipsed component. Vast component-to-component differences in physical properties would be expected to lead to inconsistencies which were not observed.

The approximate modeling of thermal-infrared beaming is expected to be uncritical, given the small phase angle ($<\unit{10}{\degree}$) at which our observations took place.


\subsection{Summary}
\label{sect:Patroclus:summary}

We report the first thermal-infrared observations of an eclipsing binary asteroid system, the Trojan (617) Patroclus.
A total of 18 thermal spectra ($\sim8$--\unit{33}{\micron})
were  obtained using the IRS on board the Spitzer Space Telescope, providing good temporal coverage of two mutual events but no spatial resolution.
The data were analyzed using a  thermophysical model in which the effects of the eclipse-occultation geometry, thermal conduction, and thermal-infrared beaming are taken into account.

We derive the first reliable estimate of the thermal inertia of a Trojan, $\Gamma\sim\unit{90}{\TIunit}$. This is comparable to the thermal inertia of Galilean satellites and indicative of a relatively coarse regolith.
However, our result is much larger than previous estimates of the thermal inertia of two Trojans which we found to be methodologically unreliable.

The diameters of the two components have been  determined to be
$100\pm10$ and $108\pm\unit{11}{\km}$, respectively; the accuracy of previously available estimates was overestimated. Our result implies an average mass density of $1.15\pm\unit{0.37}{\gramm\usk\cm\rpcubed}$ \citep[using the mass estimate by][]{Marchis2006}; the quoted uncertainty is conservative.


%% file: discuintro.tex
Our main result is the derivation of the thermal inertia of 5 near-Earth asteroids (NEAs) from analysis of extensive sets of thermal-infrared data using a detailed thermophysical  model (TPM).
This  is the only well-established method to measure the thermal inertia of asteroids, hence it is difficult to gage the reliability of our results in a direct way.
Indirect validation comes from studies of the physical consistency of the TPM \seesect{sect:discussion:TPM} and from studies of the consistency and accuracy of TPM-derived diameter estimates  \seesect{sect:NEA:D}.
In particular, TPM-derived diameters are found to be in excellent agreement with diameter estimates obtained using other techniques including spacecraft rendezvous, which is a valuable result in its own right.
The core section of this chapter is \sectref{sect:NEA:TI}, in which the thermal inertia of NEAs is discussed in the context of previously available information.  


%% file: discussion.tex
 \section{Thermophysical model (TPM)}
         \label{sect:discussion:TPM}

Our TPM takes explicit account of the effects of irregular shape, spin pole orientation, surface roughness, and thermal inertia. 
The model code described in \chaptref{chapt:TPM} allows for globally convex shapes, generalizations to non-convex shapes are under development (see \chaptref{chapt:TPMconcave} in the appendix).
As will be discussed in the following, our model has been shown to be applicable to near-Earth asteroids (NEAs) which are more challenging to model than larger objects such as main-belt asteroids (MBAs).
We conclude that our TPM is applicable to all asteroids.

         \subsection{Comparison with the Lagerros model}
         \label{sect:discussion:TPM:comparison}

Thermophysical processes are modeled following
\citet{LagerrosI,LagerrosIII,LagerrosIV}, who proposed the most realistic asteroid TPM currently available \seesect{sect:TPM:overview}.
A minor improvement of \citeauthor{LagerrosI}'s modeling is proposed in \sectrefpage{sect:beaming:fluxes}, where we provide an analytic expression for the multiple scattering of observable thermal flux inside craters to all orders; \citet{LagerrosIV} only considers direct emission and single scattering. The effect thereof, however, was found to be negligible for reasonable emissivity values \seefigpage{fig:beaming:Lagerros_all}.
\citet{LagerrosIV} proposed two different ways of modeling thermal conduction inside craters; we have only implemented the numerically more advantageous version, which is potentially less physical \seesect{sect:beaming:approximation}. 
\citeauthor{LagerrosIV} found good agreement between the numerical outcome of both versions for MBA parameters;
see \sectref{sect:discussion:TPM:consistency} and \ref{sect:discussion:TPM:othermodels} for further discussion in the context of NEAs.

The numerical design and implementation of our TPM code is fully independent of Lagerros'.
Numerical evaluation of partial differential equations and integrals 
inevitably involves discretization and truncation, which introduce numerical noise.
In the choice of discretization and truncation parameters, the required numerical accuracy must be weighed against the numerical effort. 
Parameters must be chosen fine enough to guarantee physical output for the purpose at hand, but not too fine in order to avoid excessively long computer run times.

No detailed information on the numerical implementation of Lagerros' model is publicly available.
It is clear, however, that numerical efficiency was an important design criterion in the implementation of
his model \citep[see][chapt.\ 3]{LagerrosThesis},
which
was primarily aimed at application to MBA data.
The typical thermal inertia of MBAs is very low, comparable to that of lunar regolith, and ground-based observations of MBAs are restricted to phase angles not largely exceeding \unit{30}{\degree}.
For these circumstances, the Lagerros model is reportedly very accurate and is used, e.g., for the calibration of space telescopes \citep{MuellerLagerros1998,MuellerLagerros2002}.

Our TPM code, on the other hand, has been
 designed and tested to be applicable to  NEAs, i.e.\  for a thermal inertia up to that of bare rock (\unit{2500}{\TIunit}) and for very large phase angles, when thermal emission emanating from large portions of the non-illuminated side is observable.
We found that a TPM code for NEAs must be designed in a way which is numerically much more expensive than for MBAs \seesect{sect:discussion:TPM:consistency}.

We conclude that our TPM code represents the first detailed TPM shown to be applicable to NEAs. Such a model is required for the determination of their thermal inertia, which is the primary aim of this thesis.

\subsection{Internal consistency}
\label{sect:discussion:TPM:consistency}

As a first step, it was
carefully verified that the output of the TPM code is internally consistent \seesect{sect:TPMconvex:validation}:
Model fluxes were seen to be in agreement with qualitative expectations (see, e.g., \figrefpage{fig:beaming:Sun@zenith} or \figrefpage{fig:thermal:TI}).
Furthermore, model fluxes were found to conserve energy provided that sufficiently fine discretization parameters are chosen. 
In particular, we found the required numerical effort (expressed in terms of a fractional accuracy goal) to increase considerably with increasing thermal inertia. We chose numerical parameters such that model fluxes conserve energy to within a few percent for the thermal inertia of bare rock (\unit{2500}{\TIunit}), better for lower thermal inertia.
Our approximate treatment of thermal conduction inside craters \seesect{sect:beaming:approximation} was found to lead to physically consistent flux values \seesect{sect:TPMconvex:validation:beaminginertia}.

\subsection{Consistency with other models}
\label{sect:discussion:TPM:othermodels}

We found TPM-generated synthetic flux values to agree with expectations based on experience with the NEATM \citep[see also \sectref{sect:NEATM}]{NEATM}.
Studies similar to those presented by \citet[chapt.\ 6]{Delbo2004} have been performed, where TPM-generated synthetic fluxes have been fitted using the NEATM.
In those studies, we found that the input diameter was reliably retrieved  and that the dependence of the model parameter $\eta$ on thermal inertia and roughness was as expected, i.e.\ that increasing thermal inertia increases $\eta$ while roughness decreases $\eta$ for low solar phase angle $\alpha$ and increases $\eta$ for large $\alpha$.

TPM results from fits to NEA data (see sections \ref{sect:Eros}--\ref{sect:WT24}) were generally found to be consistent with results obtained using simpler models. In the case of (1580) Betulia \seesect{sect:Betulia}, 
the TPM-derived diameter is $\sim\unit{25}{\%}$ larger than its NEATM-derived counterpart, barely consistent at the combined $1\sigma$ level; the TPM result has later been supported through radar-derived estimates.
In our study of the NEA (33342) 1998~WT24 \seesect{sect:WT24},
TPM-generated fluxes were seen to agree with the output of an independently developed, less detailed TPM.
In our study of the NEA (25143) Itokawa \seesect{sect:Itokawa}, we have, among other things, reanalyzed a data set which had previously been analyzed using \citeauthor{LagerrosIV}' model. Good  mutual consistency of the results was found
implying, in particular, that our approximate treatment of thermal inertia inside craters \seesect{sect:beaming:approximation} is uncritical at least for a thermal inertia up to \unit{700}{\TIunit}.

\subsection{Consistency of results with ``ground truth''}
        \label{sect:discussion:TPM:groundtruth}

Two of the NEAs studied by us, Eros \seesect{sect:Eros} and Itokawa \seesect{sect:Itokawa}, have been rendezvoused by spacecraft. In both cases, our TPM-derived diameter estimates are in excellent agreement with spacecraft results (see \sectref{sect:NEA:D} for a detailed discussion).
No independent technique for measuring thermal inertia has been established so far, hence it is more difficult to gage the accuracy of our thermal-inertia results. At least in the case of Eros, however, our results are in excellent agreement with qualitative expectations based on a geological interpretation of the surface makeup, while our results for Itokawa are subject to discussion \seesect{sect:NEA:TI}.
In the case of the MBA Lutetia \seesect{sect:Lutetia}, ground truth will become available after the Rosetta flyby in 2010.

\subsection{Model applicability}
        \label{sect:discussion:TPM:constrain}

Compared to simpler models, TPMs inevitably contain a larger number of free parameters,
such that a larger set of thermal-infrared data is required to constrain the model parameters in a meaningful way.
Furthermore, some knowledge on the global shape and spin state is required, which must be obtained using other techniques.

Typically, very little is known about NEA shapes and spin states \seesect{sect:intro:shape}, while thermal-infrared observations are notoriously difficult \seesect{sect:thermal:observability}.
Therefore, one has little choice but to use a ``simple'' thermal model
for the analysis of thermal-infrared observations of all but the best studied asteroids. 
This has imposed significant difficulties on studies of the thermal inertia of NEAs so far.

There is, however, an ever-increasing number of  objects with known shape and spin state, including many NEAs; optical telescope systems which are currently being built promise to increase their number by the thousands in the near future \citep{Durech2005}.
At the same time, progress in infrared detector technology and the sophistication of modern space telescopes including the Spitzer Space Telescope \seechapt{chapt:SST} render high-quality spectrophotometric or spectroscopic thermal-infrared observations of  faint asteroids  much more feasible than in the past.

\paragraph{Required data quality}
We have found that 
determination of thermal inertia imposes generally more stringent requirements on the data quality than estimation of diameter and albedo.
This is nicely exemplified in our analysis of  four different sets of thermal-infrared data of (25143) Itokawa \seesect{sect:Itokawa}, where even the two data sets which did not significantly constrain the thermal inertia allowed the diameter to be determined to within \unit{10}{\%} of the Hayabusa result or better.

\paragraph{Accuracy of thermal-inertia results}
The analysis of thermal-infrared observations using a TPM is the only 
 established technique to measure the thermal inertia of individual asteroids.
The use of simpler models entails significant systematic uncertainties; see, e.g., previous estimates of the thermal inertia of Itokawa (\sectref{sect:Itokawa}) or Betulia (\sectref{sect:Betulia}), which were found to be flawed.
While this fact underlines the importance of our studies, it also prohibits methodologically independent cross-checks on our thermal-inertia results.
Our diameter and albedo results are in excellent agreement with estimates obtained using other techniques such as spacecraft imaging \seesect{sect:NEA:D}, inspiring trust into the accuracy of our thermal-inertia results. 
Note that in the case of (1580) Betulia \seesect{sect:Betulia} our results for diameter and albedo were in contrast with previous estimates dating from the 1970s. However, after the publication of our results \citep{Betulia}, our results were independently confirmed based on reanalyses of old data and newly obtained radar observations \citep{Magri2007}.

\subsection{One-dimensional heat conduction}
\label{sect:discussion:TPM:onedimensional}

Our model neglects the effects of lateral heat conduction. 
This is justified for shape models where the typical linear dimensions of facets are large compared to the thermal skin depth, which ranges between millimeters up to $\sim\unit{50}{\cm}$ depending on the surface material.
We would therefore expect our TPM code to be applicable to all shape models with a resolution at the meter scale or coarser, comprising practically all currently available asteroid shape models.%
\footnote{ An exception may be high-resolution version of the Hayabusa-derived shape model of  Itokawa with more than 3 million facets over a body with an effective diameter around \unit{0.32}{\km}, corresponding to an average facet size around \unit{0.1}{\metre\squared}.}

Any realistic shape model for very small objects below some \unit{10}{\metre} in diameter must have very small facets, such that our TPM would not be applicable to such small objects.
No information on the shape and spin state of such small objects is currently available.

\section{Accuracy of TPM-derived diameter estimates}
        \label{sect:NEA:D}

The accuracy of TPM-derived diameter estimates for asteroids 
 depends critically on the quality and extensiveness of the available thermal-infrared database and on the quality of the used model of the asteroid shape and spin state.
However, the assumptions and approximations made in the thermophysical modeling add to the error budget.
We expect our modeling to be much more realistic than that of ``simple'' thermal models and would therefore expect considerably smaller systematic modeling uncertainties. The latter may nevertheless be the dominant source of uncertainty if
both an extensive set of high-quality data and an accurate shape model are available.

Generally speaking, the requirements for a reasonably accurate determination of the diameter are less stringent than for constraining the thermal inertia
\seesect{sect:discussion:TPM:constrain}%
---this is the physical basis of the applicability of ``simple'' thermal models.

\paragraph{NEAs rendezvoused by spacecraft}
To study the systematic diameter uncertainty inherent in our modeling, we found it instructive to 
determine the diameter of 
those two NEAs whose diameter, shape, and spin state have so far been determined very accurately during a spacecraft rendezvous, namely (433) Eros and (25143) Itokawa.
We have analyzed thermal-infrared observations of both asteroids (the data were partially obtained by us) using our TPM.
Our model neglects the effect of concavities beyond the size scale of single facets (i.e.\ beaming due to cratering)
although major global-scale concavities are present in the shape of both asteroids (see, e.g., \figrefpage{fig:Eros:shape} and \figrefpage{fig:intro:ito}).
Most thermal-infrared observations of these objects were obtained at moderate solar phase angles around \unit{30}{\degree}, although our Itokawa database contains 5 data points obtained at \unit{108}{\degree}.
Shadowing effects due to large-scale surface profile would be expected to be more important at larger phase angles, potentially leading to large diameter uncertainties.

Nevertheless, our diameter estimate for Eros is 
 within \unit{$\sim5$}{\%} of the spacecraft-derived result \seesect{sect:Eros}. Our best estimate of Itokawa's diameter \seesect{sect:Itokawa} is within the \unit{2}{\%} uncertainty range of the spacecraft-derived value.
While these two studies alone do not allow one to constrain the systematic uncertainty in a statistically significant way, 
they do show that TPM-derived diameters are potentially very accurate.
A fractional diameter uncertainty of \unit{10}{\%} appears to be a conservative upper limit on the systematic  uncertainty inherent in our  TPM.
We speculate that the actual diameter accuracy may  be better, but more than two ``ground truth'' values would be required to confirm this.
Upcoming space missions to NEAs \seesect{sect:intro:spacecraft} are expected to provide further accurate diameter estimates, which will allow our analysis to be refined.

\paragraph{Large phase angle}
One may expect the diameter accuracy to 
decrease with increasing solar phase angle at which observations have been obtained.
Our study of the NEA 
1998~WT24 \seesect{sect:WT24} 
is reassuring in this respect,
where all thermal-infrared data were
obtained at phase angles above \unit{60}{\degree}. Several independent estimates of the object's diameter are available \seetable{table:WT24:D}, all of which are consistent with our result at the \unit{10}{\%} level despite the large phase angle.

\paragraph{Radar}
The analysis of radar echoes of asteroids is a well established technique to measure their diameters \citep[see][]{Ostro2002} that is completely independent of ours.
Radar-derived diameters of our targets Itokawa, Betulia, YORP, and Lutetia have been published (see the discussion in sections \ref{sect:Itokawa}, \ref{sect:Betulia}, \ref{sect:PH5}, and \ref{sect:Lutetia}, respectively).
In each case, diameter estimates are mutually consistent 
at the combined $1\sigma$ level.
However, the nominal radar-derived diameters are consistently \unit{$\sim15$}{\%} larger than the nominal TPM-derived diameters.
It remains to be studied whether or not there is a statistically significant diameter difference.
In the case of Itokawa, the Hayabusa result nicely confirms our estimate but is $\sim 1.3 \sigma$ below the nominal radar-derived result by \citet{Ostro2005}.
No such ``ground truth'' is available for Betulia and YORP;
we note that these objects display concavities which are relatively larger than on Itokawa, potentially lowering the accuracy of our results in their case \seesect{sect:NEA:concave}.
In the case of Lutetia, ground truth will become available after the Rosetta flyby in 2010.

\paragraph{Flux calibration}
We note that diameter estimates based on thermal observations are in principle susceptible to uncertainties in the absolute flux calibration, which used to plague early thermal-infrared studies. 
Since the work of, e.g., \citet{CohenVII,CohenX} this uncertainty has been reduced drastically;
\citet{Reach2005}, e.g., report an uncertainty of \unit{3}{\%} in the absolute flux calibration of the IRAC camera  \seesect{sect:IRAC:general} on board the Spitzer Space Telescope.
Since asteroid flux levels scale with the projected  area, the corresponding diameter uncertainty is only \unit{1.5}{\%}.

 \section{Thermal inertia of NEAs}
         \label{sect:NEA:TI}
         \input{NEAprops}

%% file: NEAprops.tex
\begin{table}
  \caption[Summary of all thermal-inertia measurements of NEAs currently available.]{Summary of all thermal-inertia measurements of NEAs currently available. 
As detailed in \sectref{sect:Eros} and \ref{sect:Itokawa}, our thermal-inertia measurements of Eros and Itokawa are expected to supersede earlier determinations by \citet{LebofskyRieke1979} and \citet{MuellerItokawa}, respectively \citep[our Itokawa result is partially based on data reported in][]{MuellerItokawa}.
\citeauthor{LebofskyRieke1979}'s estimate of Eros' dimensions was converted by us
into a volume-equivalent diameter using  \eqref{eq:Deff}; the spacecraft-derived value is \unit{16.9}{\km}. 
}
  \label{table:NEA:TI}
  \centering
\small{
  \begin{tabular}{llll}
\toprule
NEA & Diameter & Thermal inertia & Source \\
     & (\km)    & (\TIunit) & \\
\midrule
(433) Eros & $17.8\pm1.8$ & $150\pm40$ & \Sectref{sect:Eros} \\
(25143) Itokawa & $0.32\pm0.03$ & $700\pm100$ & \Sectref{sect:Itokawa} \\
(1580) Betulia & $4.57\pm0.46$ & $180\pm50$ & \Sectref{sect:Betulia} \\
(54509) YORP & $0.092\pm0.010$ & 200--1200 & \Sectref{sect:PH5} \\
(33342) 1998 WT24 & $0.35\pm0.04$ & $200\pm100$ & \Sectref{sect:WT24} \\
(433) Eros & (24.8) & 140--280 & \citet{LebofskyRieke1979} \\
2002 NY40 & $0.28\pm0.03$ & 100--1000 & \citet{Mueller2004} \\
(25143) Itokawa & $0.32\pm0.03$ & $750\pm250$ & \citet{MuellerItokawa}\\
\bottomrule
  \end{tabular}
}
\end{table}

From TPM-analyses of extensive sets of thermal-infrared data, 
we have determined the thermal inertia of 5 NEAs, thus increasing the total number of NEAs with measured thermal inertia to 6.
For two of our targets,
we refine previously available estimates, no reliable estimates have been available for
the remaining three. The diameter range spanned by our targets is 0.1--\unit{17}{\km} \seetable{table:NEA:TI}.
%
%
As will be discussed in the following, our results allow the first firm conclusions to be drawn on the typical thermal inertia of NEAs.


        \subsection{Thermal-inertia results in context}
        \label{sect:NEA:props}

\paragraph{Previously available information}
Before the start of the work towards this thesis, the thermal inertia of a single NEA, (433) Eros, had been measured
\citep[see \sectref{sect:Eros} for a discussion]{LebofskyRieke1979}.
In the meantime, two more  thermal-inertia measurements of NEAs have been reported \citep{Mueller2004,MuellerItokawa}, 
based on the \citeauthor{LagerrosI} TPM (see also \sectref{sect:discussion:TPM:comparison}).
Some indirect and consequently unreliable inference on the thermal inertia of NEAs was drawn from results of simple thermal models (see \sectref{sect:intro:TI} for an overview). In the cases of (1580) Betulia \seesect{sect:Betulia} and (25143) Itokawa \seesect{sect:Itokawa}, we found such estimates to be flawed.

From a measurement of the Yarkovsky-induced orbital drift of the NEA (6489) Golevka (effective diameter \unit{$\sim0.53$}{\km}), \citet{YarkoGolevka} derive a thermal conductivity  around \unit{0.01}{\kappaunit} depending on the unknown bulk mass density of the object.
They note that values in excess of \unit{0.1}{\kappaunit} would also result in a good fit to their data.
The corresponding  thermal-inertia values are%
\footnote{ The  heat capacity assumed by \citet{YarkoGolevka} is not stated explicitly, but quoted by \citet{Bottke2006} to be \unit{680}{\joule\usk\power{\kg}{-1}\power{\kelvin}{-1}}. The quoted thermal conductivity corresponds to a  near-surface mass density of \unit{1700}{\kg\usk\power{\metre}{-3}}.}
\unit{$\sim100$}{\TIunit} and \unit{$>340$}{\TIunit}.
They rejected the latter as ``unrealistically high'', which appears unjustified in the light of our results \seetable{table:NEA:TI}.


\paragraph{Typical thermal inertia of NEAs}
The  weighted average of the 6 thermal-inertia values reported in \tableref{table:NEA:TI} (disregarding the superseded previously available estimates for Eros and Itokawa) is \unit{212}{\TIunit} (weighted by absolute uncertainty) 
or \unit{400}{\TIunit} (weighted by fractional  uncertainty).
We conclude that the typical thermal inertia of NEAs is 
moderately high, roughly \unit{300}{\TIunit}, with a significant scatter exceeding a factor of two.

Such values  are more than an order of magnitude above typical values for large MBAs \citep[around \unit{15}{\TIunit}, see][]{MuellerLagerros1998}, yet around an order of magnitude below that of bare rock on Earth \seetablepage{table:thermalproperties}. 
In particular, our results display an apparent trend of increasing thermal inertia with decreasing asteroid size \seesect{sect:NEA:TI:trend}.
See \sectref{sect:NEA:geology} for a discussion of implications on the surface structure of NEAs.

\paragraph{Yarkovsky effect}
Our thermal-inertia result is of immediate relevance for studies of the important Yarkovsky effect \seesect{sect:intro:Yarko} which is governed by thermal inertia.
Our results imply that 
for model calculations of the Yarkovsky or YORP effects
a typical thermal inertia  around \unit{300}{\TIunit} should be assumed, 
more than an order of magnitude higher than that derived for MBAs.
This corresponds to a thermal conductivity of 
\unit{$\sim0.08$}{\kappaunit} assuming a heat capacity of \unit{680}{\joule\usk\power{\kg}{-1}\power{\kelvin}{-1}} and a near-surface bulk density of \unit{1700}{\kg\usk\power{\metre}{-3}}.

The implications of our thermal-inertia result on the magnitude of the Yarkovsky effect for individual asteroids depends on other 
parameters such as the heliocentric distance, orbital eccentricity, and spin axis obliquity; for plausible parameters, the Yarkovsky effect is a strong function of thermal inertia \citep[e.g.][]{Bottke2006}.
See \citet[Fig.\ 5]{Vokrouhlicky2000} for the thermal-conductivity dependence of the Yarkovsky-induced orbital drift of several NEAs;
see \citet[Fig.\ 5]{Morbidelli2003} for an analogous plot for spherical objects in the inner main belt.

Our results are expected to enable a more accurate assessment of the impact hazard posed by individual objects such as (29075) 1950~DA, the object with the currently highest known Earth impact probability \seesect{sect:intro:impact},
where the orbital uncertainties are dominated by the lack of knowledge on physical parameters governing the Yarkovsky effect \citep{Giorgini2002}.

\paragraph{Recent confirmation by \citet{Delbo2007}}
Our result for the typical thermal inertia of NEAs has recently been confirmed in a complementary study by \citet{Delbo2007}.%
\footnote{ I am a coauthor of that paper.}
While herein we determine the thermal inertia of individual NEAs (and are therefore limited to a small  number of objects for which information on shape and spin state is available and require extensive sets of thermal-infrared data for each object), \citeauthor{Delbo2007}\ consider the much larger sample
of  NEAs with published multi-wavelength thermal-infrared observations. 
No attempt was made at constraining the thermal inertia of individual
objects
 but rather the ensemble-average thermal inertia was determined.
They obtain $200\pm\unit{40}{\TIunit}$ for the average thermal inertia of an ensemble of NEAs clustering around \unit{1}{\km} in diameter, in excellent agreement with our average value of 
\unit{300}{\TIunit}, keeping the large scatter of roughly a factor of two in mind.

While the 
lack of knowledge on important physical properties of their individual targets might induce a significant systematic uncertainty in the 
\cite{Delbo2007} result,
our result alone may be thought to lack statistical significance, being based on a sample of only 6 NEAs.
The excellent mutual consistency of the results of these two complementary thermal-inertia studies
greatly supports the validity of both.

        \subsection{Thermal inertia correlates with size}
                \label{sect:NEA:TI:trend}

\begin{figure}
   \centering
\includegraphics[width=0.8\linewidth]{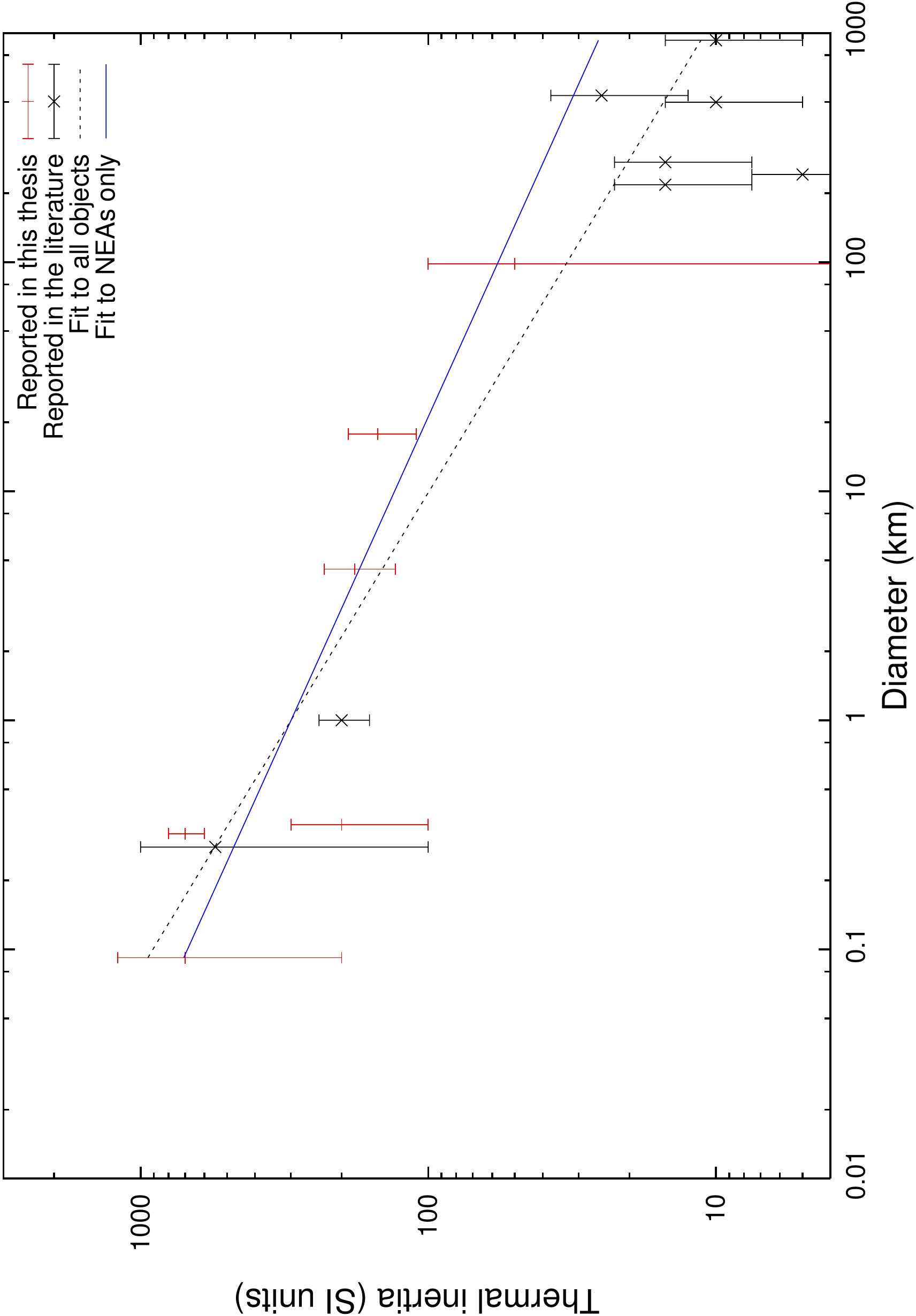}
\caption{Results of all currently available measurements of the thermal inertia of asteroids. See text for references. Previous results for Eros and Itokawa, which have been superseded in this thesis, are omitted. The dashed black line is the power-law correlation for all asteroids reported by \citet{Delbo2007}, the solid blue line is its counterpart considering only NEAs.}
\label{fig:NEA:TI}
\end{figure}

There is an intriguing correlation between the thermal inertia 
and diameter of asteroids \seefig{fig:NEA:TI}.
The available thermal-inertia measurements of asteroids are:%
\footnote{ Our result for the thermal inertia of (617) Patroclus \seesect{sect:Patroclus}, a Trojan, is disregarded. It is unclear at present whether or not the surface structure of Trojans is comparable to that of inner-Solar-System asteroids.}
\begin{itemize}
\item those reported in \tablerefpage{table:NEA:TI}
\item the result by \citet{Delbo2007} discussed above
\item the thermal inertias of  (1) Ceres, (2) Pallas, (3) Juno, (4) Vesta, and (532) Herculina reported by \citet{MuellerLagerros1998}
\item the thermal inertia of (65) Cybele reported by \citet{MuellerBlommaert2004}
\item the thermal inertia of (21) Lutetia reported in \sectref{sect:Lutetia} of this thesis.
\end{itemize}
As can be seen in \figref{fig:NEA:TI},
all values appear to scatter around a straight line on a log-log scale, indicating that thermal inertia $\Gamma$ and diameter $D$ are related through a power law $\Gamma\propto D^{-\xi}$. 
From an analysis of all results depicted in \figref{fig:NEA:TI},%
\footnote{ Except for (54509) YORP \seesect{sect:PH5}, which corresponds to the leftmost data point in \figref{fig:NEA:TI}, and is not yet published. Note the consistency of that point with the extrapolation of the correlation found. \label{foot:TI:PH5}}
\citet{Delbo2007} determined the exponent
$\xi$ to be $\sim0.48$ if all data are considered, and $\xi\sim0.36$ if only NEAs are considered.

As discussed by \citeauthor{Delbo2007}, the 
apparent thermal-inertia dichotomy between large MBAs and small NEAs is partially caused by the difference in heliocentric distance and correspondingly in temperature; 
as will be discussed in \sectref{sect:NEA:T3}, the thermal inertia of regolith would be expected to depend on the temperature $T$ like $T^{3/2}$, leading to a dependence on heliocentric distance $r$ proportional to $r^{-3/4}$.
Correcting the available thermal-inertia results for this effect reduces the thermal-inertia contrast somewhat, but does not remove it, yielding a smaller exponent of $\xi\sim0.37$ for the size-dependent thermal inertia of the entire ensemble.

We conclude that there appears to be a significant correlation of thermal inertia with size, although more data would be valuable to confirm the trend.

\paragraph{Size dependence of the Yarkovsky effect}
If thermal inertia were size independent, the Yarkovsky force would be proportional to $D^2$. Since the mass scales with $D^3$, the resulting strength of the Yarkovsky effect is proportional to $D^{-1}$ \seesect{sect:intro:Yarko}.
The apparent correlation of thermal inertia and size, however, implies a weaker dependence, roughly proportional to $D^{-0.6}$ \citep[see][]{Delbo2007}.

\paragraph{Size-frequency distributions of NEAs and MBAs}
The Yarkovsky effect determines the size-dependent efficiency of the delivery of small asteroids from the main belt into near-Earth space \citep[see also \sectref{sect:intro:populations}]{Morbidelli2003}.
The size dependence of the Yarkovsky-induced orbital drift should therefore be reflected in a difference in the size-frequency distribution (SFD) of NEAs compared to MBAs in the same size range.
In particular, the weaker size dependence of the Yarkovsky effect following from our thermal-inertia results (see above)
implies that the SFD of NEAs
should be less skewed towards smaller sizes than previously assumed.

While the SFD of MBAs at typical NEA diameters is currently not well known, available observational constraints 
are in better agreement with the SFD implied by our results than with the steeper SFD 
for size-independent thermal inertia \citep[see][and references therein]{Delbo2007}.

        \subsection{Geological interpretation}
                \label{sect:NEA:geology}

                \subsubsection{Is there regolith on NEAs?}
                        \label{sect:NEA:regolith}

We find the typical thermal inertia of NEAs to be intermediate between that of lunar regolith and bare rock on Earth.
This suggests that NEAs are covered with particulate materials, where the thermal inertia increases with typical grain size \seesectpage{sect:conduction:physics}.
In particular, the observed size-dependence of thermal inertia is readily explained in terms of regolith coarseness and/or  abundance (see also \sectref{sect:NEA:size_rego}).

No convincing evidence has yet been found for very high thermal inertia ``bare-rock'' surfaces amongst NEAs.
Even the $D\sim\unit{100}{\metre}$ ultra-fast rotator (54509) YORP appears to have at least some regolith on its surface, despite the  large centrifugal force which overwhelms the surface gravity (see also \sectref{sect:NEA:barerock}).

Relations between grain size and thermal conductivity are well established for the atmospheric conditions prevailing on Earth and Mars, but not for a vacuum. 
In particular, \citet{Presley1997} found that even under the thin Martian atmosphere heat transfer  is dominated by atmospheric transfer.%
\footnote{ This is not explicitly stated by \citeauthor{Presley1997}, but implied by their extrapolation formulae for the thermal conductivity as a function of atmospheric pressure (e.g.\ their eqn.\ 17) which yield zero for zero pressure.}
Laboratory measurements similar to those reported by \citeauthor{Presley1997}\ but obtained in a higher vacuum may be required to correlate the thermal inertia of asteroids with a typical grain size.

Nevertheless, all  thermal-inertia values of NEAs measured so far \seetablepage{table:NEA:TI} are significantly larger than that of lunar regolith, implying that asteroidal regolith is coarser than lunar regolith and/or not much deeper than the thermal skin depth (which is of the \cm\ scale).
This agrees nicely with results of spacecraft observations of Eros and Itokawa, which revealed coarser-than-lunar regolith on Eros
\citep[see][]{Veverka2001a,Veverka2001b}
and coarser-than-Eros regolith on Itokawa \citep{Yano2006}.

\subsubsection{Size-dependent efficiency of regolith formation}
        \label{sect:NEA:size_rego}

While it is not yet fully understood how  regolith on asteroids forms, it is widely believed to be generated during impact processes and to be retained by asteroidal gravity \seesect{sect:intro:regolith}.
In this picture, one would expect 
small bodies with low gravity to lose most of their ejecta towards space, and the particle size distribution of the retained ejecta to be skewed towards particles with lower thermal velocity, i.e.\ towards
larger particles relative to the original ejecta distribution.
Furthermore, smaller bodies have lower collisional lifetimes, so small NEAs might be disrupted before they have built up a thick layer of regolith depending on the (largely unknown) typical timescale for regolith formation.

It is therefore natural to expect that the regolith on smaller asteroids is less abundant and coarser than on larger asteroids, leading to enhanced thermal inertia. Hence,
the existence of a size dependence of thermal inertia does not come as a big surprise as such.
However, the form and parameters of this dependence are not constrained by our currently incomplete theoretical understanding. 
It was actually widely expected that sub-\km\ NEAs should not be able to retain regolith at all. Our findings imply that Itokawa, which has been demonstrated by Hayabusa imaging  to be at least partially covered in  regolith, appears to be the rule rather than an exception.

It appears to be well established 
that physical properties of the asteroid material, chiefly its porosity, greatly influence the amount of produced ejecta, its velocity distribution, and furthermore the size-frequency distribution of ejected particles. Furthermore, a body's ability to gravitationally retain regolith depends not only on its size but also on its bulk mass density.

The size dependence of thermal inertia will serve as a valuable constraint for future modeling of impact processes on asteroids and for the physical properties of asteroids.
We speculate 
that the presence of regolith on small asteroids indicates a large porosity in their near-surface layers, which would be expected to lower ejecta velocities relative to an impact into solid material \seesect{sect:intro:regolith}.

Furthermore, one may  expect 
the correlation between size and thermal inertia to differ between members of different taxonomic classes, which appear to display different mineralogy and different bulk mass density.
We speculate that 
this is the reason for
the large difference in thermal inertia found between the NEAs Itokawa (S type) and 1998~WT24 (E type), which are roughly equal in size \seetablepage{table:NEA:TI}.
However, significantly more data are required to confirm this expectation.

Finding possible taxonomy-dependent differences in thermal inertia 
might stimulate further progress in the modeling of impact processes,
which is of crucial importance for many aspects of planetary research including age determination of planetary surfaces and the assessment of the hazard due to impacts on Earth.

                \subsubsection{What is the ``bare rock'' on NEAs?}
                        \label{sect:NEA:barerock}

All of our NEA targets display a thermal inertia much below that of bare rock on Earth, which is  \unit{$\sim2500$}{\TIunit} \citep[see, e.g., \tablerefpage{table:thermalproperties} or][]{Jakosky1986}.
This is rather surprising, in particular for (54509) YORP and (25143) Itokawa, which one may expect to display a larger thermal inertia than actually found.

\paragraph{(54509) YORP}
YORP is an ultrafast rotator with a diameter of only $\sim\unit{0.1}{\km}$ \seesect{sect:PH5}.
Its fast spin rate of \unit{$\sim12$}{\minute} implies that the centrifugal force overwhelms gravity on most of its surface, hence dust cannot be retained by gravity except for small parts of the surface close to the rotational poles.
Nevertheless, our preliminary result
for its thermal inertia is only 200--\unit{1200}{\TIunit}, much below that of bare rock but consistent with the size dependence obtained from our remaining results \citep[see \figrefpage{fig:NEA:TI}, note that the best-fit straight line therein was derived by][\emph{without} our YORP result]{Delbo2007}, which we interpret in terms of regolith coarseness and/or abundance (see above).
We caution, however, that our result for YORP is preliminary \seesect{sect:PH5:discussion}.

\paragraph{(25143) Itokawa}
We found Itokawa, the target of the Japanese rendezvous mission Hayabusa, to have a thermal inertia of $700\pm\unit{100}{\TIunit}$, refining the estimate by \citet[$750\pm\unit{250}{\TIunit}$, see \sectref{sect:Itokawa}]{MuellerItokawa}.
However, Hayabusa imaging clearly demonstrates that \unit{$\sim80$}{\%} of the surface are dominated by boulders and only \unit{$\sim20$}{\%} are covered in coarse regolith.

\paragraph{Tentative interpretation in terms of a thin dust coating}
We speculate that boulders on Itokawa are (partially?) covered with a thin layer of thermally insulating material such as dust, potentially through cohesion. Such a coating would significantly reduce the thermal inertia provided its thickness is a non-negligible fraction of its thermal skin depth ($\sim\unit{5}{\milli\metre}$ for fine dust on Itokawa).
To the best of our knowledge, such a dust coating is neither indicated nor ruled out by Hayabusa imaging results published so far.
All Hayabusa data have recently (24 April 2007) been made publicly available; we hope they will shed light on the issue at hand.
Cohesion between regolith particles has frequently been discussed in the context of NEAs \citep[e.g.][]{Cheng2004b,Colwell2005,Starukhina2005}; cohesive attraction appears to be stronger than gravitational attraction for typical NEAs but depends critically on the (largely unknown) particle size. Fine dust is more cohesive than coarse dust. 
This may explain why
fine dust apparently sticks to boulders, despite the fact that coarser pebble-sized grains were found to be mobile by \citet{Miyamoto2007}.

On YORP, dust is  destabilized by the centrifugal force, which overwhelms gravity.
It can be shown, however, that the former is only of the order of \unit{$10^{-3}$}{\metre\usk\power{\second}{-2}}, the latter is still smaller.
In the light of the above, 
it appears plausible that cohesion can stabilize fine dust on the surface of YORP.
We also note that, due to the fast spin rate, the thermal skin depth is much lower than on other objects (smaller by a factor of $\sim\sqrt{120}\sim11$ compared to the skin depths for  spin period of \unit{24}{\hour} quoted in \tablerefpage{table:thermalproperties}), 
hence a dust ``thickness'' of a fraction of a millimeter is sufficient to reduce the thermal inertia appreciably.

\paragraph{Alternative interpretation in terms of porosity}
\citet{MuellerItokawa} speculate that the thermal inertia of Itokawa may indicate porous rock (see also \sectrefpage{sect:ito:discussion}).
This conforms with the fact that NEAs appear to be typically under-dense. They are thus assumed to contain major voids \seesect{sect:intro:internal}, although virtually nothing is known about the size scale of those voids.
If the lower-than-expected thermal inertia of Itokawa and YORP is indeed due to porosity of boulders exposed at the surface, 
this would imply that near-surface pores are small compared to the thermal skin depth for bare rock (i.e.\ up to a few \centi\metre---see \tablerefpage{table:thermalproperties}).
Such information would be  relevant for impact modeling.
However, it seems to us
that an unrealistically large porosity would be required to reduce the thermal inertia by a factor of 2500/700.

                \subsubsection{Comparison with Martian satellites}
                        \label{sect:NEA:Mars}

It is instructive to compare our asteroid thermal-inertia results  with thermal-inertia values derived for the Martian satellites, Deimos and Phobos, which are spectroscopically similar to asteroids \citep[e.g.][]{Rivkin2002} 
and are widely believed to be captured asteroids.
Deimos and Phobos bracket Eros in size and are  intermediate between NEAs and MBAs in terms of heliocentric distance.

From Viking observations, Deimos and Phobos are known to be covered with regolith;  \citet{Lunine1982} report thermal-inertia measurements  resulting in  
 25--\unit{84}{\TIunit} for Deimos ($D\sim\unit{12.6}{\km}$)
and 
38--\unit{67}{\TIunit} for Phobos ($D\sim\unit{22.4}{\km}$).
The latter is consistent  with the independent thermal-inertia estimate of 20--\unit{40}{\TIunit} by \citet{Kuehrt1992}
based on infrared observations of Phobos obtained with the Soviet Phobos-2 spacecraft.

To facilitate comparison of these values (obtained at a heliocentric distance of \unit{$\sim1.5$}{\AU}) with values obtained in the hotter thermal environment of  near-Earth space,
they must be multiplied by $1.5^{3/4}\sim1.4$, 
resulting in 
35--\unit{118}{\TIunit} for Deimos
and
28--\unit{94}{\TIunit} for Phobos.
While these values may be larger than the lunar value, they are somewhat below the thermal inertia of Eros and significantly below that of smaller NEAs.

We conclude that the Martian satellites appear to have a finer regolith than asteroids in their size range (e.g.\ Eros), possibly as fine as lunar regolith.
In their case, regolith formation  may be aided by the gravitational influence of Mars, which would be expected to influence the  ejecta dynamics appreciably.
Additionally, impacts on Mars produce fine ejecta which may be captured by its satellites;
this is supported by recent Mars Express results obtained with the High Resolution Stereo Camera HRSC,
 which indicate that Phobos captured groove-forming Martian ejecta
\citep{Murray2006}.
Phobos continues to be observed with HRSC \citep{Oberst2006}.

Another possibility to explain the difference in thermal inertia between Eros and the Martian satellites is their different composition: 
Eros is a silicaceous S-type asteroid, while 
D or T-type asteroids provide the closest spectral match  to the Martian satellites \citep{Rivkin2002}.
Our sample of NEA targets does not contain D or T-type asteroids, but a member of the 
 spectrally similar C class, (1580) Betulia.
Betulia, with a diameter of \unit{$\sim40$}{\%} of Deimos',
has  a thermal inertia of $180\pm\unit{50}{\TIunit}$, significantly above that of the Martian satellites.

        \subsection{Possible improvements to the TPM}
        \label{sect:NEA:futuremodeling}

                 \subsubsection{Non-homogeneous thermal properties}
                 \label{sect:NEA:T3}

In our TPM, all thermal properties are assumed to be homogeneous over the asteroid surface. Furthermore, it is assumed that they do not vary with depth nor temperature.
It may be instructive to relax some of these modeling assumptions.

\paragraph{Thermal inertia variegation}

As discussed in sections \ref{sect:Itokawa} and \ref{sect:NEA:barerock},
Itokawa displays a dichotomy between rocky ``rough'' and regolith-dominated ``smooth'' terrains,
although it is not clear whether the boulders in the rough terrain are ``bare'' or covered with a thin coating of dust.
Exposed boulders would  display a dramatic contrast in thermal inertia relative to regolith,
while already a thin dust coating would reduce the contrast appreciably.

This may be studied using a generalized TPM, in which the surface is decomposed into disjoint units with constant thermal inertia over each unit but possible differences in thermal inertia from unit to unit.
A natural choice of such units on Itokawa's surface would be the smooth and rough terrains, which are reportedly well distinguishable on the shape model by 
\citet{Demura2006}.

In the case of Itokawa, the dichotomy between  smooth and rough terrains 
is caused by mobility of regolith, which concentrates in the minima of the combined gravitational-centrifugal potential \citep{Fujiwara2006,Miyamoto2007}.
This may be expected to be generic to small NEAs.
It is possible to determine the gravitational-centrifugal potential of other NEAs if a detailed shape model  is available; e.g.\ the paper describing Betulia's radar-derived shape \citep{Magri2007} contains an extensive discussion of local  minima and maxima of the potential.
The study of such objects may  benefit from a generalized TPM which allows variegation of thermal inertia.
Note that, while the total number of small NEAs well-studied at radar or thermal-infrared wavelengths is small compared to the total number of NEAs, these two samples overlap strongly 
since they are essentially drawn from the much smaller sample of NEAs which made very close approaches with the Earth within the past few years.

A further application of thermal-inertia variegation may be generalized models of roughness at a small scale, e.g.\ by means of craters as in our present TPM.
It may prove fruitful to assign different thermal-inertia values to facets of different slopes, with a natural first choice being to distinguish between facets with slopes above and below reasonable estimates for the angle of friction of regolith.

\paragraph{Two-layer model}
In order to test our hypothesis of a thin dust coating on YORP and on the exposed boulders on Itokawa \seesect{sect:NEA:barerock}, a TPM should be developed in which two horizontal layers of different thermal inertia are considered. 
Similar models have been used before, e.g.\ to analyze eclipse observations of the Galilean satellites \citep{MorrisonCruikshank1973} or for model calculations of the Yarkovsky effect \citep[e.g.][]{Vokrouhlicky1999}.

While we caution that such a model is probably poorly constrained in practical application to  thermal-infrared data, we expect it to be useful
for theoretical studies. 
In such studies, 
our current TPM would be used to fit synthetic flux values generated 
for bare rock covered with dust coatings of different thickness.
The resulting effective thermal inertia as a function of dust thickness would aid the interpretation of thermal inertia results.

\paragraph{Temperature dependence}
In principle, all thermal properties, such as thermal conduction, heat capacity, and also surface bulk density, vary with temperature $T$, leading to a temperature dependence of the thermal inertia, which is neglected in our model. 
Temperature dependence of thermal inertia would be expected to  influence diurnal surface-temperature distributions directly, but also indirectly through its influence on the sub-surface temperature profile.

While one may expect the heat capacity and bulk density to be negligibly weak functions of $T$ for the conditions relevant to our purposes,%
\footnote{ \citet{Ghosh1999} discuss  the temperature dependence of the heat capacity. Their findings (see their Fig.\ 1) imply that for temperatures of 300--\unit{400}{\kelvin} the heat capacity is constant to within \unit{10}{\%} or better for plausible asteroid materials, inducing thermal-inertia variability of \unit{5}{\%} at most ($\Gamma\propto\sqrt{\kappa}$).}
 the thermal conduction may vary greatly with $T$ depending on the heat transfer mechanism (see also the discussion in \sectref{sect:conduction:physics}).
Purely radiative heat transfer, in particular, is characterized by a thermal conductivity proportional to $T^3$, implying a thermal inertia proportional to $T^{3/2}$.
Purely conductive heat transfer, on the other hand, leads to a virtually constant thermal conductivity.
From Apollo-era studies it is well known that both heat-transfer mechanisms are relevant in the lunar regolith (while the third basic heat transfer mechanism, convection, is clearly irrelevant on atmosphereless bodies).

The relatively large thermal-inertia values found by us on NEA surfaces argue in favor of a conduction-dominated heat transfer,
and therefore appear to justify our modeling assumption of temperature-independent thermal inertia in hindsight.
It would be instructive, nevertheless, to double-check our results using a model with, e.g., $T^3$ thermal conductivity or a mixed model, where thermal conductivity is a sum of a constant term and a $T^3$ term.
We caution that 
the thermal-infrared  data typically available for NEAs 
would be expected to be insufficient to constrain an additional fit parameter, such as the relative importance of the two heat transfer mechanisms at a given temperature.
Furthermore, 
to enable meaningful comparison of results obtained from TPMs 
with different temperature-dependence of thermal inertia,
a suitable reference temperature must be defined.

                 \subsubsection{Non-convex-shape TPM for NEAs}
                 \label{sect:NEA:concave}

Our thermal-inertia results for NEAs
are based on the TPM described in \chaptref{chapt:TPM}, in which the asteroid shape is considered to be convex and neither shadowing nor mutual heating are considered outside craters.
If large-scale concavities are present, neglecting them may cause the
model temperature distribution to deviate severely from the physical temperature distribution, and may lead to flawed estimates of diameter and thermal inertia.
Note that most
 available asteroid shape models are convex;  in particular, shape models derived from the inversion of optical photometry are convex by design \seesect{sect:intro:shape}. 


As discussed in \sectref{sect:NEA:D}, our diameter results for (433) Eros and (25143) Itokawa are in excellent agreement with spacecraft results despite several prominent concavities on their surfaces.
However, in the cases of (1580) Betulia and (54509) YORP 
our diameter estimates are somewhat below radar-derived estimates, barely within the combined range of uncertainty.
While we caution that the reason for this, if any, may also be on the radar side \seesect{sect:NEA:D}, 
we note that our observations took place at large solar phase angles of \unit{53}{\degree} (Betulia) and \unit{59}{\degree} (YORP).
The radar-derived shape models of both Betulia and YORP display extremely large concavities, with the major concavity on Betulia having a diameter comparable to the asteroid's radius.
The difference in nominal diameter estimates is therefore consistent with the expectation that neglecting shadowing effects at large phase angles leads to an underestimation of diameter.
Furthermore, we appear to have observed features in the thermal lightcurve which are not well reproduced by our convex-shape TPM (see \sectref{sect:Betulia:discussion} and \sectref{sect:PH5:discussion}).

It seems worthwhile first to  study the available data  using a non-convex-shape TPM, in which shadowing is considered but mutual heating is not, before developing a more general TPM, in which mutual heating among facets is considered.

                 \subsubsection{Brute-force modeling of positive relief such as boulders}
                 \label{sect:NEA:boulders}

As is common practice, we model thermal-infrared beaming by adding 
synthetic ``craters'' to the asteroid surface, i.e.\ indentations which take the shape of sections of hemispheres \seesect{sect:TPM:beaming}.
One may expect this to be a fair approximation for the surfaces of our Moon, Mercury, or other large atmosphereless bodies which are well known to be densely covered with impact craters, which typically have the shape of subdued hemispheres.
However, very little is currently known about asteroid surfaces
due to the scarcity of available high-resolution spacecraft imaging data.
Itokawa, the only sub-\km\ asteroid to be scrutinized by spacecraft so far, was found to be virtually devoid of craters \citep[see also the discussion in \sectref{sect:ito:discussion}]{Saito2006} but most of its surface is dominated by boulders.

Generally, one might expect the thermal effect of positive relief features to be quite different from that of indentations, particularly for observations at large phase angles, when shadowing effects are relatively more important.
Thermophysical models with positive rather than negative surface relief  have, however, only rarely been considered in the literature \citep[see][and references therein for a rarely used exception using random surfaces]{LagerrosIV}. In particular, no model is known to us in which boulders are specifically considered.

It appears to be worthwhile  to develop and study a generalized TPM for NEAs
in which roughness is modeled by adding boulders to the surface rather than craters.
Suitable geometric boulder models may be hemispherical bulges (if this enables analytical calculations similar to those presented in \sectref{sect:TPM:beaming}) or, alternatively, cuboids
 which would be advantageous for the numerical implementation.
Such a study could be based on a general TPM for non-convex shapes which includes the effects of mutual shadowing and heating (see the discussion above).

The main task remaining to be solved would then be 
to computer-generate very detailed ``shape models,''
where a suitable distribution of boulders 
is explicitly placed on the surface of a pre-existing shape model.
This approach causes the number of surface facets, with which the numerical effort scales, to increase significantly. 
On the other hand, one thus avoids the computationally very expensive calculation of ``crater'' fluxes.
It remains to be studied whether or not such a ``brute-force'' modeling of boulders is feasible with currently available computers, possibly after optimizing the TPM code for numerical efficiency.

A test case for such modeling would be to use the 
shape models of Eros and Itokawa with the highest resolution available, which have 200,700 facets in the case of Eros, and  3,145,728 facets in the case of Itokawa.

%% file: conclusions.tex
\chapter{Conclusions}
         \label{chapt:conclusions}

\section{Thermal modeling of near-Earth asteroids (NEAs)}

\paragraph{\bf We have developed and tested a detailed thermophysical model applicable to NEAs.}
Effects of convex irregular shape, spin state, surface roughness, and thermal inertia are explicitly taken into account.
The model is applicable to all asteroids except for objects at the meter scale or smaller.
This is the first such model shown to be applicable to NEAs. 

\paragraph{\bf A model like ours is required for reliable determination of thermal inertia.} 
Moreover, it enables more accurate size and albedo determination compared to less detailed thermal models, which are frequently used.

 \section{Thermal inertia of NEAs}
 \label{sect:conclusions:TI}

\paragraph{\bf We have doubled the number of NEAs with measured thermal inertia.}
We have determined the thermal inertia of  5 NEAs. For 2 of these, we refine previously available estimates while the remaining 3 increase the total number of NEAs with measured thermal inertia from 3 to 6.
%
For each object, we have also determined its size and albedo, and have constrained its surface mineralogy (taxonomic type). Our NEA targets range between 0.1 and \unit{17}{\km} in diameter.

\paragraph{\bf The typical thermal inertia of NEAs, which was previously unknown, is around \unit{300}{\TIunit}.}
The corresponding value for large main-belt asteroids (MBAs) is 
smaller by more than an order of magnitude, 
indicating significant differences in surface structure.
Our result has recently been confirmed in a complementary study by \citet{Delbo2007}. The corresponding  thermal conductivity  is \unit{0.08}{\kappaunit}.

\paragraph{\bf Our results allow more realistic model calculations of the Yarkovsky effect, which is important in the assessment of the impact hazard.}
Thermal inertia governs the Yarkovsky effect, a non-gravitational force known to influence the orbits of asteroids below some \unit{20}{\km} in diameter significantly.
Uncertainties in the strength of the Yarkovsky effect dominate uncertainties
 in the assessment of the risk posed by objects such as (29075) 1950~DA, the asteroid with the largest currently known probability of impacting Earth.

\paragraph{\bf The thermal inertia of asteroids correlates with size.}
Using our results, \citet{Delbo2007} found a power law relating the thermal inertia and diameter of asteroids.
From this, they deduce a modified size dependence of the Yarkovsky effect
and draw conclusions on differences between the size-frequency distributions of NEAs and similarly sized MBAs.

\paragraph{\bf There is regolith on sub-\km\ NEAs.}
The thermal inertia of NEAs is intermediate between that of lunar regolith and bare rock on Earth, indicating the presence of coarse regolith.
Asteroid regolith is believed to be gravitationally retained collisional debris. With decreasing asteroid mass, the formation of regolith should be less efficient and skewed towards coarse grains, consistent with our findings.
Our quantitative results may  lead to an improved understanding of regolith formation through impact processes.

\paragraph{\bf All NEAs studied by us have a thermal inertia significantly below that of bare rock.}
This is surprising, particularly so for two of our targets:
\begin{itemize}
\item (54509) YORP with a diameter of  $\sim\unit{0.1}{\km}$ and an ultrashort spin period of only $\sim\unit{12}{\minute}$; the centrifugal force overwhelms gravity on most of its surface and would be expected to destabilize any regolith
\item (25143) Itokawa, the target of a rendezvous with the Hayabusa spacecraft in 2005, was seen to be predominantly covered with boulders 
\end{itemize}
We tentatively explain the reduced thermal inertia with  a thin (\milli\metre\ scale) coating of particulate material, which may be stabilized by cohesion. We caution that further study is required.
Alternatively, both objects may display an extremely large near-surface porosity at the \milli\metre--\cm\ scale, but it remains to be studied whether realistic porosity models can explain the observed reduction in thermal inertia.
In the case of Itokawa, 
both theories may be testable on the basis of obtained Hayabusa imagery, which was only partially analyzed so far.

\section{Thermal inertia of an eclipsing binary}

\paragraph{\bf We have pioneered thermal-infrared observations of eclipsing binary asteroids.}
We have clearly detected the thermal response of the Trojan binary  (617) Patroclus to mutual events, where respectively one component shadowed the other.
This allowed us to determine their thermal inertia  in a uniquely direct way, in addition to a more reliable diameter estimate.

\paragraph{\bf The thermal inertia of Patroclus is around \unit{90}{\TIunit},}
indicating a cover of  relatively coarse regolith.
No reliable information on the thermal inertia of Trojans has been available beforehand.


\section{Physical characterization of spacecraft targets}

\paragraph{\bf Rosetta flyby target (21) Lutetia}
We have found the diameter of Lutetia to be $98.3\pm\unit{5.9}{\km}$ and its geometric albedo to be $\pv=0.208\pm0.025$, consistent with an M-type classification and  with previous diameter estimates.
In the past few years, spectroscopic results have been published which indicate a C-type-like surface.
We can now rule out a low albedo typical of a C-type classification.
Rosetta will fly by Lutetia in 2010.

\paragraph{\bf Potential spacecraft target (10302) 1989~ML}
The NEA (10302) 1989~ML is among the most favorable spacecraft targets in terms of energy and flight time required to reach it.
It has been taken as a working target for phase-A studies of the ESA mission Don Quijote, although virtually nothing was previously known about its physical properties.
On the basis of Spitzer observations, for which we have been awarded Director's Discretionary Time, we have determined its diameter to be $0.28\pm\unit{0.05}{\km}$ and its albedo to be $\pv=0.37\pm0.15$. Combining our results with optical and near-infrared data we conclude that 1989~ML is an E-type object---note that only 4 E-type NEAs were known beforehand.
Most probably, our results  imply that 1989~ML is not a suitable target for Don Quijote.

\paragraph{\bf Thermal-infrared characterization of potential spacecraft targets is time efficient.}
Only $\sim\unit{30}{\minute}$ of telescope time at the \unit{3.0}{\metre} IRTF were required for our Lutetia observations, and only \unit{1.2}{\hour} at the Spitzer Space Telescope in the case of 1989~ML, despite the faintness of the latter.

\chapter{Future work}
\label{chapt:future}

\section{Thermal inertia measurements}

\paragraph{\bf More NEA measurements  are required.} 
While we consider our result for the typical thermal inertia of NEAs to be well established, its apparent size dependence
may require more data points to be confirmed.

\paragraph{\bf  M-type asteroids}
Judging from  near-infrared spectroscopy \citep{Rivkin2000}
and radar measurements \citep[e.g.][]{Magri1999}, some M-type asteroids appear to be metallic, while others appear  to be non-metallic.
Metal is an excellent thermal conductor, 
potentially leading to an enhanced thermal inertia.
We have performed thermal-infrared observations  of M-type MBAs at the IRTF. A preliminary analysis indicates that metallic M types 
have indeed a larger thermal inertia than 
their non-metallic counterparts, but further study is required.

\paragraph{\bf Does thermal inertia correlate with taxonomic type?}
In general, the efficiency of regolith formation and the thermal properties of regolith may be a function of mineralogy, which may translate into a dependence of thermal inertia on taxonomic type.
This is supported by our preliminary M-type results (see above) and the difference in thermal inertia between the two NEAs 1998~WT24 (E type) and Itokawa (S type), which are roughly equal in size.
Any such taxonomy dependence would not only enable more accurate model calculations of the Yarkovsky effect but also inform future modeling of regolith formation through impact processes.

\paragraph{\bf Small MBAs} 
The diameter distribution of asteroids with measured thermal inertia displays a dichotomy between relatively small NEAs and much larger MBAs.
Measurements of the thermal inertia of smaller MBAs, with diameters of 10--\unit{100}{\km}, would be required to fill the gap. In particular it would be instructive to compare the thermal inertia of MBAs of Eros' size with that of Eros. 

\paragraph{\bf Karin family}
We have performed Spitzer observations of 17 members of the intriguing  Karin family of MBAs, which was formed in a catastrophic collisional event only $5.8\pm\unit{0.2}{\mega\yr}$ ago.
The data analysis is ongoing (due to calibration problems which we hope are now resolved). We expect to determine the typical thermal inertia of our targets, which range between $<2$ and \unit{$\sim14$}{\km} in diameter.

\paragraph{\bf Kuiper belt objects (KBOs)}
Very little is currently known about the thermal inertia of KBOs. 
Due to their large heliocentric distance, their thermal emission peaks at wavelengths which are inaccessible from the ground, requiring the use of airborne (SOFIA) or space-based telescopes (Spitzer, Herschel).
Knowledge about the thermal inertia of  KBOs would aid the accurate determination of KBO sizes and albedos  and would provide information on the surface particle grain size which is difficult to obtain otherwise.


\section{Binary asteroids}

\paragraph{\bf Mass densities from TPM-derived diameters}
The elusive mass density of asteroids can be determined in the case of binary systems  from accurate determinations of  the mutual orbit (providing the mass) and diameter measurements.
TPM-derived diameters promise to reduce the usually large  uncertainties, which are very sensitive to diameter uncertainty  ($\rho\propto D^{-3}$).
An ongoing collaboration has been established with a group of  researchers (based in Paris and Berkeley) who determine mutual orbits through  high-angular-resolution optical observations.
We have been awarded observing time with Spitzer in cycle IV (starting summer 2007) to characterize 26 binary asteroid systems.

\paragraph{\bf More eclipses}
We aim at further thermal-infrared observations of eclipsing binaries. 
The number of suitable targets (with well determined mutual orbit to allow precise prediction of eclipse events and later thermophysical modeling) is  limited but increasing;
potential targets have been identified in collaboration with our partners. 


\section{Improvements to the thermophysical model (TPM)}

\paragraph{\bf TPM for concave shapes} Our TPM for  non-convex shapes 
does not yet include the effect of mutual heating beyond craters and 
is not well tested, yet. 
Furthermore, its practical applicability is limited by its current numerical inefficiency. Numerically more efficient algorithms have been developed but not yet implemented.

\paragraph{\bf Generalized models of asteroid surface roughness}
An important application of the general non-convex-shape TPM to be developed would be thermal modeling of boulders on NEA surfaces.
In the current TPM,   surface roughness is modeled in terms of craters (negative relief) but,
judging from Hayabusa imaging of (25143) Itokawa, 
boulders (positive relief) appear to be more abundant on small asteroids.
This may have a significant influence on the temperature distribution, particularly so at large solar phase angles, when shadowing effects are important.
We propose a ``brute-force'' model of boulders, whereby cuboids are added to each facet of asteroid shape models.


%% file: concaveintro.tex
In this chapter,  generalizations of our TPM \seechapt{chapt:TPM} are discussed 
to account for the effects of shape non-convexity beyond that of cratering.
On globally non-convex shapes, facets may shadow one another and additionally exchange energy through direct and indirect radiation.

The work reported in this chapter is in progress.
A first model variant has been developed in which shadowing is taken into account but mutual heating  is not; 
detailed tests  are  on-going.
This model variant aims primarily at simulations of eclipsing binary systems 
and has been used 
 to analyze Spitzer observations of the binary system (617) Patroclus during two mutual events \seesect{sect:Patroclus}.

A  more general
model variant, in which also mutual heating of facets is taken into account, has been partially designed  but not yet implemented.


%% file: concave.tex
\section{Physical background}
        \label{sect:TPMconcave:phys}

Most available asteroid shape models  are intrinsically convex
such that the ``convex-shape'' TPM is applicable \seechapt{chapt:TPM}.
Several important asteroids, however, have well determined non-convex shapes, e.g.\ the spacecraft target NEAs (433) Eros and (25143) Itokawa (see also \figrefpage{fig:intro:ito}).
Also, doubly tidally locked binary asteroid systems can  be thermally modeled as a single, non-convex object \seesect{sect:TPMconcave:binary}.

Non-convexity of a body's shape adds to the complexity of modeling its thermal emission because facets may eclipse and/or occult one another (i.e.\ obstruct the line of sight towards the Sun and/or the observer, respectively), and furthermore they can directly and indirectly exchange energy by scattering and reabsorption of optical and thermal radiation, leading to mutual heating.

Mutual heating couples the temperatures on different facets  and thus significantly increases the difficulty of determining surface temperatures.
Shadowing is easier to model, requiring only  the directions towards the Sun and the observer to be checked for possible obstructions before fluxes are calculated.

We have designed, implemented, and partially tested a TPM variant for non-convex bodies where shadowing effects are considered and mutual heating is neglected. 
We have partially designed, but not yet implemented,  a variant in which mutual heating is fully taken into account.
Mutual heating may be negligible if the global surface concavities are  shallow such that only moderate fractions of the solid angle above the facets' local horizons are covered by other facets \citep[see also][]{LagerrosIII}.
Primarily, however, the model described herein is designed to model 
the thermal effects of eclipses in
binary asteroid systems.
The binary system is described as a single rigid body with a shape consisting of two disjoint parts.
For eclipsing binaries, we expect the effect of mutual heating to be negligible relative to that of shadowing.

In the following, both model variants shall be described.
As in the case of the convex model, geometric and thermal aspects are separated as far as possible. An auxiliary software was developed to convert asteroid shape models into a TPM-specific \code{.concave} format containing all geometric information  required for thermal modeling.

\subsection{Geometric aspects}
\label{sect:TPMconcave:geometry}

While on a convex body
all facet-specific geometric information 
required to determine the thermal flux emanating from it is its outbound surface-normal vector, more geometric information is required to
determine whether or not a facet is obstructed by another facet, and furthermore to calculate the mutual heating among them.
To model obstructions, it is required to know for each facet  which other facets are visible from it and what their relative position is.
To model mutual heating, it is required to know the \emph{view factors} of all visible facets, a measure of the solid angle under which facets are visible to one another (view factors have  been discussed in the context of hemispherical craters, see \eqrefpage{eq:beaming:viewfactor_general}).

The surface temperature on a given facet is coupled 
to the surface temperatures of all facets 
in its \emph{thermal unit}, which we define as the set of all facets with which it can exchange energy through direct or indirect radiation.
Each asteroid shape model can be uniquely decomposed into disjoint thermal units. If lateral heat conduction can be neglected, temperatures inside one thermal unit are independent of temperatures inside any other thermal unit.
For a convex object, each thermal unit consists of  a single surface element, whereas a concavity in an otherwise convex object forms a contiguous thermal unit.
In general, not all facets belonging to a thermal unit are directly visible to one another; e.g.\ a pyramid on a large sphere belongs to a thermal unit containing all points visible from the peak of the pyramid, although points on one side of the pyramid do not see points on the other side.

In an auxiliary program, asteroid shape models are read in (the same input formats are supported as for the convex model), all required geometric information is calculated and stored in \code{.concave} files, which can be read in by the non-convex TPM code.
To generate a \code{.concave} file, outbound surface-normal vectors and the intrinsic diameter are determined like for a convex object (see \sectref{sect:TPM:dA} and \ref{sect:TPM:intrinsicD}).%
\footnote{ Note that it is assumed in our algorithms  that the origin of the coordinate system is connected to all vertices by straight lines which lie entirely within the object (see footnote \ref{foot:convex_concave} on \pageref{foot:convex_concave}).
Judging from visual inspection of computer renderings, this appears to be the case for all shape models considered in this thesis.}
Then, all facets are checked whether they are visible from one another.
To be visible from one another
\begin{enumerate}
\item  facets must be tilted towards one another, i.e.\ each must be above the other facet's local horizon. This is determined from the sign of the scalar product of the connecting vector and the outbound normal vectors.
\item additionally, the line of sight between the two facets must  not be obstructed by another facet. Only facets above either horizon can do that.
\end{enumerate}
To determine whether a facet $a$ obstructs a line from facet $b$, 
one determines the intersection of that line with the plane containing facet $a$.%
\footnote{ We here only consider lines originating at the midpoint of facet $b$. This avoids model complications due to finite facet size, such as partially obstructed facets. 
If the line of sight between two facets is partially obstructed by a third facet, our approach risks defining one facet to be ``visible'' from the other, but not vice versa. To avoid this, visibility is only checked in one direction and the result is applied in both directions.}
The vector of the intersection point can be uniquely decomposed as 
\begin{equation}
  \label{eq:TPMconcave:obstruct}
  \vec{v_1} + x\cdot\left(\vec{v_2}-\vec{v_1}\right) + y\cdot\left(\vec{v_3}-\vec{v_1}\right)
\end{equation}
($\vec{v_i}$ denote the vertices of facet $a$),
facet $b$ is obstructed by facet $a$ if $x+y\leq 1$. This routine is also used in the proper TPM code to determine whether facets are shadowed.

Thermal units are built up recursively by adding a facet which has not yet been associated with a thermal unit. The routine to add facets cycles through the list of facets visible from the added facet, determines whether the visible facet is already contained in the thermal unit and adds it otherwise (recursion).

\subsection{Doubly tidally locked binaries as special non-convex bodies}
        \label{sect:TPMconcave:binary}

For the non-convex TPM code, it is not necessary to assume that the asteroid shape be contiguous.
A rigid body consisting of two disjoint pieces is equivalent to a doubly tidally locked binary system  which is at rest in a  co-rotating coordinate system (i.e.\ the two spin periods are synchronized with the period of the circular mutual orbit, and both spin axes are perpendicular on the latter).
While tidal friction is known to attract all binary systems towards this asymptotic attractor state, many binary systems are  not  doubly tidally locked; this variant of the TPM is not applicable to the latter.
It was primarily designed for planning and analyzing our Spitzer observations of (617) Patroclus, which appears to be doubly tidally locked  \seesect{sect:Patroclus}.

It is straightforward to model the observable effects of eclipse and occultation events using a non-convex TPM code in which shadowing is accounted for.
Both components are assumed to be convex. It is plausible that mutual heating from component to component can be neglected 
because their distance
 is typically much larger than either component's radius.

An auxiliary program has been developed to 
generate \code{.concave} files describing binary systems. 
To this end, two arbitrary \code{OBJ} shape files are read in, typically triangulated spheres. Both can be stretched (e.g.\ to obtain ellipsoids),  one can be rescaled to obtain different-size components. 
After determining the outbound surface-normal vectors on each component, they are placed at a user-specified mutual distance.
Both components are shifted by a distance proportional to the other component's volume, such that the $z$ axis (i.e.\ 
the spin axis of the mutual orbit in the body-fixed system)
runs through the center of mass. It is tacitly assumed that the center of mass of either component coincides with the origin of the respective shape model's coordinate system, and that the mass density is homogeneous throughout the system.

All geometric quantities, such as lists of shadowers, are determined like they would be determined for other shape files.
If both components are convex, 
the resulting binary shape has one 
 large thermal unit containing the two hemispheres facing one another, while each facet on the two opposite hemispheres forms its own thermal unit.

The intrinsic volume-equivalent diameter (which is stored in the \code{.concave} file) is determined from the sum of the individual components' volumes, where stretching and rescaling factors are taken into account.
Note that in this special case the algorithm discussed in \sectref{sect:TPM:intrinsicD} to calculate the volume of a shape model would fail.

For a binary system, the two usual diameter definitions (area-equivalent or volume-equivalent; see \sectref{sect:thermal:size-albedo}) generally lead to significantly different diameter values.
For a system composed of two spheres with diameters $D_1$ and $D_2$, the total area-equivalent diameter $D_\Area$ equals $\sqrt{D_1{}^2+D_2{}^2}$, whereas the total volume-equivalent diameter $D_V$ equals $\sqrt[3]{D_1{}^3+D_2{}^3}$.

It must be noted that our TPM uses
$D_\Area$ to 
relate the $H$ value with \pv\ (because $H$ is based on area-dependent optical magnitudes), while $D_V$ is used to scale model fluxes.
Therefore, for 
given $H$ and  \pv\ values, model fluxes must be multiplied with
\begin{equation}
  \label{eq:TPMconcave:rescale}
  \frac{f_\text{``true''}}{f_\text{model}} = \left(\frac{D_V}{D_\Area}\right)^2 = \frac{\left(D_1{}^3+D_2{}^3\right)^{2/3}}{D_1{}^2+D_2{}^2}.
\end{equation}
Alternatively, the TPM code outputs ``true'' flux values if it is given an $H$ value which is offset by 
$2.5 \log_{10} (f_\text{``true''}/f_\text{model})$ from the $H$ value quoted in the literature.

\subsection{Temperature distribution without mutual heating}
\label{sect:TPMconcave:Tshadowingonly}

If mutual heating can be neglected, then the temperatures on individual facets are independent from one another. 
We can continue to use \eqrefpage{eq:boundarycondition_dimensionless} to determine surface temperatures on smooth facets, with the sole change that $\mu_S$ is redefined to vanish when the facet is eclipsed. In the absence of thermal inertia, in particular, $T=\sqrt[1/4]{\mu_S}\ \TSS$.
Analogously, $\mu_O$ is defined to vanish when the facet is occulted such that observable flux contributions from a facet continue to be proportional to $\mu_O$.

\subsection{Temperature distribution with mutual heating}
\label{sect:TPMconcave:Tmutualheating}

The effect of mutual heating  on smooth facets inside a thermal unit can in principle be determined 
along the lines discussed in \sectrefpage{sect:TPM:beaming} for facets inside hemispherical craters.
However, since view factors  are no longer  constant, no general analytic solutions can be found.

The first step towards a solution is to determine the radiation field at optical wavelengths, $J_V$, for all time steps by numerically solving \eqref{eq:beaming:optical1} (with suitably defined $m_S(\vec{r})$ to take shadowing into account). Since asteroid Bond albedos are typically small, the perturbative approach \eqref{eq:beaming:optical:direct} would be expected to converge rapidly. The integral therein is replaced by a sum over all triangular facets inside the thermal unit.

The temperature distribution then follows from  \eqref{eq:beaming:T1general} combined with \eqref{eq:beaming:T2general}, which must be solved together with the heat diffusion equation \eqref{eq:heatconduction_dimensionless} below each facet.
A TPM with mutual heating has not yet been implemented, we plan to tackle the temperature distribution using the following algorithm: 
After a suitable initialization of the temperature profile below each facet,%
\footnote{ It should be verified at the code validation stage that model fluxes are independent of the particular initialization chosen; a good first estimate may lead to significant improvements in convergence speed.}
the system is propagated one time step further by iterating the following steps for each facet:
\begin{enumerate}
\item calculate the sub-surface temperature profile. 
If the heat diffusion is discretized in a fully explicit way, sub-surface temperatures for the new time step are fully determined by the old temperatures.
\item determine the amount of absorbed thermal flux emanating from other facets. Since the current surface temperatures are not yet known, approximate them with values from the previous time step. Recursively add higher scattering orders until convergence is reached to within a user-specified  accuracy goal. Alternatively, cut off after a certain number of scattering orders (for $\epsilon=0.9$, the reflectivity equals $1-\epsilon=0.1$, so 
the third scattering order may already be insignificant).
\item from this, combined with the pre-computed optical flux at the current time step and the current sub-surface temperature, calculate the current surface temperature from a discretization of the surface boundary condition (\eqref{eq:beaming:T2general}).
\end{enumerate}
This requires 
that the  temperature profiles of all facets must be calculated (and therefore kept in RAM) simultaneously.
It remains to be studied at the code validation phase whether or not the approximation made in step 2 introduces significant systematic errors. 
A suitable test method might be to compare the mutual heating terms determined in step 2 with mutual heating terms which would result from the resulting surface temperatures.

After temperatures have been determined, observable fluxes remain to be calculated by summing up the Planck contributions from the individual facets taking account of obstructions of the observer's line of sight  and multiple scattering. Due to the smallness of the infrared reflectivity ($1-\epsilon\sim 0.1$), it is probably sufficient to truncate after the single-scattering order (see also \figrefpage{fig:beaming:Lagerros_all}).

\subsection{Beaming}
\label{sect:TPMconcave:beaming}

\paragraph{Without mutual heating}

If mutual heating by other facets is neglected, the 
temperature distribution inside a hemispherical crater 
can in principle be determined in a way which is analogous to that
 used for the convex TPM \seesect{sect:TPM:beaming}, 
provided that $m_S(\vec{r})$ is set to zero when the crater is eclipsed by another facet.

If thermal inertia is neglected, an eclipse cools the crater down to $\unit{0}{\kelvin}$ instantaneously. An explicit numerical model of thermal conduction inside craters would be relatively easy to generalize such that it allows for the effect of eclipsing by other facets.
Our approximative treatment of thermal conduction inside craters on convex bodies \seesect{sect:beaming:approximation}, however, is not readily generalized to allow for eclipses.

Instead, we have reverted to an approximation similar to one proposed by \citet{Spencer1989,Spencer1990}, where beaming is crudely taken into account by multiplying surface temperatures with a globally constant factor $\eta^{-1/4}$. This is analogous to the NEATM $\eta$. Note, however, that here $\eta$ solely reflects the effects of thermal-infrared beaming, whereas in the NEATM $\eta$ reflects the \emph{combined} effect of beaming and thermal inertia.
Here, surface temperatures are first calculated 
for smooth facets, taking thermal conduction and shadowing fully into account.
%
We expect this approximation  to be appropriate for small phase angles and not too large values of thermal inertia, hence it should be feasible
to model thermal-infrared observations of a Trojan binary such as (617) Patroclus, which is observed at small phase angles and is expected to be covered with regolith.

\paragraph{With mutual heating}
When hemispherical craters are added to the smooth facets within a thermal unit, the list of integral equations to be solved grows further. 
Inside a crater, 
the temperature distribution becomes more complex since there are additional sources of incoming flux to be considered.
%
%
Additionally, the facet's thermal emission gains a different directional characteristics, influencing the mutual heating of facets.
The number of integral equations which in principle all have to be solved simultaneously quickly becomes prohibitively large. A suitable approximation appears to be in order.
To the best of our knowledge, no detailed modeling of this problem is available in the literature, \citet{LagerrosIII} proposed a model in which beaming is only applied to the direct component.

It seems difficult to make use of the high symmetry of hemispherical craters in the presence of mutual heating by other facets (and their craters). It may be advantageous to give up the distinction between global shape and local craters, but rather to model roughness 
by modifying the shape model itself, 
i.e.\ by replacing planar facets with suitable rough facets (which are themselves triangulated). This 
opens a conceptually easy road to much more general thermal models of asteroid surface roughness allowing, in particular, for positive surface relief such as boulders.
See also the discussion in \sectrefpage{sect:NEA:boulders}.

The implementation of  such  roughness models is beyond our current scope. It is clear, however, that a TPM code suitable for this task must be optimized for numerical performance, chiefly because the number of facets contained in thermal units grows vastly if small-scale roughness is added, requiring relatively large amounts of RAM and CPU time (see below).

\section{Implementation}
       \label{sect:TPMconcave:implementation}

Our implementation of the non-convex TPM is based on that of the convex TPM discussed in \sectref{sect:TPMconvex:implementation}.
While the current version, which neglects mutual heating, 
is a helpful tool in itself, great care was taken to implement it in a way which makes it easy to add mutual heating at a later stage.

The class structure of the convex model is largely preserved. The abstract base classes \code{asteroid} and \code{ThermalModel} are reused without any modification, the definition of  \code{TriangulatedConcave} differs marginally from that of \code{TriangulatedConvex}.
An important new  class is \code{ThermalUnit},
containing a list of pointers to \code{FacetConcave} objects.
 The latter contain all required precomputed geometric information for one facet, such as a list of visible facets and view factors.
The routine \code{fluxModFactors} (defined within the scope of \code{ThermalModelConcave}) is passed a reference to a \code{ThermalUnit} object and determines the observable flux contribution from that unit.
Flux contributions from all thermal units are summed up and converted into physical units by the calling routine within the scope of \code{TriangulatedConcave}.

\code{ThermalInertiaOnlyConcave} objects calculate temperatures and observable fluxes emanating from all facets within the thermal unit, taking shadowing effects (both towards the Sun and the observer) into account. 
To solve the heat diffusion equation, the same algorithm is used as in the convex model. As in the convex model, 
a vector containing $\mu_S$ for each time step is calculated
before the diffusion equation is solved; in doing so, shadowing is accounted for.
The algorithm to determine whether a facet is shadowed is identical to that described in \sectref{sect:TPMconcave:geometry} (\eqrefpage{eq:TPMconcave:obstruct}).
Typically, identical discretization steps can be chosen as in the convex-model case. 
To model eclipse events on binary systems, 
it is important to consider the duration of eclipse events relative to the rotation period, since it may require a finer time resolution than is usual
\seesect{sect:Patroclus:modeling}.

Anticipating that in a  more general  TPM surface temperatures will be coupled through mutual heating, the
code is designed 
in a way that
 keeps the  thermal profiles of all facets belonging to a  thermal unit  simultaneously  in the computer's RAM for all time steps.
The number $N$ of \code{double} variables required is
\begin{equation}
  \label{eq:TPMconcave:RAM}
  N=\code{nTime * nZ * facets.size()},
\end{equation}
\code{facets.size()} denotes the number of facets contained in the thermal unit.
Typical values for a binary during eclipse are $\code{facets.size()} \sim 2500$, $\code{nTime}=1000$, and $\code{nZ}= 30$, totaling to 75 million \code{double} variables requiring \unit{600}{\mega\text{B}} of RAM. 
This has a critical impact on the numerical efficiency of the program unless it  runs on a machine with significantly more RAM available (typically used machines, however, had only \unit{512}{\mega\text{B}} of RAM).

We note that knowledge of the temperature profile at \emph{all} time steps is not required for the primary purposes of our model, although it proved helpful in the validation phase.
Consequently, the RAM demand could be cut down by a factor of $\code{nTime}/2$ if only two time steps would be stored simultaneously.%
\footnote{ In addition to a vector containing surface temperatures for all time steps, which is currently not needed because surface temperatures are stored as zeroth element of the profile.}
This algorithm has been adapted from the convex TPM, where only a single facet is handled at any time and therefore RAM is not a matter of concern. 

As discussed in \sectref{sect:TPMconcave:beaming}, we account for thermal-infrared beaming by multiplying temperatures with a globally constant factor $\eta^{-1/4}$. 
This is implemented in the class \code{ThermalInertiaEta} which inherits from \code{ThermalInertiaOnlyConcave}. All temperatures are rescaled by overwriting the routine \code{TSS} (defined within the scope of \code{ThermalModel}): the value returned by the base class is multiplied by $\eta^{-1/4}$ which is a constant within the scope of \code{ThermalInertiaEta}.

Functions on the main-routine level are adapted from the convex model, which determine $\chi^2$ and the best-fit \pv\ for a given data set and scan the parameter space spanned by thermal inertia and $\eta$. Additionally, time-resolved model spectra can be output for any combination of thermal parameters.

        	\section{Validation}
                        \label{sect:TPMconcave:validation}
The model variant discussed in this chapter is still under development. Various validation test have  been successfully performed, however.

\begin{figure}
\centering
   \includegraphics[width=0.6\linewidth]{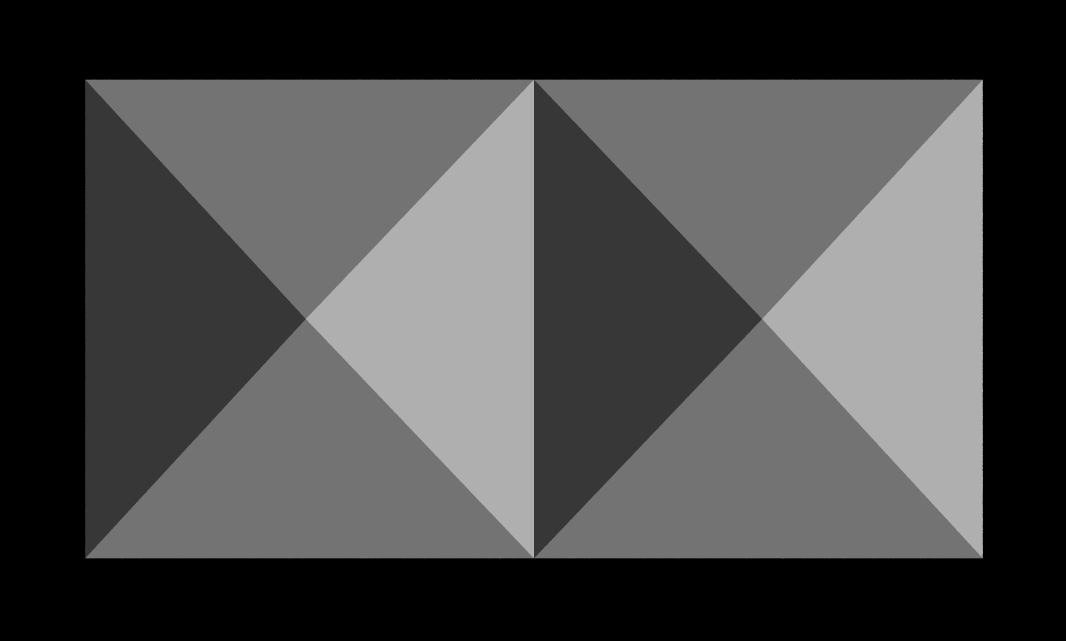}
\caption[Model figure used to verify that visible and invisible facets are correctly identified.]{One of the model figures  used to verify that visible and invisible facets are correctly identified. It consists of two adjacent pyramids. If viewed from above, two facets see one another, the remaining six are isolated. Viewed from below, each inverse pyramid forms a thermal unit in which each facet sees the remaining three.}
\label{fig:TPMconcave:doublesink}
\end{figure}

\paragraph{Geometry}
The tool \code{OBJ2concave} to generate \code{.concave} files has been tested intensively to verify that it correctly distinguishes visible from invisible facets.
To this end,  model figures such as that depicted in \figref{fig:TPMconcave:doublesink} were used or more complicated variants thereof.

Furthermore, the tool \code{binary} which generates binary \code{.concave} files has been checked by manual inspection of  simple output files (e.g.\ a model binary consisting of two tetrahedra).

\paragraph{Thermal physics}

The concave TPM was verified to reproduce convex-model fluxes for spherical shapes.
For a non-eclipsing binary viewed and irradiated pole on, the model flux equals the sum of the two individual spherical components, provided the 
diameter conversion  (\eqrefpage{eq:TPMconcave:rescale})
is applied; the validity of the latter has  been  numerically verified for a large range of diameter ratios.

The treatment of eclipses and occultations was qualitatively validated by calculating model fluxes for a synthetic binary system consisting of two equally sized spheres with the spin axis perpendicular on the viewing plane, leading to total eclipse and occultation events.
There are two eclipses and occultations per mutual orbit, when respectively one component obstructs the other's line of sight towards the Sun or the observer, respectively.
Without thermal inertia, eclipses have the same observable effect as occultations, both leading to a peak flux drop of \unit{50}{\%}.
For increasing values of thermal inertia, the relative depth of the eclipse event decreases (shadowed regions gradually cool down but their thermal emission is not ``switched off'' immediately) while occultation events, which are purely geometric in nature, remain unchanged. 
Finally, for very large thermal inertia values, eclipse events become virtually invisible in the thermal infrared.

There appear to be flaws in the model code, however.
Assuming the radar-derived shape model of (54509) YORP by \citet{Taylor2007}, which contains major concavities, and the observing geometry of our Spitzer observations of this object \seesect{sect:PH5}, synthetic model fluxes output by the non-convex TPM are $\sim2$ times larger than convex-TPM fluxes. 
This is inconsistent with the expectation that
 shadowing effects  should \emph{reduce} model fluxes rather than \emph{increase} them.
While the convex-shape TPM is well tested, the non-convex TPM is not; additionally, ``convex'' fluxes are more consistent with expectations based on the NEATM. It is not clear at present what causes this discrepancy. Further testing of the non-convex TPM is required.


%% file: lib.tex



%% file: acknow.tex
\chapter{Acknowledgments}
I am deeply indebted to you, \textbf{Alan Harris,}
for sparking my interest in this fascinating field and for your continued and never-ceasing support.  I could not  dream of a better thesis adviser, thank you for everything! I do hope our collaboration will go on, despite the slight increase in geographic distance from across-the-hall to trans-Atlantic!

It was a pleasure working in the Asteroids and Comets group.
I wish to thank 
\textbf{Ekkehard \Kuehrt} for enlightening discussions about thermal physics and for his support throughout my time at DLR Berlin including, but not restricted to,  generous support for conference travel. 
I wish to thank
\textbf{Prof.\ Spohn and Prof.\ Jaumann} for providing me with the opportunity to work in the DLR Institute of Planetary Research and for their support of my work.

Thank you and Mahlzeit to the  ``lunch club'': \textbf{Gerhard Hahn, Stefano Mottola, Uri Carsenty, Detlef de Niem, and Nikolaos Gortsas,} not only for being pleasant fellow eaters but also for inspiring me in many ways.
Thanks go to my office mates, \textbf{Frank Trauthan and Kay Lingenauber,} for a fair bit of enjoyable time, whenever our working hours overlapped.
Special thanks to \textbf{Katrin Stephan,} for patient proof reading 
of this (much too long) dissertation and for late night discussions in the empty institute building.

Mille grazie, \textbf{Marco \Delbo,} for helping getting me started on asteroids and data reduction, for lively and helpful discussions, and for our great co-operation on telescope time proposals, among other things. Don't forget that I'm in for the next round of Döner and beer! 

D\v{e}kuji,
\textbf{Franck Marchis,} for inviting me over to Berkeley and for your hospitality! Thanks also to \textbf{Josh Emery} for giving me the opportunity to give a talk at the SETI institute.
Thank you, Danke, and Merci, \textbf{Daniel Hestroffer,} for truly Babylonian discussions (not only in pubs) all over the world.
Thank you, \textbf{Alan Fitzsimmons,} for our Spitzer collaboration and for your support with PPARC!
Thank you, \textbf{John Stansberry,} for inviting me over to Arizona, so I get to see some low-thermal-inertia regolith!

Mahalo nui loa to all the \textbf{IRTF people,} most notably to \textbf{Bobby Bus,} for making my stay at \Hawaii{} unforgettable and for their continued support
after I had switched (alas!) to remote observing from Berlin. Specific thanks  go to the \textbf{telescope operators,} to \textbf{Eric Volquardsen,} and to \textbf{Eric Tollestrup.}

Many thanks to you helpful people of the \textbf{Spitzer team!} The \textbf{helpdesk,} in particular, kept on impressing me with their incredibly fast and helpful replies to my many questions.

\bigskip

\foreignlanguage{ngerman}{%
\textbf{Monika, Stephan, Kadda, und Maria: }
Ich danke Euch für Eure Hilfe und Unterstützung, an guten und an schlimmen Tagen; ich bin froh und dankbar, dass wir uns haben!
}
Finalement, merci \`a toi, \textbf{Marie,} pour me faire savoir ce que c'est, le bonheur. Je t'aime!

\rule{0.3\textwidth}{0.5pt} \\

This work is partially based on observations made with the \textbf{Spitzer Space Telescope,} which is operated by the \textbf{Jet Propulsion Laboratory, California Institute of Technology} under a contract with \textbf{NASA.}

The author is visiting astronomer (if virtually so for most of the time) at the \textbf{Infrared Telescope Facility,} which is operated by the \textbf{University of Hawaii} under Cooperative Agreement no.\ NCC 5-538 with the \textbf{National Aeronautics and Space Administration, Office of Space Science, Planetary Astronomy Program.}

I gratefully acknowledge partial financial support by \textbf{Deut\-sche For\-schungs\-ge\-mein\-schaft}
and 
generous travel grants by the American Astronomical Society \textbf{Division for Planetary Sciences DPS} (DPS meetings 2003 and 2005), 
by the \textbf{Japan Aerospace Exploration Agency JAXA} (1st Hayabusa symposium 2004), and by 
the \textbf{International Astronomical Union IAU} (ACM meeting 2005).


%% file: zusammenfassung.tex
\onehalfspacing
%
Das Thema der vorliegenden Dissertation ist die physikalische Charakterisierung von Asteroiden, wobei das Hauptaugenmerk auf der thermischen Trägheit erdnaher Asteroiden (NEAs) liegt.
Thermische Trägheit bestimmt die Stärke des Yarkovsky-Effekts, eines nichtgravitativen Effekts der die Umlaufbahnen kleiner Asteroiden mit Durchmessern bis  zu $\sim\unit{20}{\km}$ maßgeblich beeinflusst.
Für die Einschätzung des mit Erdeinschlägen von NEAs verbundenen Risikos ist der Yarkovsky-Effekt wesentlich.
Allerdings war  über die thermische Trägheit kleiner Asteroiden, insbesondere von NEAs, bislang nur sehr wenig bekannt.

Beobachtungen und theoretische Arbeiten wurden durchgeführt.
Die thermische Emission von Asteroiden wurde im mittleren infraroten Wellenlängenbereich (5~-- \unit{35}{\micron}) beobachtet. Diese Beobachtungen wurden mit dem "`Spitzer Space Telescope"' und dem \unit{3,0}{\metre} Infrarotteleskop der NASA (IRTF) durchgeführt, darunter  die ersten von Berlin aus ferngesteuerten IRTF-Beobachtungen.
Ein detailliertes thermophysikalisches Modell (TPM) wurde entwickelt und intensiv getestet; dies ist das erste TPM, 
dessen Anwendbarkeit auf NEAs gezeigt wurde.

Unser Hauptergebnis ist die Bestimmung der thermischen Trägheit von 5 NEAs; die Gesamtzahl der NEAs mit gemessener thermischer Trägheit beträgt nun 6 einschließlich der von uns betrachteten Objekte.
Für zwei von diesen verbessern wir in der Literatur verfügbare Abschätzungen, für die restlichen drei lagen keine verlässlichen Abschätzungen vor. Die Durchmesser der betrachteten NEAs liegen zwischen 0,1 und \unit{17}{\km}.

Unsere Ergebnisse erlauben eine erste Abschätzung der typischen thermischen Trägheit 
erdnaher Asteroiden:
Diese beträgt etwa \unit{300}{\TIunit} (entsprechend einer thermischen Leitfähigkeit von \unit{0,08}{\kappaunit})%
, mehr als eine Größenordnung über der typischen thermischen Trägheit großer Haupt\-gür\-tel\-asteroiden (MBAs).
Durchmesser und thermische Trägheit 
scheinen negativ zu korrelieren.

Unter Benutzung dieser Ergebnisse wurde von Kollegen die Größenabhängigkeit des Yar\-kovsky-Effekts neu bestimmt. Dies erlaubte ihnen, Unterschiede in der Größenverteilung von NEAs und MBAs ähnlicher Größe zu erklären.

Thermische Trägheit ist ein empfindlicher Indikator für feinkörniges Oberflächenmaterial, dies wird z.B. in der Marsgeologie häufig genutzt.
Die thermische Trägheit von NEAs liegt zwischen der von Mondregolith und irdischem Felsgestein. Dies deutet darauf hin, dass sogar die kleinsten betrachteten Asteroiden mit grobem Regolith bedeckt sind, in Übereinstimmung mit
 Nahaufnahmen des \unit{0,32}{\km} großen NEA (25143) Itokawa, die im Jahr 2005 durch die Raumsonde Hayabusa gewonnen wurden.

Die gefundene Korrelation zwischen thermischer Trägheit und Durchmesser lässt auf eine mit abnehmender Objektgröße  grobkörniger und/oder dünner werdende Regolithschicht schließen.
Voraussichtlich ermöglicht dies ein verbessertes Verständnis von Vorgängen, die für die Regolithbildung bedeutsam sind, z.B. von Impaktprozessen.

Zum ersten Mal wurde ein Verfinsterungsereignis in einem Doppelasteroiden-System im thermischen Infrarot beobachtet: Dazu wurde das Doppelsystem (617) Patroclus, ein Trojaner, mit Spitzer beobachtet.
Wie gezeigt wird, erlauben solche Beobachtungen eine einzigartig direkte Bestimmung der thermischen Trägheit. 
Unser Ergebnis ist die erste zuverlässige Bestimmung der thermischen Trägheit eines Trojaners.

Weiterhin wurden zwei künftige Raumsonden-Ziele untersucht: (21) Lutetia und (10302) 1989 ML. Die Größe und Albedo beider Objekte wurde bestimmt, die mögliche Oberflächenmineralogie eingeschränkt. Unsere Ergebnisse für 1989 ML sind von Bedeutung für die gegenwärtige Planung der ESA-Mission Don Quijote.


%% file: cv_online.tex
\chapter{Publikationen / Selbständigkeitserklärung}

\subsection*{Publikationen}
\foreignlanguage{USenglish}{
\begin{itemize}
\item 
\journalinpressref{\Delbo, M., dell'Oro, A., Harris, A.W., Mottola, S., Mueller, M.}{2007}{Thermal inertia of near-Earth asteroids and implications for the magnitude of the Yarkovsky effect}{\Icarus}
\item 
\journalref{Hahn, G., Mottola, S., Sen, A.K., Harris, A.W., Kührt, E., Mueller, M.}{2006}{Photometry of Karin Family Asteroids}{\BASI}{34}{393--399}
\item
\journalref{Harris, A.W., Mueller, M., \Delbo, M., Bus, S.J.}{2005}{The surface properties of small asteroids: Peculiar Betulia---A case study}{\Icarus}{179}{95--108}
\item 
\journalref{Harris, A.W., Mueller, M., \Delbo, M., Bus, S.J.}{2007}{Physical Characterization of the Potentially Hazardous High-Albedo Asteroid (33342) 1998 WT24 from Thermal-Infrared Observations}{\Icarus}{188}{414--424}
\item
\journalinpressref{Kaasalainen, M.\ and 13 colleagues, including M.\ Mueller,}{submitted in 2004}{Photometric observations 2001--2004 and modeling of (25143) Itokawa}{ASP Conference Series}
\item
\journalinpressref{Mueller, M., \Delbo, M., di Martino, M., Harris, A.W., Kaasalainen, M., Bus, S.J.}{submitted in 2004}{Indications for regolith on Itokawa from thermal-infrared observations}{ASP Conference Series}
\item 
\journalref{Mueller, M., Harris, A.W., Bus, S.J., Hora, J.L., Kassis, M., Adams, J.D.}{2006} {The size and albedo of Rosetta fly-by target 21 Lutetia from new IRTF measurements and thermal modeling} {\AandA}{447}{1153--1158}
\item
\journalref{Mueller, M., Harris, A.W., Fitzsimmons, A.}{2007} {Size, Albedo, and Taxonomic Type of Potential Spacecraft Target Asteroid (10302) 1989 ML}{\Icarus}{187}{611--615} 
\end{itemize}
}

\subsection*{Selbständigkeitserklärung}
Hiermit erkläre ich, die vorliegende Dissertation selbständig verfasst und keine weiteren Hilfmittel und Hilfen neben den angegebenen benutzt zu haben. 
In der Darstellung vorveröffentlichter Ergebnisse (siehe oben) wurden Beiträge von Koautoren nur in Ausnahmefällen  aufgenommen und ausdrücklich als solche gekennzeichnet.
\\
\\
Berlin, 21. Mai 2007
\\
\\
\\ 
\\ 
Michael Müller
